\documentclass[12pt]{article}
\usepackage{amsmath}
\usepackage{amsfonts}
\usepackage{amssymb}
\usepackage{graphicx}
\usepackage{enumerate}
\usepackage{adjustbox}
\usepackage{caption}
\usepackage{subcaption}
\usepackage{multirow}
\usepackage[margin=1in]{geometry}
\usepackage[compact]{titlesec}
\usepackage[utf8]{inputenc}
\usepackage{listings}
\usepackage{color}

\usepackage[hidelinks]{hyperref}
\usepackage{authblk}
\usepackage{float}
\usepackage{booktabs}
\usepackage{array}
\usepackage{natbib}

\usepackage{paralist}
\usepackage{url} %

\begin{document}

\bibliographystyle{plainnat}

\def\spacingset#1{\renewcommand{\baselinestretch}%
{#1}\small\normalsize} \spacingset{1}

{

\title{ \textbf{
Short-term forecasting of wildfire spread: A network epidemiology approach}}

  \author[1]{Indrila Ganguly}
  \author[2]{Muhammad Ali}
  \author[3]{Swarnali Sanyal}
  \author[2]{Viney Aneja}
  \author[4]{Srijan Sengupta}
  \affil[1]{Department of Experimental Statistics, Louisiana State University}

  \affil[2]{Department of Marine, Earth, and Atmospheric Sciences, North Carolina State University}
  \affil[3]{Department of Atmospheric Sciences, University of Illinois at Urbana--Champaign}
   \affil[4]{Department of Statistics, North Carolina State University}
  \date{}
  \maketitle
}

\bigskip

\noindent%

\spacingset{1.1} %

\begin{abstract}
Wildfire spread poses substantial environmental and public-health risks, motivating interpretable models for short-term forecasting. We develop a statistical framework that combines cellular automata with ideas from network epidemiology to model wildfire evolution across a spatial lattice. Each grid cell is classified as available, burning, or consumed. State transitions distinguish spread from burning neighbors, intrinsic ignition, and cessation of burning, with transition rates linked to meteorological and environmental covariates. A likelihood-based estimation procedure yields transition-specific covariate effects and probabilistic forecasts of cell states. We assess forecasting performance in a simulation study and in applications to the 2018 California wildfires and the 2019--2020 Australian wildfires. We also compare the method with a published forecasting approach using the 2017 Haypress fire. The results show strong short-term discrimination in many settings, with reduced accuracy at longer forecast horizons and during abrupt fire expansion. The framework provides an interpretable basis for studying wildfire dynamics and identifies opportunities to improve ignition forecasts through richer spatial and observation models.
\end{abstract}

\section{Introduction}
\label{sec:intro}
Climate change influences wildfire risk and behavior through changes in temperature, moisture, and other environmental conditions \citep{fried2008predicting,westerling2008climate}. Severe wildfire seasons in the western United States, Australia, the Mediterranean, and the Arctic have drawn attention to the need for models that characterize fire spread and its consequences \citep{shvidenko2013climate,perestrelo2021modelling,damany2022australian,akdemir2022estimating}. The 2019--2020 Australian fire season provides a prominent example of the widespread environmental impacts associated with extreme wildfire activity \citep{akdemir2022estimating,damany2022australian}.

Wildfire smoke also poses substantial public-health risks. Exposure to wildfire-related particulate matter has been associated with adverse respiratory and cardiovascular outcomes \citep{chen2021cardiovascular,matz2020health,aguilera2021wildfire}. Wildfires release carbon dioxide and other gases in addition to particulate matter \citep{shiraishi2021estimation,akdemir2022estimating}, and long-term residential wildfire exposure has been associated with increased incidence of some cancers \citep{korsiak2022long}. These findings motivate efforts to characterize the spatial and temporal extent of wildfire activity.

Forecasts of wildfire extent can inform assessments of smoke emissions and exposure. For example, \citet{akdemir2022estimating} use burned-area information to estimate emissions of $\mathrm{PM}_{2.5}$ and ammonia. Forecasting the evolution of active wildfires can therefore support downstream emissions modeling, as well as the planning and deployment of fire-suppression resources.

In recent years, a wide range of approaches have been developed for modeling wildfire behavior. Classical wildfire models include empirical and quasi-empirical rate-of-spread models, physics-based models, and simulation frameworks \citep{sullivan2009wildlanda,sullivan2009wildlandb,sullivan2009wildlandc}. Fire-growth simulation models such as FARSITE integrate established models of surface and crown fire behavior, fire acceleration, spotting, and fuel moisture, and use
vector propagation to simulate the spatial and temporal expansion of fire perimeters across a landscape \citep{finney1998farsite}. Coupled atmosphere--fire
models, in contrast, represent interactions between wildfire behavior and atmospheric dynamics \citep{clark1996coupled,coen2020computational}. Another commonly used class of approaches is based on cellular automata (CA) \citep{von1966theory}, in which the state of a spatial cell evolves
according to the states of neighboring cells
\citep{karafyllidis1997model,encinas2007simulation,alexandridis2008cellular,perestrelo2021modelling}. Level-set methods \citep{osher1988fronts} provide another framework for
representing the evolution of wildfire perimeters, in which the fire front is represented implicitly and allowed to evolve over space and time. These
methods have been used in both simulation and statistical modeling of
fire-front propagation
\citep{yoo2022bayesian,mallet2009modeling,lautenberger2013wildland,
zhai2020learning}. Among these, \citet{yoo2022bayesian} develop a Bayesian
spatio-temporal level-set model that allows the rate of fire-front evolution to vary over space and time and demonstrate the approach using observed wildfire data. While such models provide useful representations of wildfire behavior, they primarily characterize the evolution of the fire perimeter rather than probabilistic forecasting of the state of an entire spatial region.

A complementary statistical literature focuses on wildfire occurrence and
burned area at broader spatial and temporal scales. For example,
\citet{joseph2019spatiotemporal} develop spatiotemporal Bayesian models for wildfire frequency and size across the contiguous United States, combining
zero-inflated count and burned-area models to predict the occurrence of
extreme wildfires as functions of meteorological and anthropogenic
covariates, while \citet{opitz2020point} develop a Bayesian spatiotemporal log-Gaussian Cox point-process model for wildfire occurrences in the French Mediterranean basin, incorporating land-use proxies for human activity together with climatic and environmental covariates. \citet{pimont2021prediction} develop the ``Firelihood'' framework, a two-component Bayesian marked point-process model in which wildfire occurrences and fire sizes are modeled jointly, with fire-weather and forest area as key explanatory variables. \citet{castel2023disentangling} extend this framework by incorporating additional non-climatic drivers and a
more detailed spatio-temporal component to study variation in wildfire
activity across southeastern France. Other recent work includes joint models linking wildfire frequency and size
\citep{becker2022assessing,koh2023spatiotemporal}, multistage statistical
and machine-learning approaches using sparse SPDE-based spatial models for extreme wildfire frequencies and sizes \citep{cisneros2023combined}, and partially interpretable neural-network methods for modeling wildfire extremes \citep{richards2026regression}. Relatedly, \citet{lawler2024anthropogenic} develop a Bayesian hierarchical
model for wildfire counts and burned areas across the contiguous United States using meteorological, fire-weather, and anthropogenic covariates. They find that individual meteorological covariates are more informative for wildfire occurrence, whereas the Energy Release Component (ERC) and Fire Weather Index (FWI) provide greater predictive information for burned area. Their objective is to characterize large-scale patterns of wildfire occurrence and severity.

A more directly related forecasting approach is the deep hierarchical
generalized model of \citet{bradley2023deep}, which incorporates cellular-automata dynamics and is illustrated using the 2017 Haypress wildfire. Related work by \citet{zhou2026multiscale} jointly models remotely sensed wildfire activity and population change at multiple spatial scales, with an emphasis on large-scale spatial association.

In this work, we develop a stochastic framework for modeling and forecasting the short-term spatial evolution of an active wildfire. The proposed model represents wildfire evolution through three distinct mechanisms: spread from
neighboring burning cells, intrinsic ignition not attributed to neighboring fire, and transition out of the burning state. The corresponding transition rates depend on meteorological and spatial covariates, allowing the effects of these variables to be interpreted separately for the different mechanisms of wildfire evolution. This work draws on the network-based modeling framework for infectious disease transmission introduced in \citet{bu2021likelihood} and further developed in \citet{bu2022likelihood}.

Wildfire spread has a useful analogy with infectious disease transmission: a burning cell can ignite neighboring available cells, just as an infected individual can transmit infection through contacts. Cessation of burning is analogous to recovery, although the physical mechanisms differ. We combine the susceptible--infected--recovered (SIR) framework \citep{kermack1927contribution} with a two-dimensional cellular automaton \citep{packard1985two}, incorporating time-varying and time-invariant covariates into the transition rates. The lattice supplies the contact structure, while the covariates describe heterogeneity in spread, ignition, and recovery.

The proposed framework addresses a different problem from the
large-scale occurrence and burned-area models described above, while being more directly comparable to the cellular-automata-based forecasting approach of \citet{bradley2023deep}. Our formulation is highly interpretable and explicitly separates neighboring spread, intrinsic ignition, and recovery, providing
transition-specific interpretations of environmental covariates. A direct
forecasting comparison with \citet{bradley2023deep} using the 2017 Haypress fire is provided in Section~\ref{sec:haypress}.

The remainder of the paper is organized as follows. Section~\ref{sec:method} introduces the model, parameter estimation, and uncertainty quantification. Section~\ref{sec:simulation} evaluates forecasting performance using simulated data. Section~\ref{sec:real} presents applications to the California and Australian wildfires and a comparison with an existing method using the Haypress fire. Section~\ref{sec:discussion} discusses limitations and directions for future work.

\section{Methodology}
\label{sec:method}
\subsection{Background}
\begin{figure}[H]
    	\centering
    	\includegraphics[height=5cm,keepaspectratio]{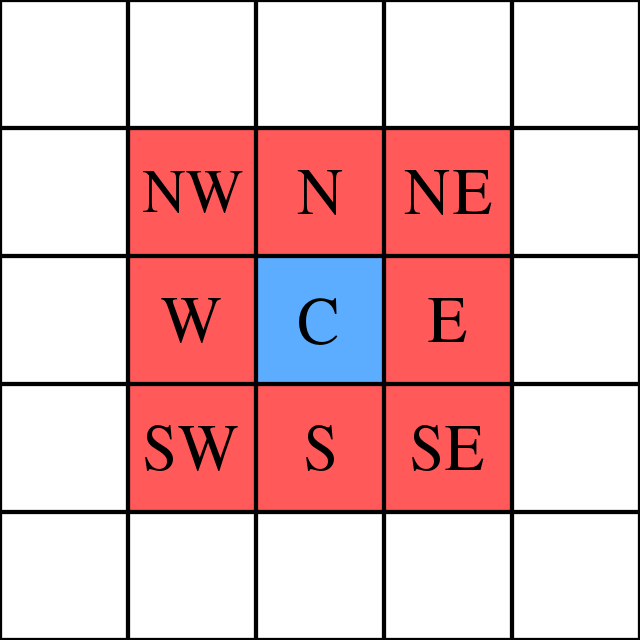}
        \caption{Representation of a Moore neighborhood (Source:\citet{enwiki:1128664897}) where the center grid marked in blue has contact with its eight neighbors marked in red.}
        \label{Fig1}
    \end{figure}

We represent the region of interest as a two-dimensional rectangular lattice containing $N$ grid cells. The objective is to forecast cell states over the next few observation times, such as the next several hours or days. Each interior cell has eight adjacent cells---north, south, east, west, and the four diagonals---forming a Moore neighborhood \citep{packard1985two}; boundary cells have fewer neighbors. If the central cell has coordinates $(0,0)$, the neighboring offsets are $\{(1,0),(-1,0),(0,1),(0,-1),(1,1),(1,-1),(-1,1),(-1,-1)\}$. Figure~\ref{Fig1} illustrates this structure. Each cell is assumed to be homogeneous and occupies exactly one of three mutually exclusive states:
\begin{itemize}
    \item $A$ (Available): The grid has not yet been affected by the wildfire, and contains biomass that is \textit{available} for burning.  We assign this state the value `0'.
    \item $B$ (Burning): The grid is actively burning at the current time point. We assign this state the value `1'.
    \item $C$ (Consumed): The biomass in this grid has already been consumed by the fire and this grid is assumed to remain in this state for the foreseeable future. We assign this state the value `-1'.

\end{itemize}

The states $A$, $B$, and $C$ are analogous to the susceptible, infected, and recovered compartments in network epidemiology \citep{bu2021likelihood,bu2022likelihood}. Here, ``recovery'' denotes cessation of burning, rather than ecological regeneration.

Covariates are treated as homogeneous within each cell. A cell in state $A$ may remain available or transition to $B$, while a cell in state $B$ may continue burning or transition to $C$. State $C$ is absorbing over the modeled period. This is an approximation: at coarse spatial resolutions, a cell may retain unburned biomass after a fire passes through part of it, and subsequent burning can violate the absorbing-state assumption.

Ignition in an available cell can arise through two mechanisms: spread from a burning neighbor or ignition not attributed to neighboring fire. Meteorological and environmental covariates can affect both mechanisms. A burning cell subsequently transitions to the consumed state when it ceases to burn.

Lattice-based models provide a natural representation of local fire propagation. In many cellular automata, transition rules are specified or calibrated to reproduce observed behavior. Our formulation expresses these transitions through covariate-dependent rates and a likelihood-based estimation procedure, allowing us to estimate transition-specific associations and construct probabilistic forecasts.

\subsection{Model}
\label{model}

Following \citet{bu2021likelihood,bu2022likelihood}, we model wildfire dynamics as a continuous-time Markov process conditional on the covariates. Available cells are subject to two competing ignition hazards, arising from burning neighbors and intrinsic factors; burning cells are subject to a recovery hazard. With fixed rates, waiting times are exponential. When covariates or neighboring states change, the hazards change accordingly, and waiting-time distributions are determined by the integrated hazards \citep{guttorp2018stochastic}.

Let $i$ and $j$ denote a $B$ and $A$ grid respectively such that $i$ and $j$ are neighbors. Let $\beta_{ijt}$ denote the rate of fire spread from $i$ to $j$ at time $t$, $\xi_{jt}$ denote the rate of ignition at time $t$ due to intrinsic factors only, and $\gamma_{it}$ denote the recovery rate of grid $i$ at time $t$. Let $Z_t$ denote the state of the entire process at time $t$. Then, the transition probabilities over an infinitesimal time interval $h$ are given by:\begin{align} &P(j \ \ \text{starts burning because of } i \text{ at } t+h|Z_t)=\beta_{ijt}h+o(h) \label{eq_burn}\\ & P(i \text{ recovers at } t+h|Z_t)=\gamma_{it} h+o(h) \label{eq_recovery}\\ &P(j \ \ \text{starts burning due to an intrinsic factor}\text{ at } t+h|Z_t)=\xi_{jt}h+o(h) \label{eq_burn_intrinsic}\end{align} where $h\downarrow 0$. We model the fire spread rate $\beta_{ijt}$ as \begin{equation}\log \beta_{ijt}= \log \beta + y_{it}\eta+ v_{ijt}\epsilon+x_{jt}'b \end{equation} where \begin{itemize}
    \item $\beta$ is the baseline rate of fire transmission between a $B$ and $A$ grid per contact.
    \item $x_{jt}=(x_{1jt}, x_{2jt}, \ldots, x_{pjt})'$ accounts for those $p$ covariates on $j$ at time $t$ that affect the fire spread.
    \item $y_{it}$ denotes the intensity of the fire at grid $i$ at time $t$, for $i \in \mathcal{N}_j$.
    \item $v_{ijt}$ denotes the dot product between
    (a) the wind vector (magnitude and direction) at $i$ and (b) the directional vector (unit length) from $i$ to $j$, for $i \in \mathcal{N}_j$.
\end{itemize}

\noindent In a similar fashion, the recovery rate $\gamma_{jt}$ is modeled as \begin{equation}
    \log \gamma_{jt}= \log \gamma + w_{jt}'\alpha \end{equation} where \begin{itemize}
    \item $\gamma$ is the baseline rate of recovery from fire.
    \item $w_{jt}$ accounts for those covariates on $j$ at time $t$ that affect the fire recovery.
    \end{itemize}

\noindent The intrinsic fire ignition rate $\xi_{jt}$ is modeled as: \begin{equation}\log \xi_{jt}= \log\xi + x_{jt}'b_E\end{equation} where \begin{itemize}
    \item $\xi$ is the baseline rate of ignition due to intrinsic factors.
    \item $x_{jt}$ is the same covariate vector as above, accounting for those covariates on $j$ that affect the fire initiation.
    \end{itemize}

To define the wind term consistently, let $\mathbf{u}_{it}=(u_{it},v_{it})'$ be the wind vector at cell $i$, expressed in common spatial coordinates, and let $\mathbf{d}_{ij}$ be the unit vector pointing from $i$ to $j$. Then $v_{ijt}=\mathbf{u}_{it}'\mathbf{d}_{ij}$. For example, if $j$ lies northeast of $i$ on a square lattice, $\mathbf{d}_{ij}=(1,1)'/\sqrt{2}$. Positive values indicate a wind component directed from the burning cell toward the available cell.

The model is formulated in continuous time, whereas cell states are generally recorded at discrete observation times. Transition times are therefore interval-censored, and multiple transitions may occur between observations. In the applications, we approximate each transition time by the observation time at which the new state is first recorded. This approximation is most credible when transition probabilities within an observation interval are small. A small numerical integration step alone does not eliminate uncertainty about transitions between observations.

\subsection{Parameter estimation and uncertainty quantification}
\label{subsec: est}
Let $\theta=\{\beta,\gamma,\eta,\epsilon,b,\alpha,\xi,b_E\}$ denote the parameters. Write $Z_j(t)$ for the state of cell $j$, and let $\mathcal{N}_j$ denote its neighbors. For notational convenience, define
\[
\mathcal{B}_j(t)=\{i\in\mathcal{N}_j:Z_i(t^-)=B\},\qquad
q_{ij}(t)=\exp\{y_{it}\eta+v_{ijt}\epsilon\},\qquad
S_j(t)=\sum_{i\in\mathcal{B}_j(t)}q_{ij}(t).
\]
Thus, only burning neighbors contribute to the spread hazard. Covariates and states in event-rate terms are evaluated immediately before the event. The three conditional intensities are
\begin{align*}
\lambda_j^{(N)}(t)&=1\{Z_j(t^-)=A\}\beta e^{x_{jt}'b}S_j(t),\\
\lambda_j^{(E)}(t)&=1\{Z_j(t^-)=A\}\xi e^{x_{jt}'b_E},\\
\lambda_j^{(C)}(t)&=1\{Z_j(t^-)=B\}\gamma e^{w_{jt}'\alpha}.
\end{align*}
Let $\mathcal{J}_N$ and $\mathcal{J}_E$ be the sets of cells igniting through neighboring spread and intrinsic ignition, respectively, and let $\mathcal{J}_C$ be the set of cells that cease burning during $(0,T_{max}]$. Their sizes are $n_B$, $n_B^{(E)}$, and $n_C$. Let $t_j^{(B)}$ and $t_j^{(C)}$ denote the ignition and recovery times of cell $j$. Conditional on the initial state, the covariates, and complete transition histories including ignition mechanisms, the likelihood is
\begin{align*}
L(\theta;\mathrm{data})={}&
\prod_{j\in\mathcal{J}_N}\!\left[\beta e^{x_{jt_j^{(B)}}'b}S_j(t_j^{(B)})\right]
\prod_{j\in\mathcal{J}_E}\!\left[\xi e^{x_{jt_j^{(B)}}'b_E}\right]
\prod_{j\in\mathcal{J}_C}\!\left[\gamma e^{w_{jt_j^{(C)}}'\alpha}\right]\\
&\times\exp\left\{-\int_0^{T_{max}}\sum_{j=1}^N
\left[\lambda_j^{(N)}(t)+\lambda_j^{(E)}(t)+\lambda_j^{(C)}(t)\right]dt\right\}.
\end{align*}
This is the conditional-intensity likelihood for the state-dependent transition processes \citep{guttorp2018stochastic}. Its product form does not imply that neighboring cell histories are marginally independent. The supplementary material provides further details of the construction.

In the observed data, ignition mechanisms are not recorded. We use a deterministic attribution rule: an ignition is assigned to neighboring spread if the cell has at least one burning neighbor immediately before ignition, and to intrinsic ignition otherwise. Maximizing the complete-data likelihood after this assignment gives a working likelihood-based estimator. The rule does not identify the true ignition mechanism. In particular, intrinsic ignitions near burning cells are assigned to spread, which can distort both baseline rates and covariate coefficients. The magnitude and direction of these effects require a sensitivity analysis. With continuously observed transitions but unobserved mechanisms, the ignition contribution to the observed-data likelihood would instead involve $\lambda_j^{(N)}(t_j^{(B)})+\lambda_j^{(E)}(t_j^{(B)})$.

Writing $l(\theta)=\log L(\theta;\mathrm{data})$, the score components for the working likelihood are
\begin{align}
\frac{\partial l}{\partial\beta}
&=\frac{n_B}{\beta}-\int_0^{T_{max}}\sum_{j=1}^N
1\{Z_j(t^-)=A\}e^{x_{jt}'b}S_j(t)\,dt,\\
\frac{\partial l}{\partial b}
&=\sum_{j\in\mathcal{J}_N}x_{jt_j^{(B)}}-
\int_0^{T_{max}}\sum_{j=1}^N\lambda_j^{(N)}(t)x_{jt}\,dt.
\end{align}
For $a\in\{\eta,\epsilon\}$, set $r_{ij,\eta}(t)=y_{it}$ and $r_{ij,\epsilon}(t)=v_{ijt}$. Then
\begin{align}
\frac{\partial l}{\partial a}
={}&\sum_{j\in\mathcal{J}_N}
\frac{\sum_{i\in\mathcal{B}_j(t_j^{(B)})}q_{ij}(t_j^{(B)})r_{ij,a}(t_j^{(B)})}{S_j(t_j^{(B)})}\notag\\
&-\beta\int_0^{T_{max}}\sum_{j=1}^N1\{Z_j(t^-)=A\}e^{x_{jt}'b}
\sum_{i\in\mathcal{B}_j(t)}q_{ij}(t)r_{ij,a}(t)\,dt.
\end{align}
The remaining score components are
\begin{align}
\frac{\partial l}{\partial\gamma}
&=\frac{n_C}{\gamma}-\int_0^{T_{max}}\sum_{j=1}^N
1\{Z_j(t^-)=B\}e^{w_{jt}'\alpha}\,dt,\\
\frac{\partial l}{\partial\alpha}
&=\sum_{j\in\mathcal{J}_C}w_{jt_j^{(C)}}-
\int_0^{T_{max}}\sum_{j=1}^N\lambda_j^{(C)}(t)w_{jt}\,dt,\\
\frac{\partial l}{\partial\xi}
&=\frac{n_B^{(E)}}{\xi}-\int_0^{T_{max}}\sum_{j=1}^N
1\{Z_j(t^-)=A\}e^{x_{jt}'b_E}\,dt,\\
\frac{\partial l}{\partial b_E}
&=\sum_{j\in\mathcal{J}_E}x_{jt_j^{(B)}}-
\int_0^{T_{max}}\sum_{j=1}^N\lambda_j^{(E)}(t)x_{jt}\,dt.
\end{align}
We maximize $l(\theta)$ numerically, for example using Newton--Raphson or quasi-Newton methods, with positive baseline rates $\beta,\gamma,\xi$. Let $\hat\theta$ denote the resulting estimates. For a small interval $h$, the plug-in transition probabilities satisfy
\begin{align*}
P\{Z_j(t+h)=B\mid Z_t,\ Z_j(t)=A\}
&\approx\left[\sum_{i\in\mathcal{B}_j(t)}\hat\beta_{ijt}+\hat\xi_{jt}\right]h,\\
P\{Z_j(t+h)=C\mid Z_t,\ Z_j(t)=B\}
&\approx\hat\gamma_{jt}h.
\end{align*}
These are first-order approximations for eligible state transitions, rather than unconditional probabilities of burning or recovery. Longer-horizon forecasts require propagation of the full state distribution, and numerical steps must keep the approximated probabilities within $[0,1]$.

For a correctly specified likelihood and under suitable regularity conditions, maximum likelihood estimators are approximately normal with covariance given by the inverse Fisher information. With $l$ denoting the log-likelihood, the expected and observed information matrices are
\[
I(\theta)=-E_\theta\!\left[\frac{\partial^2l(\theta)}{\partial\theta\,\partial\theta'}\right],
\qquad
J(\hat\theta)=-\left.\frac{\partial^2l(\theta)}{\partial\theta\,\partial\theta'}\right|_{\theta=\hat\theta}.
\]
The usual model-based standard error of $\hat\theta_k$ is
$\sqrt{[J(\hat\theta)^{-1}]_{kk}}$.
For fixed ignition labels, the working log-likelihood separates into three parameter blocks, so its observed information is block diagonal. However, these standard errors condition on the attribution rule and the approximated transition times; they do not account for uncertainty in ignition mechanisms, temporal discretization, or model misspecification. The simulation study in Section~\ref{sec:simulation} evaluates forecasting performance under the specified model.

 \section{Simulation study}
 \label{sec:simulation}

 In this section, we evaluate the performance of our method in a simulation setting, and assess how well it predicts data generated from the proposed model. We consider a $50 \times 50$ square lattice representing the spatial domain.
The following parameter values are used in the simulation: $\beta=3, \eta=0.5, \gamma=0.1, \epsilon=0, b=(1,1), \alpha=(1,1)$, $\xi=0.2$, $b_E=(1,1)$. The fire intensity term $y_{it}$ is defined as follows: for grids in state $A$ or $C$, $y_{it} = 0$; for grids in state $B$, $y_{it} \sim \text{Uniform}(0,1)$. The covariates are assumed to be time-independent and are generated as:
$
x_j \sim \mathcal{N}_2\left( \begin{bmatrix} 0 \\ 0 \end{bmatrix}, \begin{bmatrix} 0.1^2 & 0 \\ 0 & 0.1^2 \end{bmatrix} \right), \quad
w_j \sim \mathcal{N}_2\left( \begin{bmatrix} 0.5 \\ 0.5 \end{bmatrix}, \begin{bmatrix} 0.1^2 & 0 \\ 0 & 0.1^2 \end{bmatrix} \right).
$
Time points at intervals of $h=0.0001$ are considered. At the initial time point, each grid has a probability of 0.01 of being in the burning state $B$, independently of others.

\begin{figure}[htbp]
    \centering
    \includegraphics[width=0.7\textwidth]{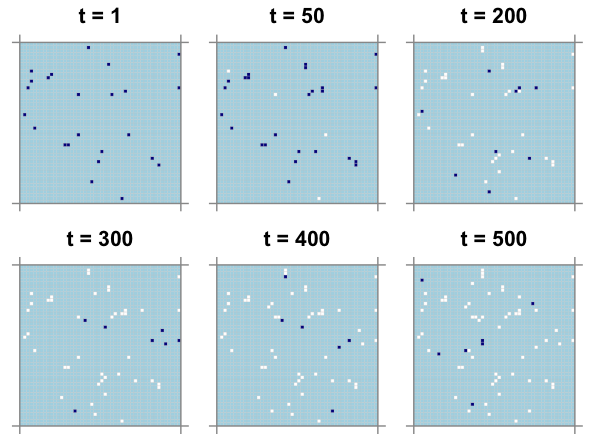}
    \caption{A simulation showing the spread of wildfire at different time points. The colors light blue, dark blue, and white denote $A$, $B$ and $C$ grids respectively.}
    \label{fig:wild_spread}
\end{figure}

Figure~\ref{fig:wild_spread} illustrates wildfire evolution under the specified simulation model. Light blue, dark blue, and white represent states $A$, $B$, and $C$, respectively. From the initial ignition sites, fire spreads to neighboring cells, while intrinsic ignitions create new burning cells elsewhere. Over time, burning cells transition to the consumed state.

\begin{figure}[htbp]
    \centering
    \includegraphics[width=0.7\textwidth]{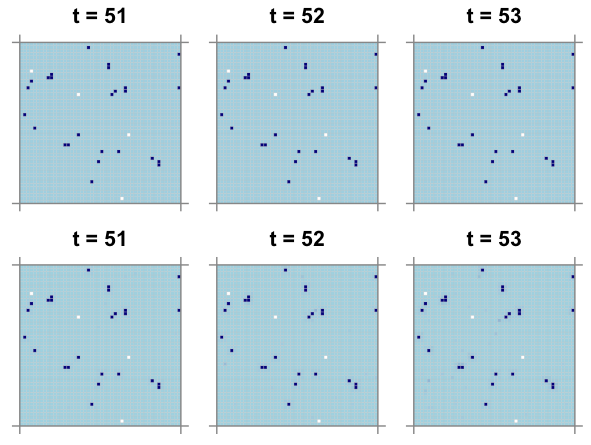}
    \caption{Comparison of Observed (top row) and Predicted Spread (bottom row) at 3 future time points based on a single simulation. }
    \label{fig:obs_v_pred50}
\end{figure}

To evaluate the forecasting capability of our model, we look at a scenario where the wildfire is observed for the first $T = 50$ time points, and forecasts are made for the next three time points. These are three simulation steps of length $h$; a correspondence with a 72-hour forecast window would require a separate calibration of the simulation time unit. The upper and lower rows of Figure \ref{fig:obs_v_pred50} depict the observed and predicted spread of wildfire respectively at the $51^{st}$, $52^{nd}$ and $53^{rd}$ time points. The plots of the predicted spread are color-coded such that the color of a grid is the RGB of a linear combination of `Light Blue', `Dark Blue' and `White' weighted by the corresponding estimated probabilities of $A$, $B$ and $C$ for that particular grid. To assess prediction accuracy, we obtain the mean areas under the receiver operating characteristic curve (AUCs) for $T=50$ based on 100 replications, and the results are shown in Table \ref{tab:auc50}. All simulations have been performed in the statistical software R (\citet{R}). AUC values close to 1 indicate strong discrimination, while values near 0.5 suggest performance no better than random guessing. Our results show that the model achieves high AUC values across all three states, demonstrating its strong discrimination under this simulation setting.

\begin{table}[htbp]
    \centering
    \begin{tabular}{|c|c|c|c|} \hline
     Prediction time point & Available ($A$) & Burning ($B$) & Consumed ($C$) \\ \hline
     $T=51$    & 0.960 (0.002) & 0.997 (0.014) & 1.000 (0.001)\\ \hline
      $T=52$   &  0.958 (0.008) & 0.992 (0.017) & 1.000 (0.001)\\  \hline
      $T=53$  &0.956 (0.012)  &0.987 (0.022) &1.000 (0.001)  \\ \hline
    \end{tabular}
    \caption{Mean AUC values (with corresponding standard deviations)}
    \label{tab:auc50}
\end{table}

The simulated results show strong discrimination at all three forecast horizons, with modest decreases in AUC for the available and burning states as the horizon increases. These results describe performance under the specified data-generating model; the applications below assess performance on observed wildfire data.

\section{Applications to the California and Australian wildfires}

\label{sec:real}
\subsection{Data sources and preprocessing}
\subsubsection{Fire data}
\label{sec:firedata}

For both wildfire datasets, we obtained active-fire detections from the Visible Infrared Imaging Radiometer Suite (VIIRS) through NASA Earthdata \citep{Earth_data}. For California, the initial collection period was May--October 2018 and included major events such as the Ranch and Carr fires \citep{California_fires}. The analysis below uses June--October 2018. We selected relatively sparsely populated counties to reduce, without eliminating, the potential influence of unmeasured suppression activities. The selected counties are Del Norte, Siskiyou, Modoc, Humboldt, Trinity, Shasta, Lassen, Tehama, Mendocino, Glenn, Butte, Plumas, and Sierra.

For the Australian Megafire, we do not restrict our analysis to specific regions. Instead, we utilize all available fire data for the period from November 1, 2019, to January 15, 2020. Unlike the California dataset, this dataset may include more densely populated areas and, consequently, more extensive fire suppression activities. The rationale for analyzing datasets from regions with potentially varying levels of human intervention is to assess the robustness and generalizability of our proposed methodology across different real-world scenarios.

For our analysis, we extract data on the following variables: `Latitude', `Longitude', `Acquisition date', `Confidence', and `Brightness'. Brief descriptions of these variables are provided below (\citet{Earth_data_details}). \begin{itemize}
    \item Latitude: Center of 1 km fire pixel, but not necessarily the actual location of the fire as one or more fires can be detected within the 1 km pixel.
    \item Longitude: Center of 1 km fire pixel, but not necessarily the actual location of the fire as one or more fires can be detected within the 1 km pixel.
    \item Acquisition date: The date on which the MODIS sensor acquired the data.
    \item Confidence: A categorical variable with three levels `High', `Nominal' and `Low' indicating the quality of fire pixels. For our purpose, we discard the data with `Low' confidence to minimize the inclusion of erroneous data.
    \item  Brightness: The brightness temperature of the fire pixel, measured in Kelvin, from MODIS Channel 21/22.
\end{itemize}

We construct the observed cell states from the retained fire detections. A cell with at least one detection at a given observation time is assigned state $B$ (value 1), and its maximum recorded brightness temperature is used as the fire-intensity proxy. In the absence of a detection, the cell is assigned state $A$ (value 0) if no earlier detection occurred during the study period, and state $C$ (value $-1$) otherwise. This coding treats nondetection as absence of active burning and does not by itself establish that biomass has been consumed. Grid construction is described below.

\subsubsection{Covariate data}

We include meteorological and environmental covariates associated with wildfire initiation and spread. Temperature, moisture availability, and precipitation can influence both fuel conditions and fire activity \citep{westerling2008climate,sazib2021leveraging}. Precipitation may reduce immediate fire risk while also promoting vegetation growth that supplies fuel in a later dry period. Wind affects directional spread \citep{beer1991interaction}, and relative humidity is associated with the extent of burning \citep{slocum2010effect}. We also include surface pressure \citep{zhang2023joint}. Fractional vegetation cover was available for the California analysis but was omitted from the Australian analysis because a comparable dataset was not obtained. Table~\ref{tab:covariates} summarizes the listed covariates and data sources.

\begin{table}[htbp]
\centering
\small
\begin{tabular}{>{\raggedright\arraybackslash}p{1.35in}>{\raggedright\arraybackslash}p{1.70in}>{\raggedright\arraybackslash}p{0.65in}>{\raggedright\arraybackslash}p{1.30in}}
\toprule
Covariate & Description and unit & California & Australia \\
\midrule
Temperature & Air temperature at 2 m (K) & NLDAS & ERA5-Land \\
Surface pressure & Pressure (Pa) & NLDAS & ERA5-Land \\
Precipitation & Hourly total ($\mathrm{kg}\,\mathrm{m}^{-2}$) & NLDAS & Google Earth repositories \\
Vegetation cover & Fraction from 0 to 1 & NLDAS & Not included \\
Soil moisture & Water in the upper 100 cm ($\mathrm{kg}\,\mathrm{m}^{-2}$) & NLDAS & Google Earth repositories \\
Relative humidity & Relative humidity (\%) & NLDAS & Global Forecast System \\
Zonal wind & East--west component at 10 m ($\mathrm{m}\,\mathrm{s}^{-1}$) & NLDAS & ERA5-Land \\
Meridional wind & North--south component at 10 m ($\mathrm{m}\,\mathrm{s}^{-1}$) & NLDAS & ERA5-Land \\
\bottomrule
\end{tabular}
\caption{Covariates and recorded data sources for the California and Australian analyses. NLDAS denotes the North American Land Data Assimilation System.}
\label{tab:covariates}
\end{table}

\subsubsection{Grid construction}
\label{subsec: grid_cal}
Grid resolution balances the within-cell homogeneity assumption against computational cost. Large cells can contain both burning and nonburning areas, whereas very small cells increase the number of states to be modeled. We use cells of approximately $4\,\mathrm{km}\times4\,\mathrm{km}$ for the California and Australian analyses. Covariates are averaged within each cell. Missing covariate values are imputed from the nearest cell with available data. State labels are assigned from the current and past detections, as described in Section~\ref{sec:firedata}.

\subsection{California wildfire}

The California analysis covers June 1--October 31, 2018. This period includes both summer fire activity and the later-season dry conditions associated with substantial wildfire risk \citep{cal_season,cal_season2}. We evaluate short-term forecasts using a sequence of expanding training windows.

\subsubsection{Parameter interpretation}
\begin{figure}[H]
    \centering
    \includegraphics[trim={0cm 0cm 0cm 0cm},clip, scale=0.24]{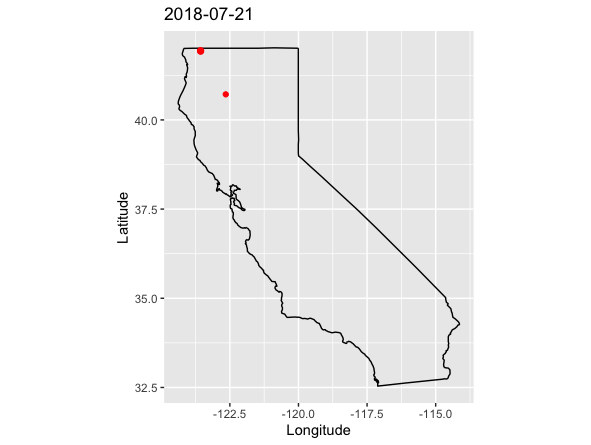}
    \hspace{-2mm}
    \includegraphics[trim={5.8cm 0cm 0cm 0cm},clip, scale=0.24]{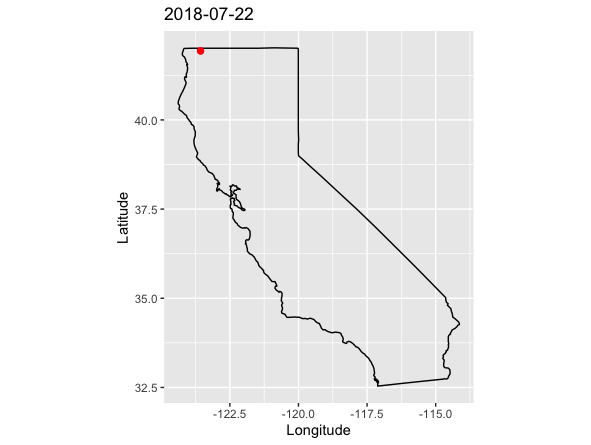}
    \hspace{-2mm}
    \includegraphics[trim={5.8cm 0cm 0cm 0cm},clip, scale=0.24]{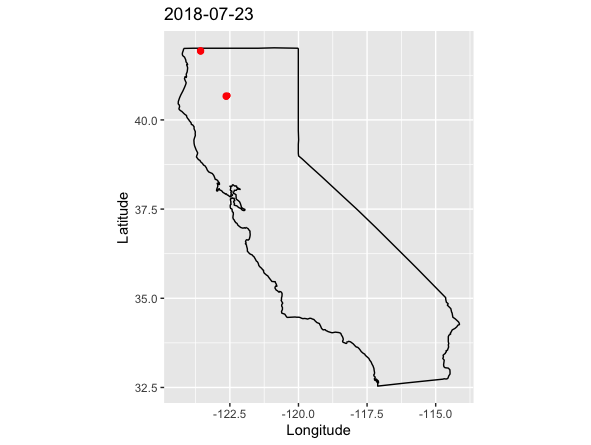}
    \hspace{-2mm}
    \includegraphics[trim={5.8cm 0cm 0cm 0cm},clip, scale=0.24]{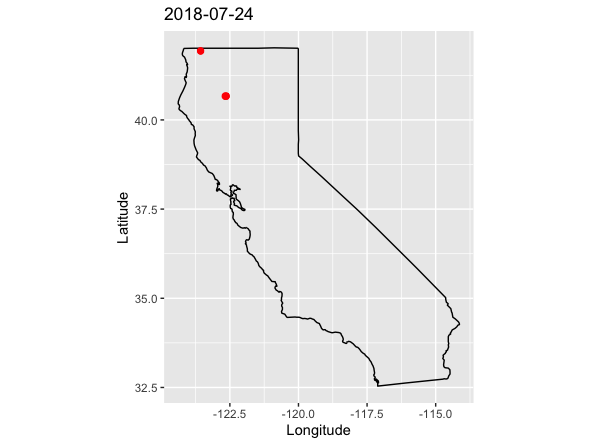}
     \includegraphics[trim={0cm 0cm 0cm 0cm},clip, scale=0.24]{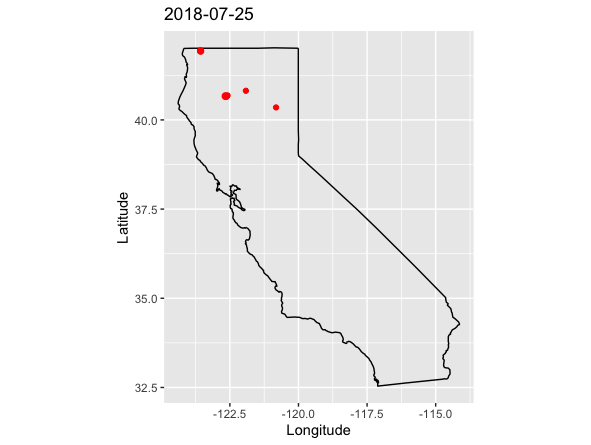}
    \hspace{-2mm}
    \includegraphics[trim={5.8cm 0cm 0cm 0cm},clip, scale=0.24]{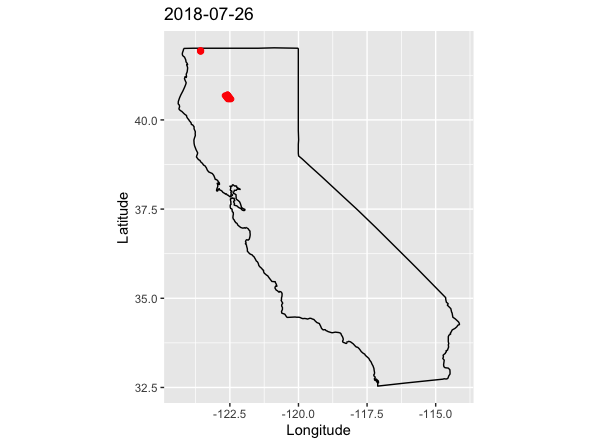}
    \hspace{-2mm}
    \includegraphics[trim={5.8cm 0cm 0cm 0cm},clip, scale=0.24]{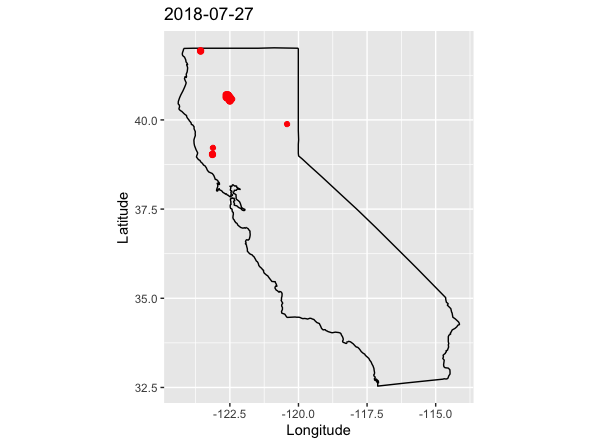}
    \hspace{-2mm}
    \includegraphics[trim={5.8cm 0cm 0cm 0cm},clip, scale=0.24]{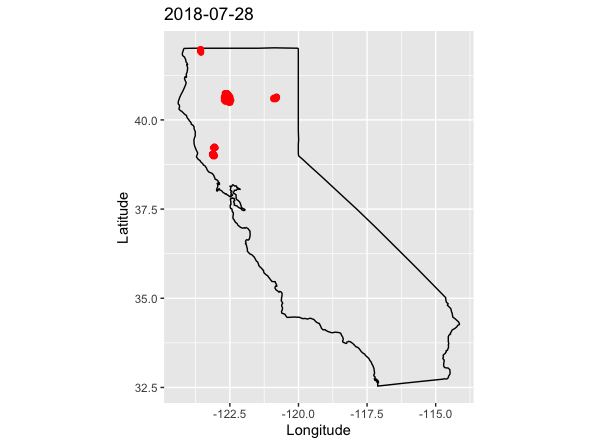}
     \includegraphics[trim={0cm 0cm 0cm 0cm},clip, scale=0.24]{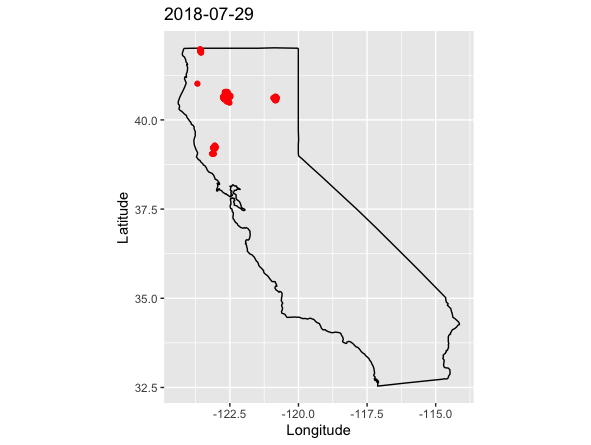}
    \hspace{-2mm}
    \includegraphics[trim={5.8cm 0cm 0cm 0cm},clip, scale=0.24]{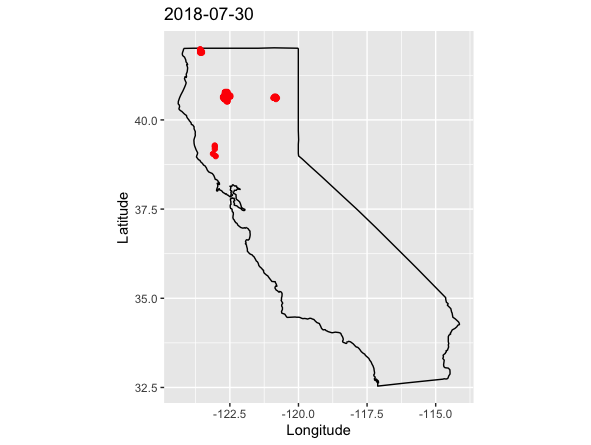}
    \hspace{-2mm}
    \includegraphics[trim={5.8cm 0cm 0cm 0cm},clip, scale=0.24]{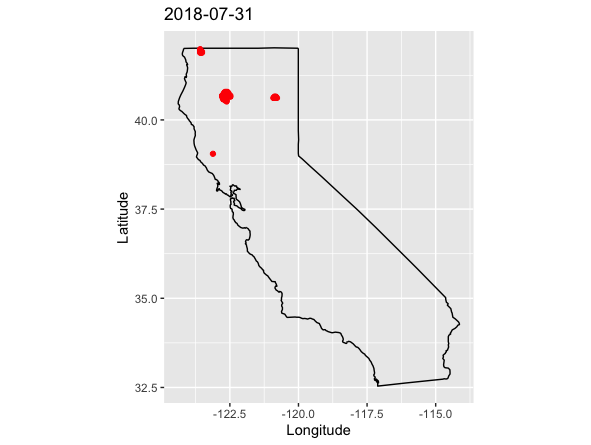}
    \hspace{-2mm}
    \includegraphics[trim={5.8cm 0cm 0cm 0cm},clip, scale=0.24]{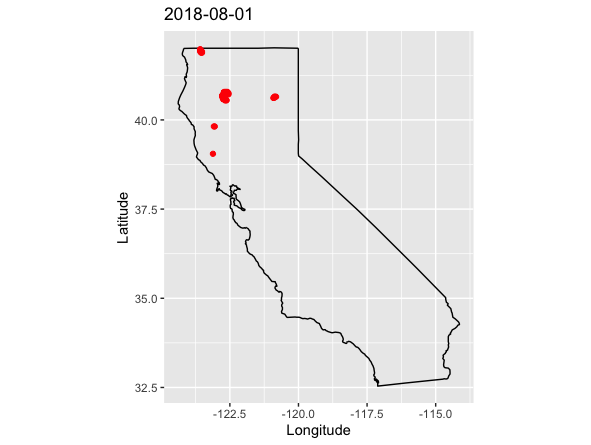}
    \caption{Progression of the fire in the chosen counties in the period July 21, 2018 to Aug 01, 2018.}
    \label{fig:cal_obs}
\end{figure}

 After the construction of the grids, our final data set consists of values on grids of 4 km by 4 km for time periods in the range of June 1, 2018 to October 31, 2018. We use the same variables as covariates in all three components of our model described previously. To provide an overview of the grid construction process and the progression of wildfire activity, Figure \ref{fig:cal_obs} illustrates the spread of fire in the selected region over a twelve-day period from July 21 to August 1. In this figure, a red dot indicates the presence of fire in a grid on a given day. In addition to visualizing fire progression, examining the estimated model parameters offers valuable insights into model performance. The estimates help interpret the influence of covariates on the transition rates between states over time.

\begin{figure}[H]
     \centering
     \includegraphics[width=\textwidth]{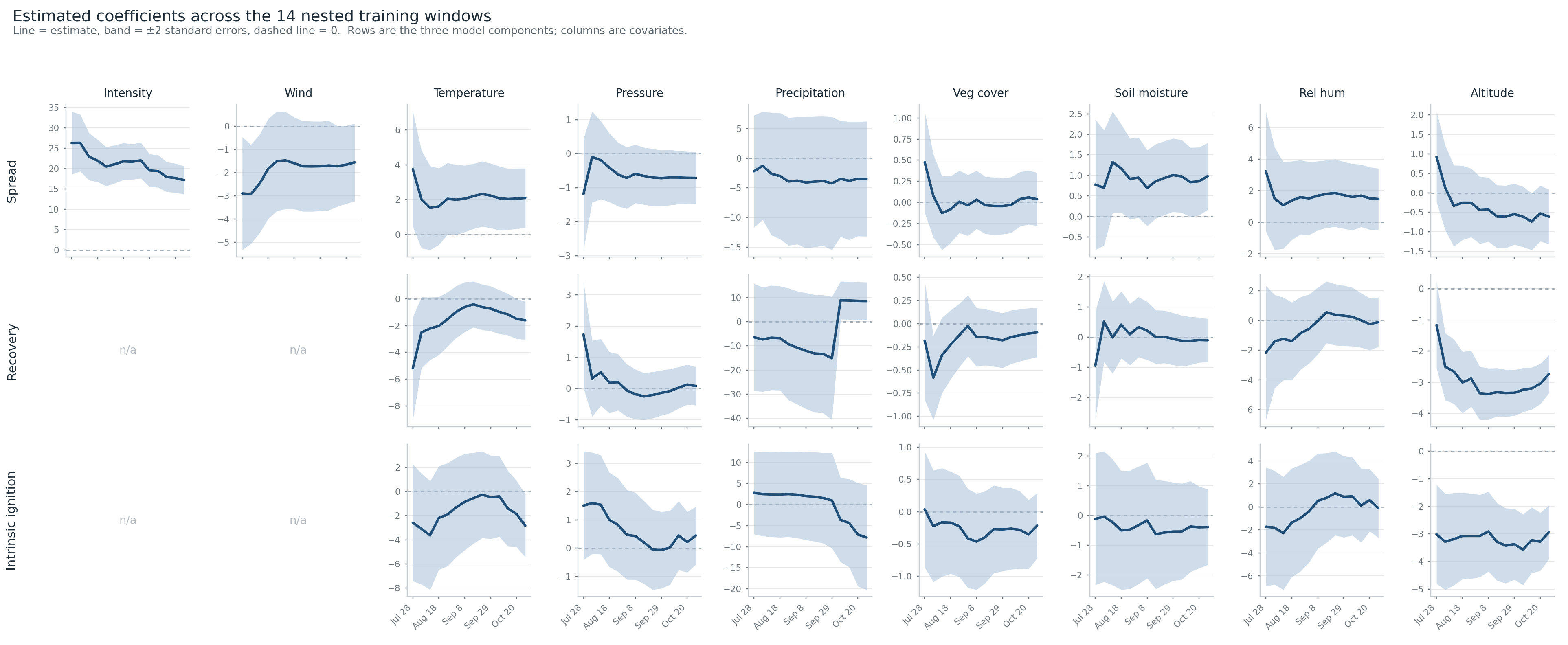}
     \caption{Estimated coefficients across the fourteen nested training windows.
     Rows correspond to the three model components: spread from burning
     neighbors, recovery, and intrinsic ignition and columns to the
     covariates. In each panel the solid line is the estimate and the shaded band
     spans two standard errors either side of it; the dashed line marks zero.
     Intensity and wind enter the spread component only, so those panels are
     empty in the lower two rows.}
     \label{fig:coef_trajectories}
\end{figure}

 The estimated coefficients and the corresponding standard errors for fire spread from neighbors, recovery from fire, and intrinsic fires respectively are reported in  Tables 1-3 in the Supplementary Materials. The parameters are estimated on nested training periods. Specifically, we begin with the interval from June 1 to July 28 and incrementally extend the training window by one week at a time (e.g., June 1 to August 4, and so on) until the end of October. These are coefficients on the log-rate scale. For a covariate with coefficient $c$, an increase of $\Delta$ on the fitted covariate scale multiplies the corresponding rate by $\exp(c\Delta)$, holding the other covariates fixed. An alternative way of expressing the rate ratio is to see how it changes when the corresponding covariate is increased by the average daily change. Suppose $z_t$ is the value of a covariate at time $t$. Then, the average daily change is given by the quantity $\frac{1}{T-1} \sum_{t=1}^{T-1} |z_{t+1} - z_t|$ where $T=153$ is the total number of time points from June 1 to October 31, 2018. Tables 4-6 in the Supplementary Materials report the rate ratio of the rates for average daily change of that covariate for fire spread from neighbors, recovery from fire, and intrinsic fires respectively. For vegetation cover, the reported rate ratio corresponds to an increase of 0.01 in the fractional cover; for soil moisture, it corresponds to an increase of $1\,\mathrm{kg}\,\mathrm{m}^{-2}$. We also plot how the coefficients evolve through the training window in Figure~\ref{fig:coef_trajectories}. From the figure, we can see that the estimates settle from about mid-August onwards with the widest intervals occurring in the earliest window, where the fewest events are available to inform them. Three effects are distinguishable from zero across essentially the whole season. Fire intensity in the spread component is positive in all fourteen windows, declining from $26.3$ to $17.2$ as the training window expands. Altitude is negative in the recovery and intrinsic ignition components, in thirteen and fourteen of the fourteen windows respectively, indicating lower fitted recovery and intrinsic ignition rates at higher elevations. Temperature is positive in the spread component and distinguishable from zero in nine of the fourteen windows. For the remaining covariates, the error bands contain zero in most or all windows. This pattern describes the evidence for individual conditional associations and does not establish their relative contributions to predictive performance.

\subsubsection{Forecasting}
For each training window, we forecast cell states over the next three days. These forecasts require covariates at future times. In our analysis, forecasted covariate values are used as proxies for the future values; the resulting fire forecasts are conditional on these inputs. Figure~\ref{fig:cal_maps} shows one example, and additional maps appear in the supplementary material. Predicted cells are colored using an RGB mixture of light blue, dark blue, and white, weighted by the probabilities of states $A$, $B$, and $C$, respectively. Table~\ref{tab:auc_cal} reports state-specific AUC values. All calculations were performed in R \citep{R}.

Each forecast set comprises 1-, 2-, and 3-day predictions from a common cutoff date. For example, forecasts for July 29--31 use training data from June 1 to July 28. The training window then expands by one week. AUC values are generally high for the available and consumed states, while performance for the burning state varies more substantially across horizons and dates (Table~\ref{tab:auc_cal}). In particular, several longer-horizon burning-state AUCs fall below 0.5.

Burning-state discrimination depends partly on the balance between persistent burning and new ignitions. A model that assigns high scores to cells already burning can perform well when those cells remain burning, but poorly when subsequent activity consists mainly of new ignitions. For the two lowest AUCs in Table~\ref{tab:auc_cal}, only 3 and 4 cells were burning, respectively, and all were new ignitions. Limited early training data, unexpected ignitions, and violations of the absorbing consumed-state assumption may also contribute. These results motivate evaluation specifically focused on newly burning cells.

Since the AUC for the Burning state is affected in this way by the composition of the burning grids, and since ignition is a rare event, we also report the precision at $k$ in Table \ref{tab:precision_pooled}. Here $k$ is the number of grids burning on the forecast day that were not burning when the forecast was made, and the precision at $k$ is the proportion of the $k$ highest ranked grids that are among them. Pooled over all fourteen training windows and the three forecast horizons, the $k$ highest ranked grids contain 48 of the 230 that ignite, a pooled precision of $48/230\approx0.209$. Table~\ref{tab:precision_pooled} reports the horizon-specific precision values and their random-ranking benchmarks. A rule that ranks grids by the number of burning neighbors and estimates no parameters achieves $0.113$ over the same set, so the fitted model identifies close to twice as many ignitions as the neighborhood structure alone.

 \begin{figure}[H]
\centering
\includegraphics[width=0.48\textwidth]{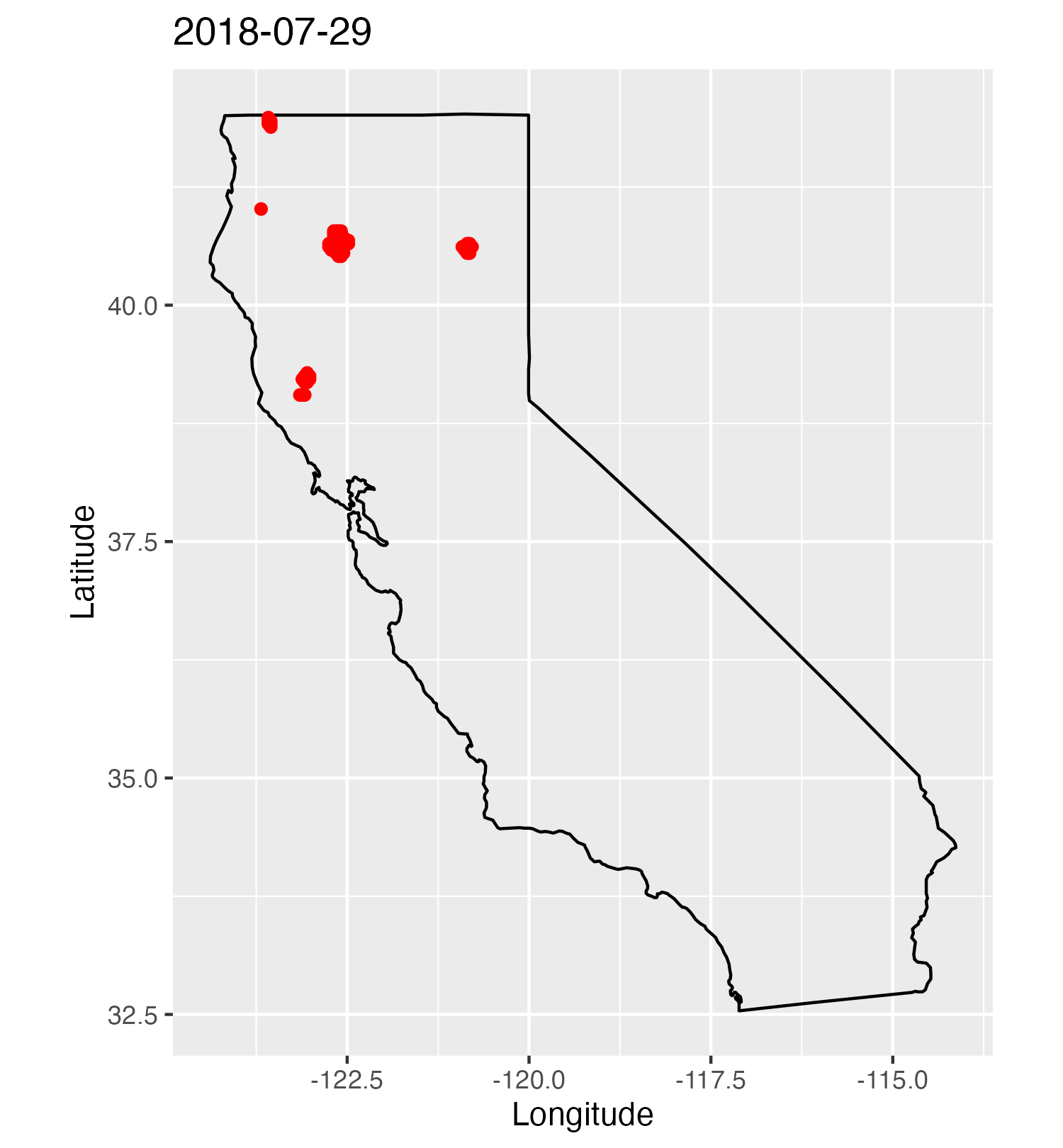}\hfill
\includegraphics[width=0.48\textwidth]{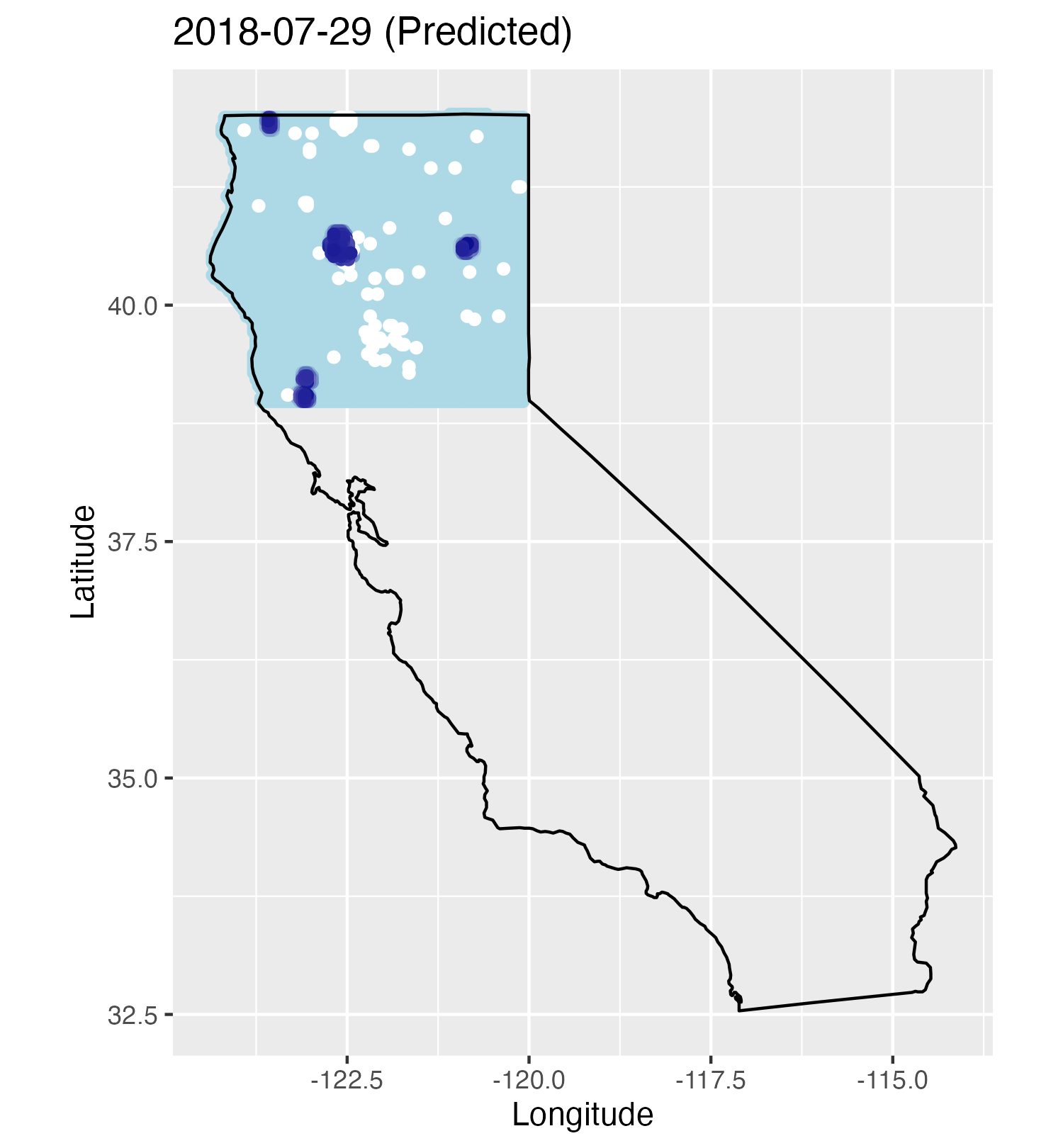}
\caption{Observed (left) and predicted (right) state of the California fire on
29 July 2018. In the observed panel a red point marks a burning cell. In
the predicted panel each cell is colored by the linear combination of its
three predicted state probabilities, with Available in light blue, Burning in
dark blue and Consumed in white.}
\label{fig:cal_maps}
\end{figure}

 \begin{table}[htbp]
     \centering
     \resizebox{\textwidth}{!}{
     \begin{tabular}{|c|ccc|ccc|ccc|}
     \hline

      Time &  A (1 step) &  B (1 step) &  C (1 step) &  A (2 step) &  B (2 step) &  C (2 step) & A (3 step) &  B (3 step) &  C (3 step)\\ \hline
 June 1-July 28 & 0.9986 & 0.9940 & 0.9996 & 0.9984 & 0.9969 & 0.9836 & 0.9925 & 0.9098 & 0.9755\\
 June 1- Aug 4 & 0.9958 & 0.9704 & 1.0000 & 0.9958 & 0.9771 & 0.9953 & 0.9875 & 0.6612 & 0.9930\\
 June 1- Aug 11 & 0.9986 & 0.9851 & 1.0000 & 0.9890 & 0.8189 & 0.9999 & 0.9891 & 0.7302 & 0.9979\\
 June 1-Aug 18 & 1.0000 & 0.9993 & 0.9999 & 0.9912 & 0.9320 & 1.0000 & 0.9872 & 0.7500 & 0.9997\\
 June 1 - Aug 25 & 1.0000 & 0.9999 & 1.0000 & 0.9990 & 0.9847 & 1.0000 & 0.9988 & 0.9979 & 0.9972\\
 June 1- Sep 1 & 1.0000 & 0.9999 & 1.0000 & 1.0000 & 0.9997 & 1.0000 & 0.9987 & 0.9631 & 0.9987\\
 June 1- Sep 8 & 1.0000 & 0.9998 & 1.0000 & 0.9992 & 0.9873 & 1.0000 & 0.9993 & 0.9117 & 0.9999\\
 June 1- Sep 15 & 0.9984 & 0.8889 & 1.0000 & 0.9979 & 0.9480 & 0.9989 & 0.9970 & 0.9254 & 0.9977\\
 June 1- Sep 22 & 0.9982 & 0.6177 & 1.0000 & 0.9975 & 0.7636 & 0.9978 & 0.9903 & 0.2245 & 0.9967\\
 June 1- Sep 29 & 1.0000 & - & 1.0000 & 1.0000 & - & 1.0000 & 0.9987 & 0.8792 & 1.0000\\
 June 1 - Oct 6 & 0.9984 & 0.8513 & 1.0000 & 0.9962 & 0.7843 & 0.9969 & 0.9941 & 0.5282 & 0.9897\\
 June 1- Oct 13 & 0.9999 & 0.9437 & 1.0000 & 0.9949 & 0.1834 & 0.9990 & 0.9929 & 0.6802 & 0.9970\\
 June 1 - Oct 20 & 1.0000 & 1.0000 & 1.0000 & 0.9884 & 0.4943 & 1.0000 & 0.9867 & 0.4324 & 0.9926\\
 June 1 - Oct 27 & 1.0000 & - & 1.0000 & 1.0000 & - & 1.0000 & 0.9958 & 0.4397 & 1.0000\\
   \hline
     \end{tabular}}
     \caption{AUC values for the California wildfires. A, B, and C denote available, burning, and consumed cells. A dash denotes an unreported AUC.}
     \label{tab:auc_cal}
 \end{table}

\begin{table}[htbp]
     \centering
     \begin{tabular}{|c|c|ccc|c|}
     \hline

      & & \multicolumn{3}{c|}{Precision@$k$} & \\
      Forecast horizon & $k$ & Proposed model & Baseline rule & Random & Lift\\ \hline
     1-step ahead & 46 & 0.261 & 0.072 & 0.00035 & 744\\
     2-step ahead & 84 & 0.215 & 0.115 & 0.00059 & 367\\
     3-step ahead & 100 & 0.183 & 0.131 & 0.00060 & 306\\
   \hline
     \end{tabular}
     \caption{Precision at $k$ for the Burning class for California wildfire, pooled across the 14 training
     windows. Here $k$ is the number of cells burning on the forecast day that were
     not burning when the forecast was made, and precision@$k$ is the proportion of
     a method's $k$ highest-ranked cells that are among them. The baseline rule
     ranks each cell by how many of its eight neighbors are burning and has no
     fitted parameters; the random column is what ranking at random achieves. Lift
     is the proposed model's precision divided by the random value.}
     \label{tab:precision_pooled}
 \end{table}

\newpage

\subsection{Australian wildfires, 2019--2020}

 Australia's vegetation and episodes of hot, dry weather contribute to substantial bushfire risk. The 2019--2020 season caused extensive environmental damage, particularly in southeastern Australia \citep{akdemir2022estimating,damany2022australian}. We analyze observations from November 1, 2019, to January 15, 2020.

\subsubsection{Parameter interpretation}

\begin{figure}[H]
    \centering
    \includegraphics[trim={0cm 0cm 1.8cm 0cm},clip, scale=0.35]{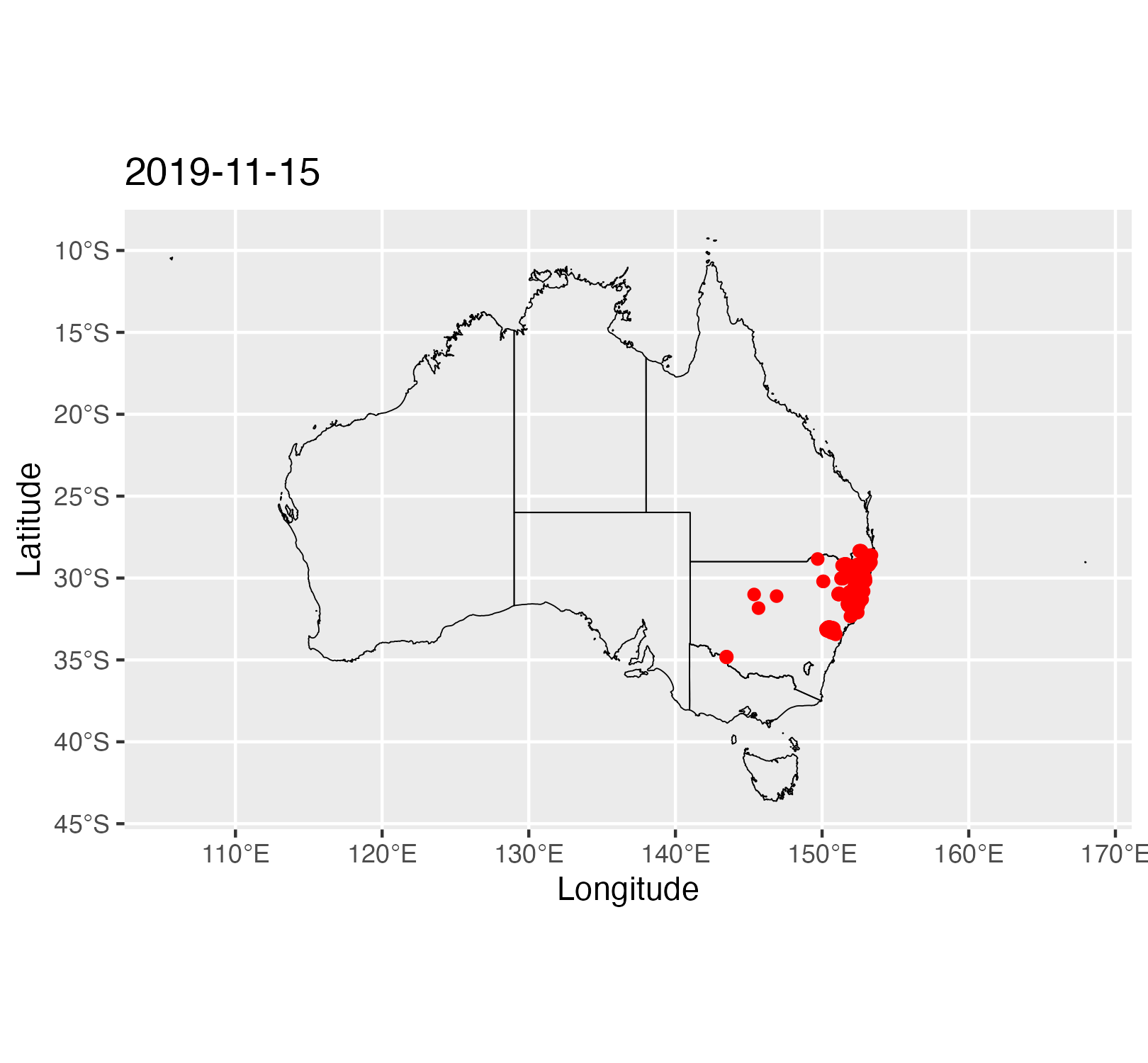}
    \hspace{-2mm}
    \includegraphics[trim={2.8cm 0cm 1.8cm 0cm},clip, scale=0.35]{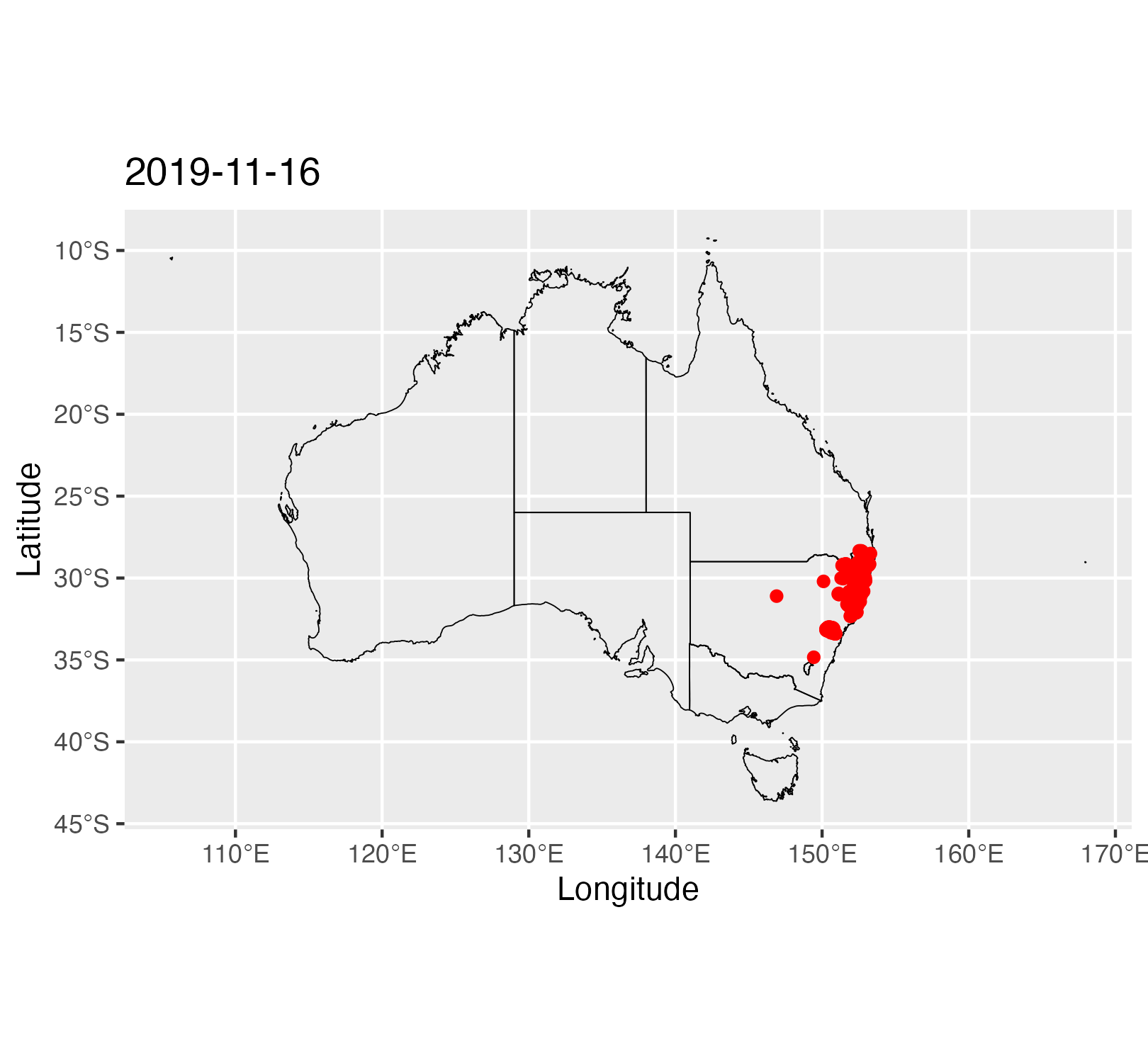}
    \hspace{-2mm}
    \includegraphics[trim={2.8cm 0cm 1.8cm 0cm},clip, scale=0.35]{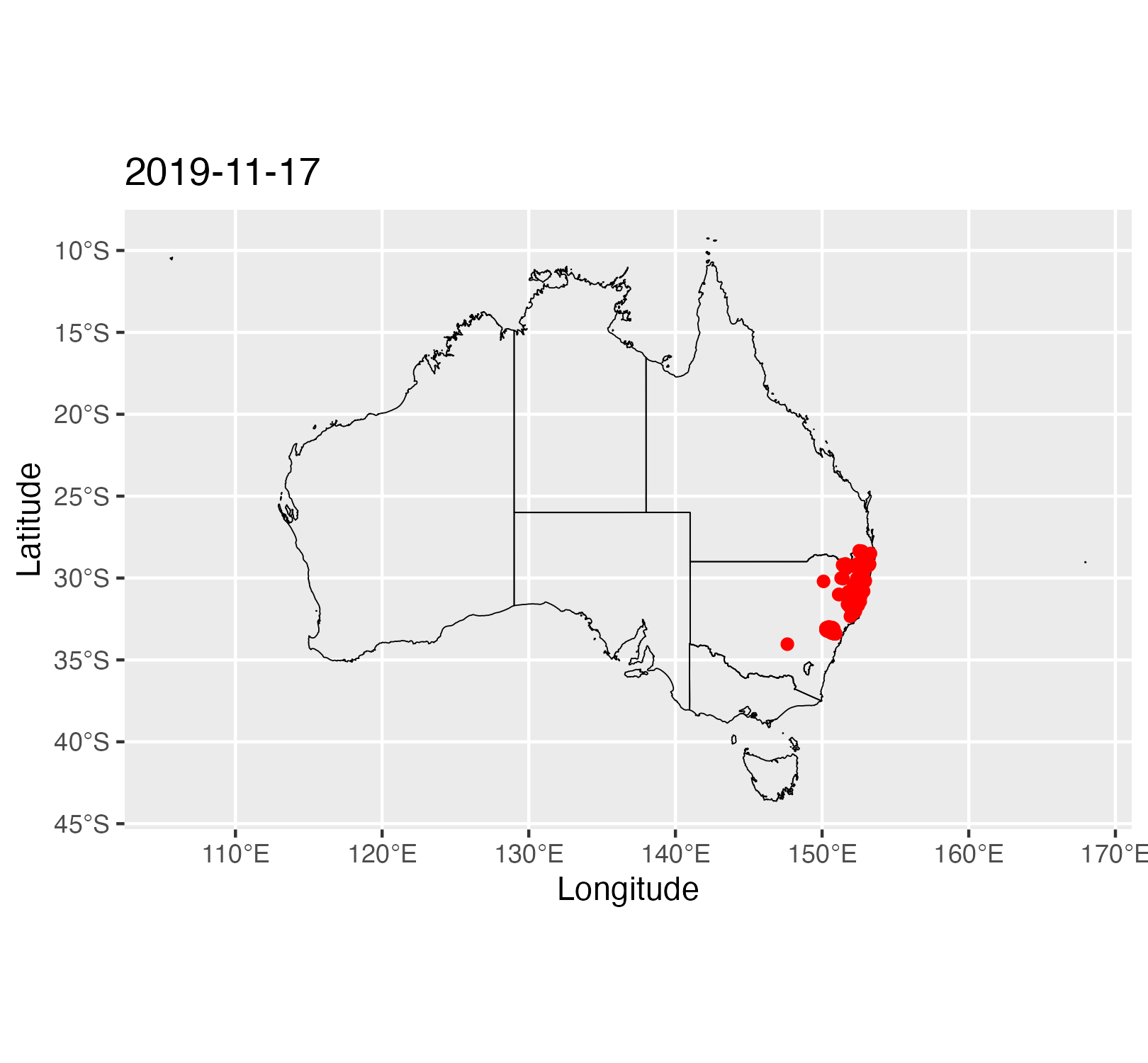}
    \hspace{-2mm}
    \includegraphics[trim={2.8cm 0cm 1.8cm 0cm},clip, scale=0.35]{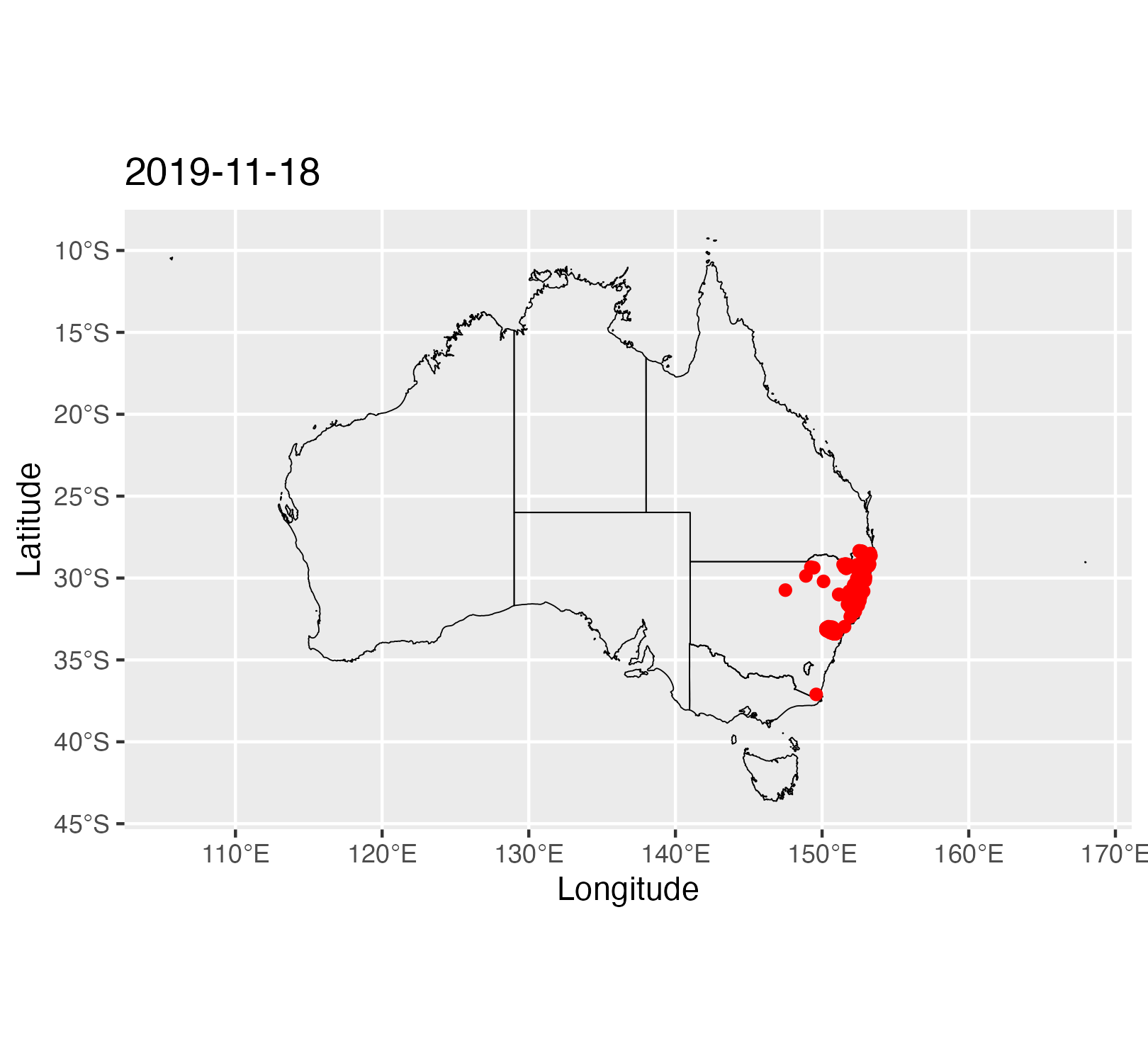}
     \includegraphics[trim={0cm 0cm 1.8cm 0cm},clip, scale=0.35]{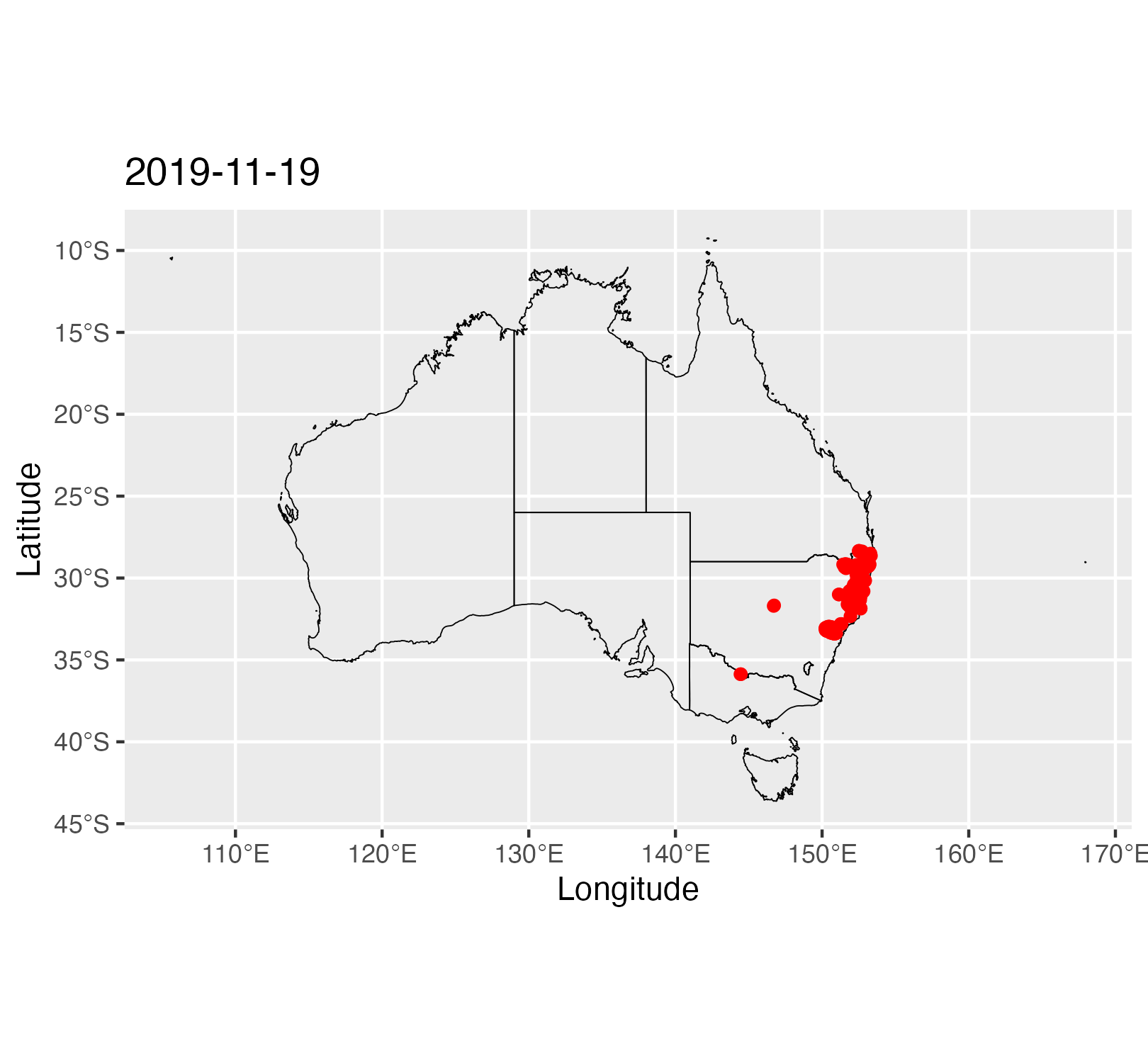}
    \hspace{-2mm}
    \includegraphics[trim={2.8cm 0cm 1.8cm 0cm},clip, scale=0.35]{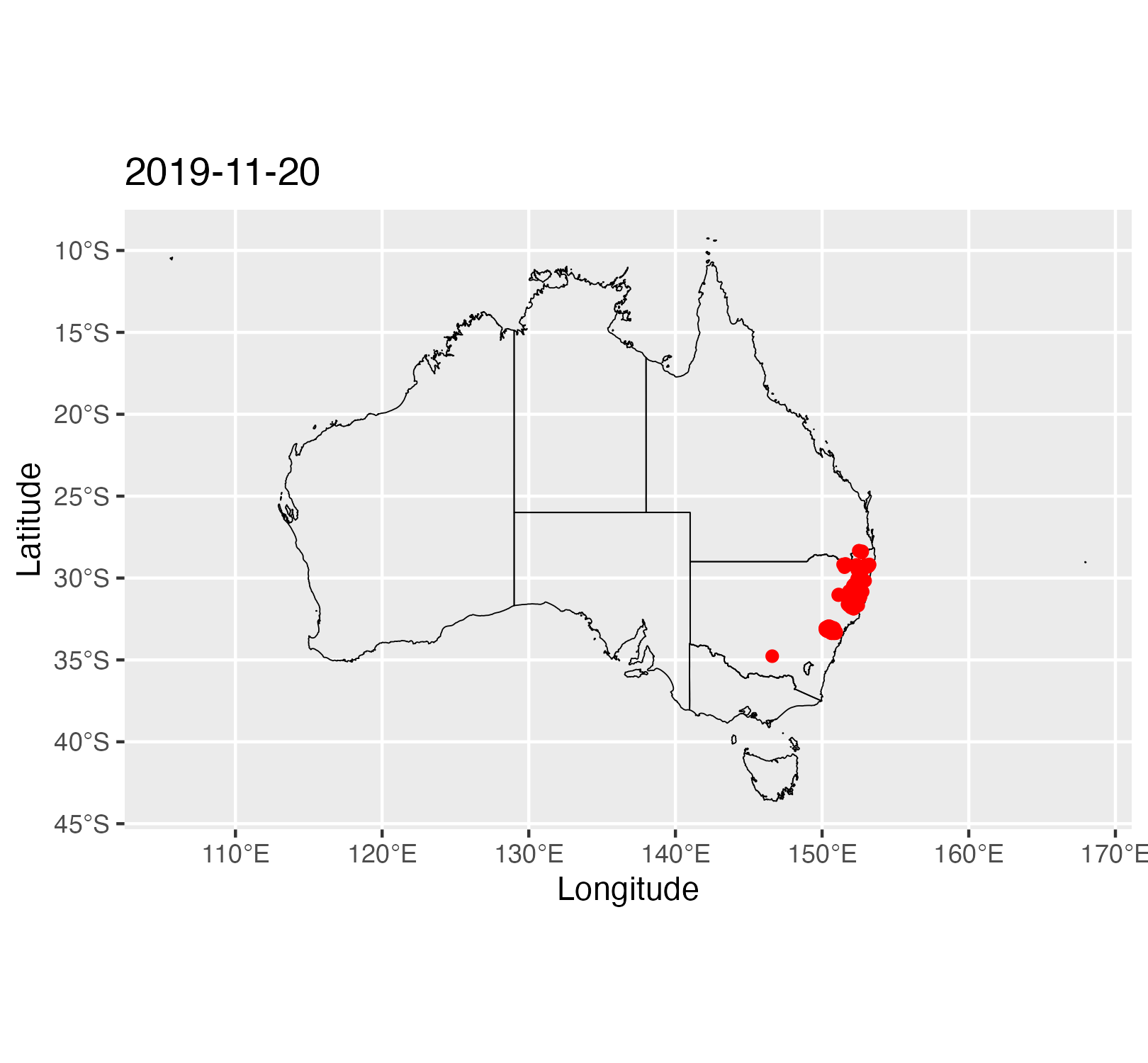}
    \hspace{-2mm}
    \includegraphics[trim={2.8cm 0cm 1.8cm 0cm},clip, scale=0.35]{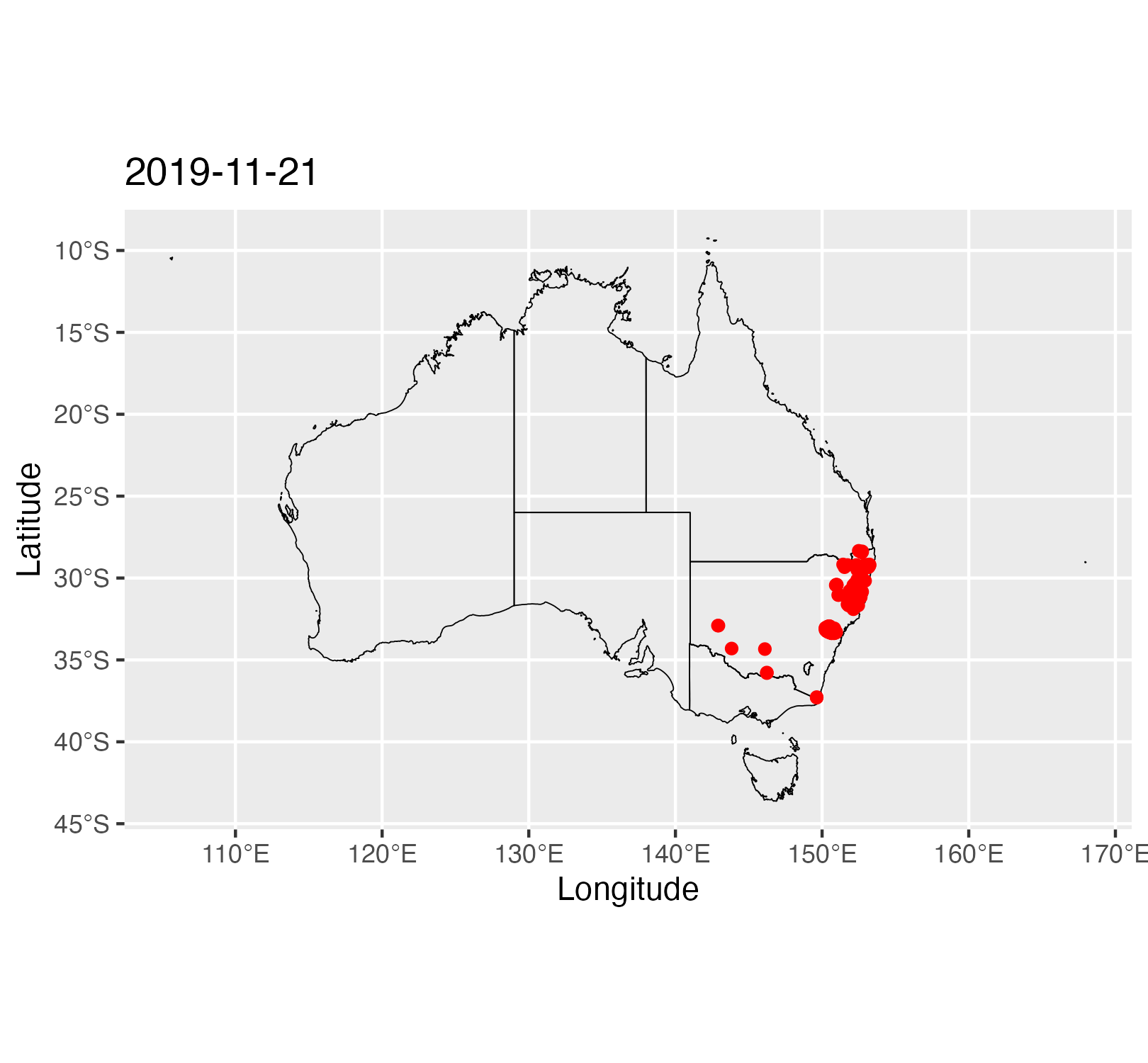}
    \hspace{-2mm}
    \includegraphics[trim={2.8cm 0cm 1.8cm 0cm},clip, scale=0.35]{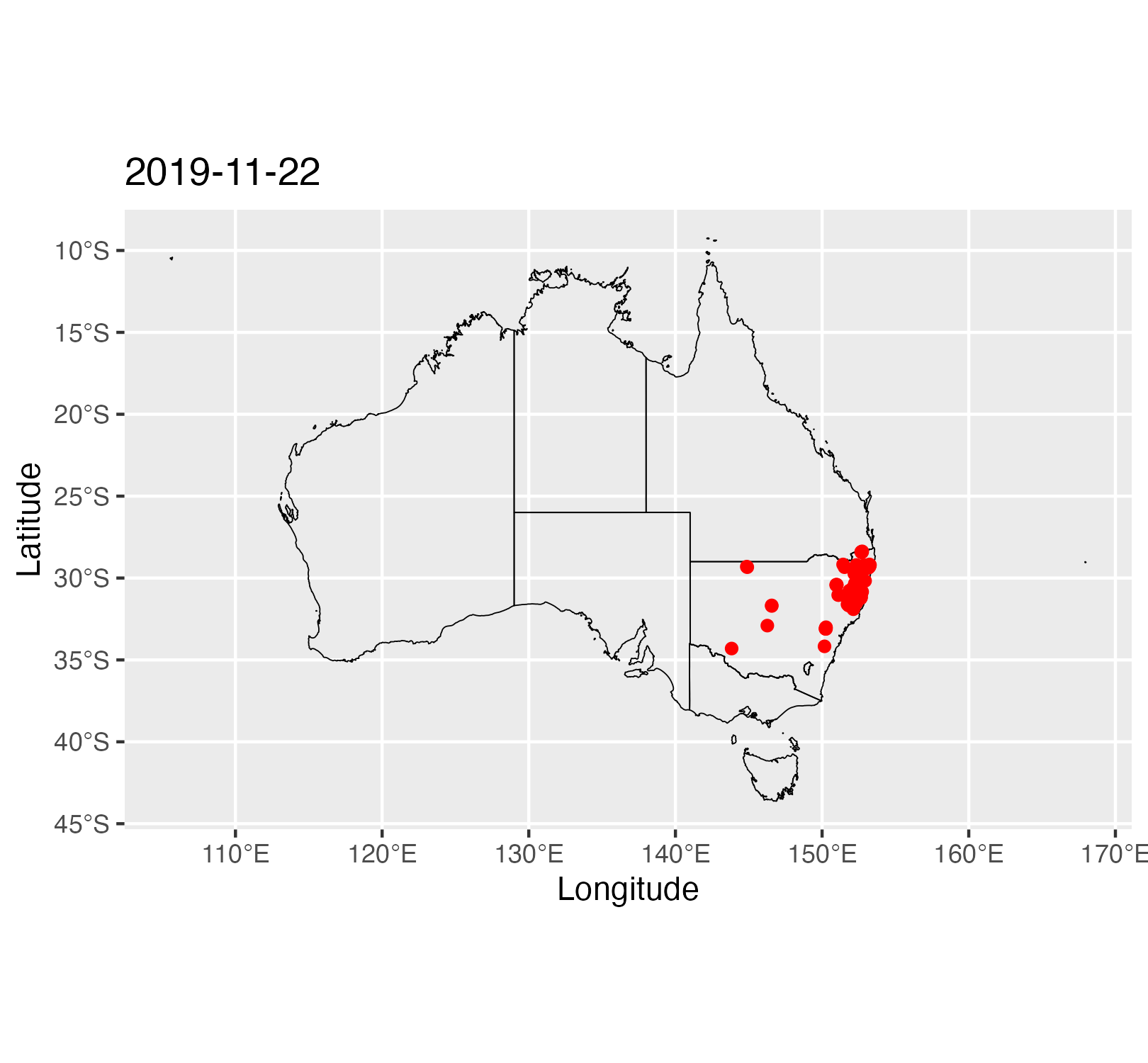}
     \includegraphics[trim={0cm 0cm 1.8cm 0cm},clip, scale=0.35]{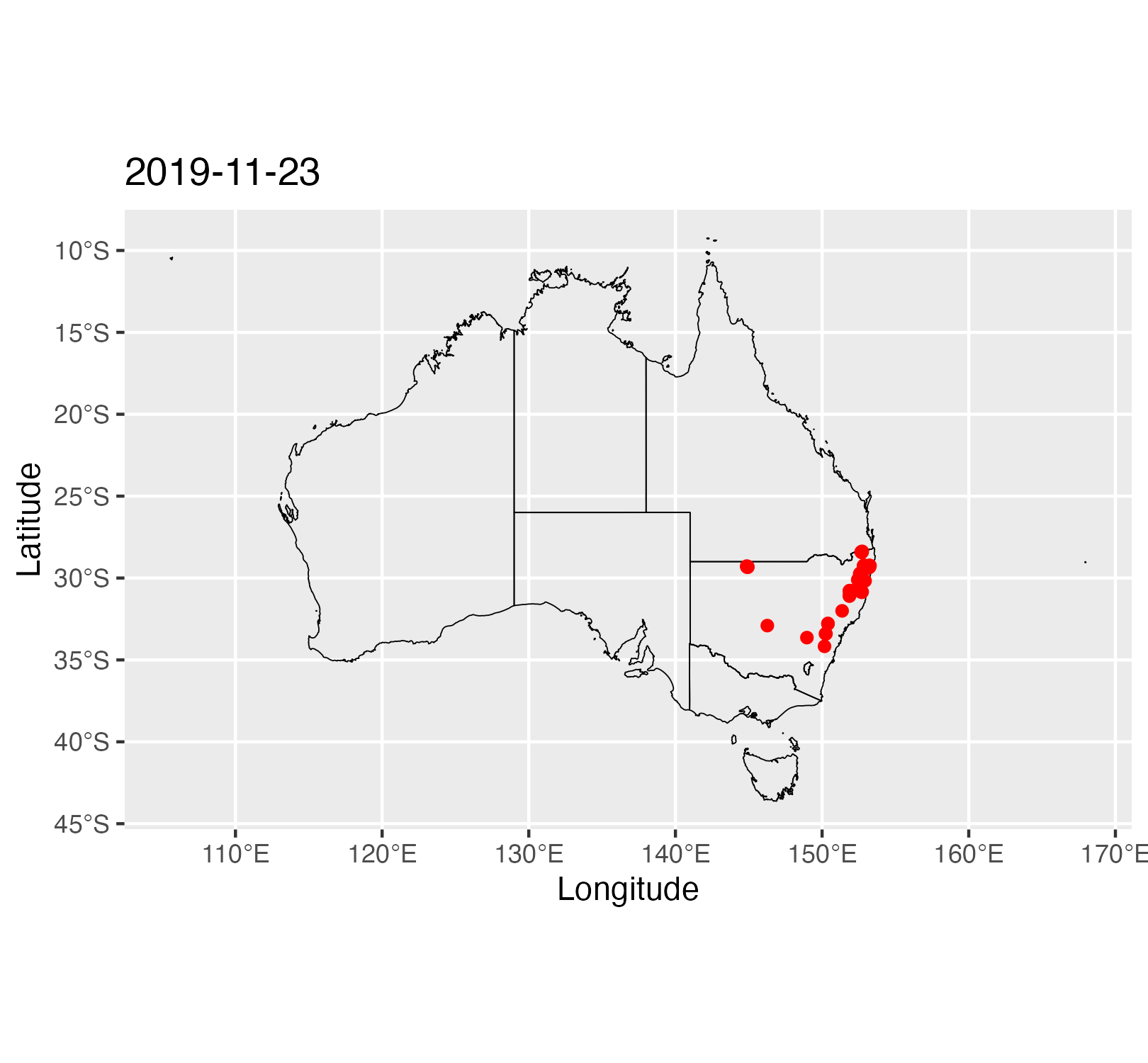}
    \hspace{-2mm}
    \includegraphics[trim={2.8cm 0cm 1.8cm 0cm},clip, scale=0.35]{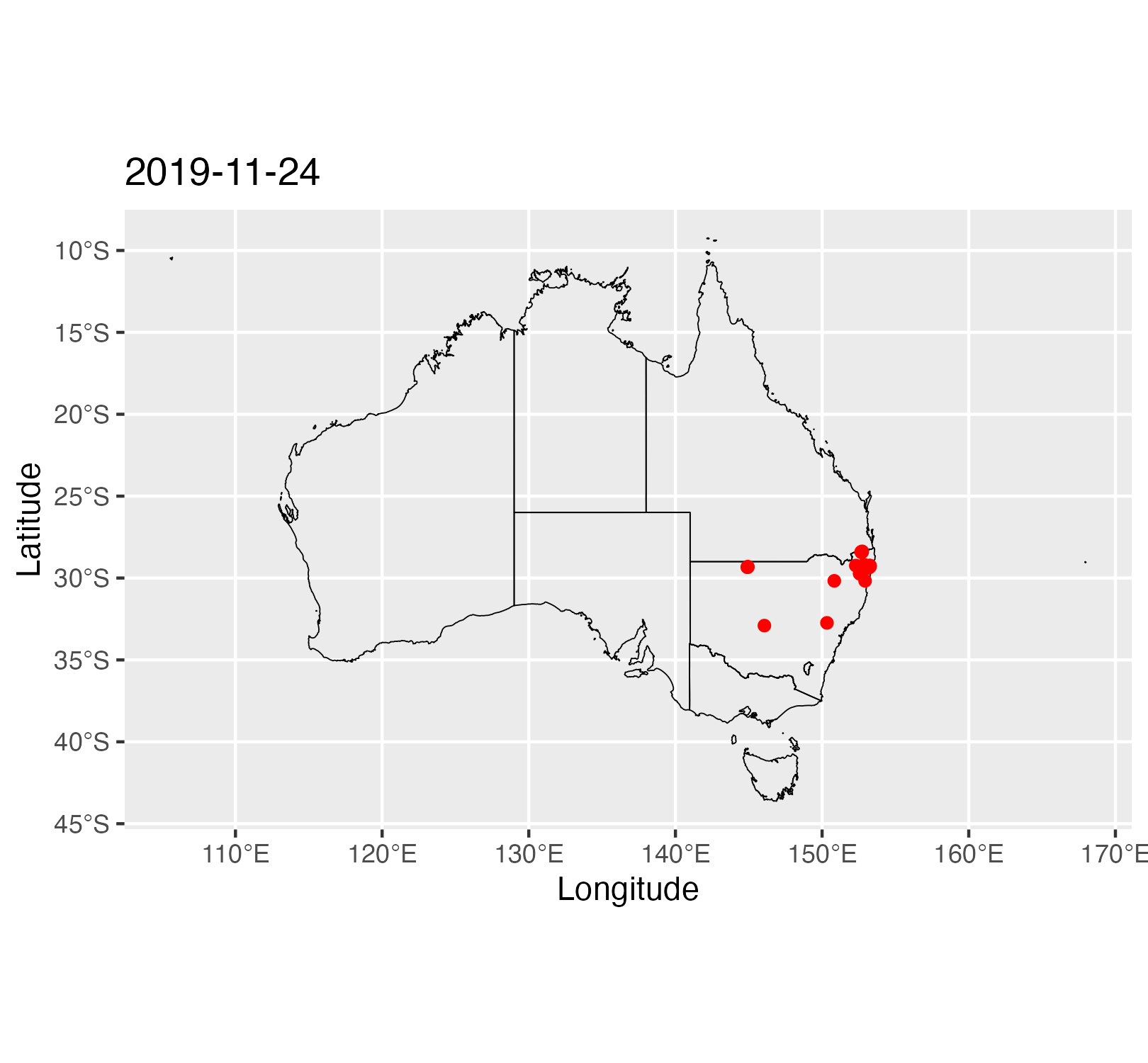}
    \hspace{-2mm}
    \includegraphics[trim={2.8cm 0cm 1.8cm 0cm},clip, scale=0.35]{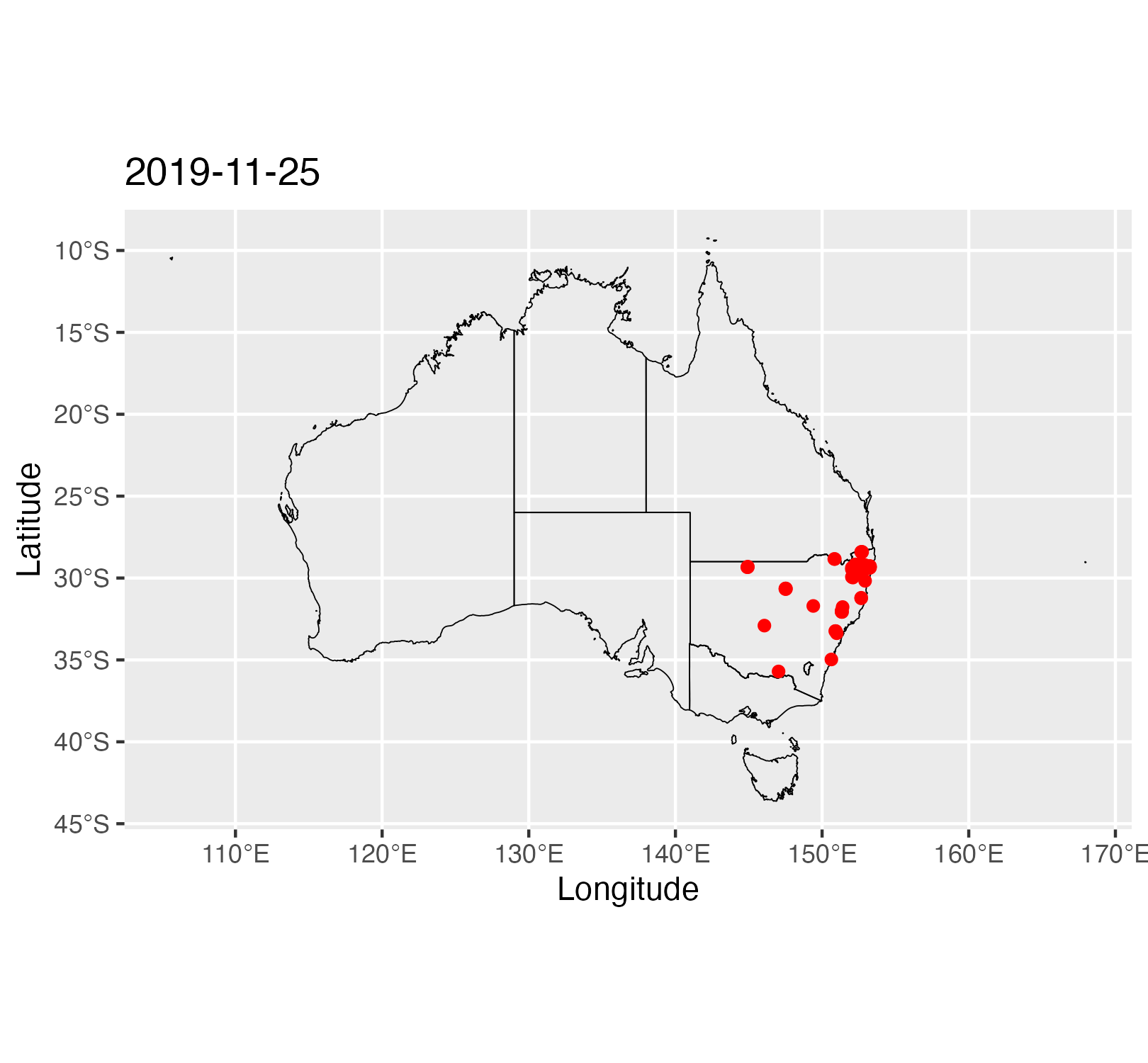}
    \hspace{-2mm}
    \includegraphics[trim={2.8cm 0cm 1.8cm 0cm},clip, scale=0.35]{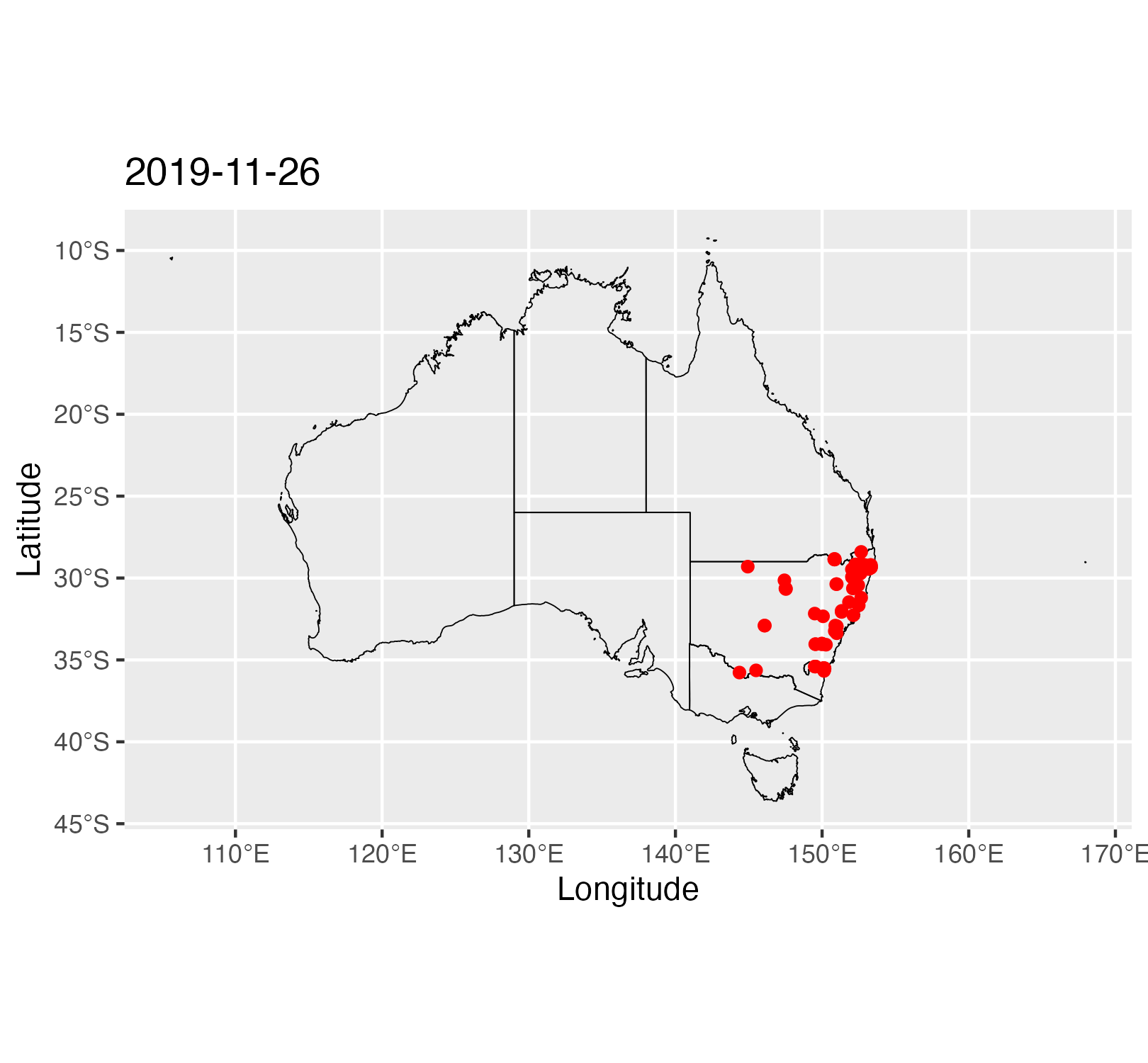}
        \caption{Progression of the fire in the period Nov 15, 2019 to Nov 26, 2019.}
    \label{fig:aus_obs}
\end{figure}

\begin{figure}[H]
\centering
\includegraphics[width=\textwidth]{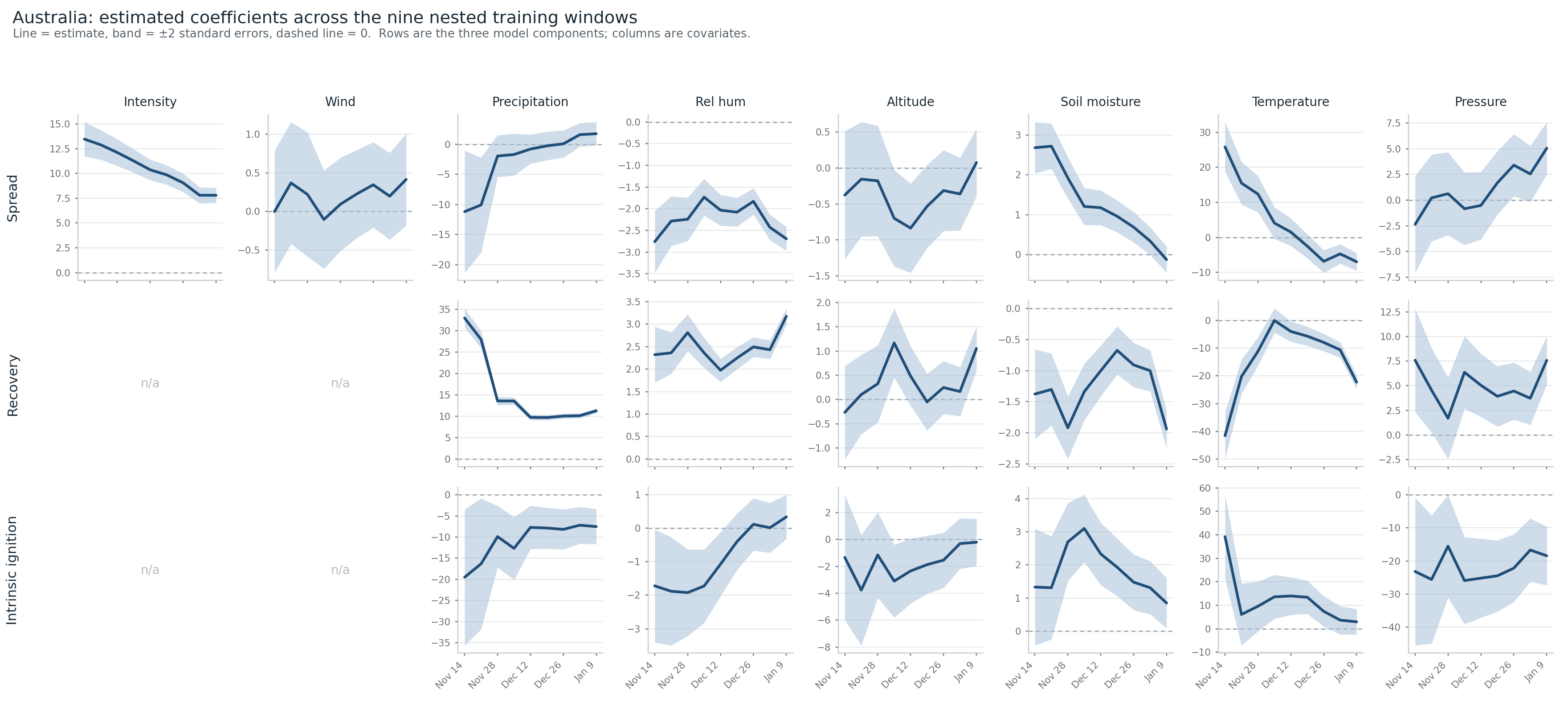}
\caption{Estimated coefficients for the Australia wildfire across the nine
nested training windows. Rows are the three model components and columns the
covariates; the solid line is the estimate, the shaded band spans two standard
errors either side, and the dashed line marks zero. Intensity and wind enter
the spread component only.}
\label{fig:coef_traj_aus}
\end{figure}

After constructing the grids, we have values of fire intensity and the covariates over the specified time period. Figure \ref{fig:aus_obs} illustrates the spread of fire in the period from Nov 15, 2019 to Nov 26, 2019. The estimated coefficients and the corresponding standard errors are provided in Tables 7-9 of the Supplementary Materials. We begin with the time period from November 1 to November 14 and incrementally extend the training window by 7 days, following the same approach used for the California dataset. To aid interpretation, we also report the relative rate ratios corresponding to the average daily change in each covariate. These are presented in Tables 10-12  of the Supplementary Materials. These scaled estimates provide a more intuitive understanding of how changes in covariates influence the transition rates in the model. Figure~\ref{fig:coef_traj_aus} shows how each
estimate evolves as the training window is extended.

Two spread coefficients are distinguishable from zero in all nine windows. Fire intensity in the neighborhood is positive throughout, declining
from $13.46$ to $7.80$ as the training window expands. Relative humidity is negative throughout, between $-1.7$ and $-2.9$, implying that spread is slower in humid conditions. Soil moisture is positive early in the season and separated from zero in seven of the nine windows. The effect of temperature is relatively unstable. Temperature is large and positive in the earliest windows and becomes small and negative later.

The recovery component has a clear and consistent interpretation. Precipitation
and relative humidity both raise the rate at which a burning cell ceases to
burn, and temperature lowers it, each in at least eight of the nine windows.
These fitted associations are consistent with faster cessation of burning under wetter, more humid, and cooler conditions. Intrinsic ignition shows the
mirror image in precipitation: the coefficient is negative in all nine windows,
indicating higher fitted intrinsic ignition rates under drier conditions, conditional on the other covariates.

Figure~\ref{fig:aus_maps} shows observed and predicted states for a
representative forecast date.

\subsubsection{Forecasting}

\begin{figure}[H]
\centering
\includegraphics[width=0.48\textwidth]{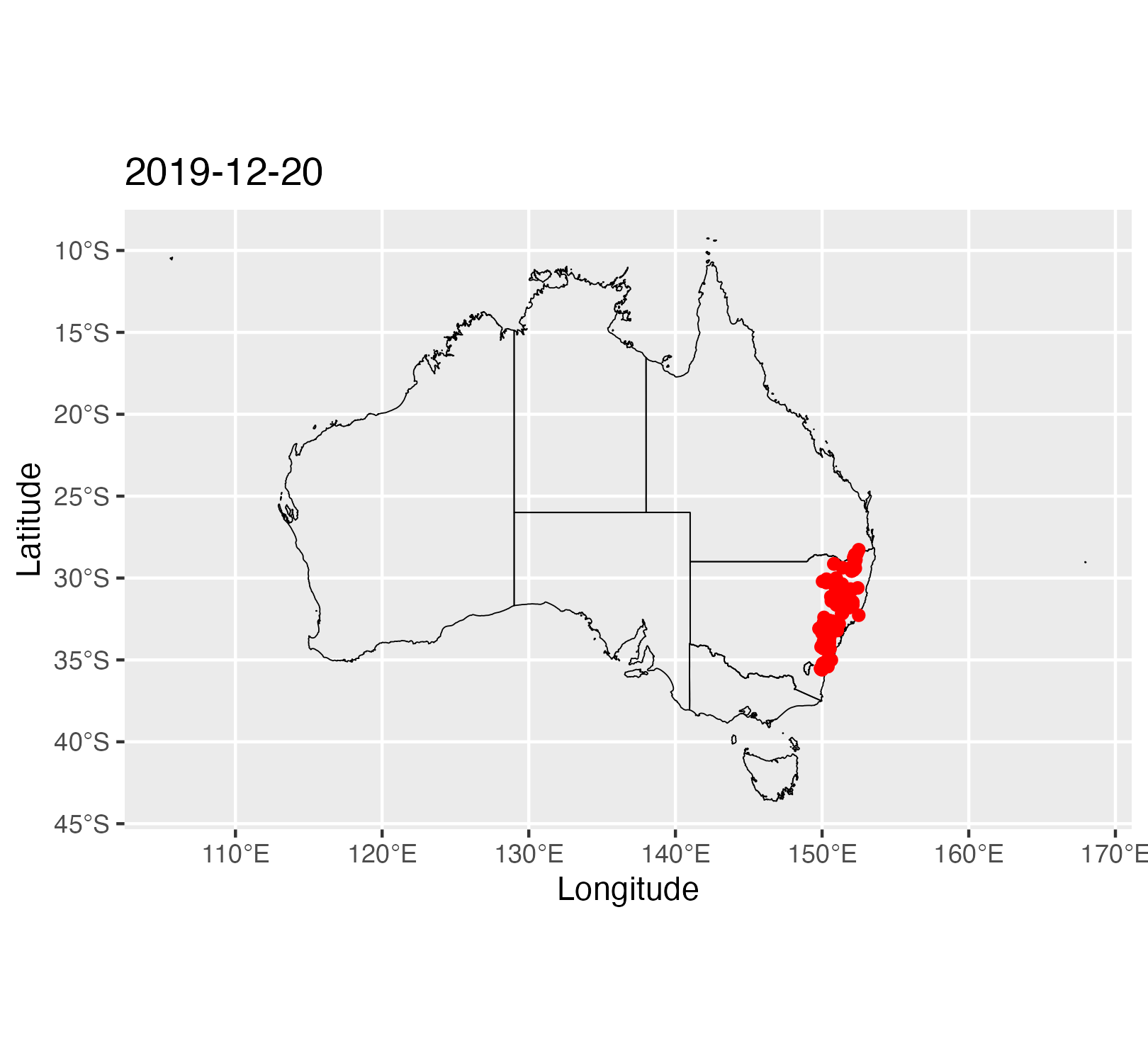}\hfill
\includegraphics[width=0.48\textwidth]{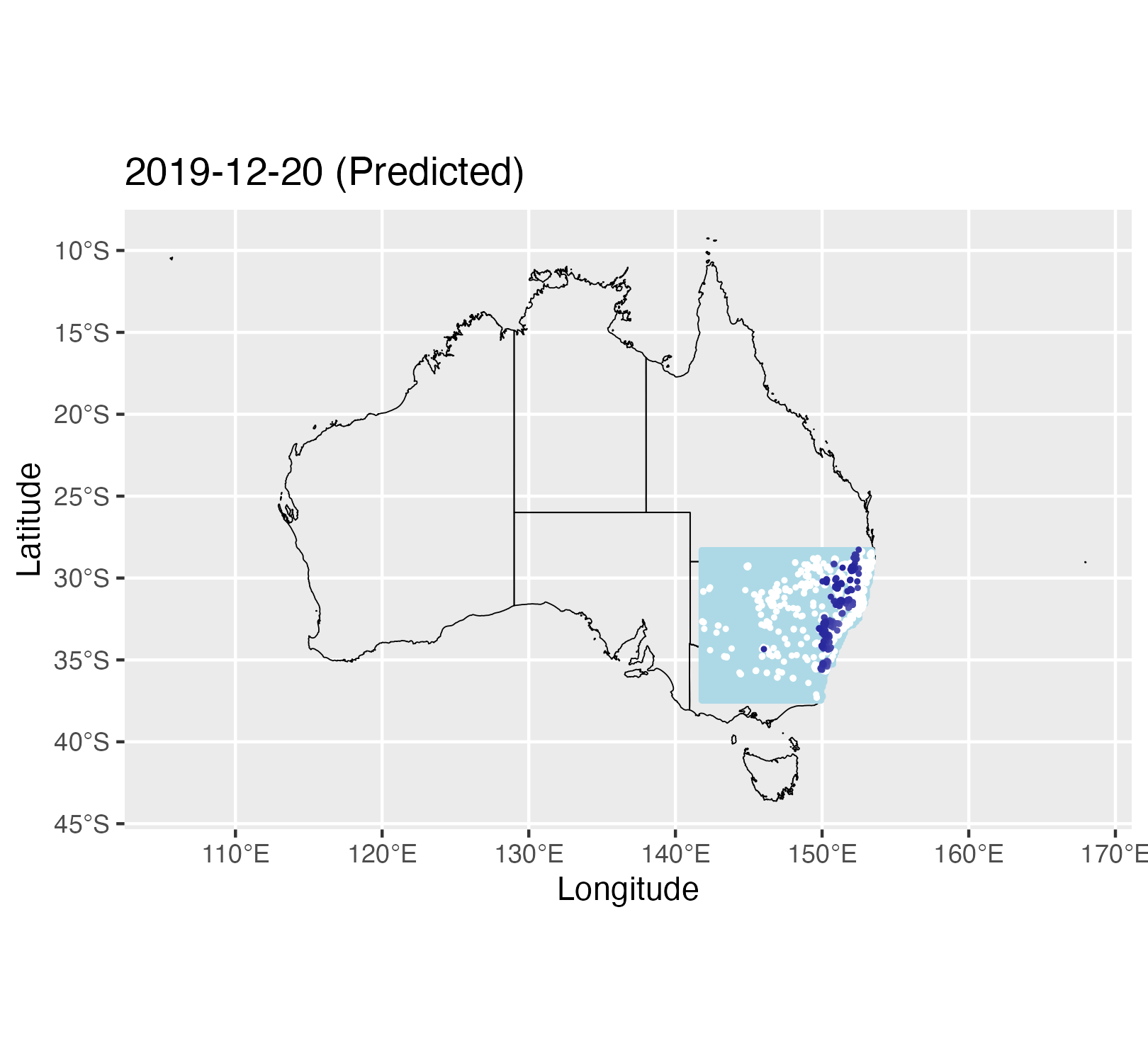}
\caption{Observed (left) and predicted (right) state of the Australian fire on
20 December 2019. In the observed panel a red point marks a burning cell. In
the predicted panel each cell is colored by the linear combination of its
three predicted state probabilities, with Available in light blue, Burning in
dark blue and Consumed in white.}
\label{fig:aus_maps}
\end{figure}

We now see how our model performs when trying to forecast fire spread for 3 consecutive future time points based on the data obtained so far. For example, we forecast fire activity for November 15, 16, and 17, 2019, using data up to November 14. The AUC values corresponding to the 1-day, 2-day and 3-day forecasts are presented in Table \ref{tab:auc_aus}. The observed vs predicted plots for all time windows considered are provided in the supplementary materials, with a specific instance provided in figure \ref{fig:aus_maps}. Similar to the California data, we observe that our model is able to capture the fire pattern reasonably well, especially for $A$ and $C$ categories. The AUC values for $B$ are relatively high for the 1-step forecasts, but the value decreases for 2 or 3-day forecasts, which is reasonable. Since ignition is a rare event, and since the AUC for the $B$ state is influenced by how many of the grids burning at the forecast time were already burning when the forecast was made, we also report the precision at $k$ in Table \ref{tab:precision_pooled_aus}. Similar to California results, $k$ is the number of grids burning on the forecast day that were not burning when the forecast was made,
and the precision at $k$ is the proportion of the $k$ highest ranked grids that are among them. Pooled over the nine training windows and the three forecast horizons, the $k$ highest ranked grids contain 987 of the 3151 grids that ignite, a precision of $0.313$ against a base rate of $0.0014$, so that the ranking is more than two hundred times richer in ignitions than one made at random. A rule that ranks grids by the number of burning neighbors and estimates no parameters achieves $0.188$ over the same set, so the fitted model identifies about $1.7$ times as many ignitions as the neighborhood structure alone.
By this measure the model performs somewhat better on the Australian data than on the
California data, where the corresponding pooled precision is $0.209$ against $0.113$ for the same
parameter-free rule.

Suppression activities may influence the Australian results, particularly in densely populated regions. Such activities were not explicitly modeled because sufficiently detailed records were unavailable. The model improves on the neighborhood baseline for the reported precision-at-$k$ summaries, but incorporating suppression and population information is a potential extension whose predictive benefit remains to be evaluated.

\begin{table}[htbp]
     \centering
     \resizebox{\textwidth}{!}{
     \begin{tabular}{|c|ccc|ccc|ccc|}
     \hline

      Time &  A (1 step) &  B (1 step) &  C (1 step) &  A (2 step) &  B (2 step) &  C (2 step) & A (3 step) &  B (3 step) &  C (3 step)\\ \hline
     Nov 1- Nov 14 & 0.9949 & 0.9817 & 0.9999 & 0.9896 & 0.9613 & 0.9889 & 0.9822 & 0.5908 & 0.9760\\
     Nov 1-Nov 21 & 0.9986 & 0.9848 & 1.0000 & 0.9956 & 0.7329 & 0.9952 & 0.9931 & 0.7056 & 0.9920\\
     Nov 1-Nov 28 & 0.9974 & 0.9281 & 1.0000 & 0.9944 & 0.7968 & 0.9923 & 0.9933 & 0.8539 & 0.9895\\
     Nov 1-Dec 5 & 0.9935 & 0.9617 & 0.9999 & 0.9871 & 0.9264 & 0.9930 & 0.9638 & 0.5309 & 0.9775\\
     Nov 1-Dec 12 & 0.9992 & 0.9573 & 1.0000 & 0.9970 & 0.9344 & 0.9993 & 0.9950 & 0.9113 & 0.9979\\
     Nov 1-Dec 19 & 0.9991 & 0.9903 & 1.0000 & 0.9971 & 0.9766 & 0.9982 & 0.9901 & 0.6884 & 0.9893\\
     Nov 1-Dec 26 & 0.9984 & 0.9540 & 1.0000 & 0.9960 & 0.9216 & 0.9995 & 0.9932 & 0.8978 & 0.9956\\
     Nov 1-Jan 2 & 0.9993 & 0.9897 & 1.0000 & 0.9787 & 0.8476 & 0.9987 & 0.9711 & 0.6994 & 0.9648\\
     Nov 1-Jan 9 & 0.9994 & 0.9564 & 1.0000 & 0.9986 & 0.9251 & 0.9976 & 0.9974 & 0.7884 & 0.9958\\
   \hline
     \end{tabular}}
     \caption{AUC values for the Australian wildfires. A, B, and C denote available, burning, and consumed cells.}
     \label{tab:auc_aus}
 \end{table}

\begin{table}[htbp]
     \centering
     \begin{tabular}{|c|c|ccc|c|}
     \hline

      & & \multicolumn{3}{c|}{Precision@$k$} & \\
      Forecast horizon & $k$ & Proposed model & Baseline rule & Random & Lift\\ \hline
     1-step ahead & 637 & 0.303 & 0.169 & 0.00085 & 355\\
     2-step ahead & 1362 & 0.339 & 0.208 & 0.00182 & 186\\
     3-step ahead & 1152 & 0.288 & 0.175 & 0.00154 & 187\\
   \hline
     \end{tabular}
     \caption{Precision at $k$ for the Burning class, pooled across the nine
     training windows, for the Australia wildfire. Here $k$ is the number of
     cells burning on the forecast day that were not burning when the forecast
     was made, and precision@$k$ is the proportion of a method's $k$
     highest-ranked cells that are among them. The baseline rule ranks each cell
     by how many of its eight neighbors are burning and has no fitted
     parameters; the random column is what ranking at random achieves. Lift is
     the proposed model's precision divided by the random value.}
     \label{tab:precision_pooled_aus}
 \end{table}

\subsection{Comparison with an existing method: the 2017 Haypress fire}
\label{sec:haypress}

The applications above compare the proposed model with a simple neighborhood rule. We next compare it with a published statistical forecasting approach. For this purpose, we consider the 2017 Haypress fire, the largest of the nineteen fires in the Orleans Complex in Siskiyou County, California. \citet{bradley2023deep} analyze this fire using a deep hierarchical generalized transformation model and report forecast AUC values at selected time points. This provides a published benchmark for evaluating our method. We reproduce their analysis setting as closely as the publicly available data permit and evaluate our model using the same forecast criterion.

\subsubsection{Setup}

The response is constructed from the GeoMAC historical fire perimeters, the same data source used by \citet{bradley2023deep}. Restricting the 2017 layer to the incident identified as ``Haypress'' yields 55 mapped perimeters, time-stamped between 11 August and 24 September 2017. \citet{bradley2023deep} report $T=52$ time points. Because several calendar days contain more than one mapped perimeter, we index time by the ordered time stamps rather than by calendar day. For each perimeter, we overlay a $100 \times 100$ lattice, matching the discretization used by \citet{bradley2023deep}. A cell is classified as Burning at time~$t$ if it lies within the perimeter mapped at that time. A cell that was burning at an earlier time but is no longer burning is classified as Consumed, with Consumed treated as an absorbing state, as described in Section~2.1.

Fire intensity data is obtained from the VIIRS active-fire product, as in Section \ref{sec:firedata}. Here, however, the VIIRS observations are re-binned directly onto the finer $118$~m lattice to match the scenario by \citet{bradley2023deep}. For burning cells without an associated intensity value, intensity is imputed using the brightness of the nearest detection available at the same time point. Meteorological covariates are obtained from gridMET at approximately $4$~km resolution and include temperature, precipitation, and relative humidity, together with the wind vector used to construct $v_{ijt}$. Elevation is obtained from a $30$~m digital elevation model. Surface pressure, vegetation cover, and soil moisture, which are available in the analyses of Sections~4.2 and~4.3, do not have directly comparable data at the resolution used here and are therefore omitted.

For evaluation, we follow the criterion used by \citet{bradley2023deep}. They compute true- and false-positive rates at locations ``that are not burning at time point 31.'' Accordingly, for a forecast from time $t$ to $t+1$, we calculate the AUC over cells that are not burning at time~$t$, using whether a cell is burning at $t+1$ as the binary outcome and its predicted probability of ignition as the score. Because our reconstruction contains 55 time points whereas the published analysis uses 52, matching the numerical indices alone does not establish that the forecast dates and evaluation cells coincide. The comparison should therefore be interpreted as approximate until those alignments are verified. This criterion differs from the state-wise AUC values reported in Tables~\ref{tab:auc_cal} and~\ref{tab:auc_aus}, which are computed over all cells. The two measures are therefore not directly comparable, and only the restricted AUC is used in the present comparison.

\subsubsection{Comparison results}

Table~\ref{tab:haypress_auc} reports results at the two forecast points for which \citet{bradley2023deep} provide AUC values: $t=31 \rightarrow 32$ and $t=51 \rightarrow 52$.

\begin{table}[htbp]
\centering
\footnotesize
\begin{tabular}{|l|r|r|r|}
\toprule
Forecast & Our model & Second Layer BHM & Third Layer BHM \\
\midrule
$t=31 \rightarrow 32$ & 0.650 & 0.65 & 0.75 \\ \hline
$t=51 \rightarrow 52$ & 0.934 & 0.79 & 0.83 \\
\bottomrule
\end{tabular}
\caption{AUC for the 2017 Haypress fire, computed over cells that are not burning at time~$t$. BHM denotes Bayesian hierarchical model. The Second Layer BHM and Third Layer BHM values are those reported by \citet{bradley2023deep}.}
\label{tab:haypress_auc}
\end{table}

At $t=51 \rightarrow 52$, when the mapped perimeter changes only modestly, the proposed model attains an AUC of $0.934$, exceeding the values of $0.79$ and $0.83$ reported for the Second Layer and Third Layer BHM, respectively. In contrast, at $t=31 \rightarrow 32$, following a much larger change in the perimeter, the proposed model achieves an AUC of $0.650$. This is essentially identical to the Second Layer BHM value of $0.65$, but below the Third Layer BHM value of $0.75$. In these two comparisons, relative performance differs with the magnitude of the observed perimeter change. In particular, \citet{bradley2023deep} obtain their largest improvement from the cellular-automata component at the large-jump forecast, whereas the proposed model performs better at the forecast time with the smaller observed change.

This difference can be understood by considering the interaction between spatial resolution and the local neighborhood structure. With cells measuring approximately $118 \times 151$~m, movement from a cell to one of its Moore neighbors corresponds to a displacement of less than approximately $0.2$~km in a single time step. The Haypress perimeters, however, can move by substantially greater distances between successive observations. To quantify this discrepancy, we calculated, for each newly ignited cell, the distance in lattice cells to the nearest cell that was already burning at the preceding time point. Overall, $59\%$ of new ignitions occur more than one cell away, with a median distance of $2$ cells. Such ignitions cannot be reached in a single synchronous update restricted to immediate neighbors. A continuous-time process can, however, propagate through multiple successive neighboring cells within one observation interval. Hence for $t=31 \rightarrow 32$, which corresponds to a larger time gap, coarse temporal observation may lead the attribution rule to label multi-step spread as intrinsic ignition.

\begin{figure}[htbp]
\centering

\begin{subfigure}[b]{0.32\textwidth}
\includegraphics[width=\textwidth]{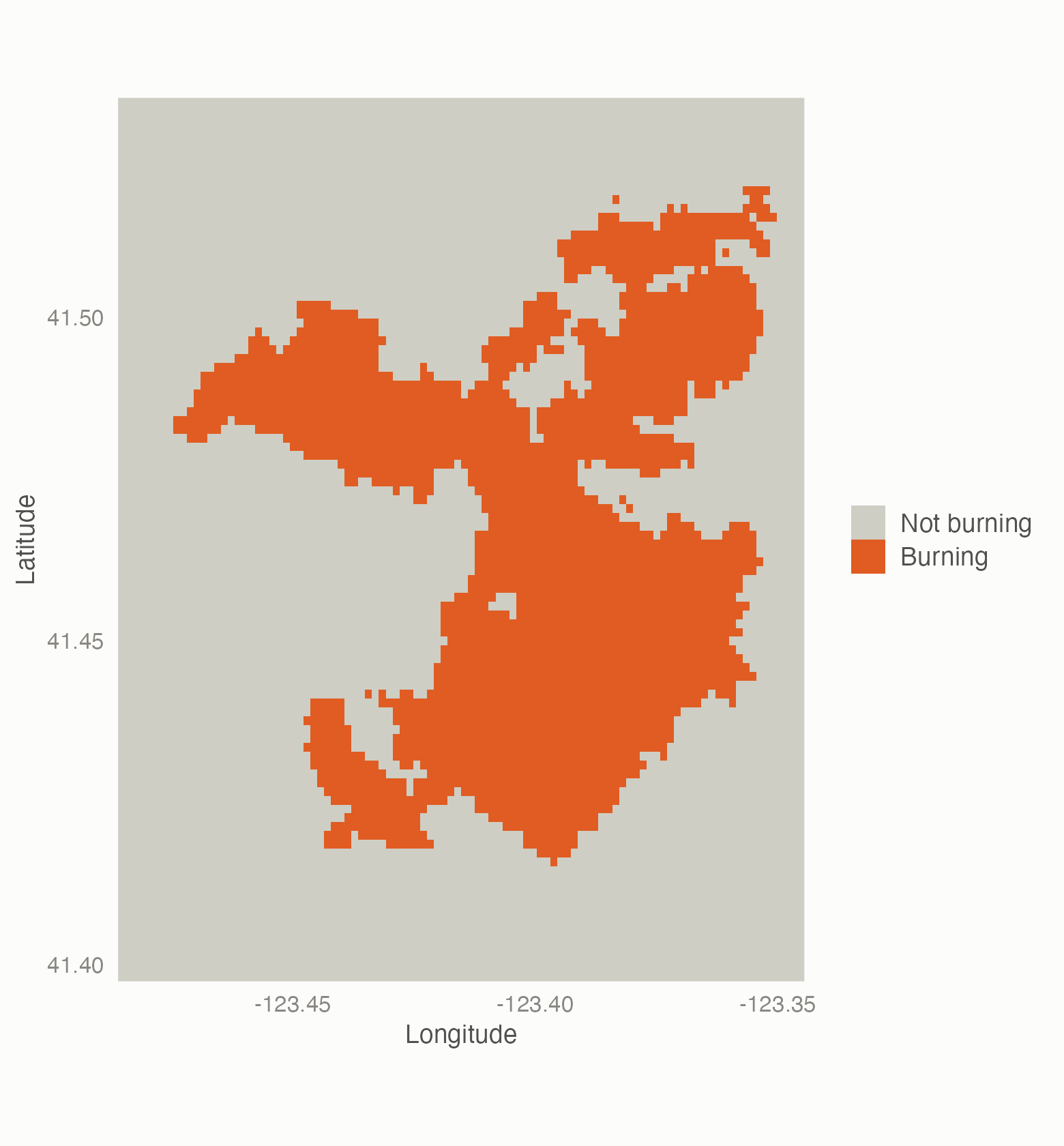}
\caption{Observed, $t=31$.}
\end{subfigure}\hfill
\begin{subfigure}[b]{0.32\textwidth}
\includegraphics[width=\textwidth]{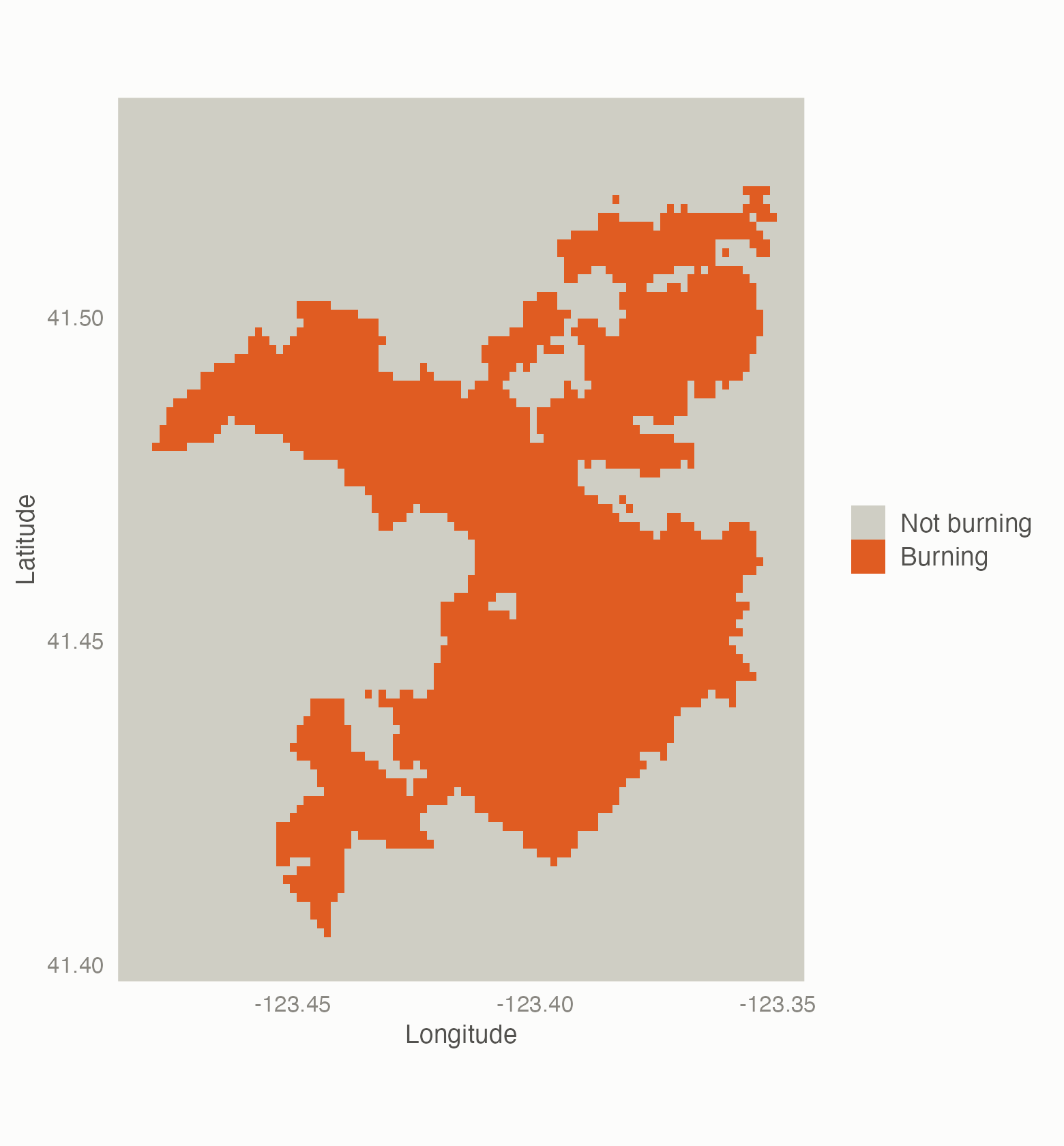}
\caption{Observed, $t=32$.}
\end{subfigure}\hfill
\begin{subfigure}[b]{0.32\textwidth}
\includegraphics[width=\textwidth]{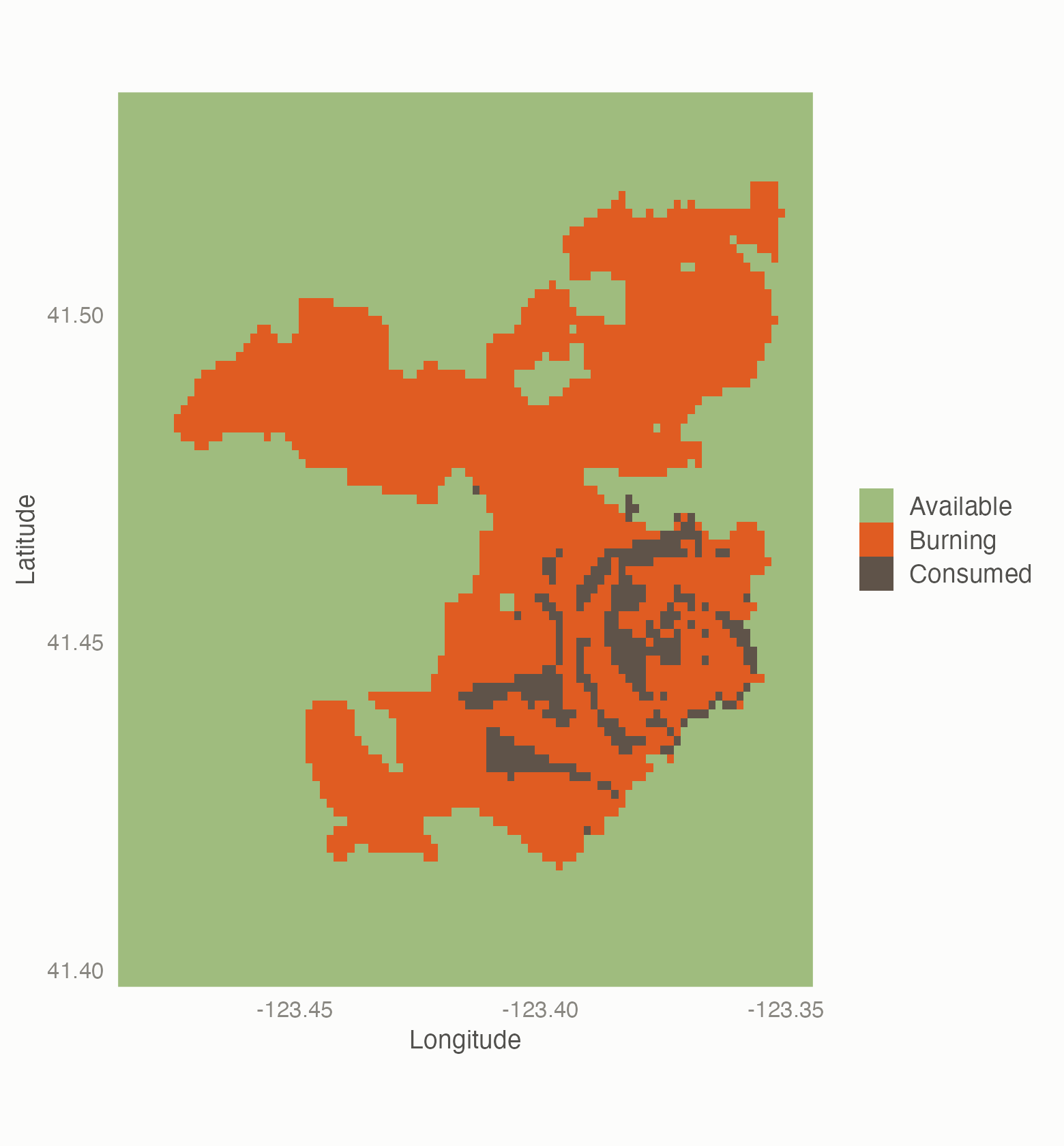}
\caption{Predicted, $t=32$.}
\end{subfigure}

\vspace{6pt}

\begin{subfigure}[b]{0.3\textwidth}
\includegraphics[width=\textwidth]{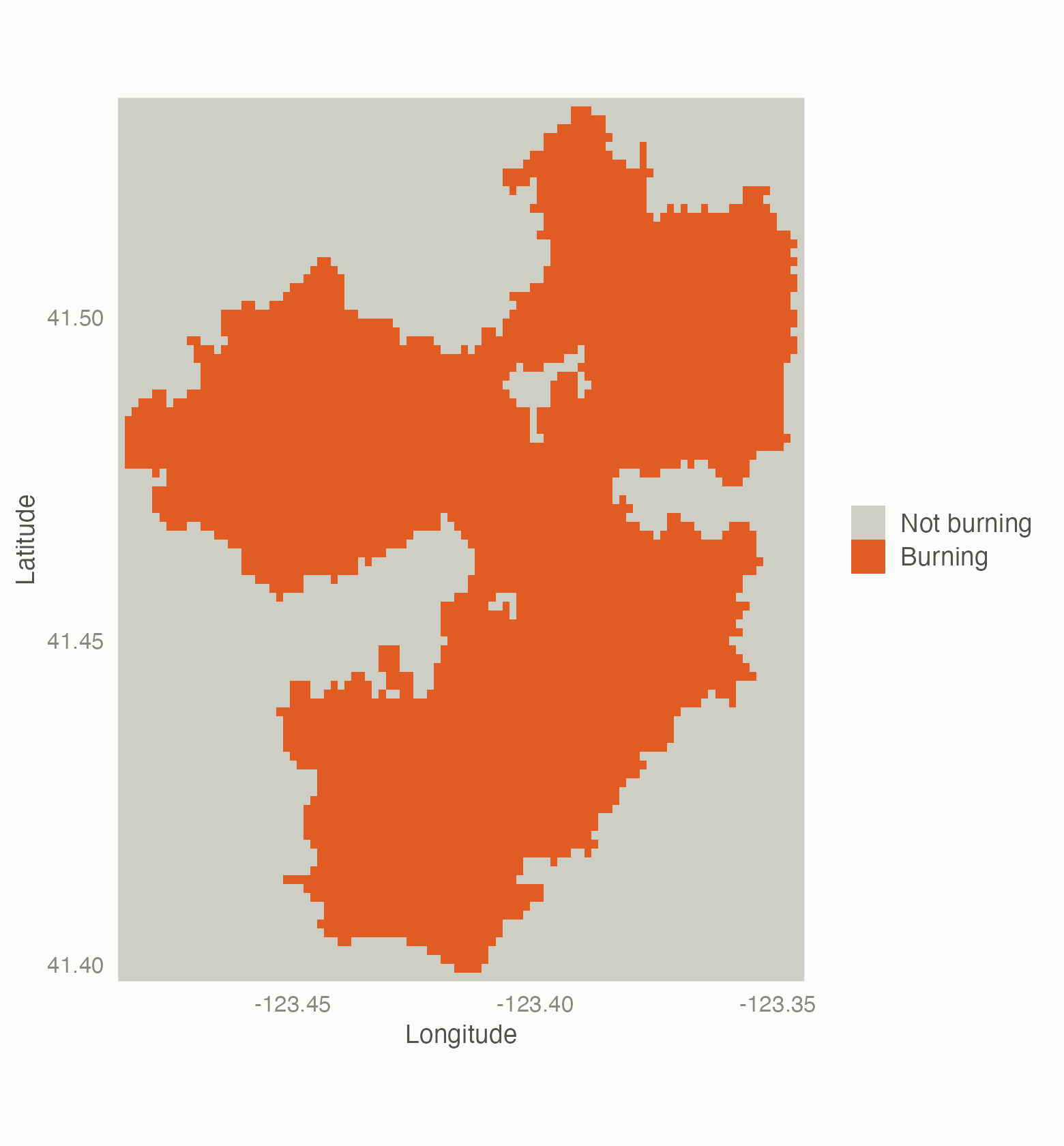}
\caption{Observed, $t=51$.}
\end{subfigure}\hfill
\begin{subfigure}[b]{0.3\textwidth}
\includegraphics[width=\textwidth]{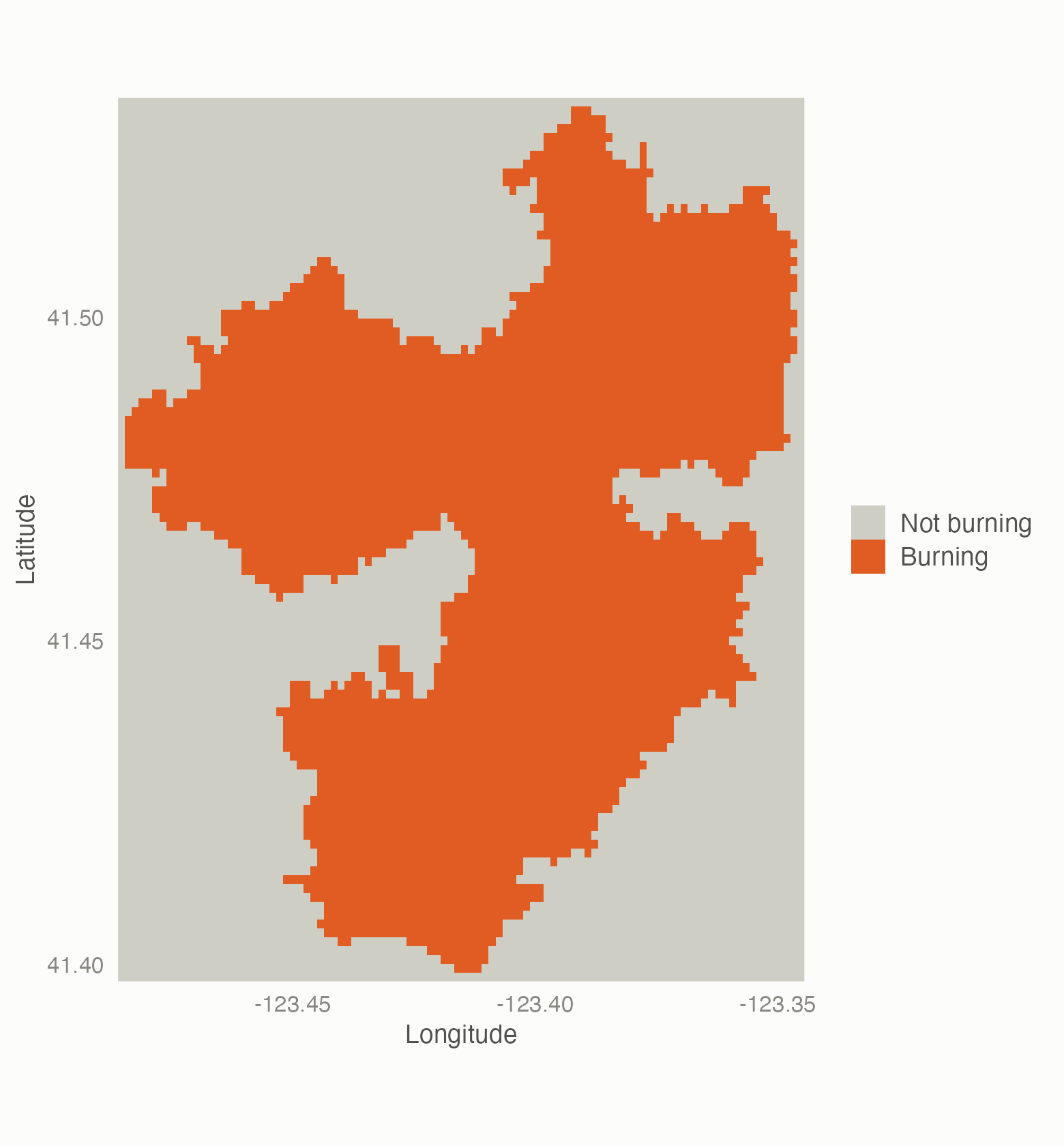}
\caption{Observed, $t=52$.}
\end{subfigure}\hfill
\begin{subfigure}[b]{0.3\textwidth}
\includegraphics[width=\textwidth]{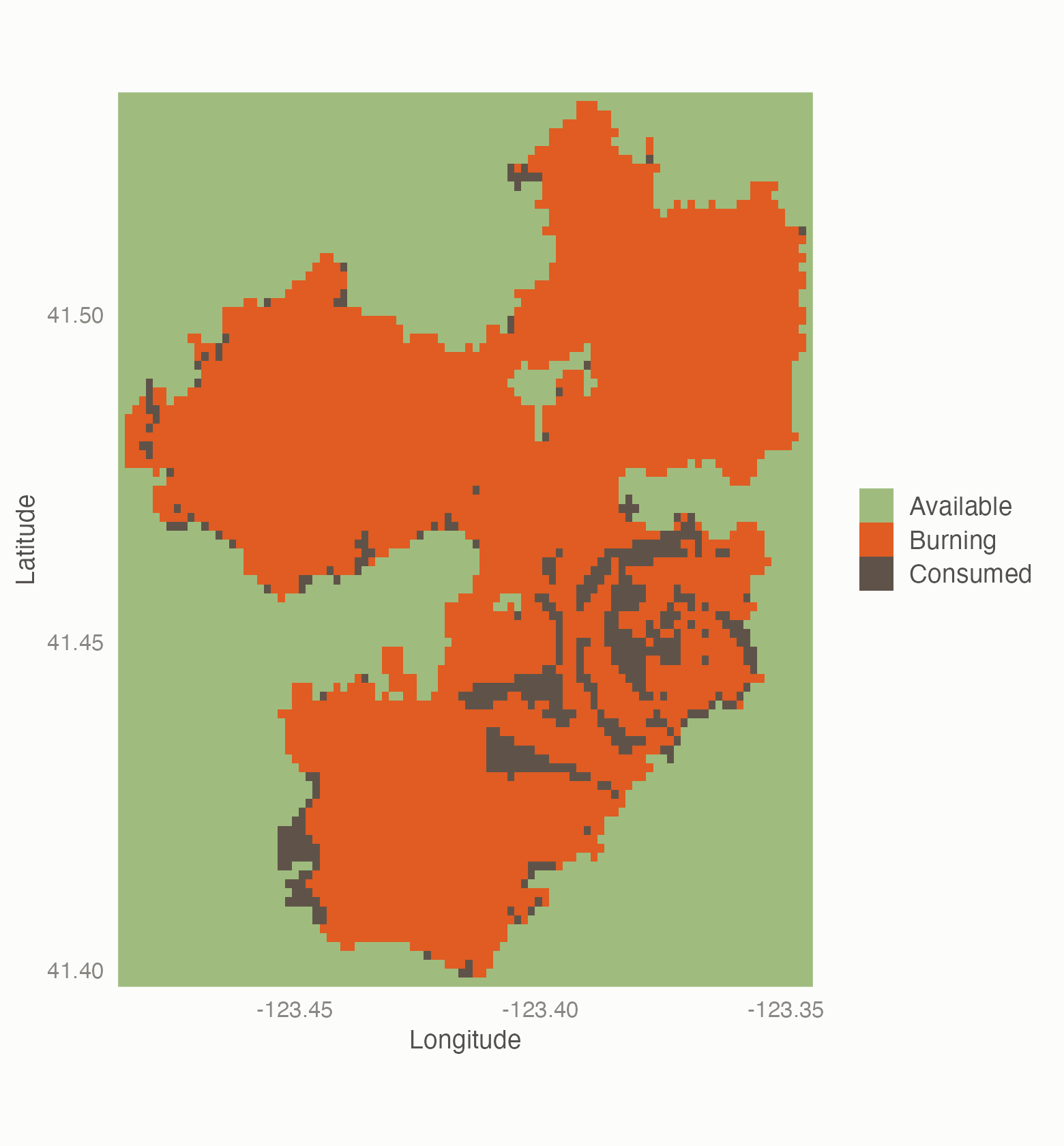}
\caption{Predicted, $t=52$.}
\end{subfigure}
\caption{Observed and predicted states of the 2017 Haypress fire at the two forecast times and the observed state in the previous time point as considered in Table~\ref{tab:haypress_auc}. The observed panels indicate whether each cell is burning, while the predicted panels show the state with the highest predicted probability.}
\label{fig:haypress}
\end{figure}

Figure~\ref{fig:haypress} compares the observed and predicted states at the two forecast times considered by \citet{bradley2023deep}. The observed panels indicate whether each cell is burning, whereas the predicted panels display the state having the highest predicted probability. At $t=52$, the predicted fire extent follows the observed perimeter closely and reproduces several unburnt regions within the fire boundary. At $t=32$, the model identifies the primary direction of spread but underestimates the magnitude of the observed expansion, consistent with the lower AUC obtained for this forecast.

\section{Discussion}
\label{sec:discussion}
We developed a likelihood-based framework for modeling wildfire evolution on a spatial lattice. The model distinguishes neighboring spread, intrinsic ignition, and cessation of burning, with covariates entering each transition rate separately. Applications to the 2018 California and 2019--2020 Australian wildfires show strong discrimination for many short-term forecasts, but performance for the burning state declines at longer horizons and during abrupt expansion. The Haypress comparison further illustrates the sensitivity of local propagation models to spatial and temporal resolution.

The framework provides interpretable conditional associations between environmental covariates and transition rates. Its cell-level state probabilities can describe spatial variation within a region and support downstream analyses of fire activity. These associations should not be interpreted as causal effects, particularly when suppression activities, fuel characteristics, or observation errors are incompletely measured.

Several limitations warrant further investigation. First, the absorbing consumed state prevents reignition. At the chosen grid resolution, unburned biomass within a partially burned cell can permit renewed burning even over short periods. An SIRS-type extension \citep{sirs}, or a model with within-cell fuel availability, could address this limitation. Efficient implementations would also permit finer grids and sensitivity analyses across spatial resolutions.

Second, ignition mechanisms are unobserved. The deterministic attribution rule in Section~\ref{subsec: est} simplifies estimation but can misclassify intrinsic ignition and spread. The reported model-based standard errors do not incorporate uncertainty in these labels or in the approximated transition times. Integrating over unobserved mechanisms and interval-censored paths would provide a more complete inferential treatment.

Third, operational forecasts require future covariate values. Errors in weather forecasts contribute uncertainty beyond that represented by plugging predicted covariates into the fitted transition rates. Measurement-error methods \citep{carroll1995measurement,robinson1986errors} and predictive distributions for uncertain covariates \citep{foster2012uncertainty} offer possible approaches for propagating this uncertainty into fire forecasts.

The Haypress analysis also highlights the interaction between the immediate Moore neighborhood and fine spatial discretization. At approximately $118\times151$~m resolution, substantial fire expansion between observations may require several successive local transitions. A single synchronous update cannot represent such propagation, even though it is possible in the underlying continuous-time process. Potential extensions include explicit simulation between observation times and distance-based neighborhoods adapted to the spatial and temporal resolution. The comparison with published forecasts also requires careful alignment of observation dates, response definitions, and evaluation cells.

Fire-suppression activities can substantially alter observed wildfire dynamics, but sufficiently detailed suppression records were not available for these analyses. The model therefore learns from fire activity under the interventions that actually occurred; it does not estimate the counterfactual spread of an uncontrolled fire. Population and other anthropogenic covariates, as considered by \citet{lawler2024anthropogenic}, may provide useful additional information, although their value for cell-level short-term forecasts requires evaluation. Analyzing California and Australia provides evidence from different settings, but does not by itself isolate the effects of population density or suppression.

Finally, satellite detections are subject to measurement error and incomplete observation. Excluding low-confidence detections reduces some false positives, but an absent detection does not necessarily establish that a cell is not burning. A model that separates latent fire states from the observation process could better represent these uncertainties and improve the interpretation of available, burning, and consumed states.

\clearpage

\bibliography{ref}
\end{document}


\bibliographystyle{apalike}

\def\spacingset#1{\renewcommand{\baselinestretch}%
{#1}\small\normalsize} \spacingset{1}

{
  \title{ \textbf{Supplementary Material for `Short-term forecasting of wildfire spread: A network epidemiology approach'}}
  \author[1]{Indrila Ganguly}
  \author[2]{Muhammad Ali}
  \author[3]{Swarnali Sanyal}
  \author[2]{Viney Aneja}
  \author[5]{Srijan Sengupta}
  \affil[1]{Department of Experimental Statistics, Louisiana State University}

  \affil[2]{Department of Marine, Earth, and Atmospheric Sciences, North Carolina State University}
  \affil[3]{Department of Atmospheric Sciences, University of Illinois at Urbana Champaign }
   \affil[5]{Department of Statistics, North Carolina State University}
  \date{}
  \maketitle
}

\bigskip

\noindent%

\spacingset{1.1} %

\section{Appendix}

 \subsection{Likelihood function construction}
\label{app:lik_construct}
We now want to show the construction of the likelihood function mentioned in Section 2. The construction of the likelihood is somewhat straightforward from Proposition 3.1 in \cite{Rasmussen2018LectureNT}. We state the above-mentioned proposition in terms of the conditional intensity of a general point process, since in our setting the rate at which a grid ignites depends on the states of its neighbors and is therefore not a deterministic function of time.

\noindent \textbf{Result 1:} Suppose we observe the time interval $[0, T)$, and $(t_1, \ldots, t_n)$ denote the arrival times of a point process on that interval. Let $\mathcal{H}_t$ denote the history of the process up to but not including time $t$, and let $\lambda^*(t)$ denote the conditional intensity function, defined as the mean number of events occurring in the interval $[t,t+h)$ given $\mathcal{H}_t$, where $h>0$ is an infinitesimal quantity. Then, the likelihood function is given by \begin{align*}
    L= \prod_{i=1}^n \lambda^*(t_i) \exp\left(-\int_{0}^T  \lambda^*(s) ds  \right)
\end{align*}
\noindent A non-homogeneous Poisson process is the special case in which $\lambda^*(t)$ does not depend on $\mathcal{H}_t$. In our setting the conditional intensity of a grid depends on the states of its neighbors at time $t-$, which are determined by the history.

\noindent Now, we would simply like to apply this result in our framework.  In our case, we observe the process till time $T_{max}$, and  $t_j^{(B)}$, $t_j^{(C)}$  denote the times an event occurs at grid `$j$'.
Thus, the likelihood is given by:

\begingroup\footnotesize
\begin{align*}
    L(\theta; \text{data}) &= \prod_{j=1}^N \Biggl[\left( \beta \exp(x_{jt_j^{(B)}}'b)\sum_{i \in \mathcal{N}_j} \exp(y_{it_j^{(B)}}\eta +v_{ijt_j^{(B)}}\epsilon )\right)^{I(j \text{ is A and ignites from a burning neighbor})} \times \\ & \left( \gamma \exp(w_{j t_j^{(C)}}' \alpha)\right)^{I(j \text{ recovers})} \left( \xi \exp(x_{j t_j^{(B)}}' b_E)\right)^{I(j \text{ is A and ignites intrinsically})} \times \\ & \exp\Biggl(- \int_0^{T_{max}} \Biggl[ \left( \beta \exp(x_{jt}'b)\sum_{i \in \mathcal{N}_j} \exp(y_{it}\eta +v_{ijt}\epsilon )\right)I(j \text{ is $A$ and has a $B$ neighbor at } t)+ \\& \left( \gamma \exp(w_{j t}' \alpha)\right) I(j \text{ is $B$ at } t)+ \left( \xi \exp(x_{j t}' b_E)\right)I( j \text{ is $A$  at } t )\Biggr] dt \Biggr) \Biggr]  \end{align*}
\endgroup
    Some straightforward simplifications lead us to the likelihood provided in Section 2.3 of the main manuscript. Thus, we get,

    \begingroup\footnotesize
\begin{align*}
 & L(\theta; \text{data})  = \beta^{n_B}\gamma^{n_C} \xi^{n_B^{(E)}}\underset{j: j \text{ is A and ignites from a burning neighbor}}{\prod} \exp ({x_{jt_j^{(B)}}'b})\sum_{i \in \mathcal{N}_j} e^{y_{it_j^{(B)}}\eta+v_{ijt_j^{(B)}}\epsilon} \\& \times  \underset{j: j \text{ is A and ignites intrinsically}}{\prod} \exp ({x_{jt_j^{(B)}}'b_E}) \times \exp \left (-\int_0^{T_{max}}\xi \sum_{j=1}^N e^{x_{jt}'b_E}1(j \text{ is $A$ at }t) dt  \right)\\ &\times \underset{j: j \text{ recovers}}{\prod} \exp ({w_{jt_j^{(C)}}'\alpha}) \times \exp \Biggl(- \int_0^{T_{max}} \bigl[\beta \sum_{j=1}^N e^{x_{jt}'b}\bigl[\sum_{i \in \mathcal{N}_j} e^{y_{it}\eta+v_{ijt}\epsilon}\bigr] I(j \text{ is $A$ and has a $B$ neighbor at } t)  \\ & +\gamma \sum_{j=1}^N e^{w_{jt}'\alpha} 1( j \text{ is $B$ at } t) \bigr]dt \Biggr)
 \end{align*}
\endgroup

The likelihood above is written for a process observed continuously, so that the transition times $t_j^{(B)}$ and $t_j^{(C)}$ are known exactly. In practice the state of each grid is recorded at regular intervals, and these times are known only to lie within one such interval. In the applications we take $t_j^{(B)}$ and $t_j^{(C)}$ to be the first interval at which the new state is recorded, which is an adequate approximation when the transition probabilities over a single interval are small.

\subsection{Parameter estimates and forecasts for California data}

\begin{table}[H]
     \centering
     \resizebox{\textwidth}{!}{
     \begin{tabular}{|c|ccccccccc|}
     \hline

      Time & Intensity & Wind & Temperature & Pressure & Precipitation & Veg cover & Soil moisture & Rel hum & Altitude\\ \hline
     June 1-July 28 & 26.27 (3.87) & -2.90 (1.22) & 3.75 (1.66) & -1.19 (0.82) & -2.18 (4.72) & 0.48 (0.30) & 0.78 (0.80) & 3.22 (1.89) & 0.92 (0.58)\\
     June 1- Aug 4 & 26.30 (3.48) & -2.93 (1.07) & 2.02 (1.40) & -0.10 (0.67) & -1.24 (4.58) & 0.08 (0.25) & 0.70 (0.70) & 1.50 (1.63) & 0.13 (0.54)\\
     June 1- Aug 11 & 22.94 (2.90) & -2.48 (1.06) & 1.52 (1.21) & -0.19 (0.57) & -2.61 (5.17) & -0.13 (0.22) & 1.33 (0.62) & 1.08 (1.37) & -0.33 (0.52)\\
     June 1-Aug 18 & 21.90 (2.59) & -1.84 (1.08) & 1.61 (1.10) & -0.41 (0.51) & -2.99 (5.33) & -0.08 (0.20) & 1.17 (0.54) & 1.38 (1.24) & -0.25 (0.48)\\
     June 1 - Aug 25 & 20.49 (2.40) & -1.51 (1.07) & 2.04 (1.03) & -0.61 (0.47) & -3.91 (5.40) & 0.01 (0.18) & 0.92 (0.49) & 1.59 (1.17) & -0.26 (0.44)\\
     June 1- Sep 1 & 21.06 (2.34) & -1.48 (1.05) & 1.99 (1.01) & -0.71 (0.45) & -3.76 (5.37) & -0.03 (0.18) & 0.95 (0.49) & 1.51 (1.15) & -0.44 (0.43)\\
     June 1- Sep 8 & 21.75 (2.26) & -1.60 (0.99) & 2.05 (0.96) & -0.59 (0.43) & -4.10 (5.53) & 0.03 (0.17) & 0.69 (0.46) & 1.68 (1.09) & -0.43 (0.41)\\
     June 1- Sep 15 & 21.66 (2.18) & -1.73 (0.97) & 2.20 (0.94) & -0.66 (0.42) & -3.94 (5.51) & -0.03 (0.17) & 0.86 (0.45) & 1.79 (1.07) & -0.61 (0.40)\\
     June 1- Sep 22 & 21.99 (2.20) & -1.73 (0.97) & 2.32 (0.94) & -0.70 (0.42) & -3.83 (5.47) & -0.04 (0.17) & 0.94 (0.45) & 1.85 (1.07) & -0.61 (0.40)\\
     June 1- Sep 29 & 19.52 (2.02) & -1.73 (0.97) & 2.23 (0.93) & -0.72 (0.41) & -4.23 (5.62) & -0.04 (0.17) & 1.01 (0.45) & 1.72 (1.06) & -0.55 (0.39)\\
     June 1 - Oct 6 & 19.36 (1.99) & -1.70 (0.96) & 2.08 (0.92) & -0.70 (0.41) & -3.46 (4.89) & -0.03 (0.17) & 0.98 (0.44) & 1.60 (1.05) & -0.61 (0.39)\\
     June 1- Oct 13 & 17.93 (1.83) & -1.73 (0.87) & 2.03 (0.88) & -0.70 (0.39) & -3.79 (5.00) & 0.04 (0.16) & 0.84 (0.42) & 1.68 (0.99) & -0.74 (0.36)\\
     June 1 - Oct 20 & 17.65 (1.79) & -1.66 (0.84) & 2.06 (0.87) & -0.71 (0.39) & -3.46 (4.83) & 0.06 (0.16) & 0.86 (0.41) & 1.52 (0.99) & -0.53 (0.36)\\
     June 1 - Oct 27 & 17.15 (1.73) & -1.56 (0.84) & 2.09 (0.85) & -0.71 (0.38) & -3.47 (4.86) & 0.04 (0.16) & 0.98 (0.41) & 1.47 (0.97) & -0.61 (0.35)\\
   \hline
     \end{tabular}}
     \caption{Spread coefficients for the California wildfire. Each cell gives the estimate with its standard error in parentheses.
      }
     \label{tab:coef_spread}
 \end{table}

\begin{table}[H]
     \centering
     \resizebox{\textwidth}{!}{
     \begin{tabular}{|c|ccccccc|}
     \hline

      Time & Temperature & Pressure & Precipitation & Veg cover & Soil moisture & Rel hum & Altitude\\ \hline
     June 1-July 28 & -5.18 (1.92) & 1.73 (0.85) & -6.42 (11.10) & -0.18 (0.32) & -0.95 (0.90) & -2.17 (2.26) & -1.16 (0.70)\\
     June 1- Aug 4 & -2.51 (1.32) & 0.33 (0.61) & -7.33 (10.79) & -0.58 (0.23) & 0.52 (0.67) & -1.41 (1.57) & -2.50 (0.54)\\
     June 1- Aug 11 & -2.22 (1.16) & 0.52 (0.54) & -6.60 (10.82) & -0.34 (0.20) & -0.01 (0.60) & -1.23 (1.39) & -2.65 (0.52)\\
     June 1-Aug 18 & -2.03 (1.08) & 0.19 (0.49) & -6.81 (10.78) & -0.23 (0.19) & 0.42 (0.56) & -1.39 (1.31) & -3.00 (0.49)\\
     June 1 - Aug 25 & -1.53 (1.01) & 0.21 (0.45) & -9.32 (11.60) & -0.13 (0.17) & 0.09 (0.51) & -0.86 (1.22) & -2.88 (0.45)\\
     June 1- Sep 1 & -0.98 (0.96) & -0.06 (0.42) & -10.76 (11.70) & -0.02 (0.16) & 0.34 (0.50) & -0.56 (1.16) & -3.35 (0.43)\\
     June 1- Sep 8 & -0.62 (0.93) & -0.18 (0.40) & -12.03 (12.00) & -0.15 (0.16) & 0.22 (0.48) & -0.02 (1.13) & -3.37 (0.41)\\
     June 1- Sep 15 & -0.42 (0.86) & -0.25 (0.37) & -13.17 (12.18) & -0.15 (0.15) & 0.01 (0.45) & 0.56 (1.04) & -3.31 (0.39)\\
     June 1- Sep 22 & -0.62 (0.85) & -0.21 (0.37) & -13.40 (12.21) & -0.16 (0.15) & 0.01 (0.44) & 0.39 (1.03) & -3.34 (0.38)\\
     June 1- Sep 29 & -0.74 (0.84) & -0.14 (0.36) & -15.11 (12.75) & -0.18 (0.15) & -0.06 (0.43) & 0.33 (1.02) & -3.34 (0.37)\\
     June 1 - Oct 6 & -0.98 (0.82) & -0.08 (0.35) & 8.89 (3.91) & -0.14 (0.14) & -0.12 (0.42) & 0.24 (0.99) & -3.24 (0.36)\\
     June 1- Oct 13 & -1.17 (0.77) & 0.03 (0.33) & 8.80 (3.92) & -0.13 (0.14) & -0.12 (0.40) & 0.02 (0.92) & -3.20 (0.34)\\
     June 1 - Oct 20 & -1.50 (0.75) & 0.13 (0.32) & 8.64 (3.93) & -0.11 (0.14) & -0.09 (0.38) & -0.24 (0.88) & -3.05 (0.32)\\
     June 1 - Oct 27 & -1.61 (0.71) & 0.08 (0.31) & 8.57 (3.90) & -0.09 (0.13) & -0.10 (0.36) & -0.11 (0.83) & -2.73 (0.31)\\
   \hline
     \end{tabular}}
     \caption{Recovery coefficients for the California wildfire. Each cell gives the estimate with its standard error in parentheses.}
     \label{tab:coef_recovery}
 \end{table}

\begin{table}[H]
     \centering
     \resizebox{\textwidth}{!}{
     \begin{tabular}{|c|ccccccc|}
     \hline

      Time & Temperature & Pressure & Precipitation & Veg cover & Soil moisture & Rel hum & Altitude\\ \hline
     June 1-July 28 & -2.59 (2.42) & 1.51 (0.96) & 2.78 (4.91) & 0.04 (0.45) & -0.12 (1.11) & -1.73 (2.58) & -3.01 (0.90)\\
     June 1- Aug 4 & -3.10 (2.30) & 1.59 (0.89) & 2.50 (4.99) & -0.22 (0.43) & -0.04 (1.10) & -1.81 (2.47) & -3.28 (0.87)\\
     June 1- Aug 11 & -3.63 (2.25) & 1.54 (0.87) & 2.42 (5.04) & -0.16 (0.42) & -0.22 (1.06) & -2.29 (2.46) & -3.18 (0.84)\\
     June 1-Aug 18 & -2.18 (2.15) & 1.00 (0.83) & 2.41 (5.09) & -0.17 (0.40) & -0.50 (1.00) & -1.37 (2.36) & -3.07 (0.78)\\
     June 1 - Aug 25 & -1.91 (2.14) & 0.82 (0.82) & 2.50 (5.07) & -0.22 (0.39) & -0.48 (0.99) & -0.98 (2.31) & -3.07 (0.77)\\
     June 1- Sep 1 & -1.31 (2.06) & 0.48 (0.79) & 2.34 (5.14) & -0.41 (0.38) & -0.33 (0.99) & -0.39 (2.21) & -3.07 (0.75)\\
     June 1- Sep 8 & -0.86 (1.99) & 0.43 (0.77) & 2.03 (5.21) & -0.46 (0.37) & -0.17 (0.97) & 0.51 (2.07) & -2.91 (0.72)\\
     June 1- Sep 15 & -0.55 (1.88) & 0.21 (0.73) & 1.86 (5.29) & -0.39 (0.36) & -0.63 (0.92) & 0.78 (1.95) & -3.29 (0.70)\\
     June 1- Sep 22 & -0.25 (1.80) & -0.05 (0.71) & 1.56 (5.38) & -0.27 (0.34) & -0.57 (0.87) & 1.17 (1.83) & -3.42 (0.68)\\
     June 1- Sep 29 & -0.46 (1.72) & -0.06 (0.68) & 0.98 (5.67) & -0.27 (0.32) & -0.54 (0.83) & 0.87 (1.76) & -3.37 (0.64)\\
     June 1 - Oct 6 & -0.40 (1.67) & 0.02 (0.65) & -3.64 (4.99) & -0.26 (0.32) & -0.54 (0.81) & 0.92 (1.70) & -3.57 (0.64)\\
     June 1- Oct 13 & -1.41 (1.57) & 0.45 (0.61) & -4.37 (5.23) & -0.28 (0.30) & -0.37 (0.76) & 0.12 (1.60) & -3.22 (0.59)\\
     June 1 - Oct 20 & -1.86 (1.37) & 0.22 (0.54) & -7.11 (6.13) & -0.35 (0.27) & -0.40 (0.69) & 0.57 (1.35) & -3.28 (0.53)\\
     June 1 - Oct 27 & -2.83 (1.31) & 0.45 (0.51) & -7.80 (6.20) & -0.22 (0.25) & -0.39 (0.64) & -0.11 (1.28) & -2.93 (0.49)\\
   \hline
     \end{tabular}}
     \caption{Intrinsic ignition coefficients for the California wildfire. Each cell gives the estimate with its standard error in parentheses.}
     \label{tab:coef_intrinsic}
 \end{table}

 \begin{table}[H]
     \centering
     \resizebox{\textwidth}{!}{
     \begin{tabular}{|c|ccccccccc|}
     \hline

      Time & Intensity & Wind & Temperature & Pressure & Precipitation & Veg cover & Soil moisture & Rel hum & Altitude\\ \hline
     June 1-July 28 & 1.01 (0.00) & 1.00 (0.00) & 1.12 (0.06) & 0.99 (0.00) & 0.99 (0.03) & 1.01 (0.00) & 1.00 (0.00) & 1.09 (0.06) & 1.00 (0.00)\\
     June 1- Aug 4 & 1.01 (0.00) & 1.00 (0.00) & 1.06 (0.05) & 1.00 (0.00) & 0.99 (0.03) & 1.00 (0.00) & 1.00 (0.00) & 1.04 (0.05) & 1.00 (0.00)\\
     June 1- Aug 11 & 1.01 (0.00) & 1.00 (0.00) & 1.05 (0.04) & 1.00 (0.00) & 0.98 (0.03) & 1.00 (0.00) & 1.01 (0.00) & 1.03 (0.04) & 1.00 (0.00)\\
     June 1-Aug 18 & 1.01 (0.00) & 1.00 (0.00) & 1.05 (0.03) & 1.00 (0.00) & 0.98 (0.03) & 1.00 (0.00) & 1.00 (0.00) & 1.04 (0.04) & 1.00 (0.00)\\
     June 1 - Aug 25 & 1.01 (0.00) & 1.00 (0.00) & 1.06 (0.03) & 1.00 (0.00) & 0.98 (0.03) & 1.00 (0.00) & 1.00 (0.00) & 1.04 (0.03) & 1.00 (0.00)\\
     June 1- Sep 1 & 1.01 (0.00) & 1.00 (0.00) & 1.06 (0.03) & 1.00 (0.00) & 0.98 (0.03) & 1.00 (0.00) & 1.00 (0.00) & 1.04 (0.03) & 1.00 (0.00)\\
     June 1- Sep 8 & 1.01 (0.00) & 1.00 (0.00) & 1.06 (0.03) & 1.00 (0.00) & 0.97 (0.03) & 1.00 (0.00) & 1.00 (0.00) & 1.05 (0.03) & 1.00 (0.00)\\
     June 1- Sep 15 & 1.01 (0.00) & 1.00 (0.00) & 1.07 (0.03) & 1.00 (0.00) & 0.98 (0.03) & 1.00 (0.00) & 1.00 (0.00) & 1.05 (0.03) & 1.00 (0.00)\\
     June 1- Sep 22 & 1.01 (0.00) & 1.00 (0.00) & 1.07 (0.03) & 1.00 (0.00) & 0.98 (0.03) & 1.00 (0.00) & 1.00 (0.00) & 1.05 (0.03) & 1.00 (0.00)\\
     June 1- Sep 29 & 1.01 (0.00) & 1.00 (0.00) & 1.07 (0.03) & 1.00 (0.00) & 0.97 (0.03) & 1.00 (0.00) & 1.00 (0.00) & 1.05 (0.03) & 1.00 (0.00)\\
     June 1 - Oct 6 & 1.01 (0.00) & 1.00 (0.00) & 1.06 (0.03) & 1.00 (0.00) & 0.98 (0.03) & 1.00 (0.00) & 1.00 (0.00) & 1.04 (0.03) & 1.00 (0.00)\\
     June 1- Oct 13 & 1.01 (0.00) & 1.00 (0.00) & 1.06 (0.03) & 1.00 (0.00) & 0.98 (0.03) & 1.00 (0.00) & 1.00 (0.00) & 1.05 (0.03) & 1.00 (0.00)\\
     June 1 - Oct 20 & 1.01 (0.00) & 1.00 (0.00) & 1.06 (0.03) & 1.00 (0.00) & 0.98 (0.03) & 1.00 (0.00) & 1.00 (0.00) & 1.04 (0.03) & 1.00 (0.00)\\
     June 1 - Oct 27 & 1.01 (0.00) & 1.00 (0.00) & 1.07 (0.03) & 1.00 (0.00) & 0.98 (0.03) & 1.00 (0.00) & 1.00 (0.00) & 1.04 (0.03) & 1.00 (0.00)\\
   \hline
     \end{tabular}}
     \caption{Estimates of rescaled parameters involved in fire spread from neighbours. Each cell gives the multiplicative change in the rate, with its standard error in parentheses; 1.00 denotes no effect. For all variables except vegetation cover and soil moisture the change is the mean absolute daily change; for vegetation cover it is 1/100th, for soil moisture 1 kg/m$^2$.}
     \label{tab:spread_scaled_california}
 \end{table}

\begin{table}[H]
     \centering
     \resizebox{\textwidth}{!}{
     \begin{tabular}{|c|ccccccc|}
     \hline

      Time & Temperature & Pressure & Precipitation & Veg cover & Soil moisture & Rel hum & Altitude\\ \hline
     June 1-July 28 & 0.85 (0.05) & 1.01 (0.00) & 0.96 (0.07) & 1.00 (0.00) & 1.00 (0.00) & 0.94 (0.06) & 1.00 (0.00)\\
     June 1- Aug 4 & 0.93 (0.04) & 1.00 (0.00) & 0.96 (0.06) & 0.99 (0.00) & 1.00 (0.00) & 0.96 (0.04) & 1.00 (0.00)\\
     June 1- Aug 11 & 0.93 (0.03) & 1.00 (0.00) & 0.96 (0.06) & 1.00 (0.00) & 1.00 (0.00) & 0.97 (0.04) & 1.00 (0.00)\\
     June 1-Aug 18 & 0.94 (0.03) & 1.00 (0.00) & 0.96 (0.06) & 1.00 (0.00) & 1.00 (0.00) & 0.96 (0.03) & 1.00 (0.00)\\
     June 1 - Aug 25 & 0.95 (0.03) & 1.00 (0.00) & 0.94 (0.07) & 1.00 (0.00) & 1.00 (0.00) & 0.98 (0.03) & 1.00 (0.00)\\
     June 1- Sep 1 & 0.97 (0.03) & 1.00 (0.00) & 0.94 (0.07) & 1.00 (0.00) & 1.00 (0.00) & 0.98 (0.03) & 1.00 (0.00)\\
     June 1- Sep 8 & 0.98 (0.03) & 1.00 (0.00) & 0.93 (0.07) & 1.00 (0.00) & 1.00 (0.00) & 1.00 (0.03) & 1.00 (0.00)\\
     June 1- Sep 15 & 0.99 (0.03) & 1.00 (0.00) & 0.92 (0.07) & 1.00 (0.00) & 1.00 (0.00) & 1.02 (0.03) & 1.00 (0.00)\\
     June 1- Sep 22 & 0.98 (0.03) & 1.00 (0.00) & 0.92 (0.07) & 1.00 (0.00) & 1.00 (0.00) & 1.01 (0.03) & 1.00 (0.00)\\
     June 1- Sep 29 & 0.98 (0.02) & 1.00 (0.00) & 0.91 (0.07) & 1.00 (0.00) & 1.00 (0.00) & 1.01 (0.03) & 1.00 (0.00)\\
     June 1 - Oct 6 & 0.97 (0.02) & 1.00 (0.00) & 1.06 (0.03) & 1.00 (0.00) & 1.00 (0.00) & 1.01 (0.03) & 1.00 (0.00)\\
     June 1- Oct 13 & 0.97 (0.02) & 1.00 (0.00) & 1.06 (0.03) & 1.00 (0.00) & 1.00 (0.00) & 1.00 (0.02) & 1.00 (0.00)\\
     June 1 - Oct 20 & 0.96 (0.02) & 1.00 (0.00) & 1.06 (0.03) & 1.00 (0.00) & 1.00 (0.00) & 0.99 (0.02) & 1.00 (0.00)\\
     June 1 - Oct 27 & 0.95 (0.02) & 1.00 (0.00) & 1.05 (0.03) & 1.00 (0.00) & 1.00 (0.00) & 1.00 (0.02) & 1.00 (0.00)\\
   \hline
     \end{tabular}}
     \caption{Estimates of rescaled parameters involved in recovery from fire. Each cell gives the multiplicative change in the rate, with its standard error in parentheses; 1.00 denotes no effect. For all variables except vegetation cover and soil moisture the change is the mean absolute daily change; for vegetation cover it is 1/100th, for soil moisture 1 kg/m$^2$.}
     \label{tab:recovery_scaled_california}
 \end{table}

\begin{table}[H]
     \centering
     \resizebox{\textwidth}{!}{
     \begin{tabular}{|c|ccccccc|}
     \hline

      Time & Temperature & Pressure & Precipitation & Veg cover & Soil moisture & Rel hum & Altitude\\ \hline
     June 1-July 28 & 0.92 (0.07) & 1.01 (0.01) & 1.02 (0.03) & 1.00 (0.01) & 1.00 (0.00) & 0.95 (0.07) & 1.00 (0.00)\\
     June 1- Aug 4 & 0.91 (0.06) & 1.01 (0.00) & 1.02 (0.03) & 1.00 (0.01) & 1.00 (0.00) & 0.95 (0.06) & 1.00 (0.00)\\
     June 1- Aug 11 & 0.90 (0.06) & 1.01 (0.00) & 1.02 (0.03) & 1.00 (0.01) & 1.00 (0.00) & 0.94 (0.06) & 1.00 (0.00)\\
     June 1-Aug 18 & 0.94 (0.06) & 1.01 (0.00) & 1.02 (0.03) & 1.00 (0.01) & 1.00 (0.00) & 0.96 (0.06) & 1.00 (0.00)\\
     June 1 - Aug 25 & 0.94 (0.06) & 1.00 (0.00) & 1.02 (0.03) & 1.00 (0.01) & 1.00 (0.00) & 0.97 (0.06) & 1.00 (0.00)\\
     June 1- Sep 1 & 0.96 (0.06) & 1.00 (0.00) & 1.01 (0.03) & 0.99 (0.01) & 1.00 (0.00) & 0.99 (0.06) & 1.00 (0.00)\\
     June 1- Sep 8 & 0.97 (0.06) & 1.00 (0.00) & 1.01 (0.03) & 0.99 (0.01) & 1.00 (0.00) & 1.01 (0.06) & 1.00 (0.00)\\
     June 1- Sep 15 & 0.98 (0.06) & 1.00 (0.00) & 1.01 (0.03) & 0.99 (0.00) & 1.00 (0.00) & 1.02 (0.05) & 1.00 (0.00)\\
     June 1- Sep 22 & 0.99 (0.05) & 1.00 (0.00) & 1.01 (0.03) & 1.00 (0.00) & 1.00 (0.00) & 1.03 (0.05) & 1.00 (0.00)\\
     June 1- Sep 29 & 0.99 (0.05) & 1.00 (0.00) & 1.01 (0.04) & 1.00 (0.00) & 1.00 (0.00) & 1.02 (0.05) & 1.00 (0.00)\\
     June 1 - Oct 6 & 0.99 (0.05) & 1.00 (0.00) & 0.98 (0.03) & 1.00 (0.00) & 1.00 (0.00) & 1.03 (0.05) & 1.00 (0.00)\\
     June 1- Oct 13 & 0.96 (0.05) & 1.00 (0.00) & 0.97 (0.03) & 1.00 (0.00) & 1.00 (0.00) & 1.00 (0.04) & 1.00 (0.00)\\
     June 1 - Oct 20 & 0.95 (0.04) & 1.00 (0.00) & 0.96 (0.04) & 1.00 (0.00) & 1.00 (0.00) & 1.02 (0.04) & 1.00 (0.00)\\
     June 1 - Oct 27 & 0.92 (0.04) & 1.00 (0.00) & 0.95 (0.04) & 1.00 (0.00) & 1.00 (0.00) & 1.00 (0.03) & 1.00 (0.00)\\
   \hline
     \end{tabular}}
     \caption{Estimates of rescaled parameters involved in intrinsic fire ignition. Each cell gives the multiplicative change in the rate, with its standard error in parentheses; 1.00 denotes no effect. For all variables except vegetation cover and soil moisture the change is the mean absolute daily change; for vegetation cover it is 1/100th, for soil moisture 1 kg/m$^2$.}
     \label{tab:intrinsic_scaled_california}
 \end{table}

\newpage

\begin{figure}[H]
     \centering
        \begin{subfigure}[b]{0.49\linewidth}
         \centering
         \includegraphics[scale=0.23]{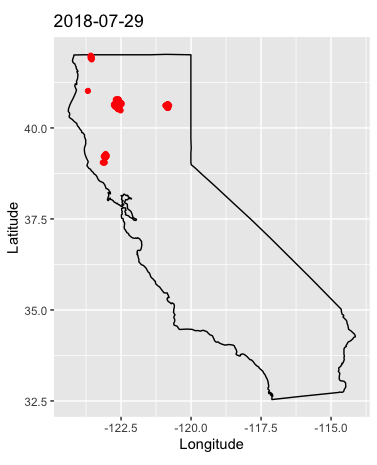}
         \includegraphics[scale=0.27]{Images/cal_pred_jul_29.png}
          \caption{July 29, 2018.}

     \end{subfigure}
       \begin{subfigure}[b]{0.49\linewidth}
         \centering
         \includegraphics[scale=0.23]{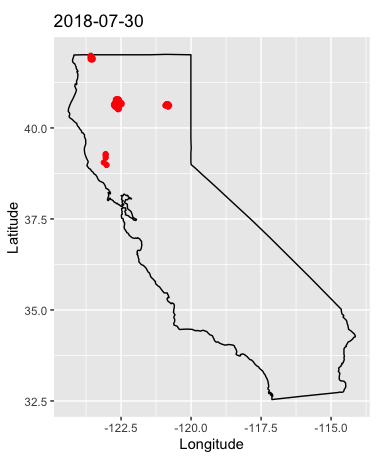}
         \includegraphics[scale=0.27]{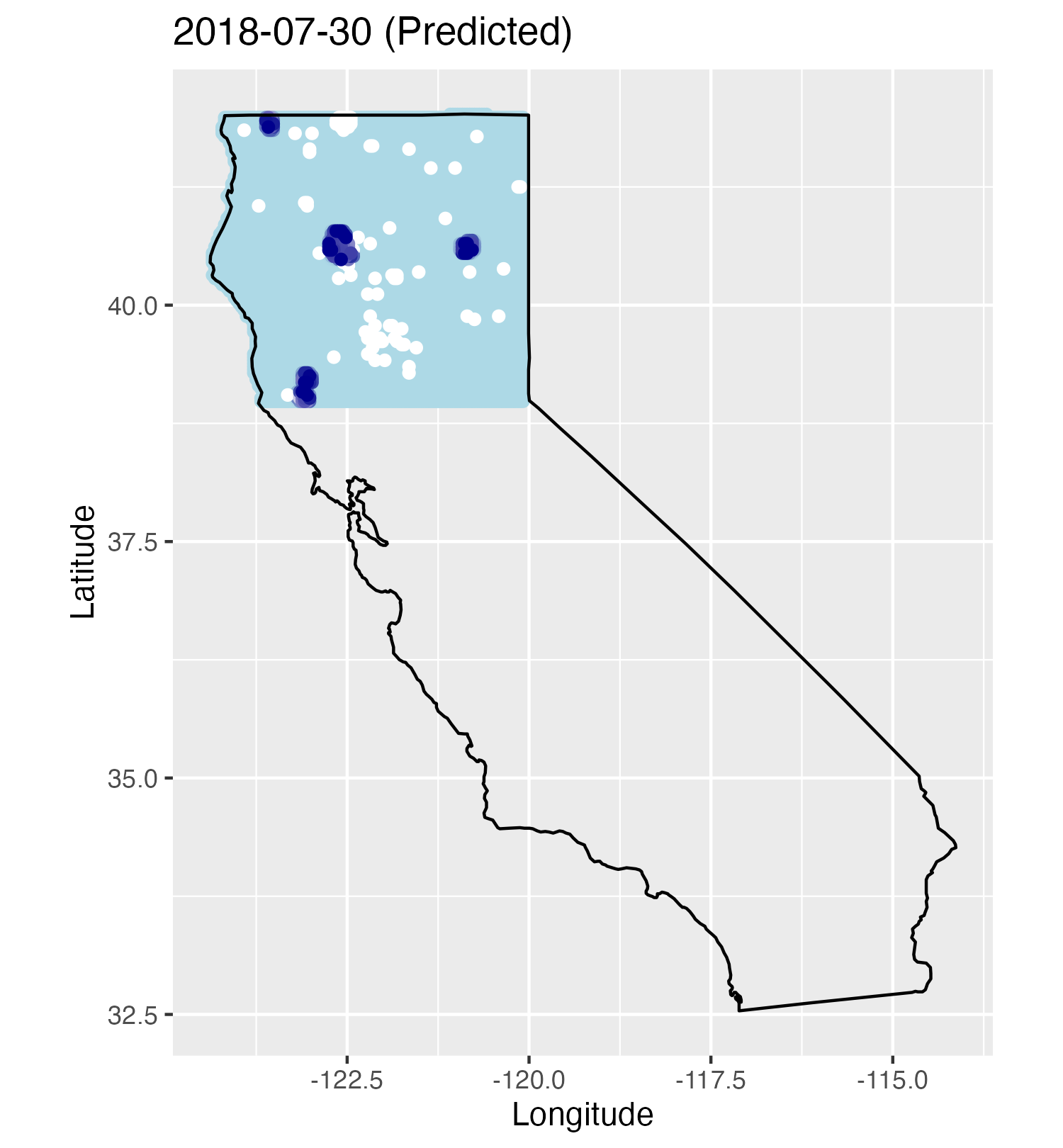}
          \caption{July 30, 2018.}
          \end{subfigure}

       \begin{subfigure}[b]{0.49\linewidth}
         \centering
         \includegraphics[scale=0.23]{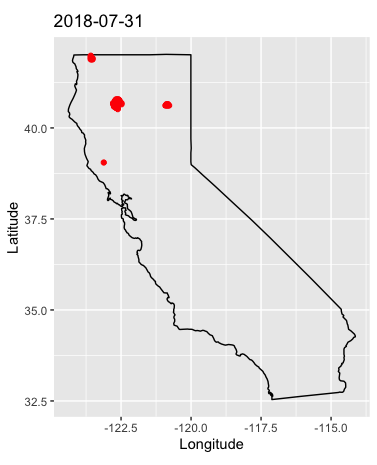}
         \includegraphics[scale=0.27]{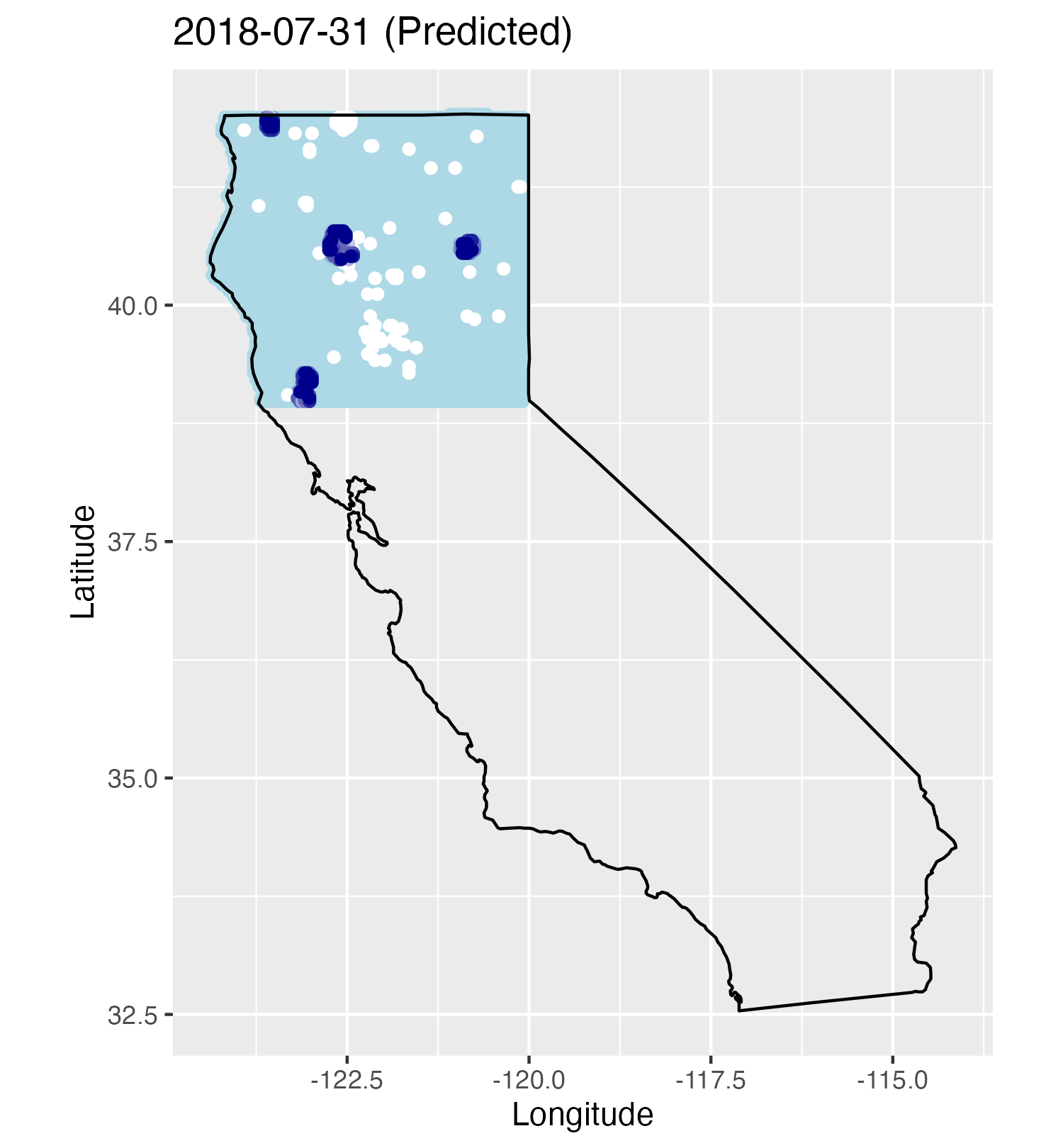}
          \caption{July 31, 2018.}

     \end{subfigure}
       \begin{subfigure}[b]{0.49\linewidth}
         \centering
         \includegraphics[scale=0.23]{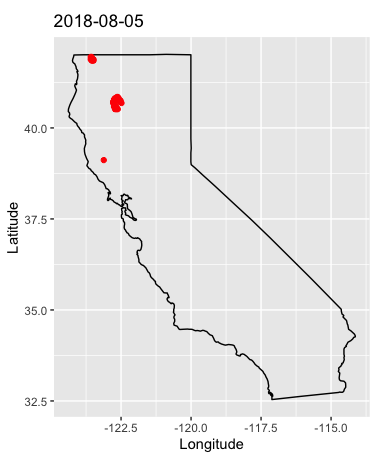}
         \includegraphics[scale=0.27]{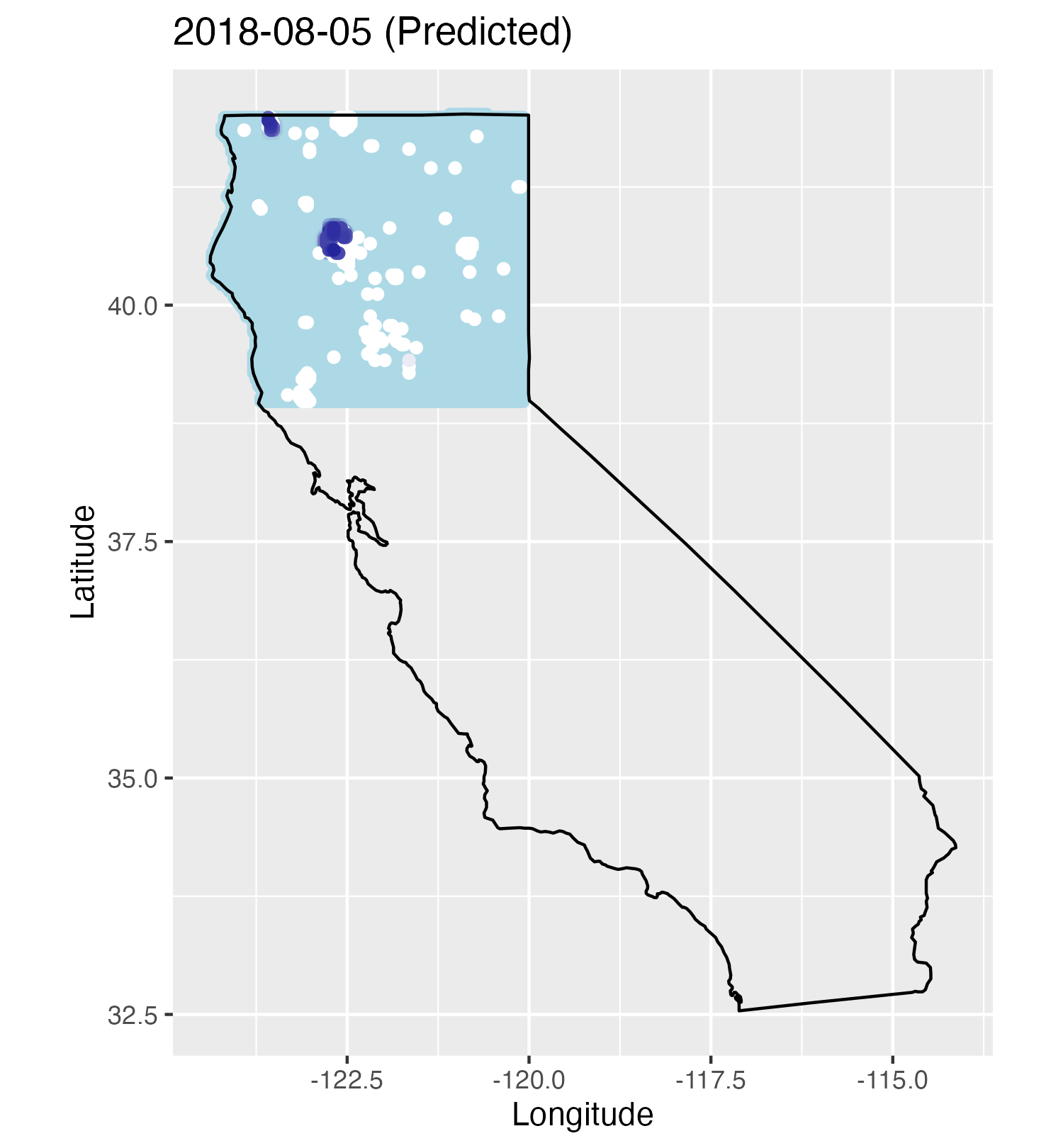}
          \caption{Aug 5, 2018.}
     \end{subfigure}

      \begin{subfigure}[b]{0.49\linewidth}
         \centering
         \includegraphics[scale=0.23]{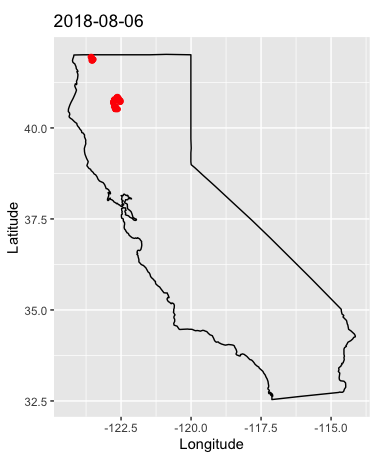}
         \includegraphics[scale=0.27]{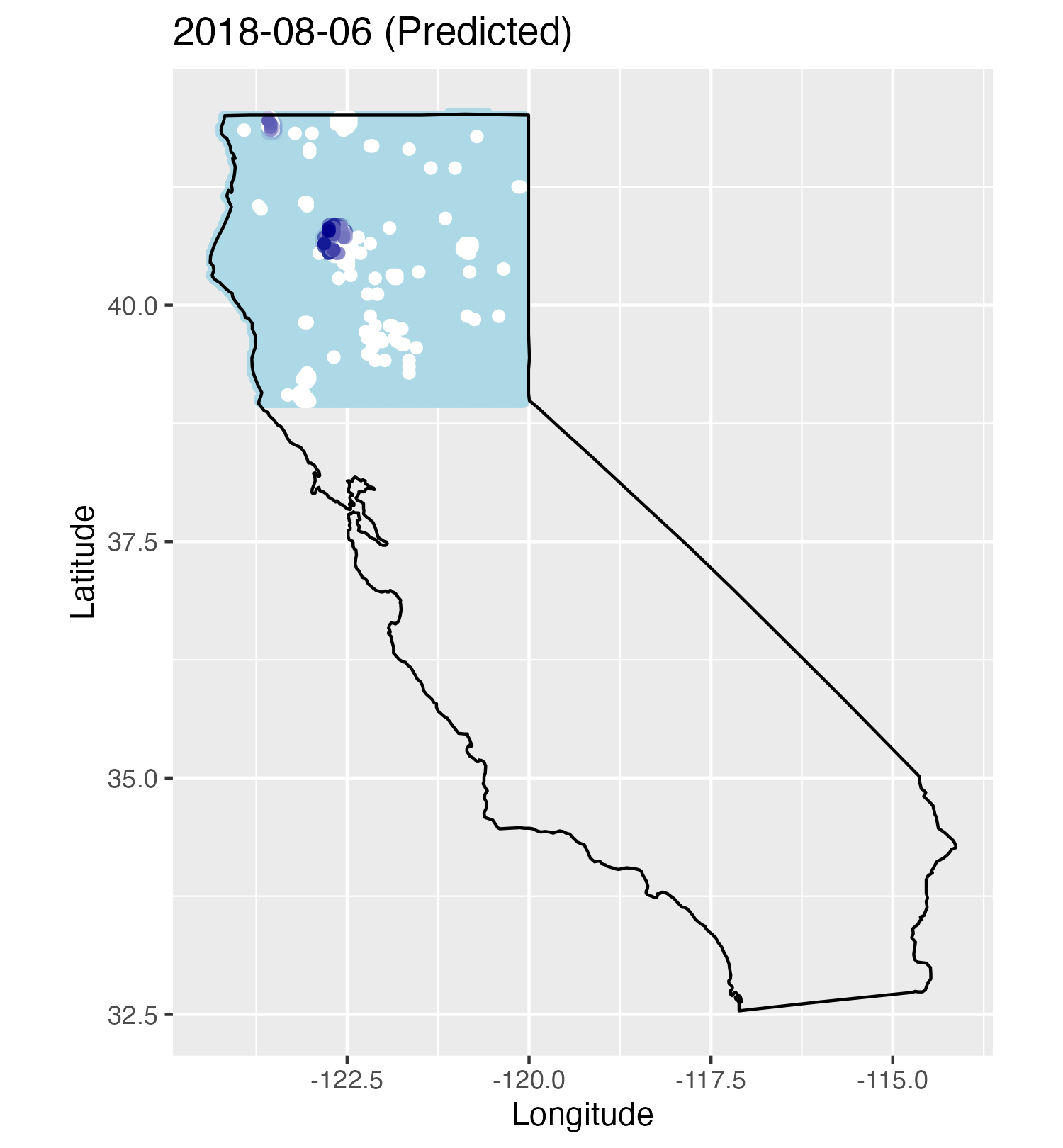}
          \caption{Aug 6, 2018.}

     \end{subfigure}
       \begin{subfigure}[b]{0.49\linewidth}
         \centering
         \includegraphics[scale=0.23]{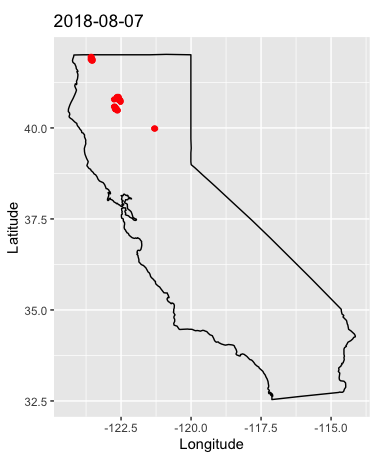}
         \includegraphics[scale=0.27]{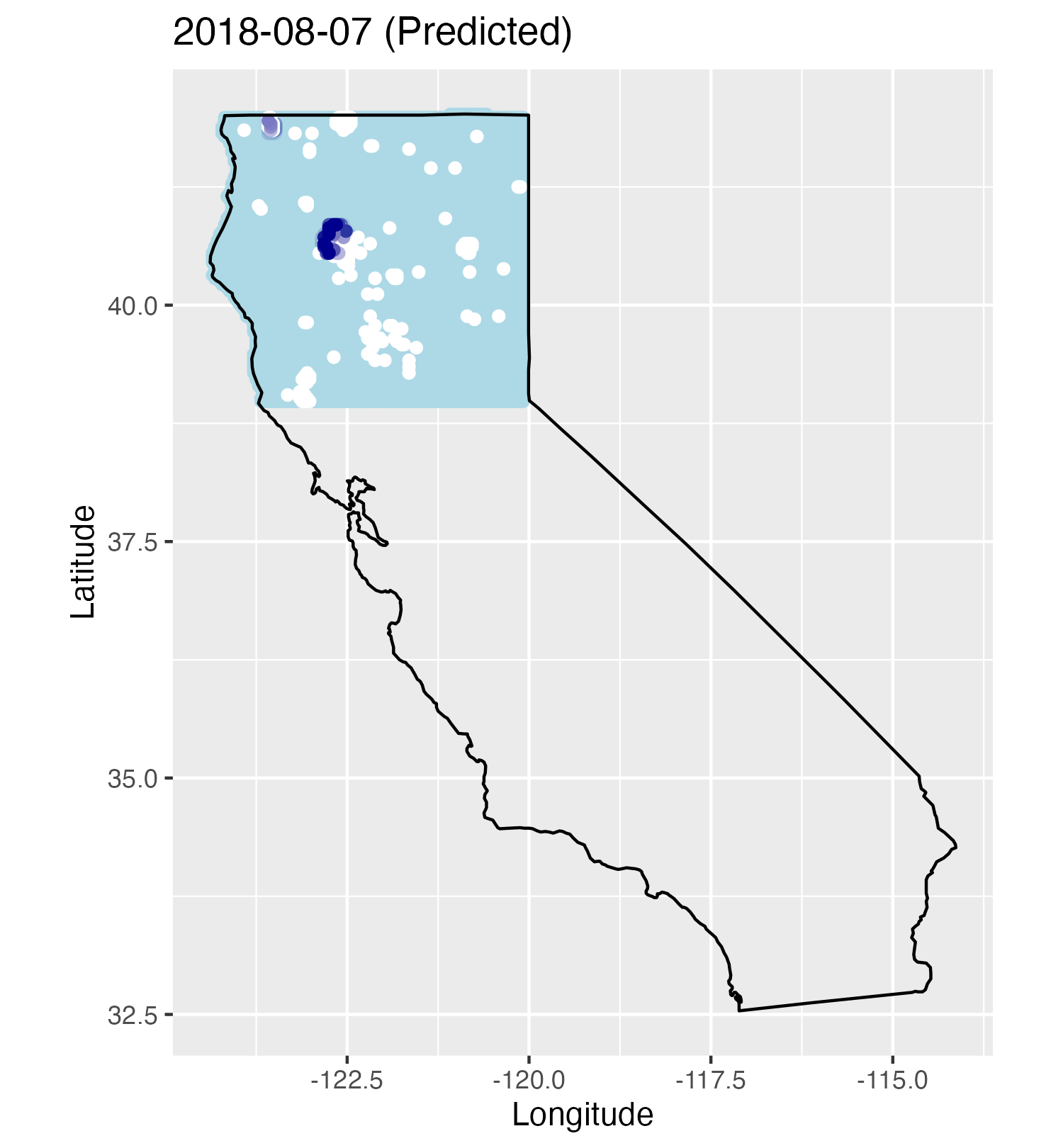}
          \caption{Aug 7, 2018.}
     \end{subfigure}

     \begin{subfigure}[b]{0.49\linewidth}
         \centering
         \includegraphics[scale=0.23]{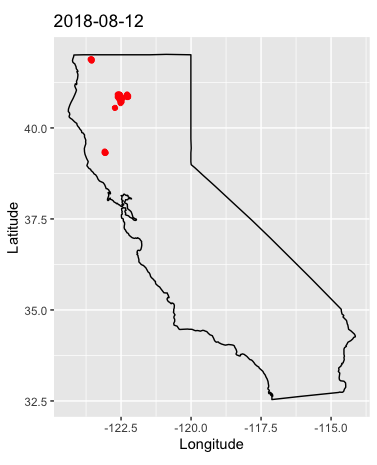}
         \includegraphics[scale=0.27]{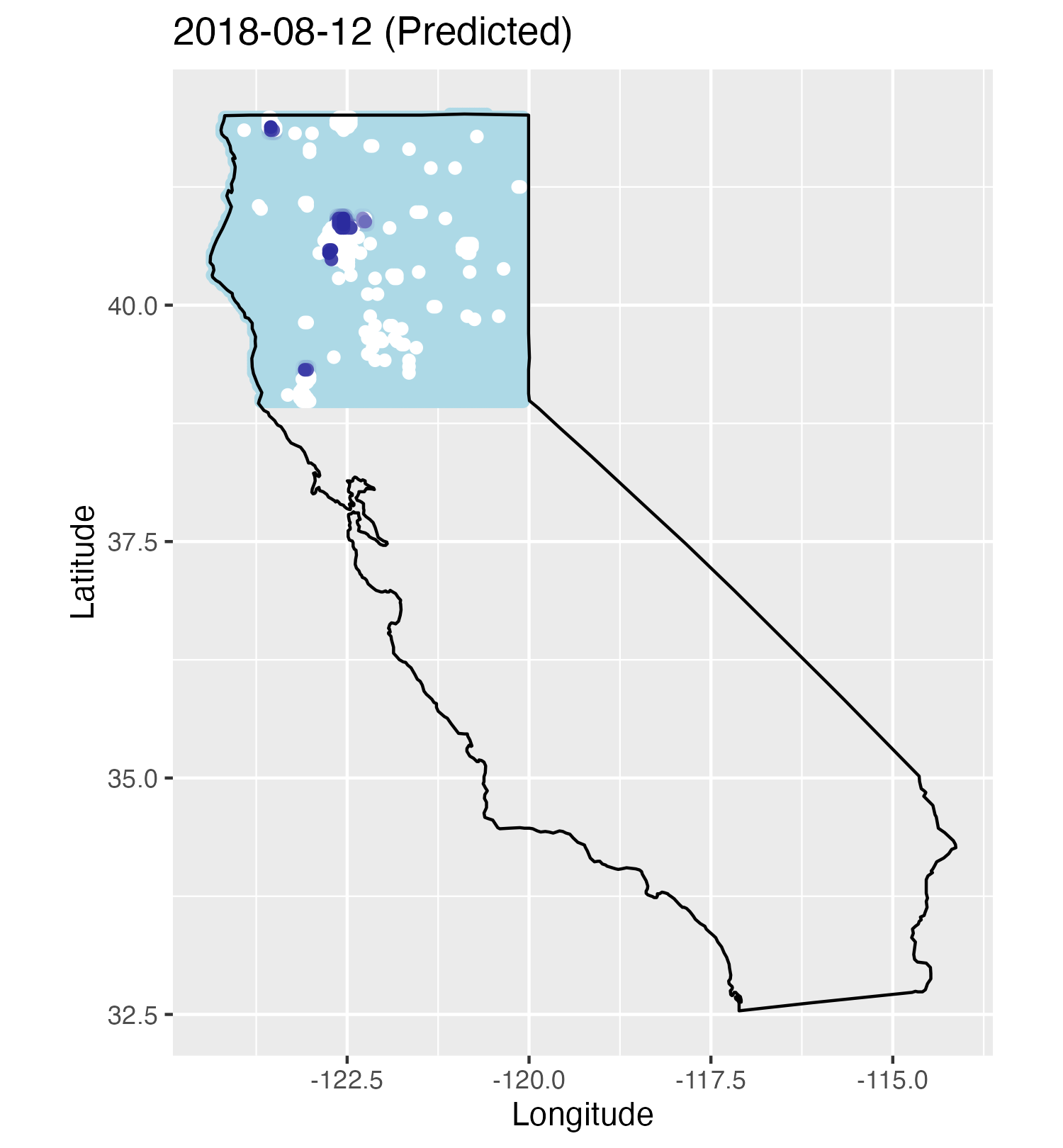}
          \caption{Aug 12, 2018.}

     \end{subfigure}
       \begin{subfigure}[b]{0.49\linewidth}
         \centering
         \includegraphics[scale=0.23]{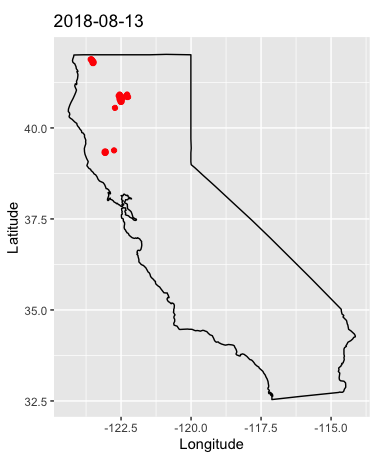}
         \includegraphics[scale=0.27]{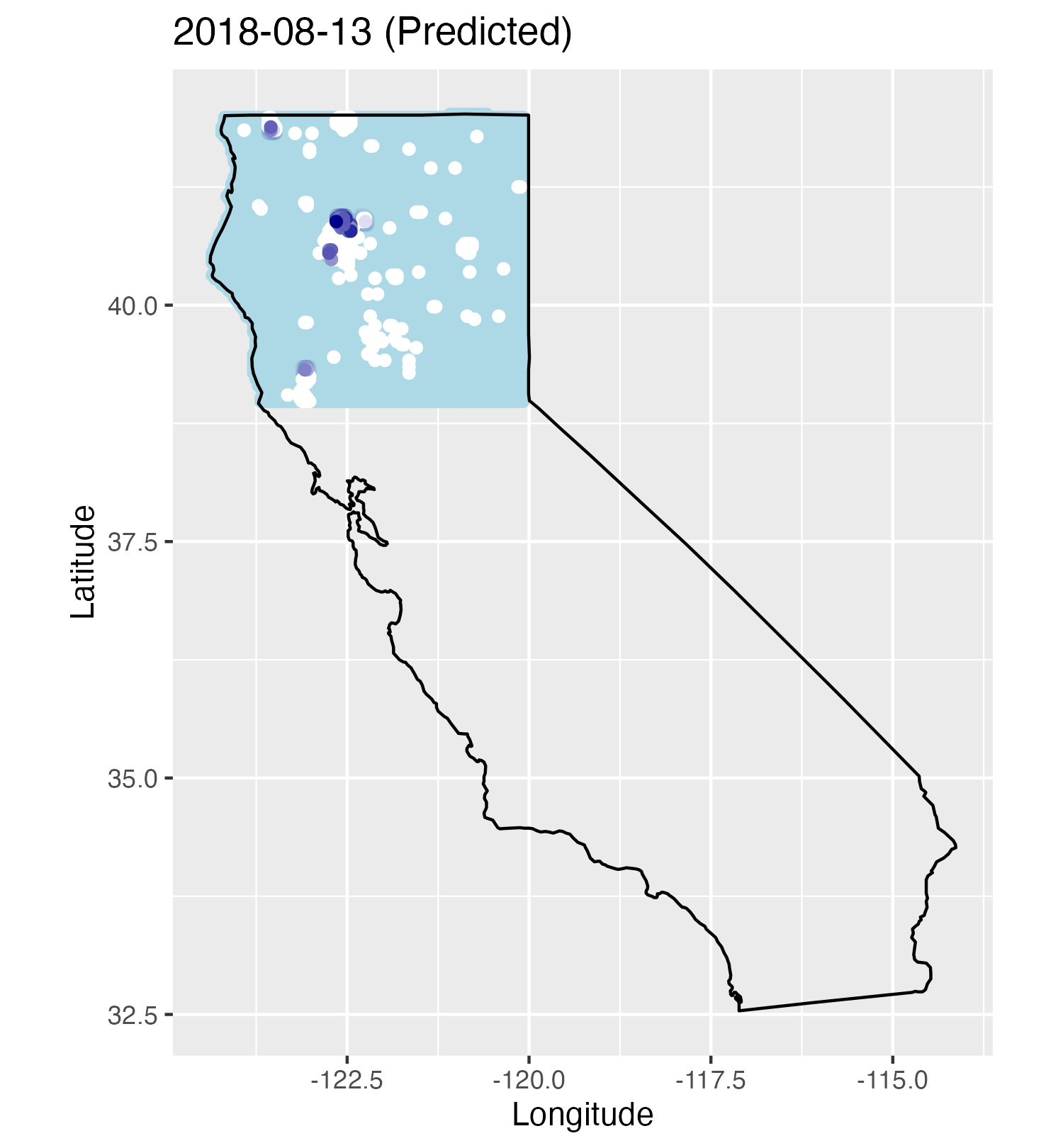}
          \caption{Aug 13, 2018.}
     \end{subfigure}

     \caption{Observed vs Predicted fire }
      \label{fig:cal_obs_vs_pred1}
\end{figure}

\begin{figure}[H]
     \centering
        \begin{subfigure}[b]{0.49\linewidth}
         \centering
         \includegraphics[scale=0.23]{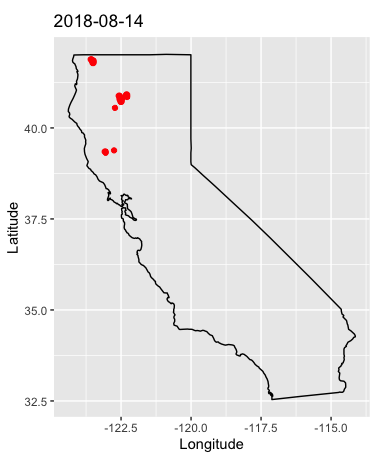}
         \includegraphics[scale=0.27]{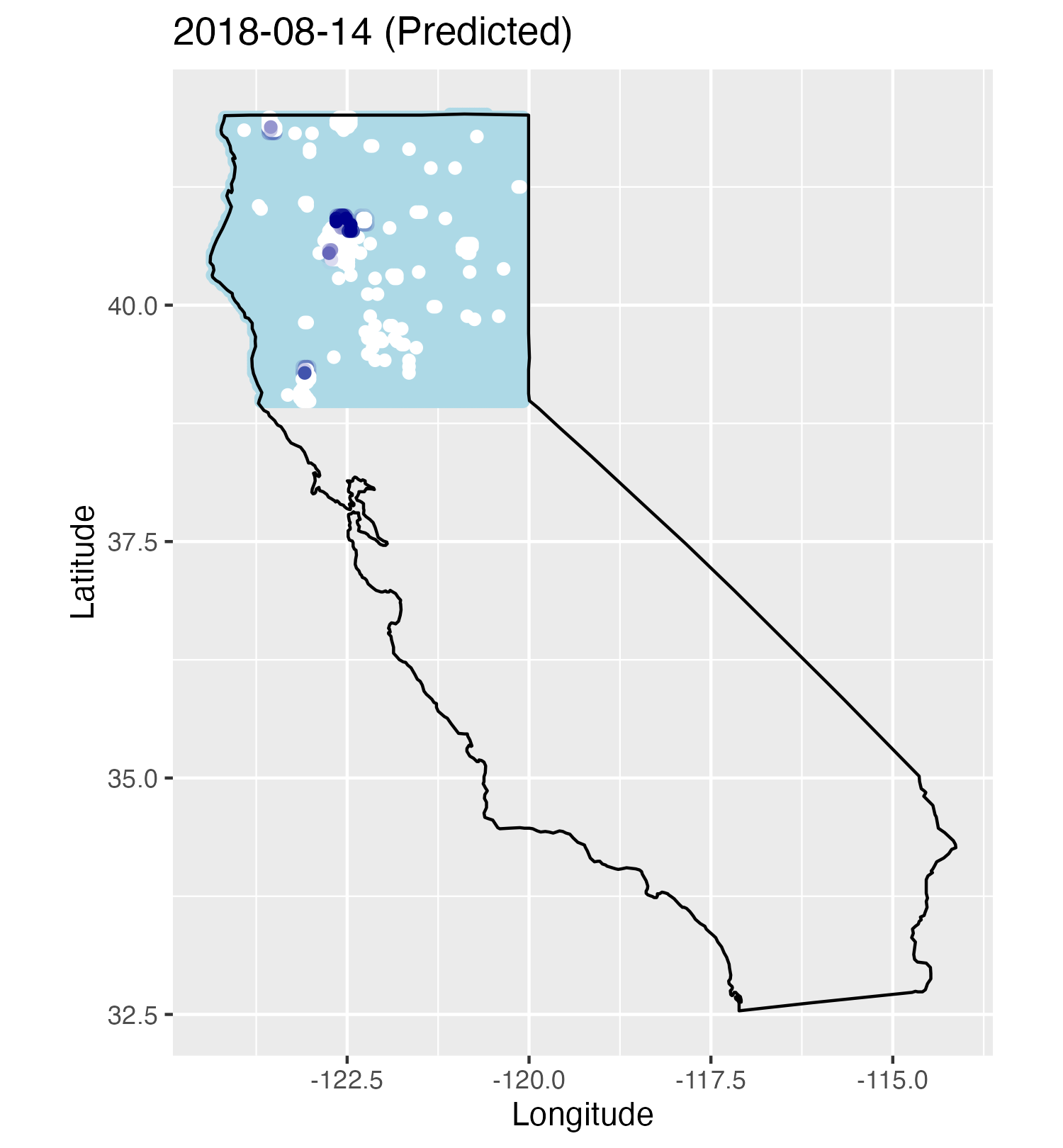}
          \caption{Aug 14, 2018.}

     \end{subfigure}
       \begin{subfigure}[b]{0.49\linewidth}
         \centering
         \includegraphics[scale=0.23]{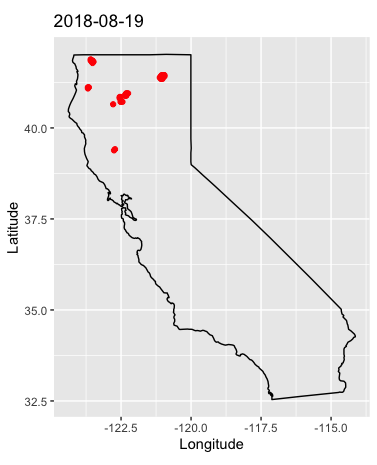}
         \includegraphics[scale=0.27]{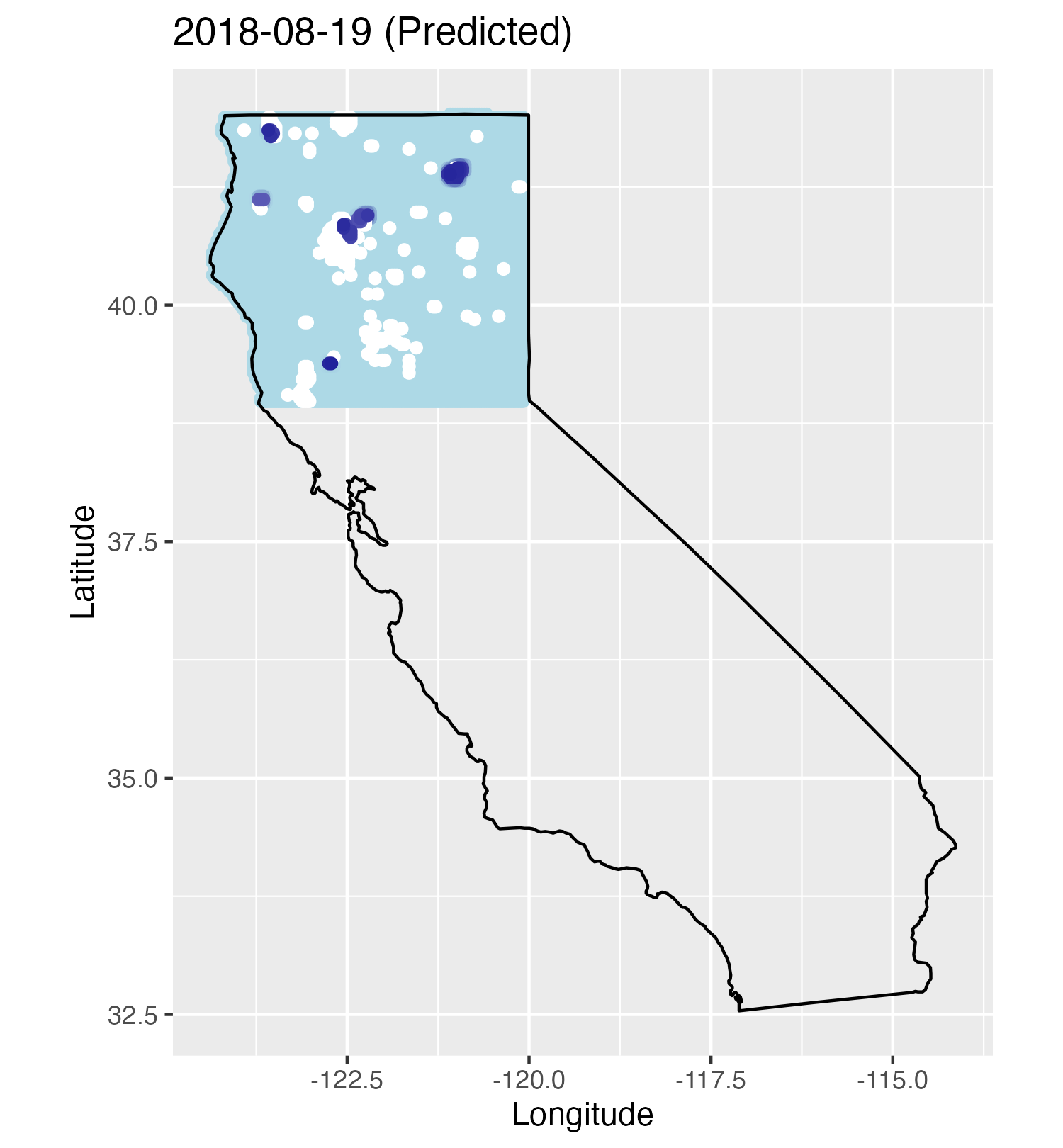}
          \caption{Aug 19, 2018.}
          \end{subfigure}

       \begin{subfigure}[b]{0.49\linewidth}
         \centering
         \includegraphics[scale=0.23]{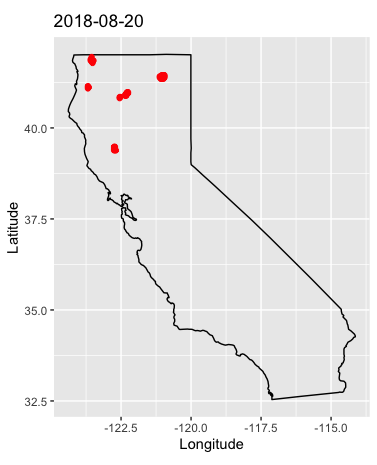}
         \includegraphics[scale=0.27]{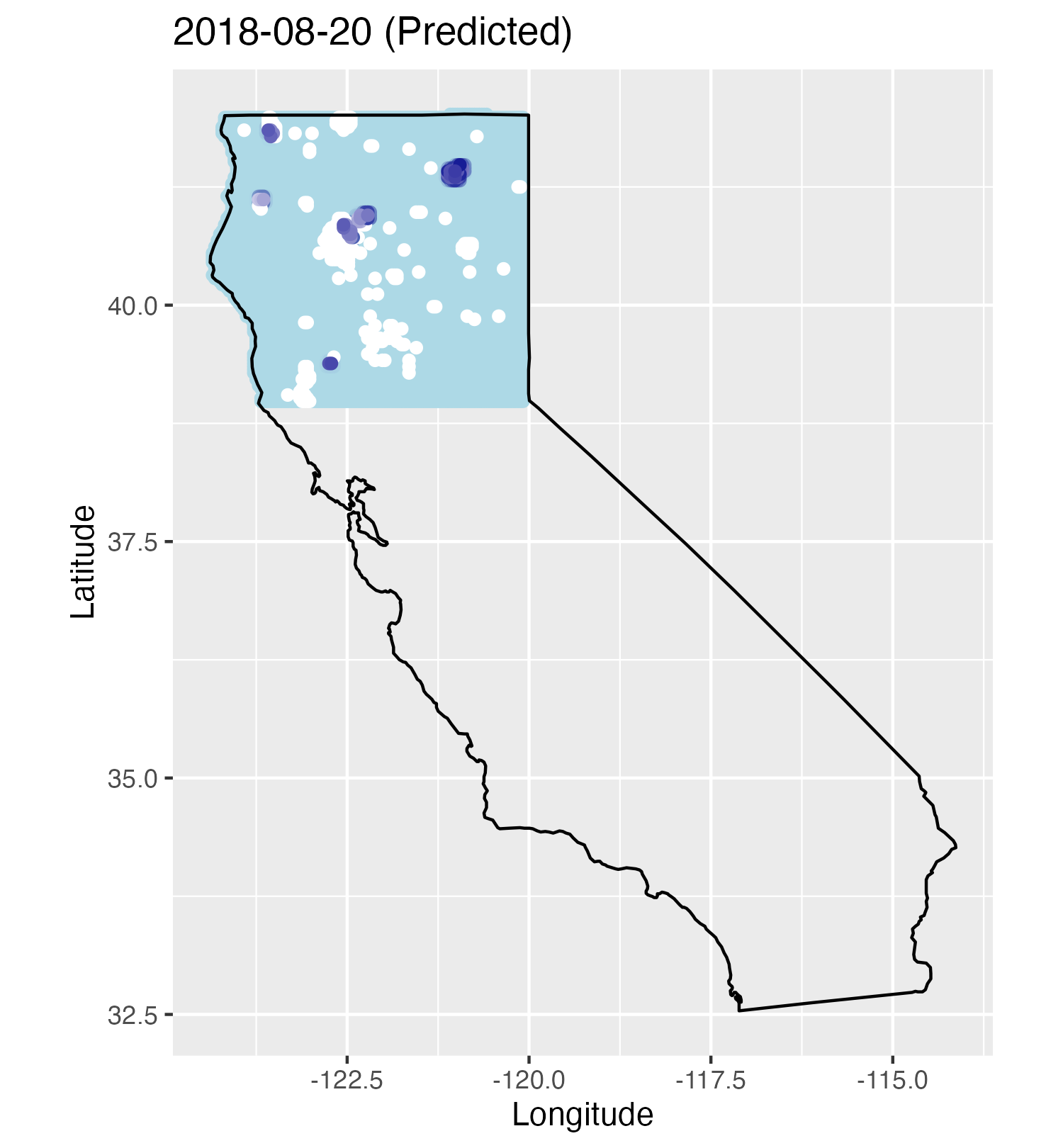}
          \caption{Aug 20, 2018.}

     \end{subfigure}
       \begin{subfigure}[b]{0.49\linewidth}
         \centering
         \includegraphics[scale=0.23]{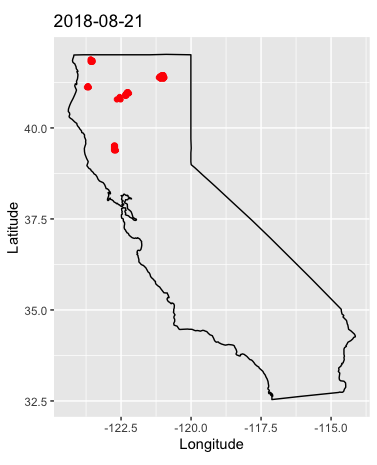}
         \includegraphics[scale=0.27]{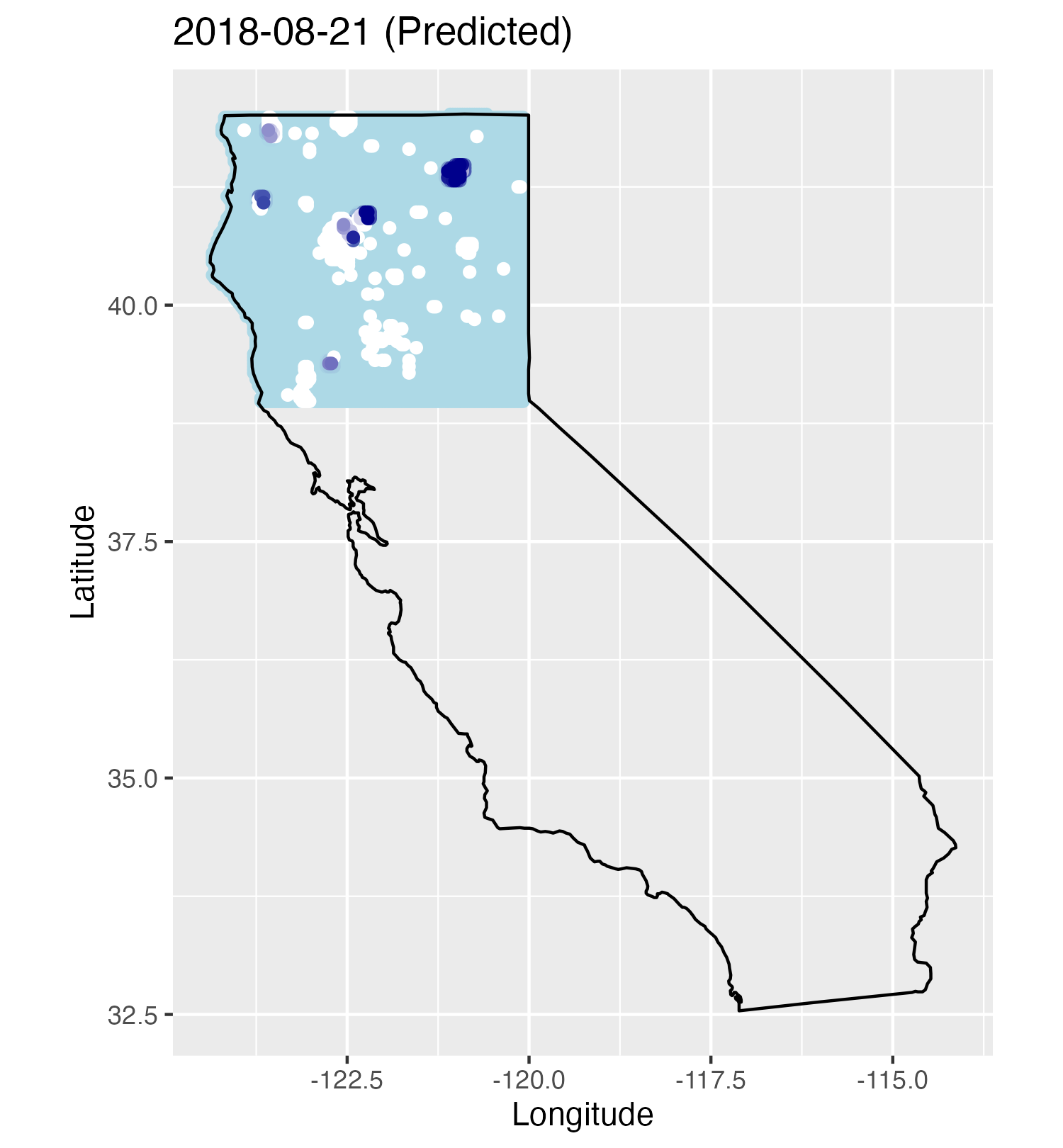}
          \caption{Aug 21, 2018.}
     \end{subfigure}

      \begin{subfigure}[b]{0.49\linewidth}
         \centering
         \includegraphics[scale=0.23]{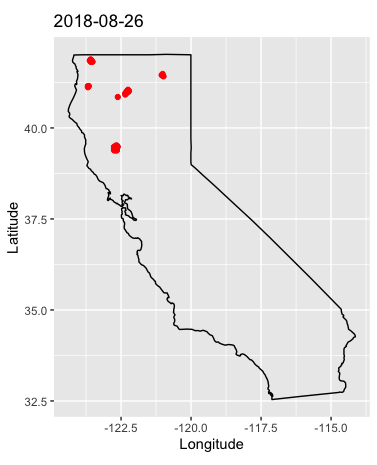}
         \includegraphics[scale=0.27]{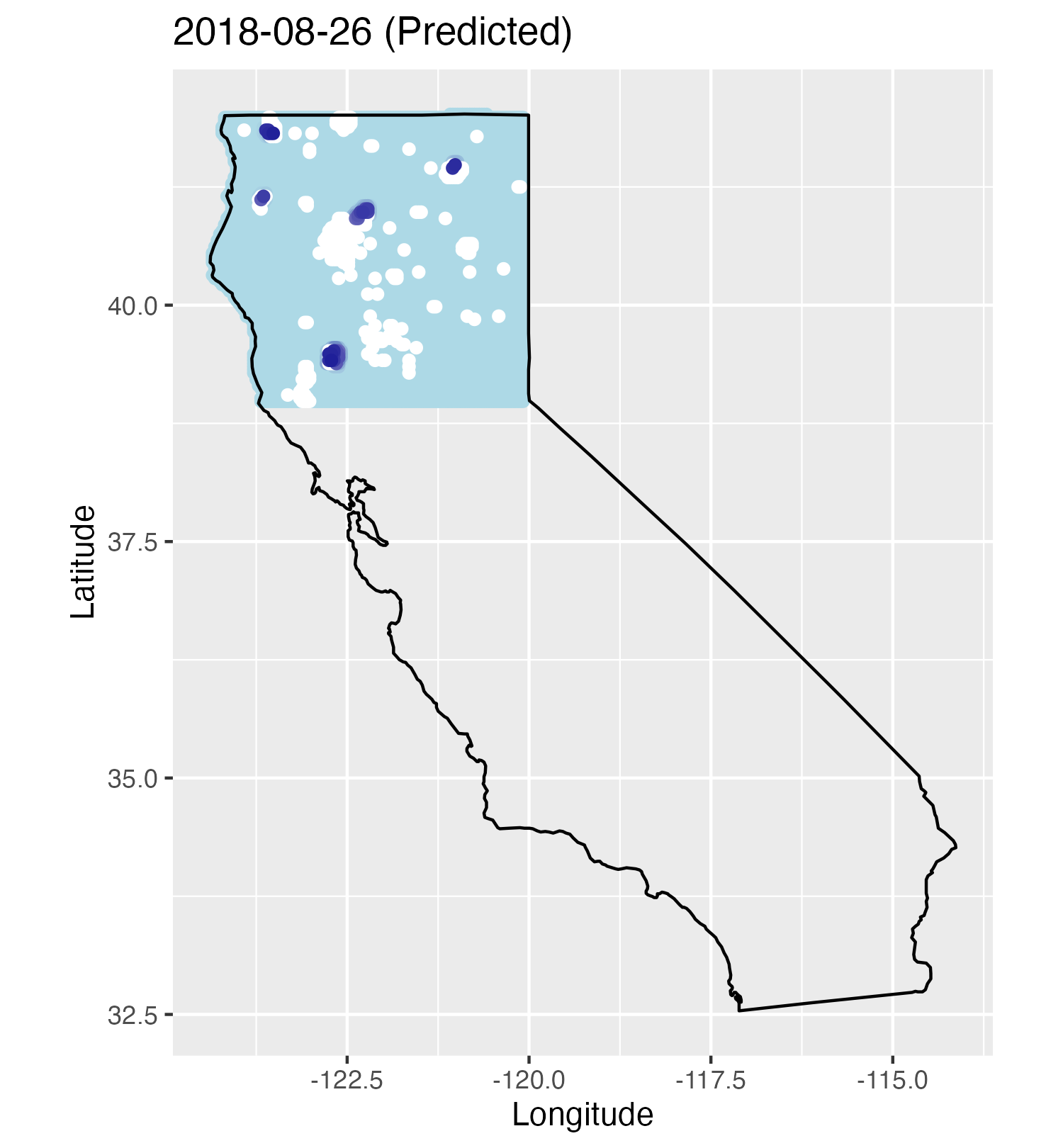}
          \caption{Aug 26, 2018.}

     \end{subfigure}
       \begin{subfigure}[b]{0.49\linewidth}
         \centering
         \includegraphics[scale=0.23]{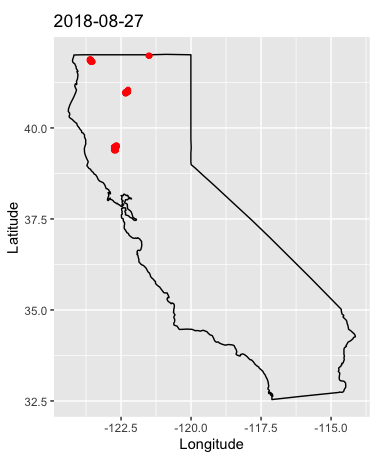}
         \includegraphics[scale=0.27]{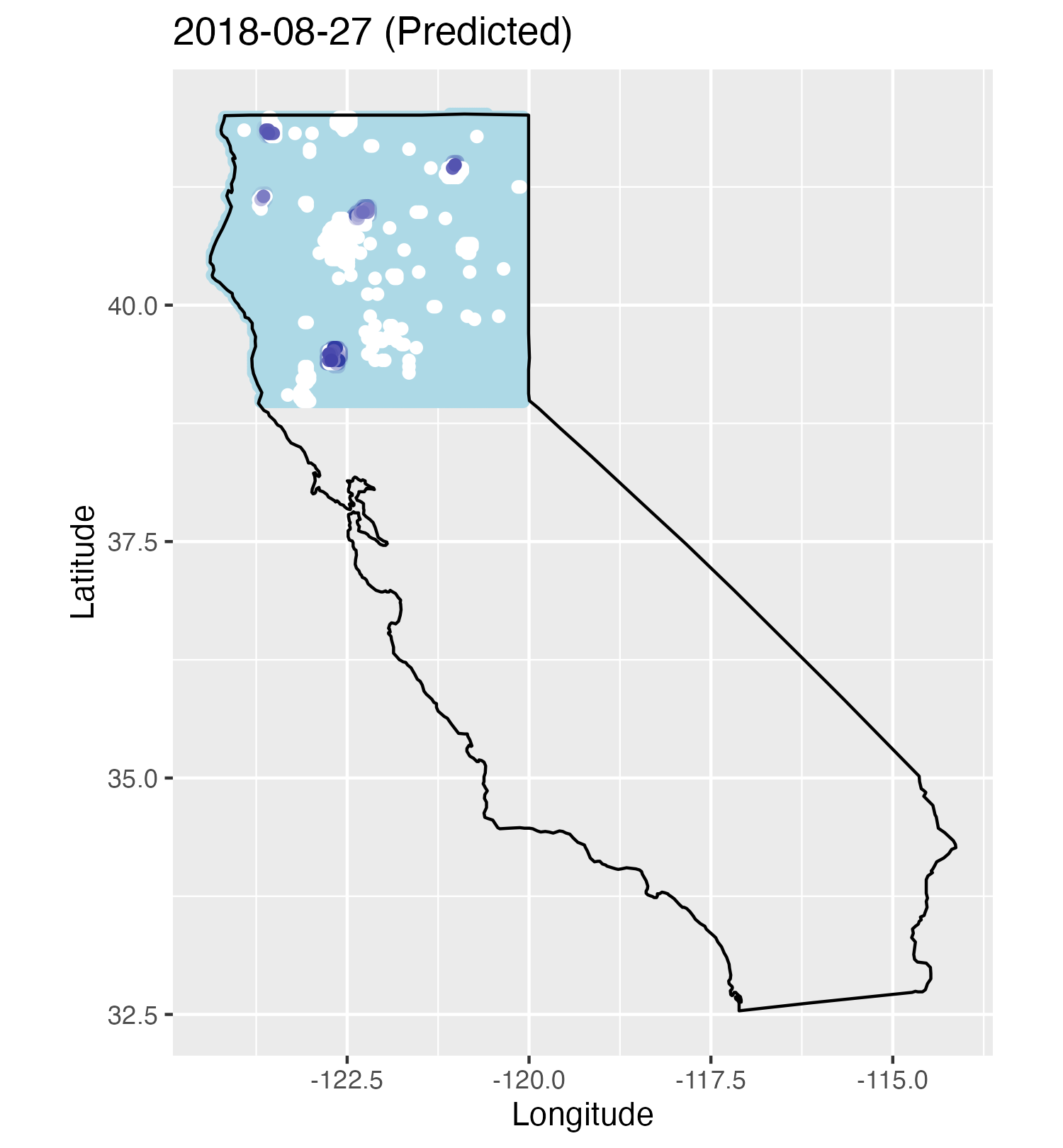}
          \caption{Aug 27, 2018.}
     \end{subfigure}

     \begin{subfigure}[b]{0.49\linewidth}
         \centering
         \includegraphics[scale=0.23]{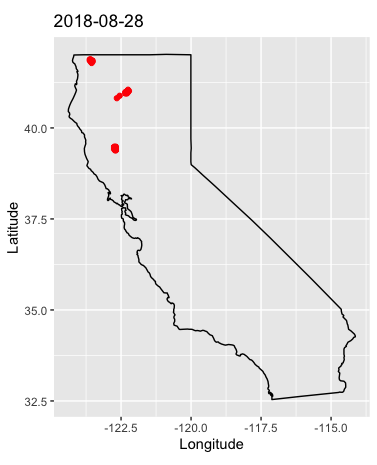}
         \includegraphics[scale=0.27]{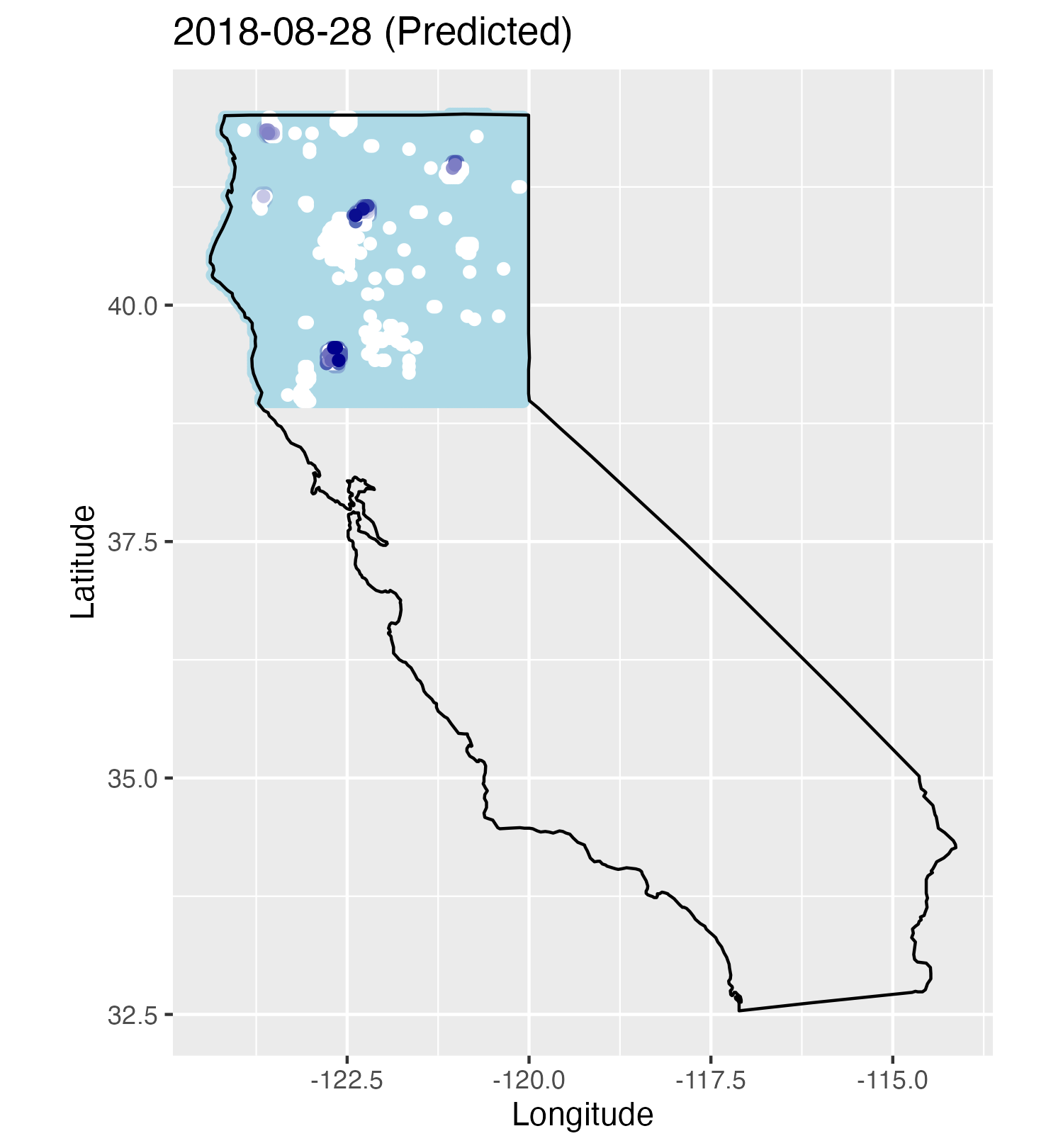}
          \caption{Aug 28, 2018.}

     \end{subfigure}
       \begin{subfigure}[b]{0.49\linewidth}
         \centering
         \includegraphics[scale=0.23]{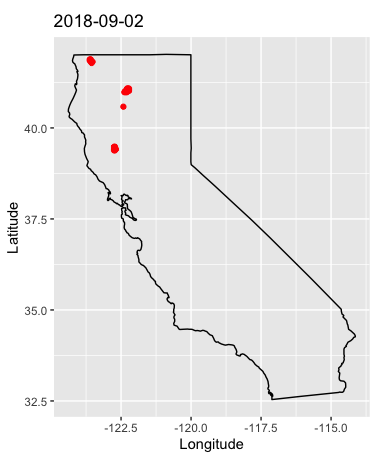}
         \includegraphics[scale=0.27]{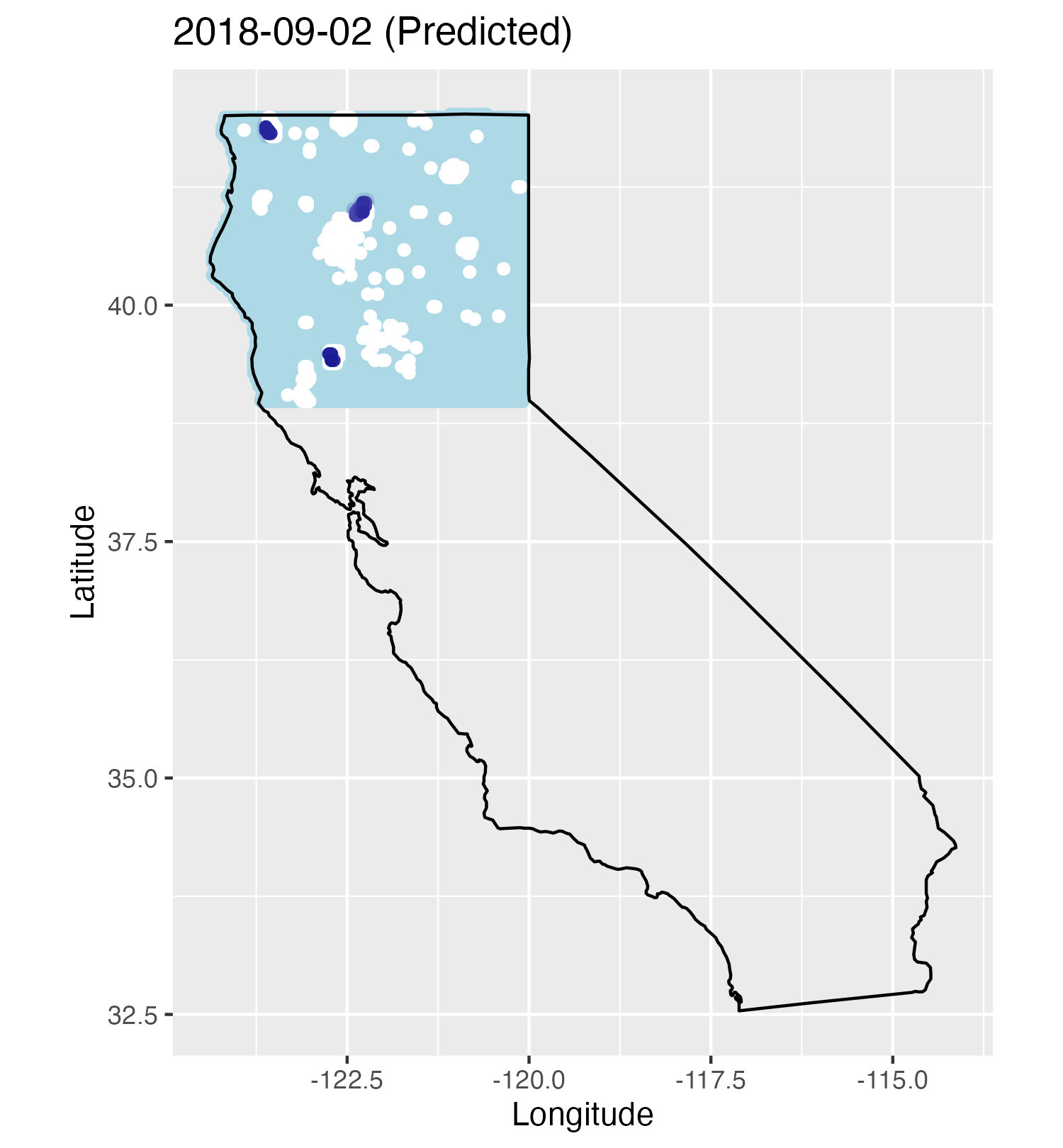}
          \caption{Sep 2, 2018.}
     \end{subfigure}

     \caption{Observed vs Predicted fire }
     \label{fig:cal_obs_vs_pred2}

\end{figure}

\begin{figure}[H]
     \centering
        \begin{subfigure}[b]{0.49\linewidth}
         \centering
         \includegraphics[scale=0.23]{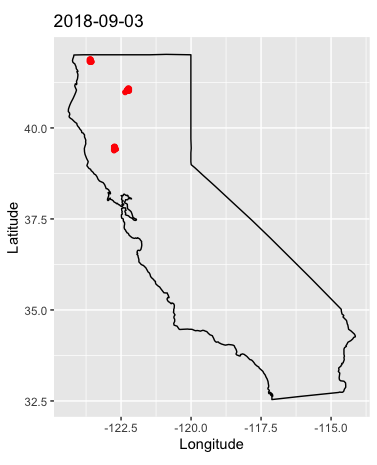}
         \includegraphics[scale=0.27]{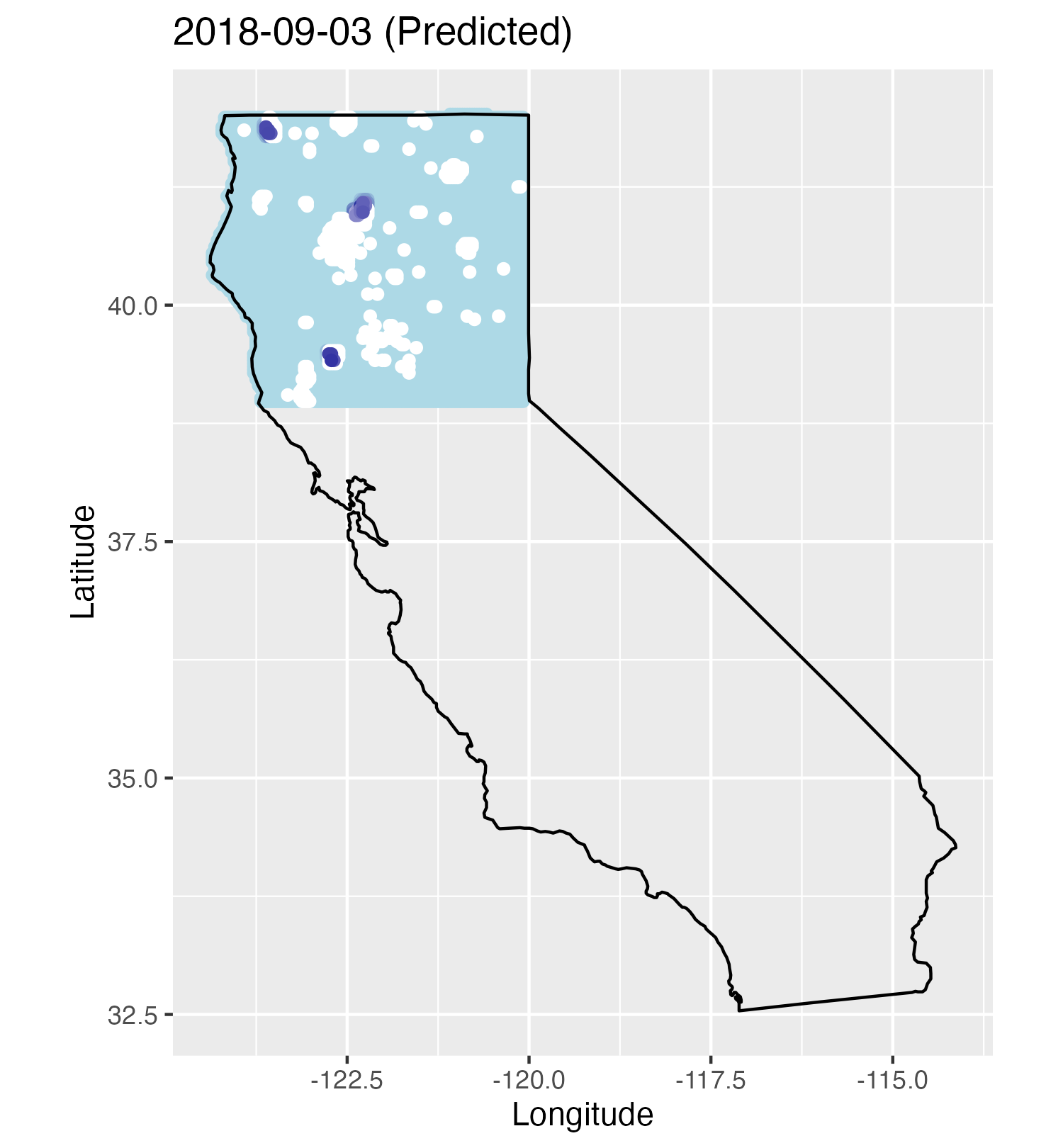}
          \caption{Sep 3, 2018.}

     \end{subfigure}
       \begin{subfigure}[b]{0.49\linewidth}
         \centering
         \includegraphics[scale=0.23]{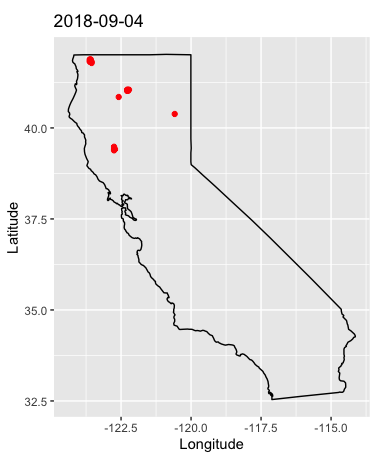}
         \includegraphics[scale=0.27]{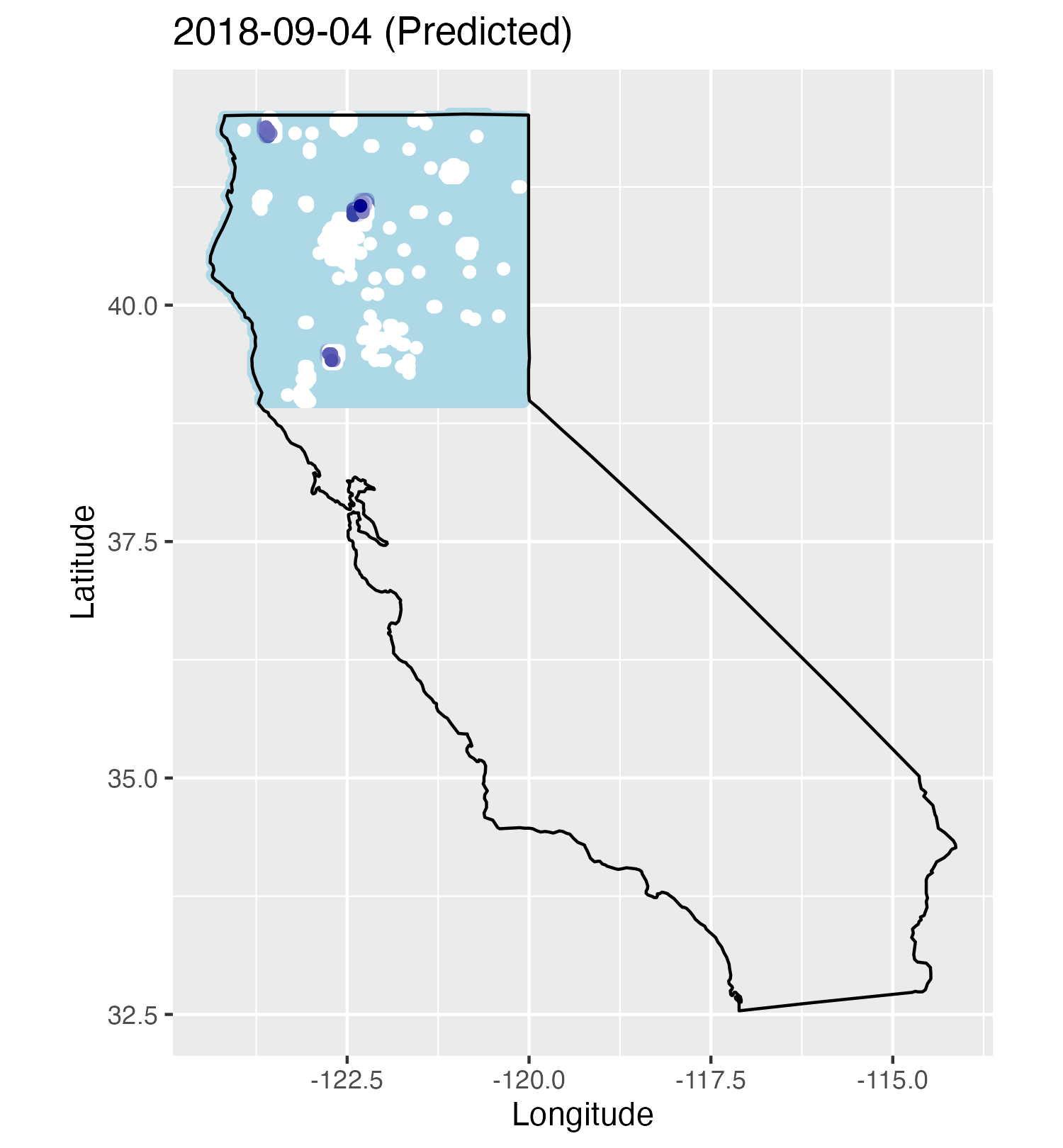}
          \caption{Sep 4, 2018.}
          \end{subfigure}

       \begin{subfigure}[b]{0.49\linewidth}
         \centering
         \includegraphics[scale=0.23]{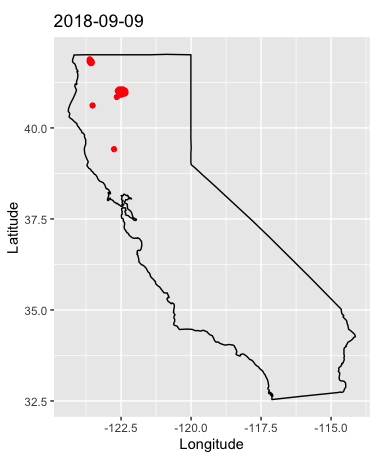}
         \includegraphics[scale=0.27]{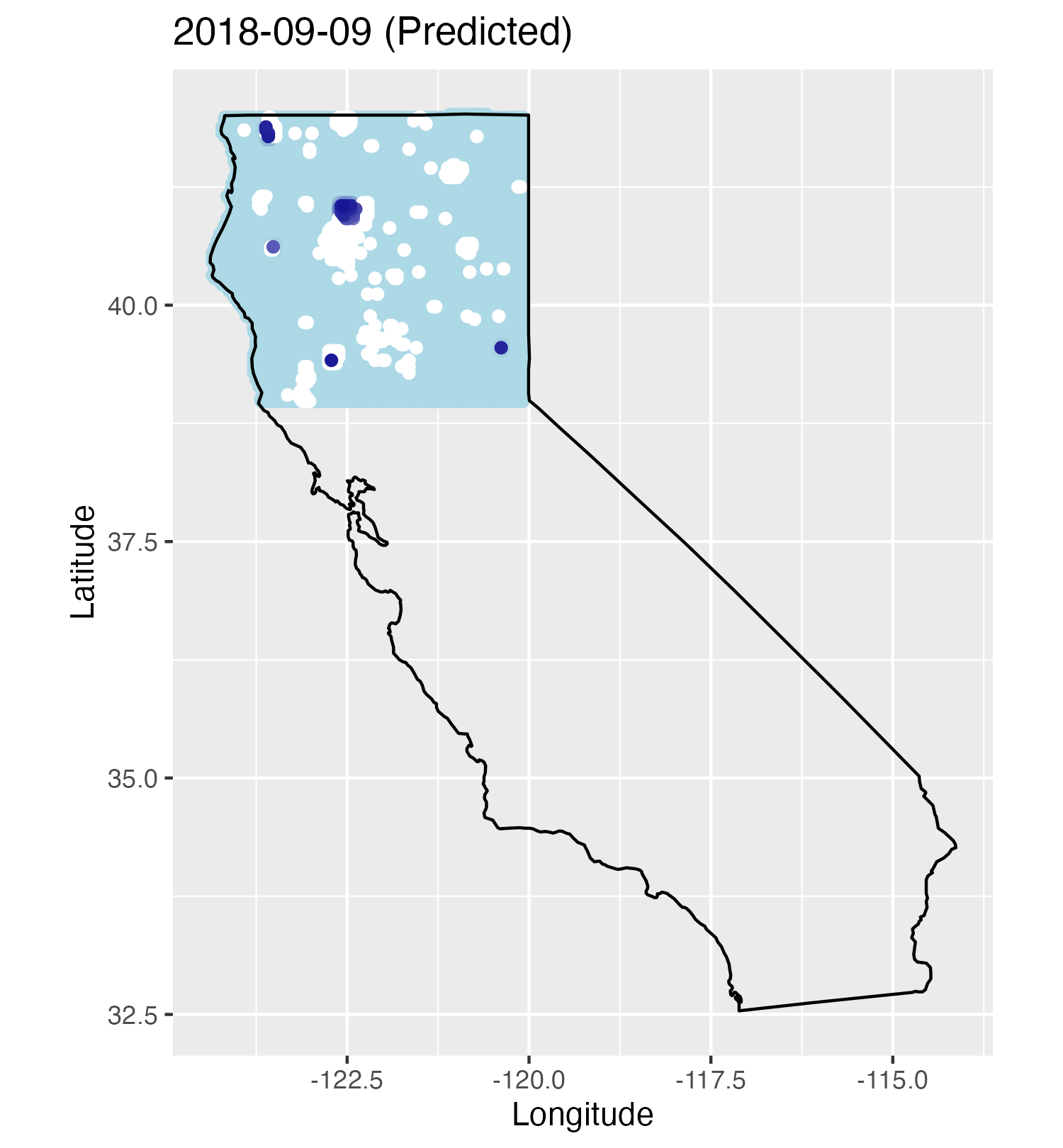}
          \caption{Sep 9, 2018.}

     \end{subfigure}
       \begin{subfigure}[b]{0.49\linewidth}
         \centering
         \includegraphics[scale=0.23]{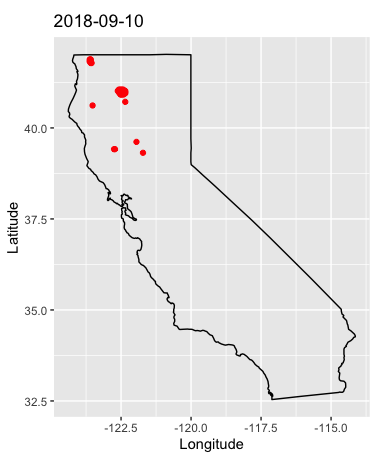}
         \includegraphics[scale=0.27]{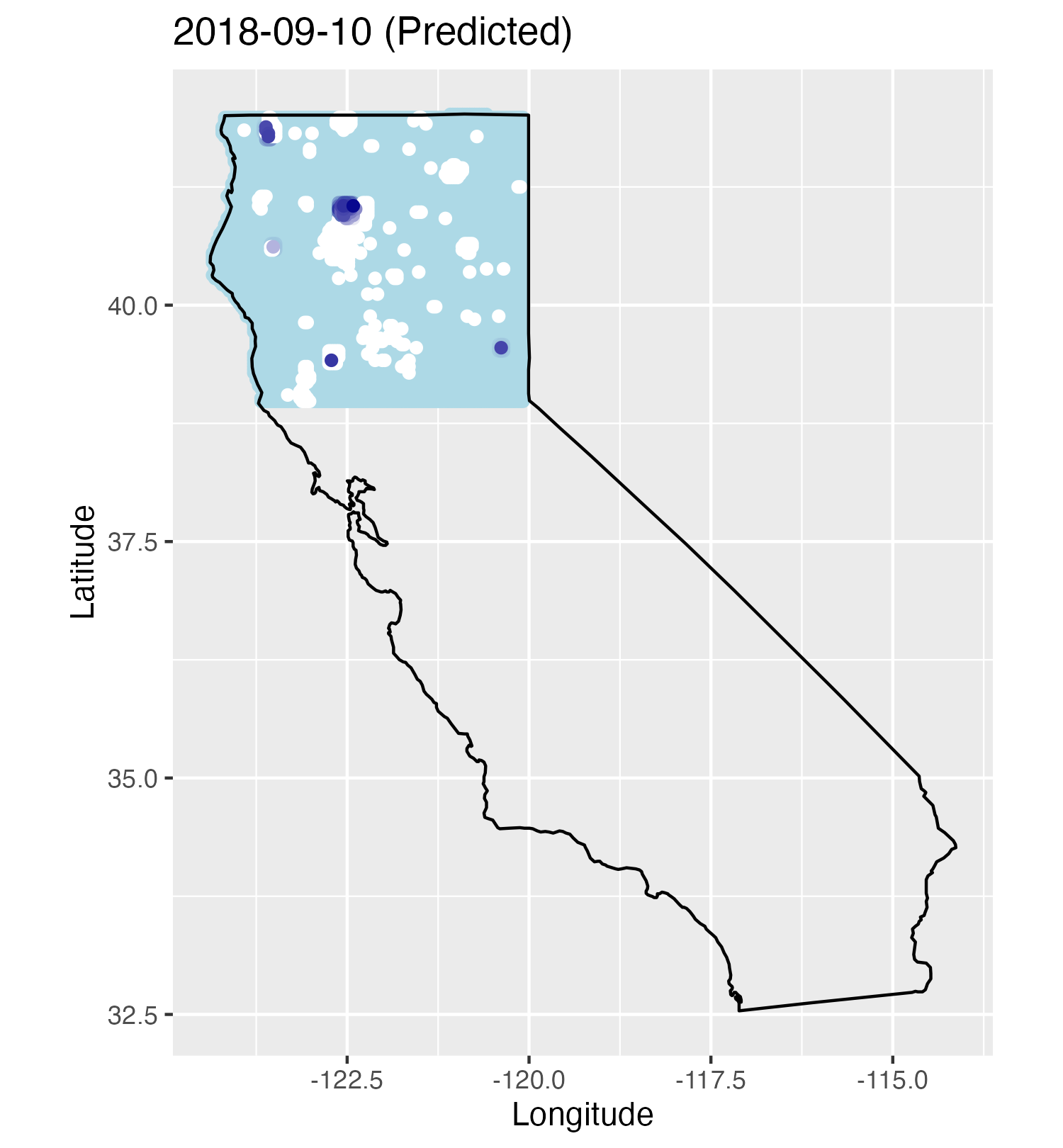}
          \caption{Sep 10, 2018.}
     \end{subfigure}

      \begin{subfigure}[b]{0.49\linewidth}
         \centering
         \includegraphics[scale=0.23]{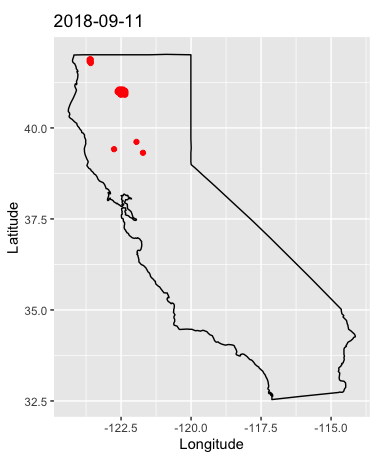}
         \includegraphics[scale=0.27]{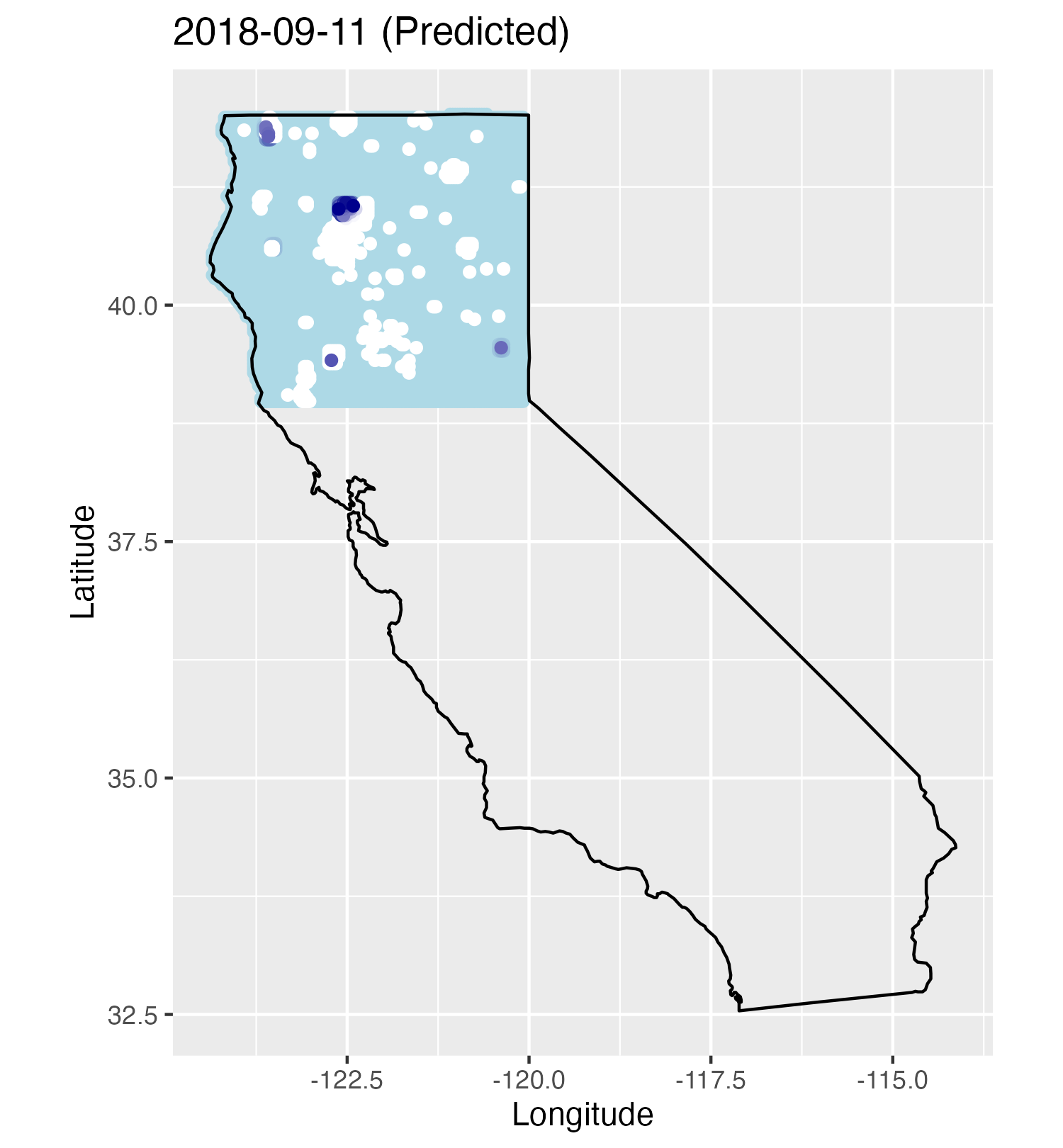}
          \caption{Sep 11, 2018.}

     \end{subfigure}
       \begin{subfigure}[b]{0.49\linewidth}
         \centering
         \includegraphics[scale=0.23]{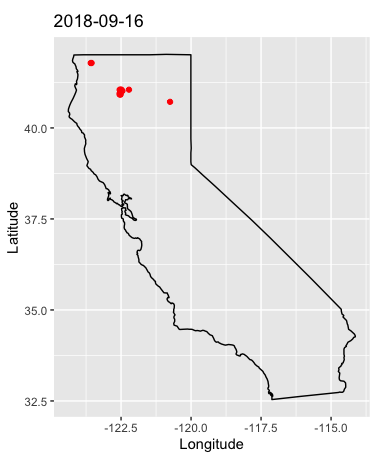}
         \includegraphics[scale=0.27]{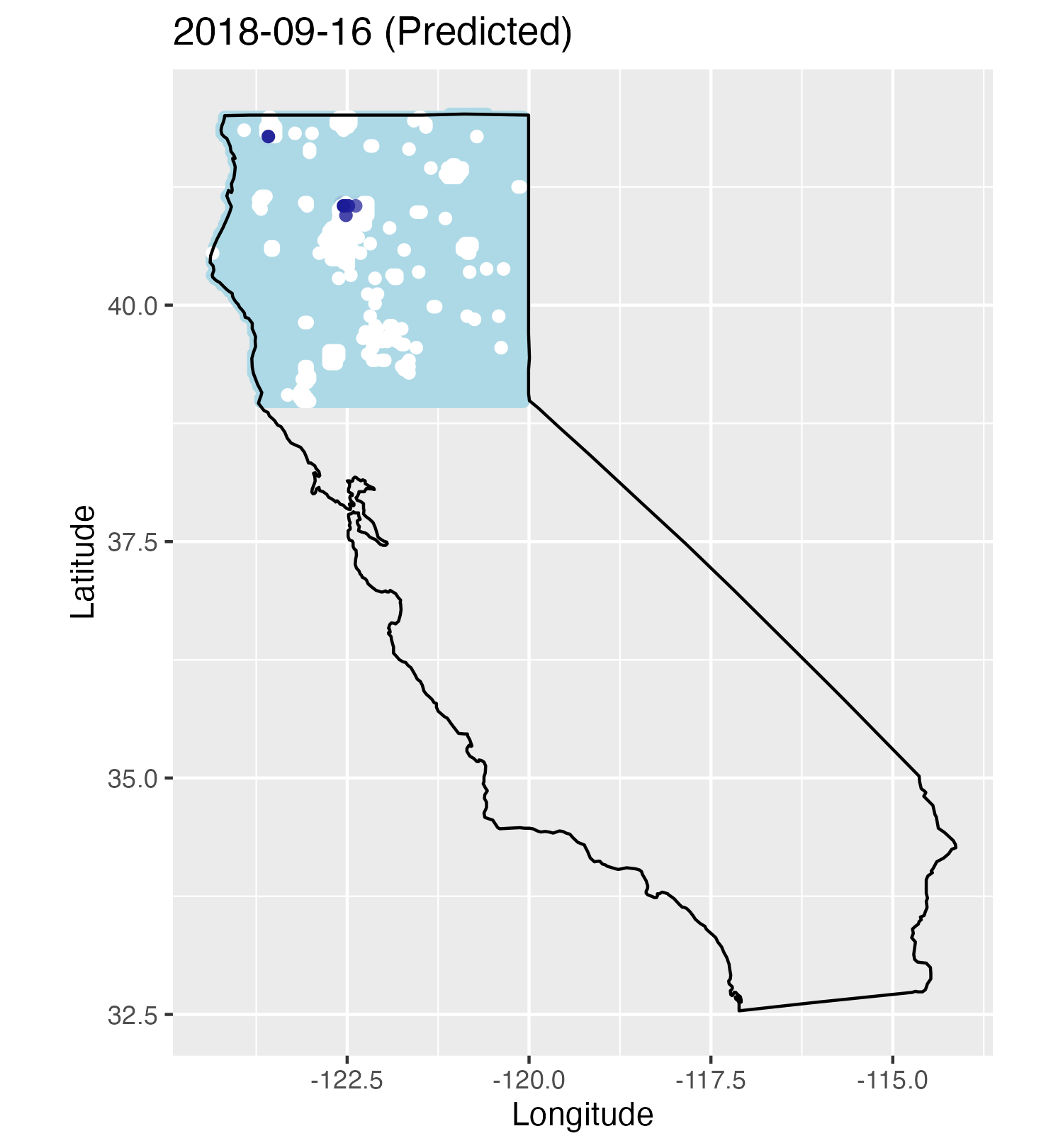}
          \caption{Sep 16, 2018.}
     \end{subfigure}

     \begin{subfigure}[b]{0.49\linewidth}
         \centering
         \includegraphics[scale=0.23]{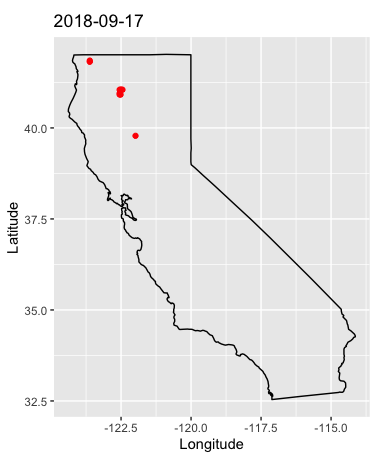}
         \includegraphics[scale=0.27]{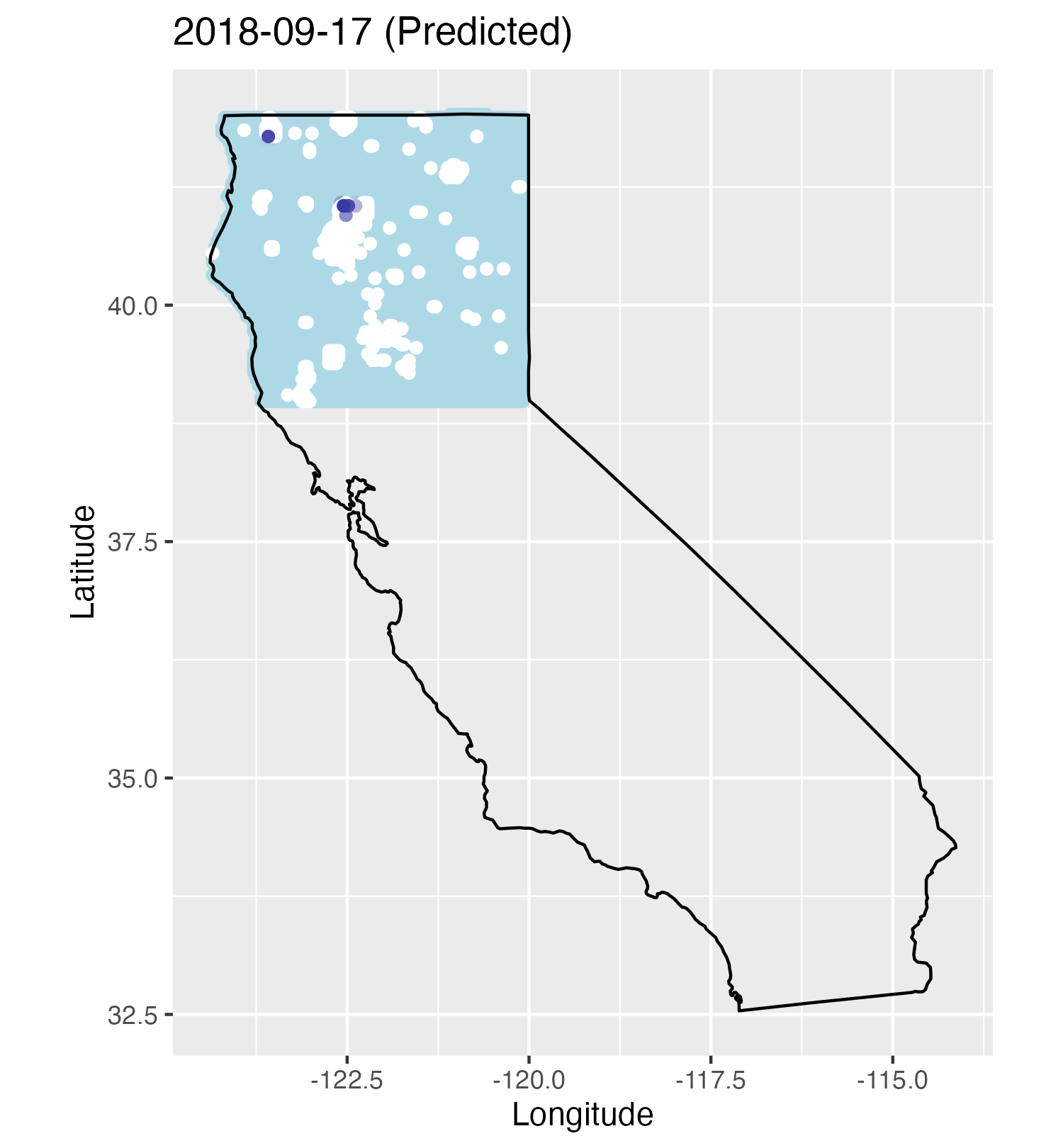}
          \caption{Sep 17, 2018.}

     \end{subfigure}
       \begin{subfigure}[b]{0.49\linewidth}
         \centering
         \includegraphics[scale=0.23]{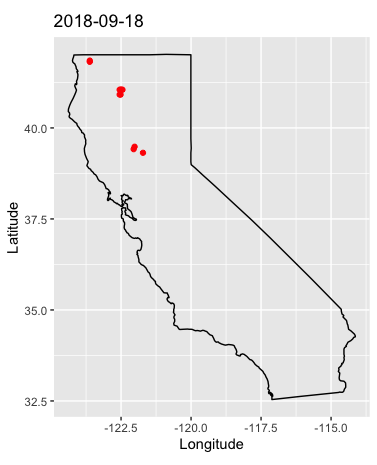}
         \includegraphics[scale=0.27]{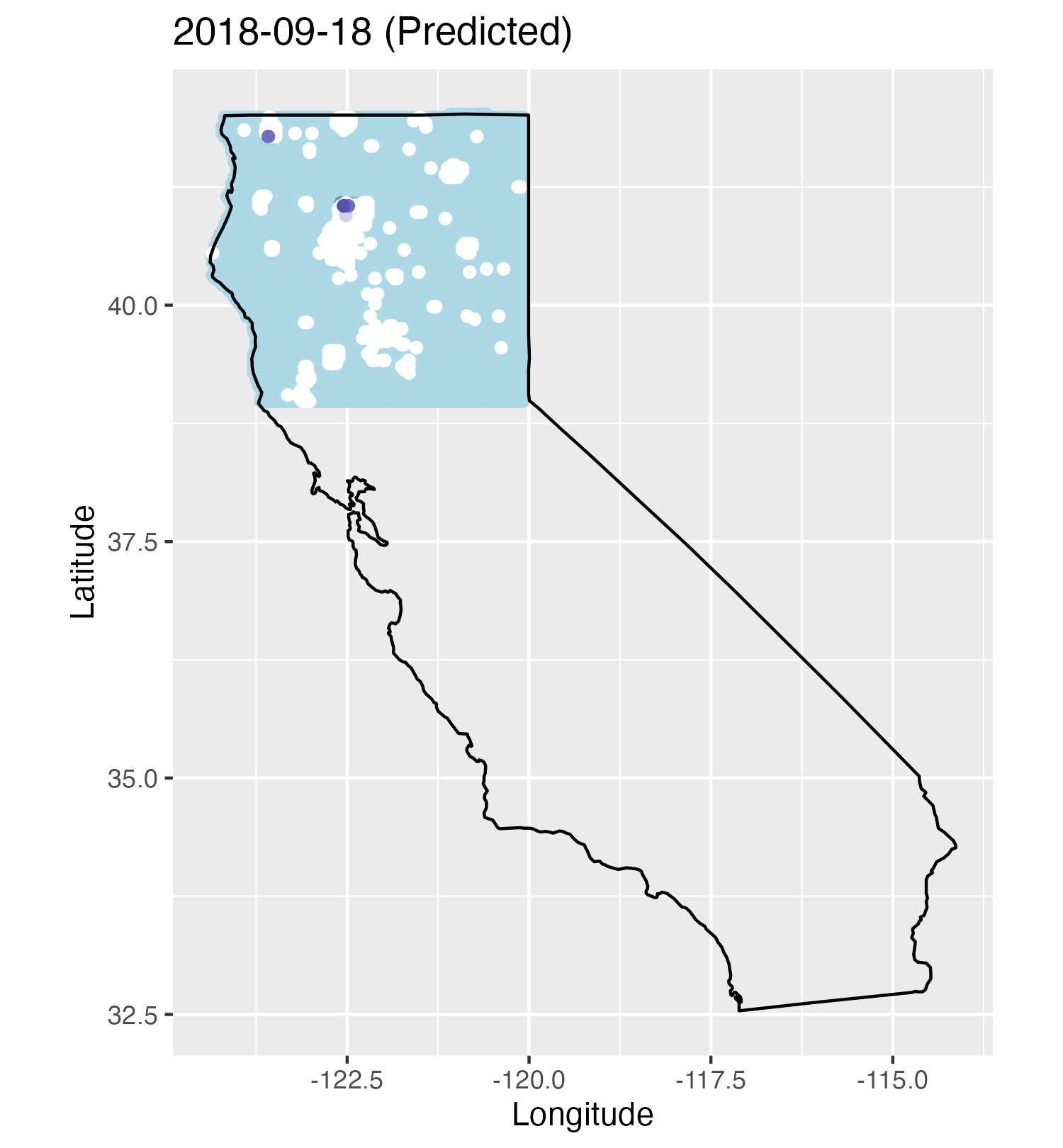}
          \caption{Sep 18, 2018.}
     \end{subfigure}

     \caption{Observed vs Predicted fire }
     \label{fig:cal_obs_vs_pred3}
\end{figure}

\begin{figure}[H]
     \centering
        \begin{subfigure}[b]{0.49\linewidth}
         \centering
         \includegraphics[scale=0.23]{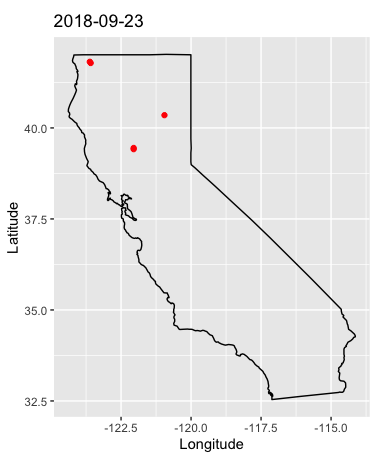}
         \includegraphics[scale=0.27]{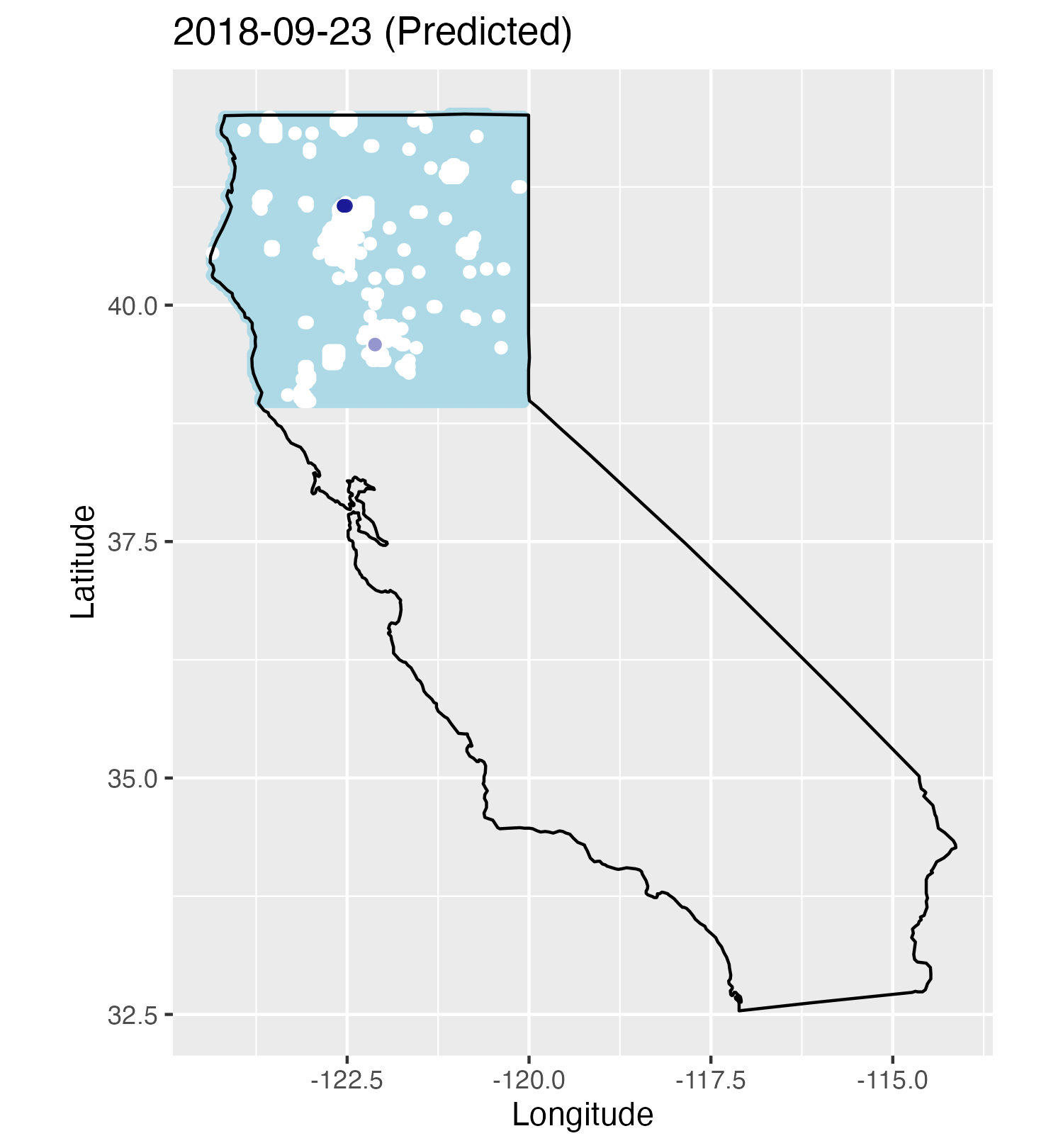}
          \caption{Sep 23, 2018.}

     \end{subfigure}
       \begin{subfigure}[b]{0.49\linewidth}
         \centering
         \includegraphics[scale=0.23]{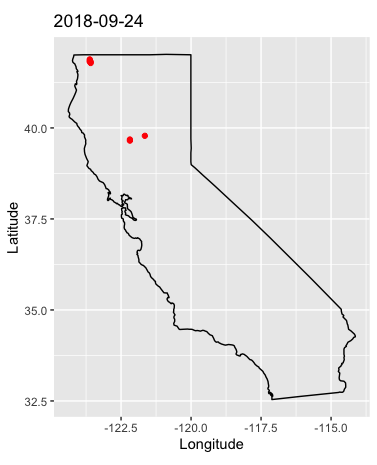}
         \includegraphics[scale=0.27]{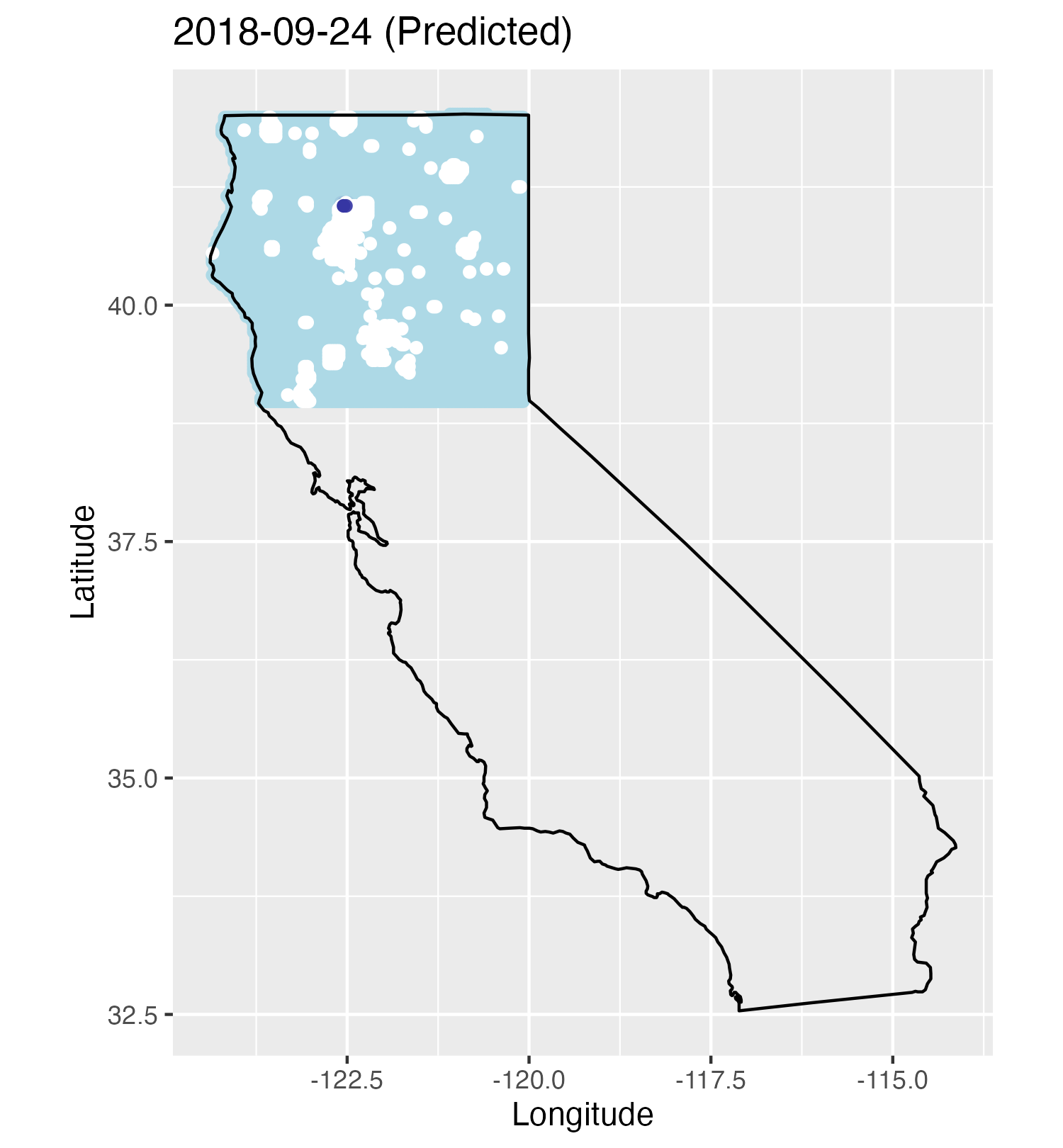}
          \caption{Sep 24, 2018.}
          \end{subfigure}

       \begin{subfigure}[b]{0.49\linewidth}
         \centering
         \includegraphics[scale=0.23]{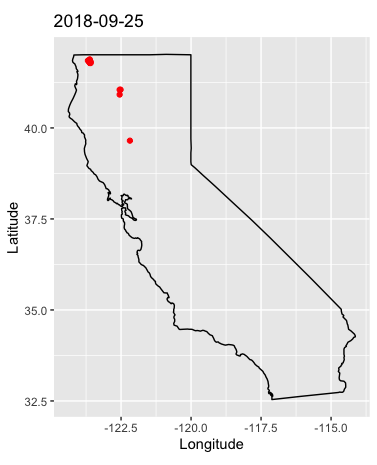}
         \includegraphics[scale=0.27]{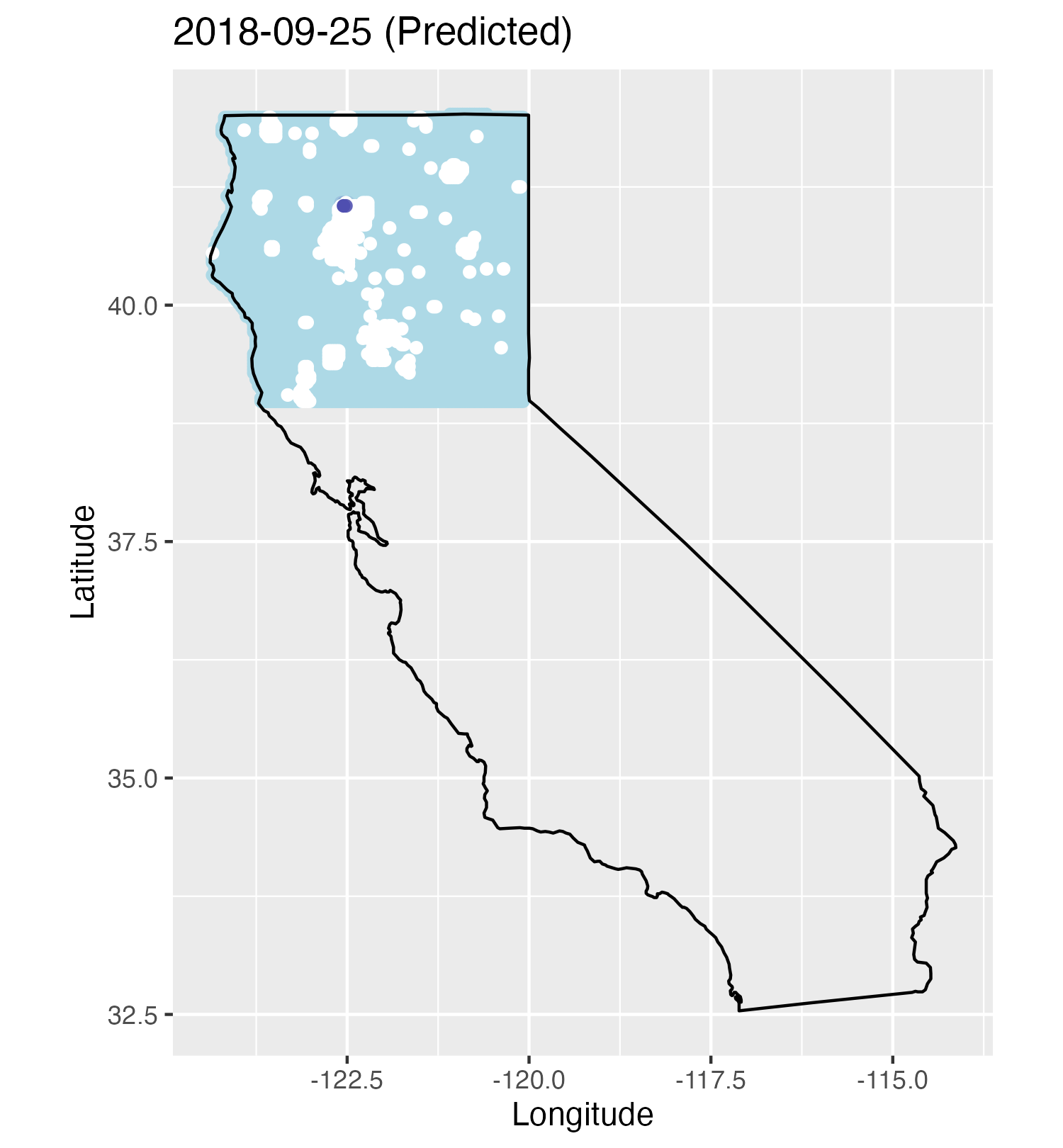}
          \caption{Sep 25, 2018.}

     \end{subfigure}
       \begin{subfigure}[b]{0.49\linewidth}
         \centering
         \includegraphics[scale=0.23]{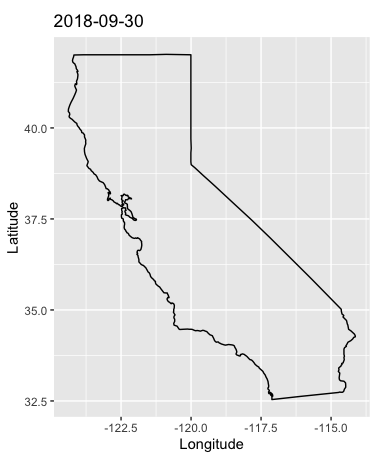}
         \includegraphics[scale=0.27]{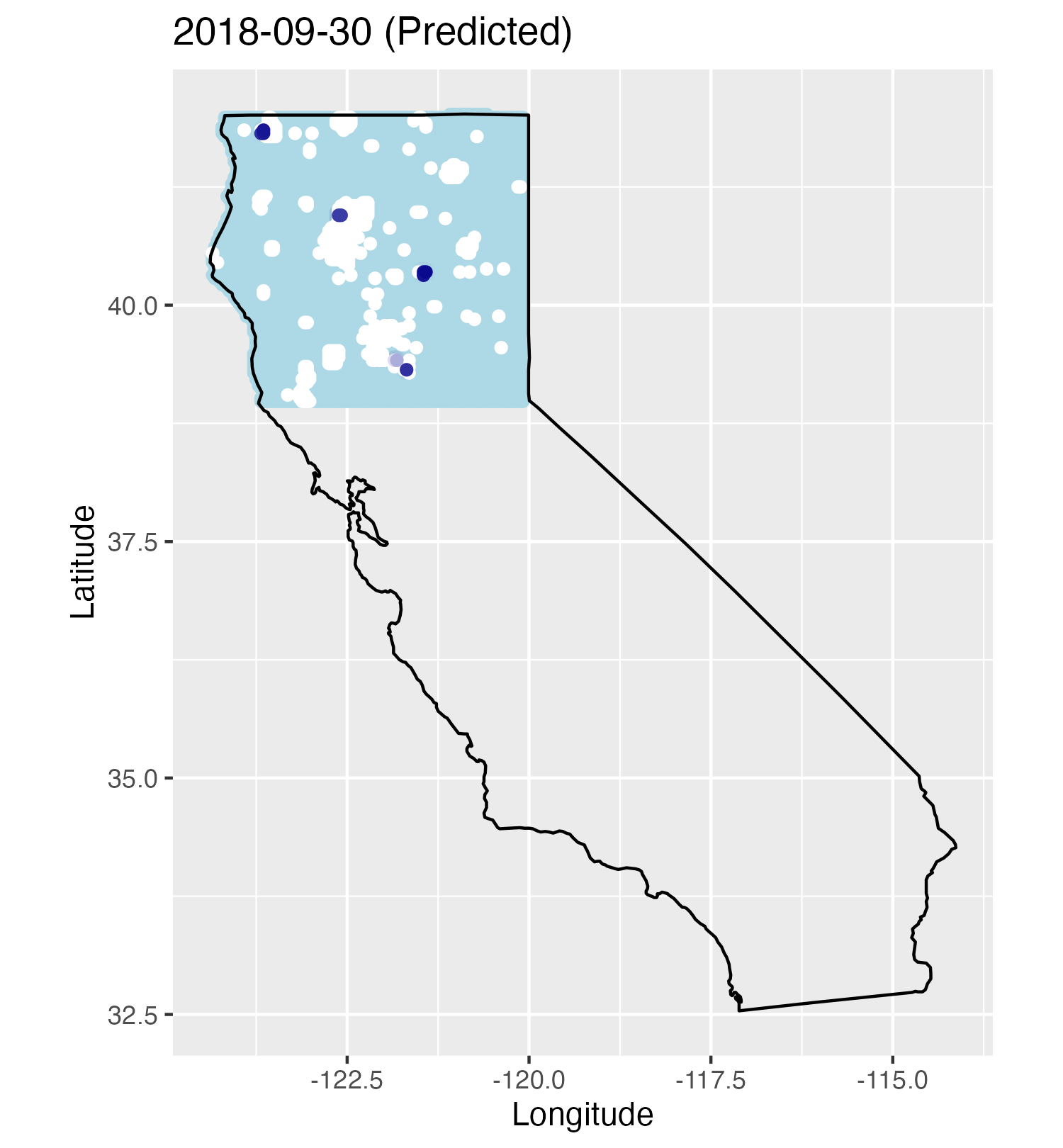}
          \caption{Sep 30, 2018.}
     \end{subfigure}

      \begin{subfigure}[b]{0.49\linewidth}
         \centering
         \includegraphics[scale=0.23]{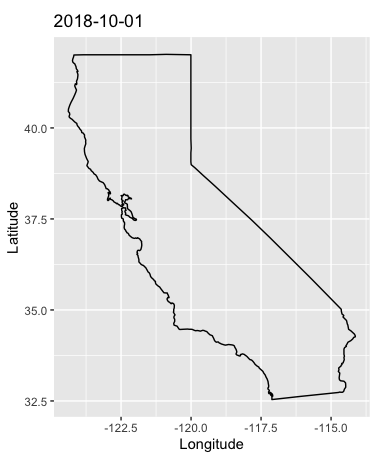}
         \includegraphics[scale=0.27]{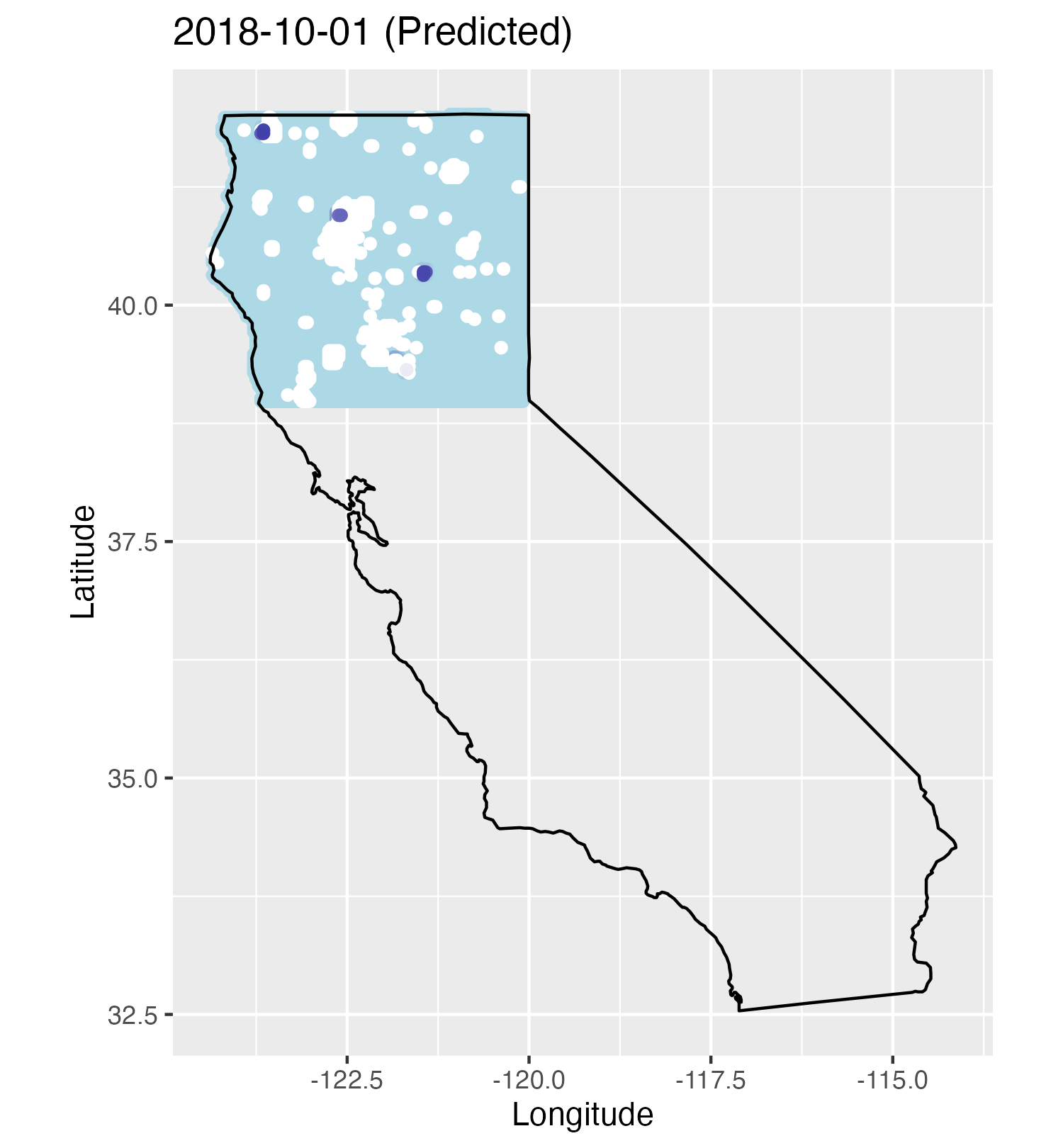}
          \caption{Oct 1, 2018.}

     \end{subfigure}
       \begin{subfigure}[b]{0.49\linewidth}
         \centering
         \includegraphics[scale=0.23]{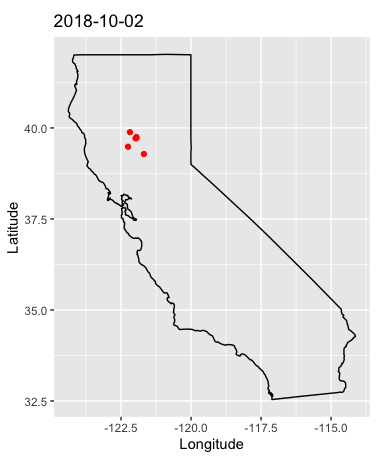}
         \includegraphics[scale=0.27]{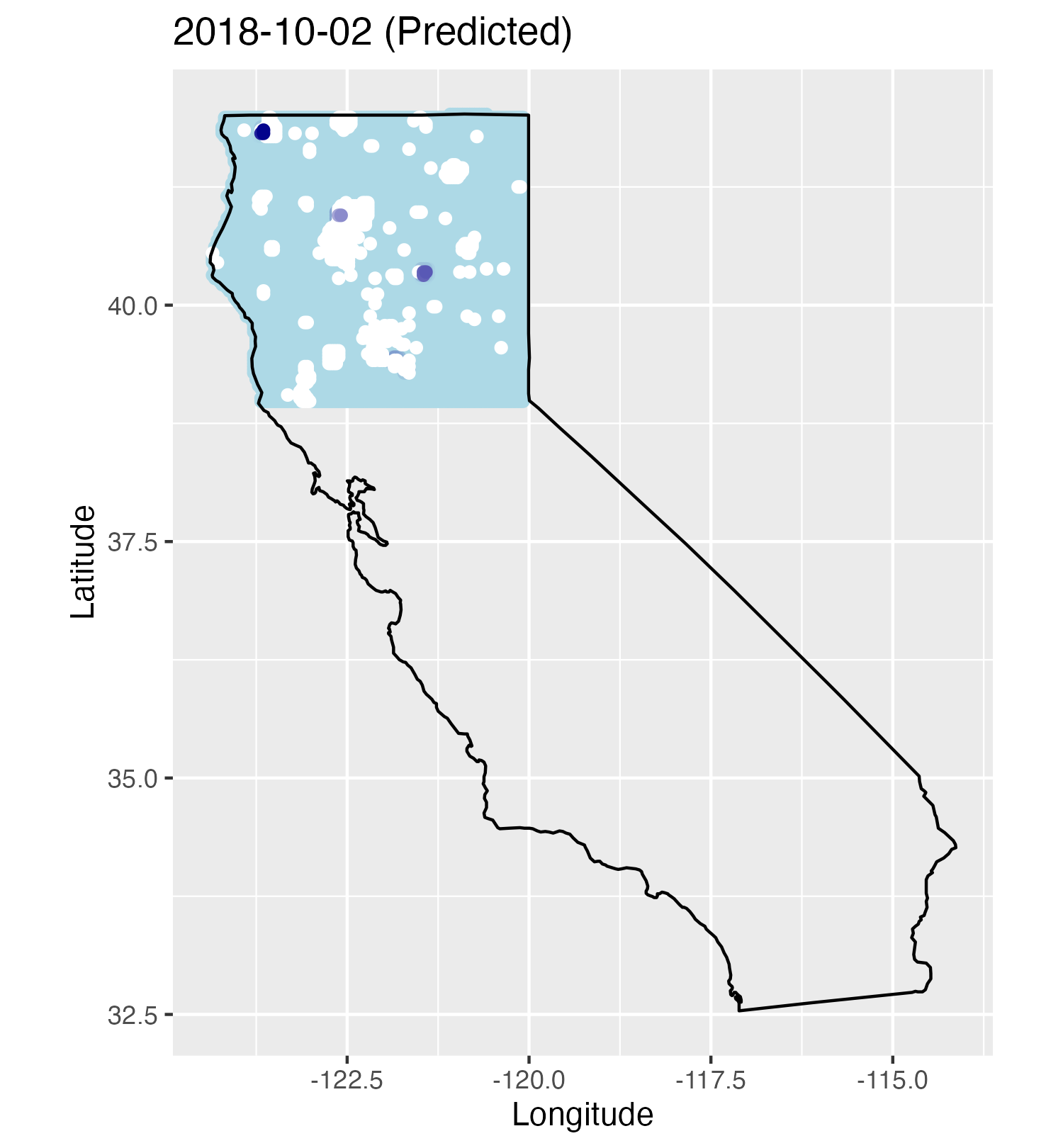}
          \caption{Oct 2, 2018.}
     \end{subfigure}

     \begin{subfigure}[b]{0.49\linewidth}
         \centering
         \includegraphics[scale=0.23]{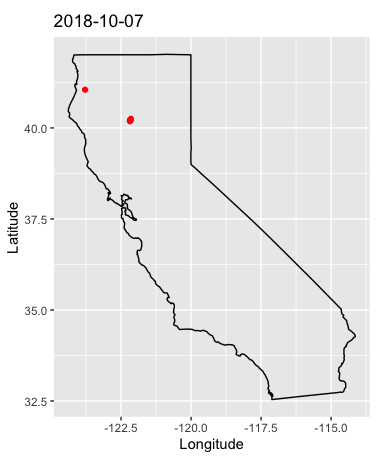}
         \includegraphics[scale=0.27]{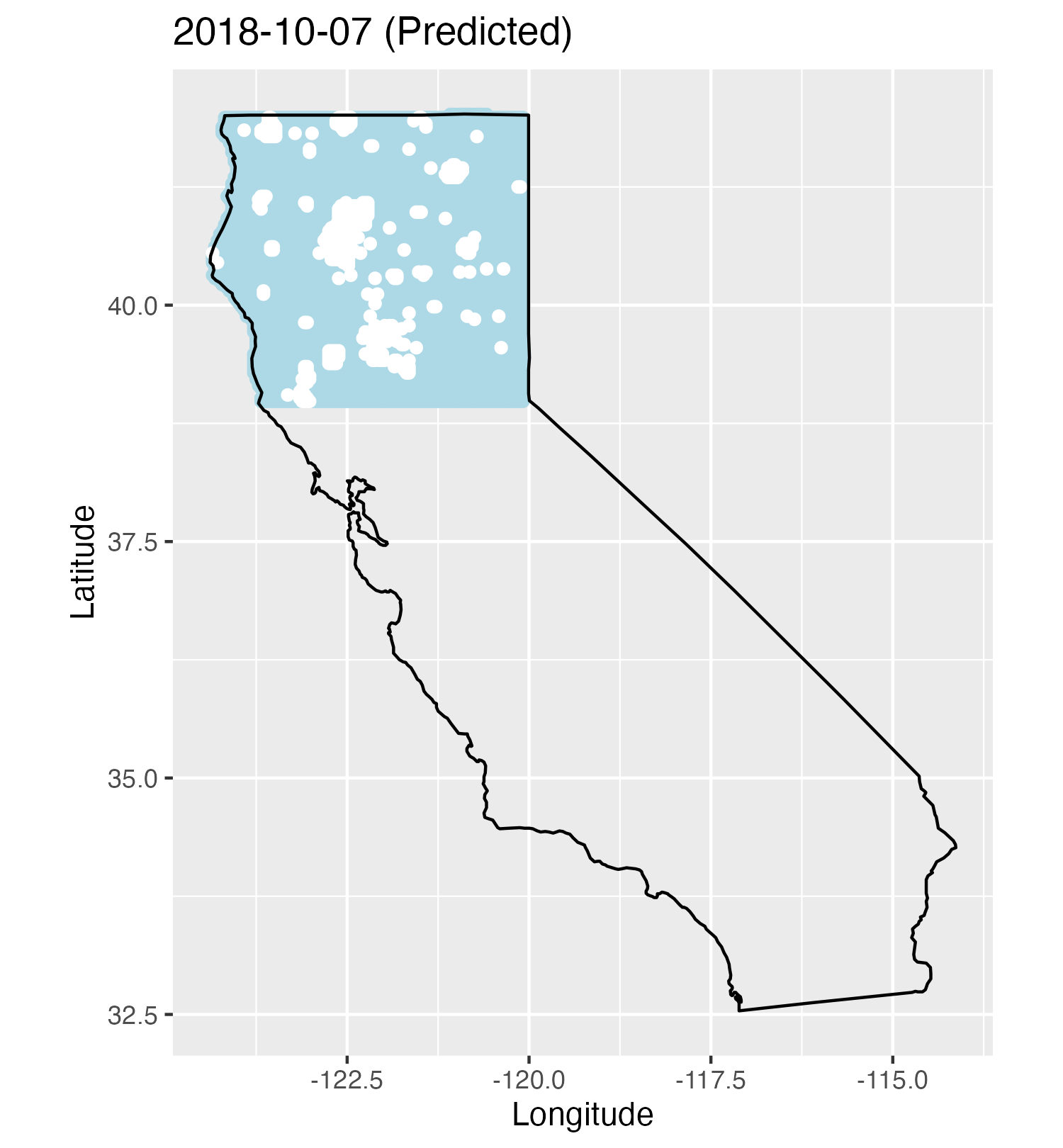}
          \caption{Oct 7, 2018.}

     \end{subfigure}
       \begin{subfigure}[b]{0.49\linewidth}
         \centering
         \includegraphics[scale=0.23]{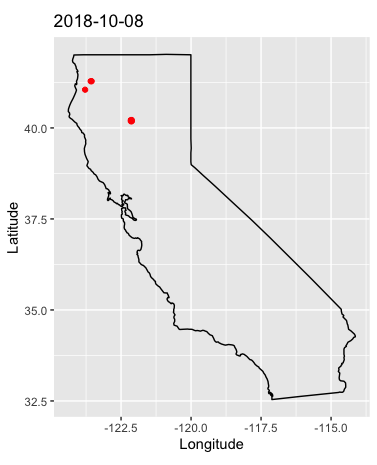}
         \includegraphics[scale=0.27]{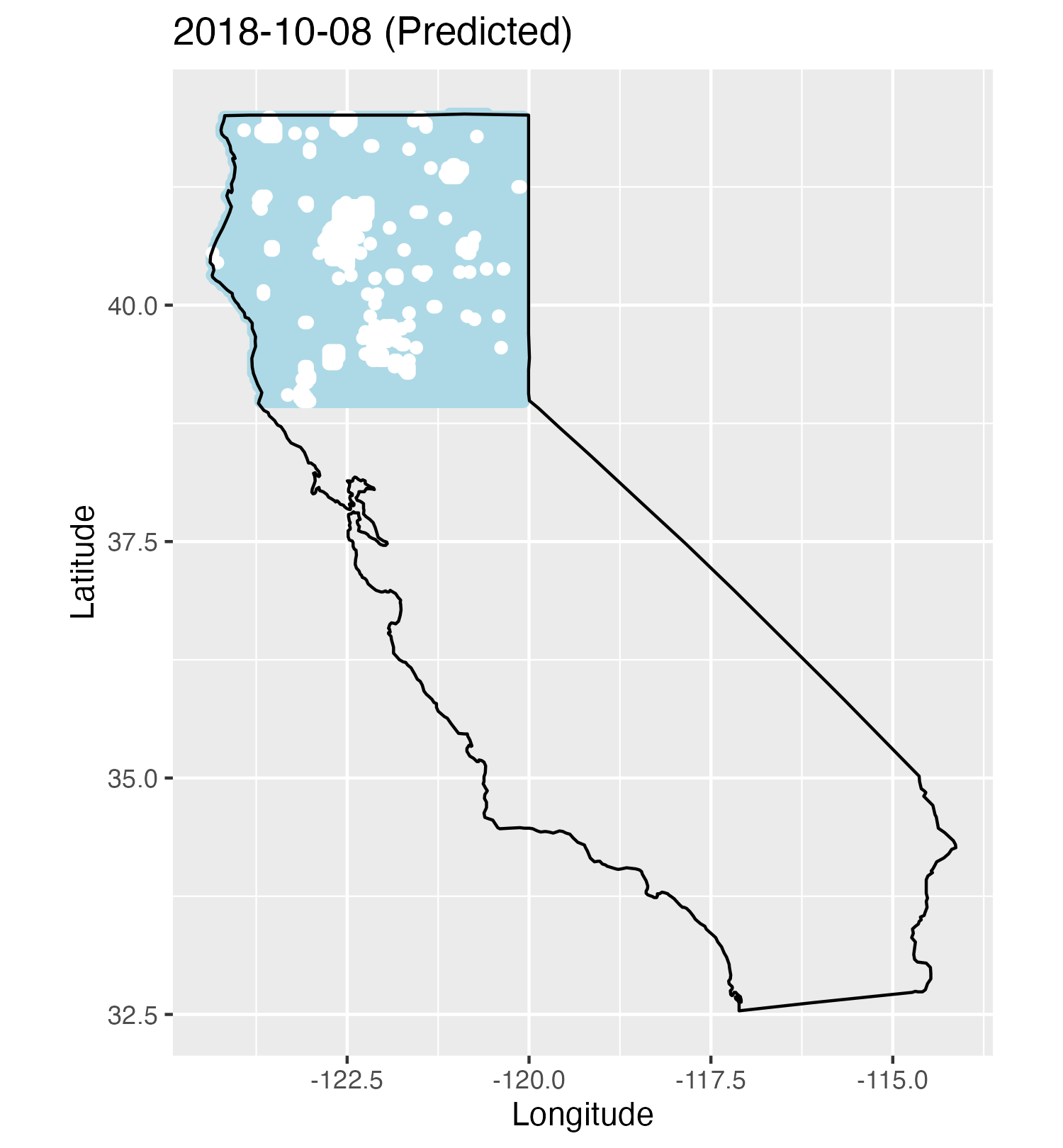}
          \caption{Oct 8, 2018.}
     \end{subfigure}

     \caption{Observed vs Predicted fire }
       \label{fig:cal_obs_vs_pred4}
\end{figure}

\begin{figure}[H]
     \centering
        \begin{subfigure}[b]{0.49\linewidth}
         \centering
         \includegraphics[scale=0.23]{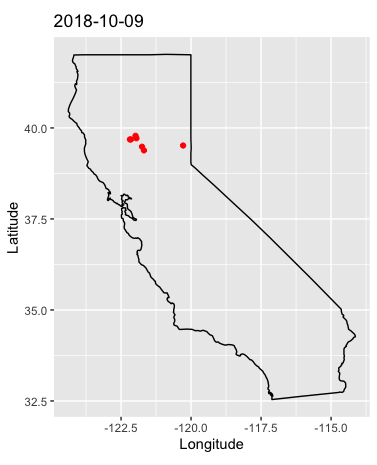}
         \includegraphics[scale=0.27]{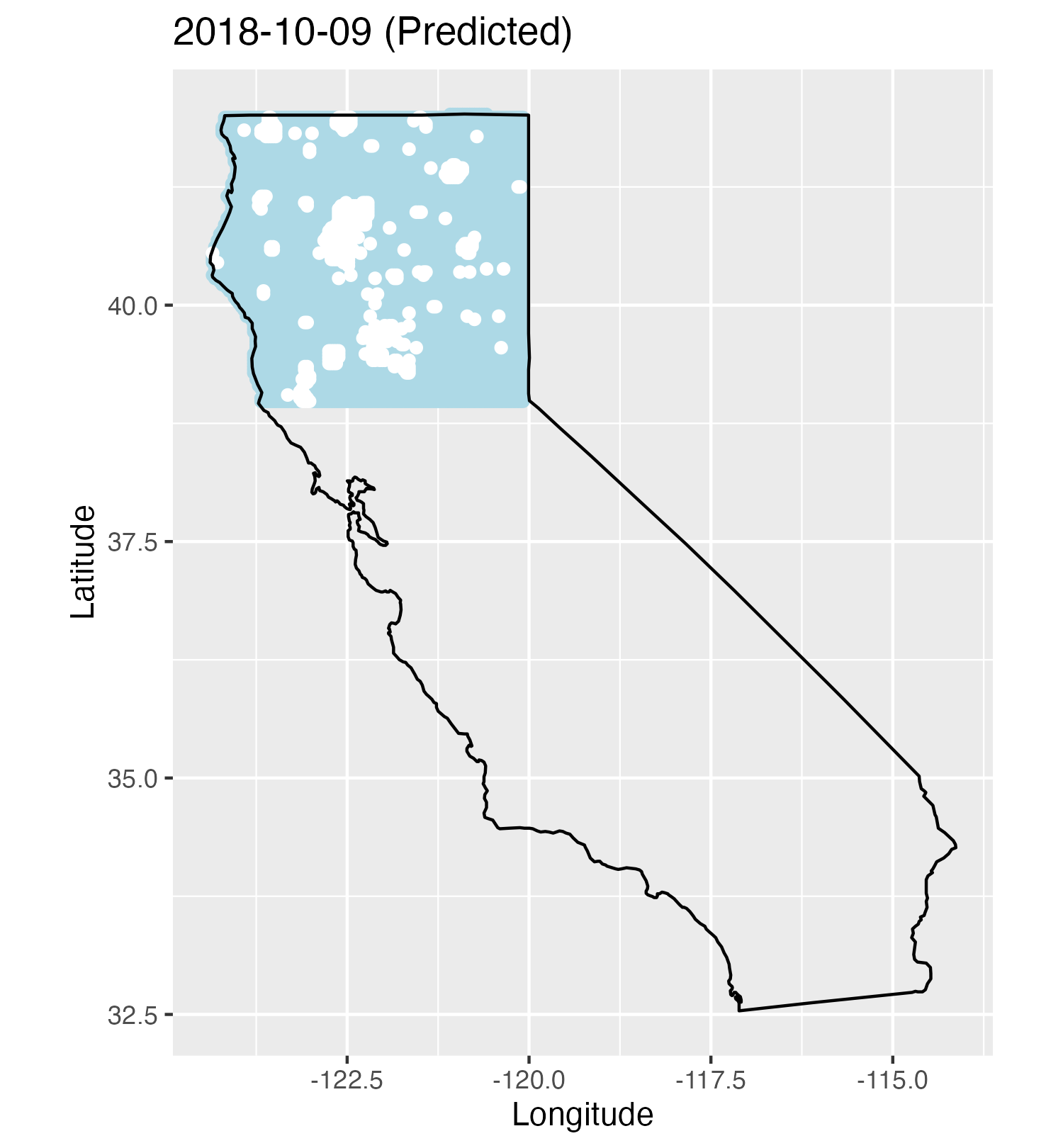}
          \caption{Oct 9, 2018.}

     \end{subfigure}
       \begin{subfigure}[b]{0.49\linewidth}
         \centering
         \includegraphics[scale=0.23]{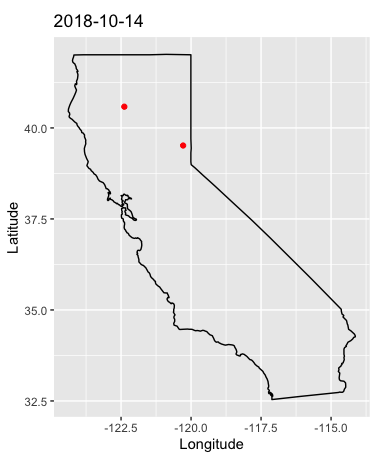}
         \includegraphics[scale=0.27]{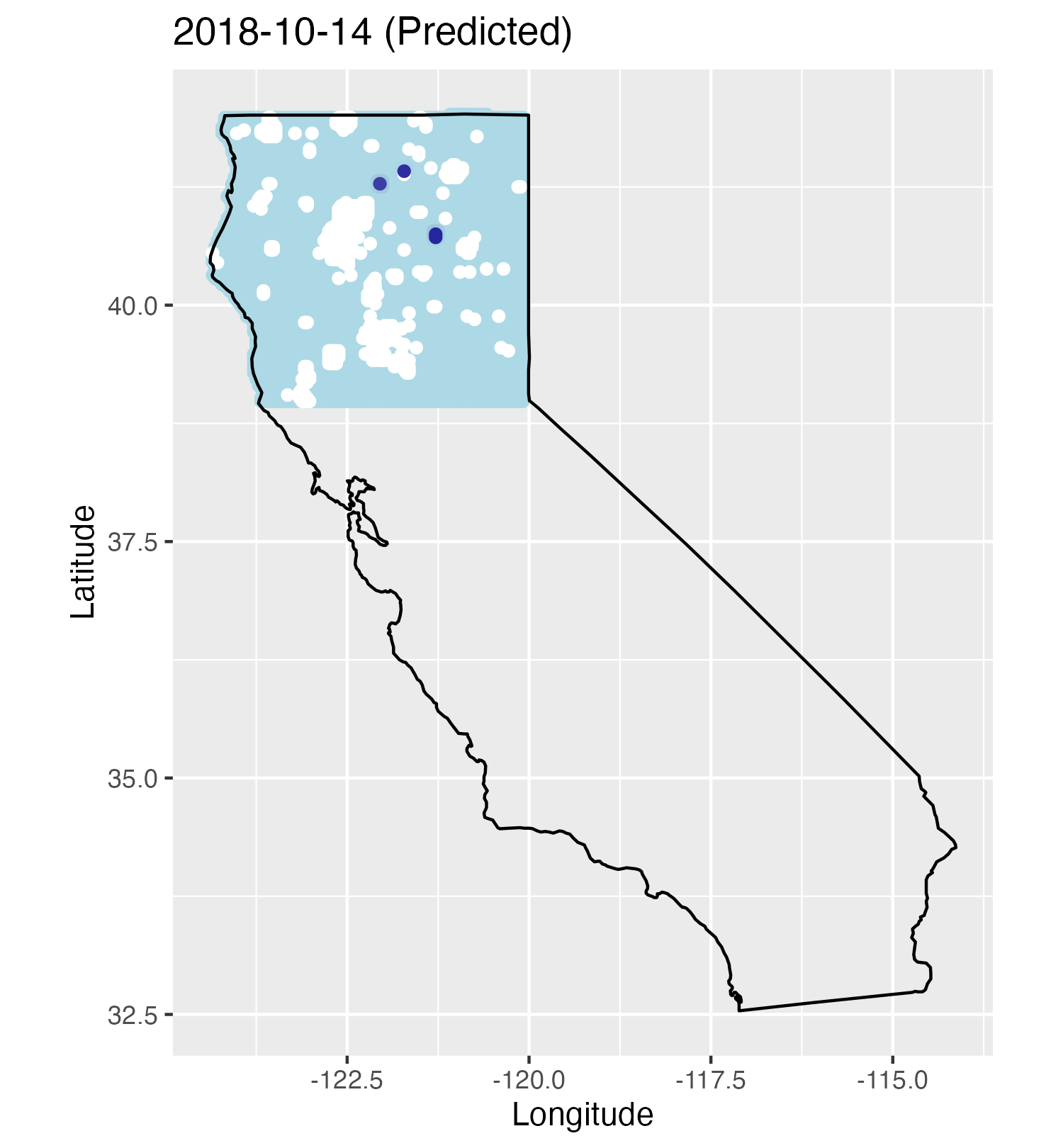}
          \caption{Oct 14, 2018.}
          \end{subfigure}

       \begin{subfigure}[b]{0.49\linewidth}
         \centering
         \includegraphics[scale=0.23]{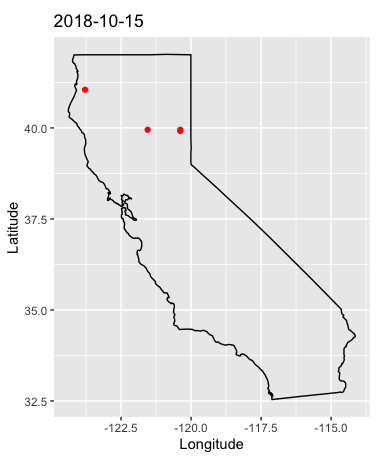}
         \includegraphics[scale=0.27]{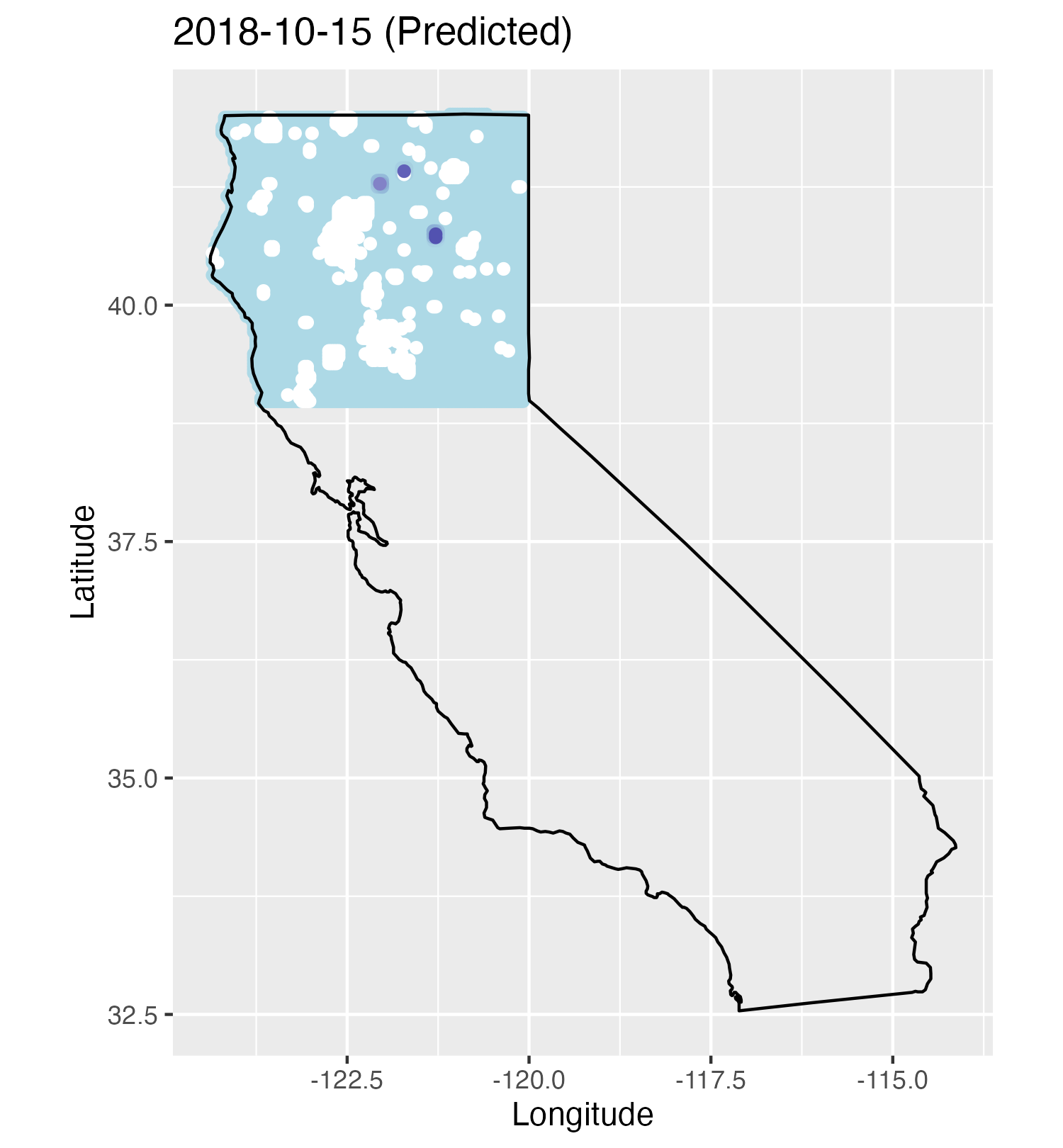}
          \caption{Oct 15, 2018.}

     \end{subfigure}
       \begin{subfigure}[b]{0.49\linewidth}
         \centering
         \includegraphics[scale=0.23]{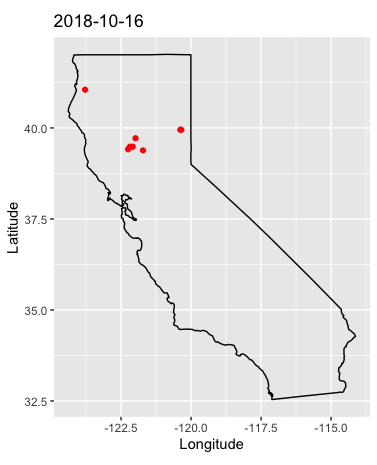}
         \includegraphics[scale=0.27]{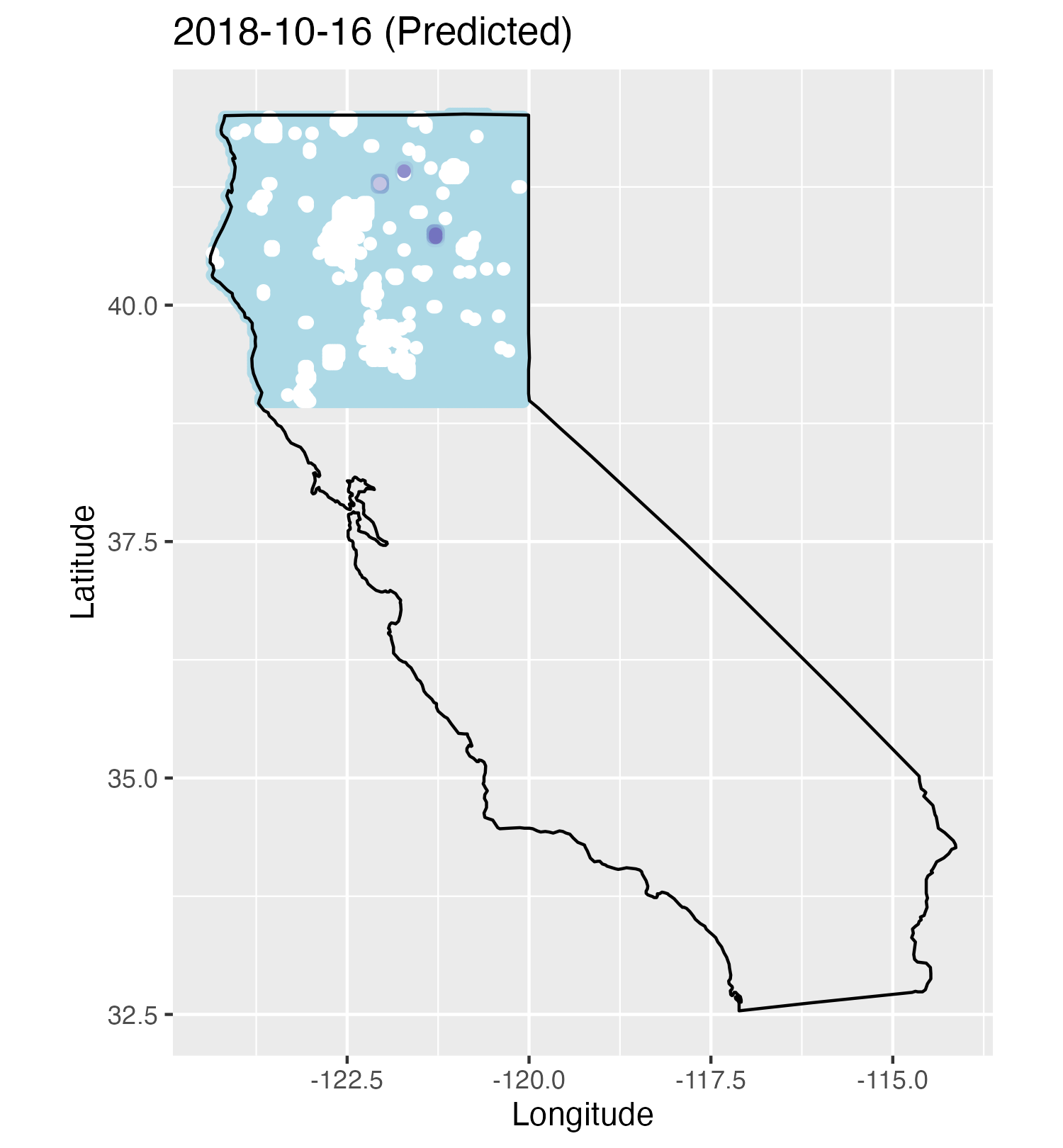}
          \caption{Oct 16, 2018.}
     \end{subfigure}

      \begin{subfigure}[b]{0.49\linewidth}
         \centering
         \includegraphics[scale=0.23]{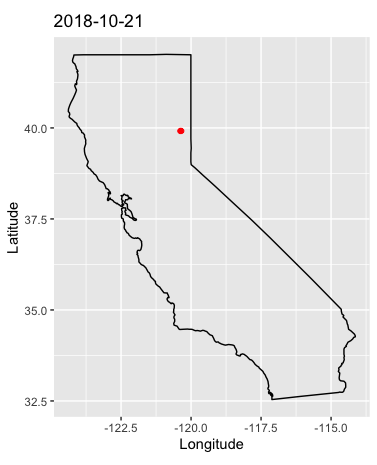}
         \includegraphics[scale=0.27]{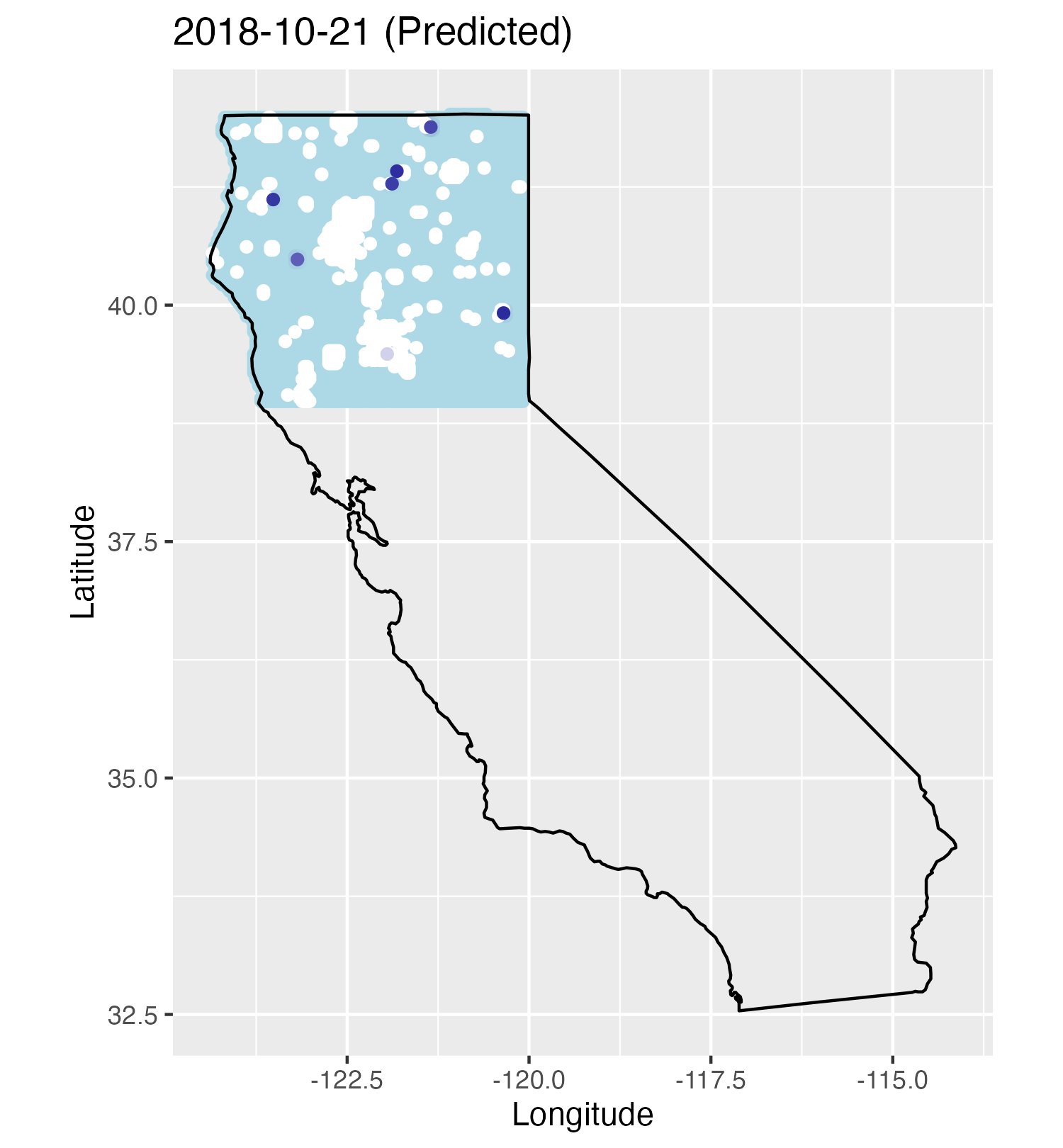}
          \caption{Oct 21, 2018.}

     \end{subfigure}
       \begin{subfigure}[b]{0.49\linewidth}
         \centering
         \includegraphics[scale=0.23]{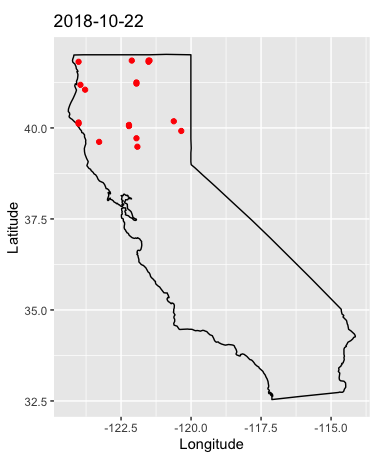}
         \includegraphics[scale=0.27]{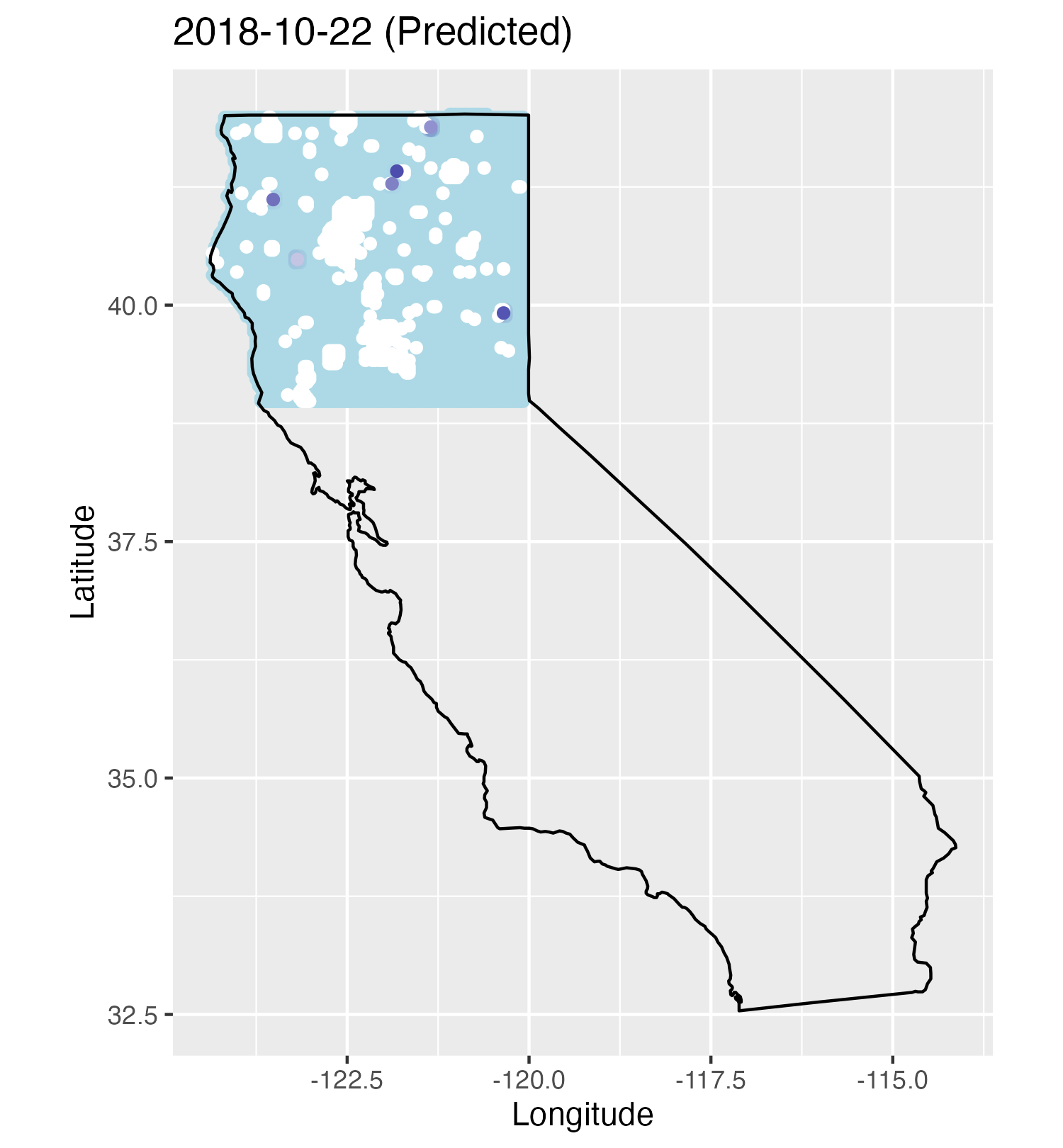}
          \caption{Oct 22, 2018.}
     \end{subfigure}

     \begin{subfigure}[b]{0.49\linewidth}
         \centering
         \includegraphics[scale=0.23]{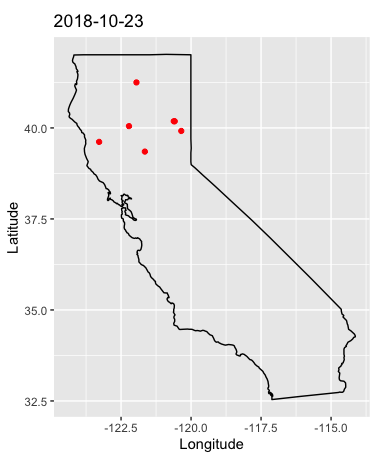}
         \includegraphics[scale=0.27]{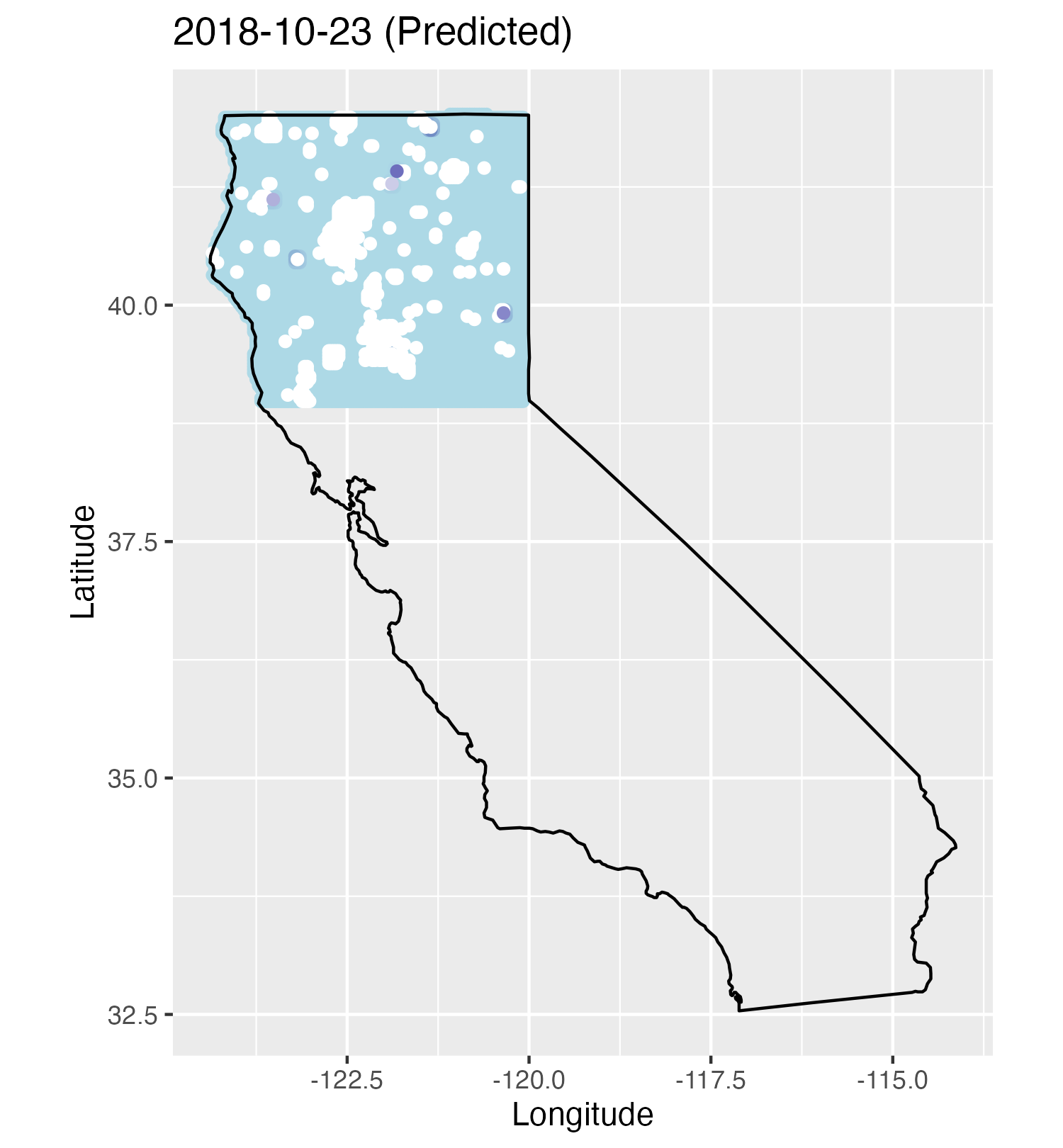}
          \caption{Oct 23, 2018.}

     \end{subfigure}
       \begin{subfigure}[b]{0.49\linewidth}
         \centering
         \includegraphics[scale=0.23]{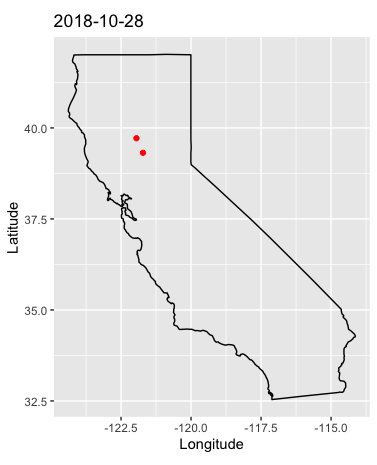}
         \includegraphics[scale=0.27]{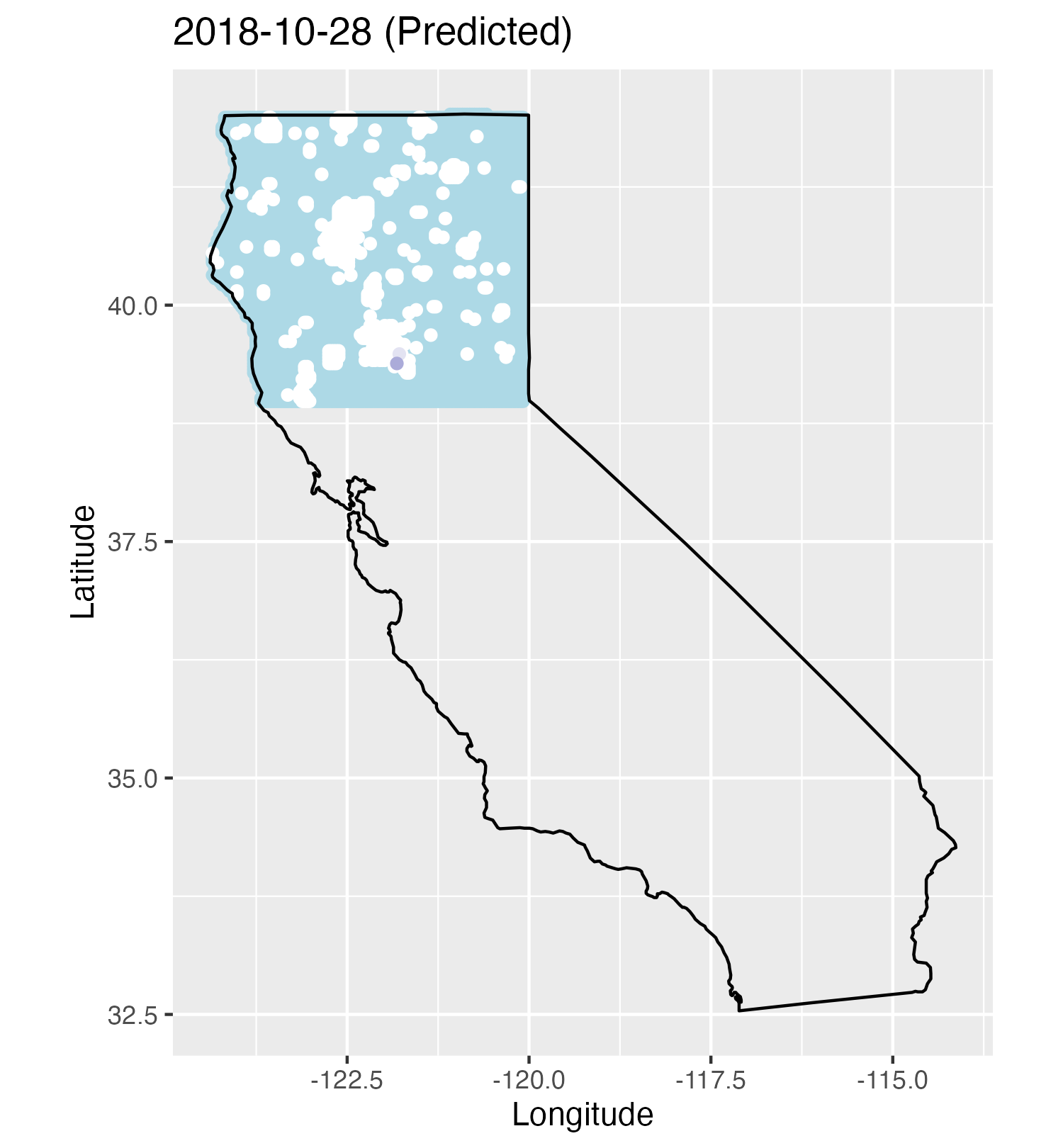}
          \caption{Oct 28, 2018.}
     \end{subfigure}

     \caption{Observed vs Predicted fire }
           \label{fig:cal_obs_vs_pred5}
\end{figure}

\begin{figure}[H]
     \centering
        \begin{subfigure}[b]{0.49\linewidth}
         \centering
         \includegraphics[scale=0.23]{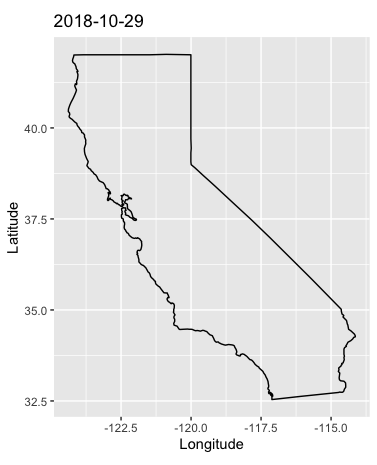}
         \includegraphics[scale=0.27]{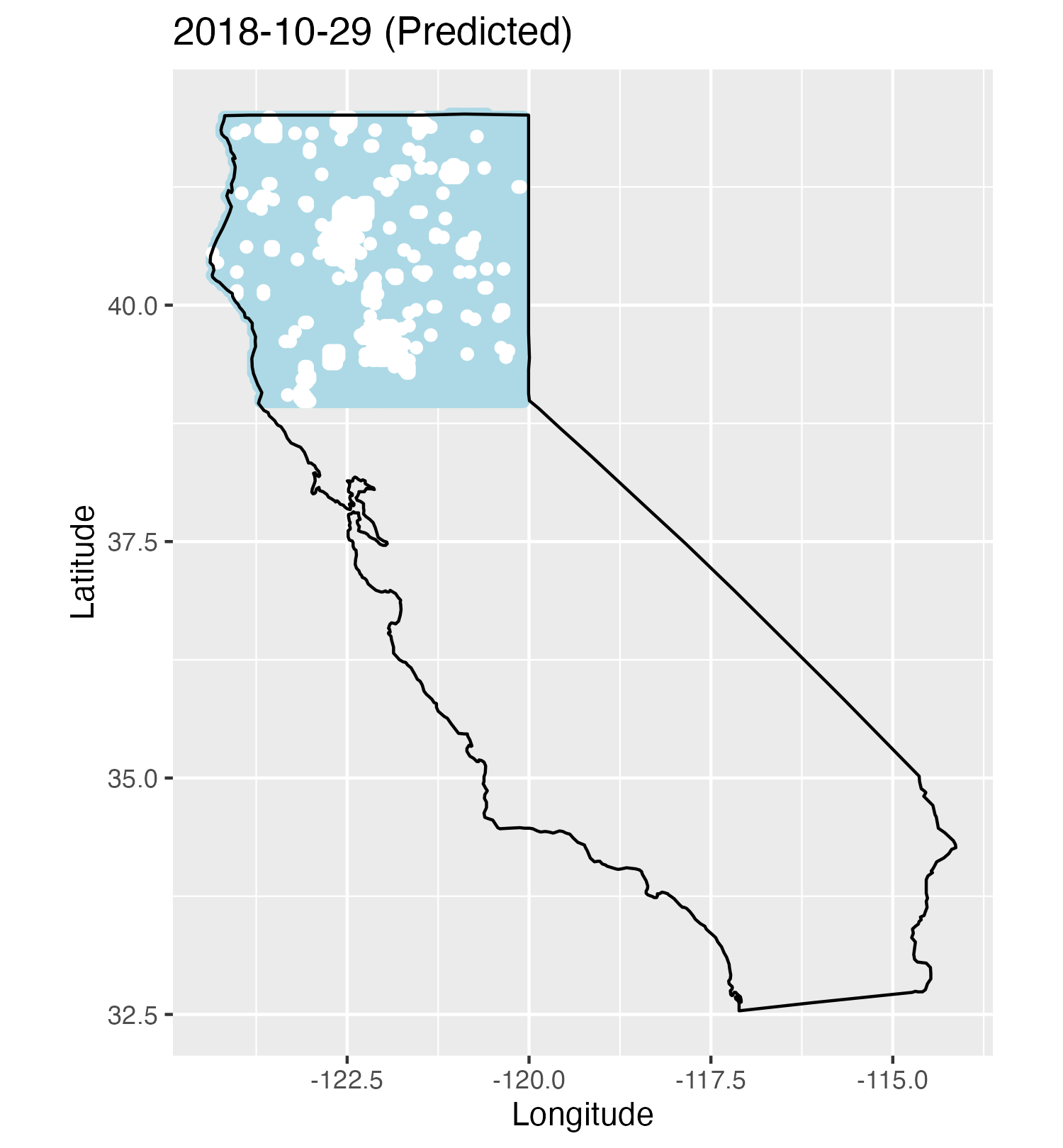}
          \caption{Oct 29, 2018.}

     \end{subfigure}
       \begin{subfigure}[b]{0.49\linewidth}
         \centering
         \includegraphics[scale=0.23]{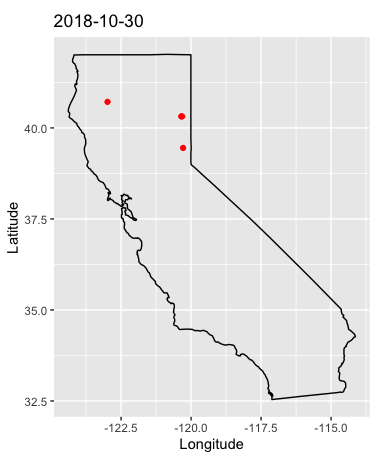}
         \includegraphics[scale=0.27]{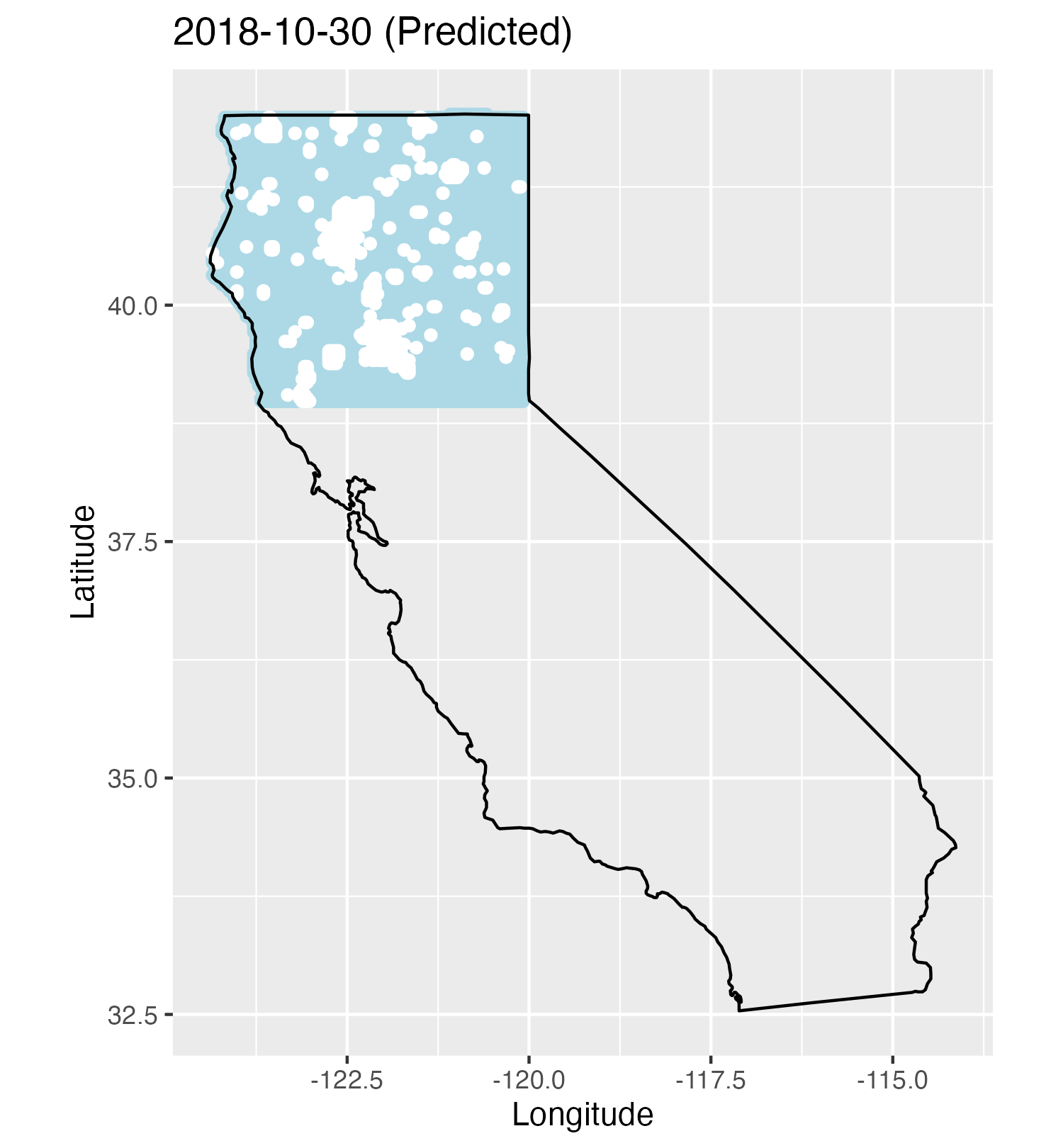}
          \caption{Oct 30, 2018.}
          \end{subfigure}
  \caption{Observed vs Predicted fire }
   \label{fig:cal_obs_vs_pred6}

\end{figure}

\subsection{Parameter estimates and forecasts for Australia data}

\begin{table}[H]
     \centering
     \resizebox{\textwidth}{!}{
     \begin{tabular}{|c|cccccccc|}
     \hline

      Time & Intensity & Wind & Precipitation & Rel hum & Altitude & Soil moisture & Temperature & Pressure\\ \hline
     Nov 1- Nov 14 & 13.46 (0.86) & -0.00 (0.40) & -11.19 (5.08) & -2.76 (0.36) & -0.38 (0.44) & 2.68 (0.32) & 25.82 (3.56) & -2.35 (2.36)\\
     Nov 1-Nov 21 & 12.88 (0.74) & 0.37 (0.40) & -10.07 (3.96) & -2.28 (0.29) & -0.16 (0.40) & 2.72 (0.29) & 15.51 (3.04) & 0.22 (2.11)\\
     Nov 1-Nov 28 & 12.11 (0.68) & 0.22 (0.40) & -1.93 (1.75) & -2.24 (0.25) & -0.18 (0.38) & 1.92 (0.26) & 12.37 (2.60) & 0.63 (2.02)\\
     Nov 1-Dec 5 & 11.26 (0.59) & -0.11 (0.32) & -1.68 (1.73) & -1.73 (0.21) & -0.70 (0.34) & 1.21 (0.23) & 4.11 (2.25) & -0.84 (1.77)\\
     Nov 1-Dec 12 & 10.37 (0.52) & 0.09 (0.30) & -0.77 (1.22) & -2.03 (0.18) & -0.84 (0.31) & 1.18 (0.22) & 1.53 (1.96) & -0.52 (1.64)\\
     Nov 1-Dec 19 & 9.86 (0.49) & 0.23 (0.29) & -0.23 (1.18) & -2.08 (0.17) & -0.53 (0.29) & 0.96 (0.20) & -2.49 (1.72) & 1.69 (1.55)\\
     Nov 1-Dec 26 & 9.06 (0.46) & 0.34 (0.28) & 0.10 (1.14) & -1.83 (0.15) & -0.31 (0.28) & 0.69 (0.19) & -6.83 (1.61) & 3.41 (1.51)\\
     Nov 1-Jan 2 & 7.80 (0.40) & 0.20 (0.28) & 1.62 (0.97) & -2.43 (0.15) & -0.36 (0.25) & 0.34 (0.17) & -4.70 (1.38) & 2.55 (1.37)\\
     Nov 1-Jan 9 & 7.80 (0.38) & 0.41 (0.30) & 1.79 (0.97) & -2.69 (0.14) & 0.07 (0.23) & -0.13 (0.16) & -6.96 (1.26) & 5.06 (1.28)\\
   \hline
     \end{tabular}}
     \caption{Estimates of the parameters involved in fire spread from neighbours for the Australia wildfire, with standard errors in parentheses.}
     \label{tab:coef_spread_aus}
 \end{table}

\begin{table}[H]
     \centering
     \resizebox{\textwidth}{!}{
     \begin{tabular}{|c|cccccc|}
     \hline

      Time & Precipitation & Rel hum & Altitude & Soil moisture & Temperature & Pressure\\ \hline
     Nov 1- Nov 14 & 32.99 (1.13) & 2.32 (0.31) & -0.26 (0.48) & -1.38 (0.36) & -41.53 (4.22) & 7.59 (2.63)\\
     Nov 1-Nov 21 & 27.97 (1.01) & 2.36 (0.23) & 0.10 (0.41) & -1.30 (0.29) & -20.21 (3.09) & 4.54 (2.14)\\
     Nov 1-Nov 28 & 13.59 (0.49) & 2.81 (0.20) & 0.32 (0.40) & -1.92 (0.25) & -11.15 (2.58) & 1.69 (2.07)\\
     Nov 1-Dec 5 & 13.59 (0.44) & 2.36 (0.17) & 1.17 (0.36) & -1.34 (0.23) & 0.00 (2.18) & 6.37 (1.85)\\
     Nov 1-Dec 12 & 9.75 (0.32) & 1.97 (0.13) & 0.48 (0.31) & -1.00 (0.20) & -3.95 (1.84) & 5.04 (1.60)\\
     Nov 1-Dec 19 & 9.71 (0.32) & 2.25 (0.12) & -0.05 (0.29) & -0.67 (0.19) & -5.64 (1.72) & 3.92 (1.54)\\
     Nov 1-Dec 26 & 10.05 (0.30) & 2.49 (0.11) & 0.25 (0.27) & -0.91 (0.18) & -7.90 (1.57) & 4.46 (1.45)\\
     Nov 1-Jan 2 & 10.14 (0.29) & 2.43 (0.10) & 0.16 (0.25) & -1.00 (0.17) & -10.60 (1.40) & 3.73 (1.35)\\
     Nov 1-Jan 9 & 11.29 (0.26) & 3.17 (0.09) & 1.05 (0.23) & -1.94 (0.15) & -22.26 (1.25) & 7.58 (1.23)\\
   \hline
     \end{tabular}}
     \caption{Estimates of the parameters involved in recovery from fire for the Australia wildfire, with standard errors in parentheses.}
     \label{tab:coef_recovery_aus}
 \end{table}

\begin{table}[H]
     \centering
     \resizebox{\textwidth}{!}{
     \begin{tabular}{|c|cccccc|}
     \hline

      Time & Precipitation & Rel hum & Altitude & Soil moisture & Temperature & Pressure\\ \hline
     Nov 1- Nov 14 & -19.50 (8.06) & -1.72 (0.84) & -1.34 (2.33) & 1.33 (0.88) & 39.23 (8.93) & -23.18 (11.17)\\
     Nov 1-Nov 21 & -16.32 (7.73) & -1.88 (0.81) & -3.76 (2.06) & 1.31 (0.78) & 6.11 (6.60) & -25.60 (9.69)\\
     Nov 1-Nov 28 & -9.89 (3.64) & -1.92 (0.64) & -1.15 (1.59) & 2.69 (0.59) & 9.58 (5.31) & -15.53 (7.75)\\
     Nov 1-Dec 5 & -12.68 (3.73) & -1.73 (0.55) & -3.09 (1.35) & 3.10 (0.51) & 13.61 (4.67) & -25.92 (6.58)\\
     Nov 1-Dec 12 & -7.71 (2.55) & -1.07 (0.48) & -2.34 (1.21) & 2.33 (0.47) & 13.96 (4.02) & -25.20 (5.96)\\
     Nov 1-Dec 19 & -7.87 (2.42) & -0.40 (0.42) & -1.87 (1.08) & 1.93 (0.43) & 13.44 (3.52) & -24.49 (5.38)\\
     Nov 1-Dec 26 & -8.18 (2.37) & 0.11 (0.39) & -1.54 (1.02) & 1.48 (0.42) & 7.42 (3.24) & -22.17 (5.12)\\
     Nov 1-Jan 2 & -7.20 (2.18) & 0.01 (0.37) & -0.30 (0.94) & 1.31 (0.40) & 3.68 (3.02) & -16.70 (4.78)\\
     Nov 1-Jan 9 & -7.52 (2.07) & 0.34 (0.33) & -0.20 (0.87) & 0.85 (0.38) & 2.97 (2.72) & -18.44 (4.44)\\
   \hline
     \end{tabular}}
     \caption{Estimates of the parameters involved in intrinsic fire ignition for the Australia wildfire, with standard errors in parentheses.}
     \label{tab:coef_intrinsic_aus}
 \end{table}

 \begin{table}[H]
     \centering
     \resizebox{\textwidth}{!}{
     \begin{tabular}{|c|cccccccc|}
     \hline

      Time & Intensity & Wind & Precipitation & Rel hum & Altitude & Soil moisture & Temperature & Pressure\\ \hline
     Nov 1- Nov 14 & 1.02 (0.00) & 1.00 (0.00) & 0.93 (0.03) & 0.87 (0.02) & 1.00 (0.00) & 1.07 (0.01) & 1.16 (0.02) & 1.00 (0.00)\\
     Nov 1-Nov 21 & 1.02 (0.00) & 1.00 (0.00) & 0.94 (0.02) & 0.90 (0.01) & 1.00 (0.00) & 1.07 (0.01) & 1.09 (0.02) & 1.00 (0.00)\\
     Nov 1-Nov 28 & 1.01 (0.00) & 1.00 (0.00) & 0.99 (0.01) & 0.90 (0.01) & 1.00 (0.00) & 1.05 (0.01) & 1.07 (0.02) & 1.00 (0.00)\\
     Nov 1-Dec 5 & 1.01 (0.00) & 1.00 (0.00) & 0.99 (0.01) & 0.92 (0.01) & 1.00 (0.00) & 1.03 (0.01) & 1.02 (0.01) & 1.00 (0.00)\\
     Nov 1-Dec 12 & 1.01 (0.00) & 1.00 (0.00) & 0.99 (0.01) & 0.91 (0.01) & 1.00 (0.00) & 1.03 (0.01) & 1.01 (0.01) & 1.00 (0.00)\\
     Nov 1-Dec 19 & 1.01 (0.00) & 1.00 (0.00) & 1.00 (0.01) & 0.90 (0.01) & 1.00 (0.00) & 1.02 (0.01) & 0.99 (0.01) & 1.00 (0.00)\\
     Nov 1-Dec 26 & 1.01 (0.00) & 1.00 (0.00) & 1.00 (0.01) & 0.92 (0.01) & 1.00 (0.00) & 1.02 (0.00) & 0.96 (0.01) & 1.01 (0.00)\\
     Nov 1-Jan 2 & 1.01 (0.00) & 1.00 (0.00) & 1.01 (0.01) & 0.89 (0.01) & 1.00 (0.00) & 1.01 (0.00) & 0.97 (0.01) & 1.00 (0.00)\\
     Nov 1-Jan 9 & 1.01 (0.00) & 1.00 (0.00) & 1.01 (0.01) & 0.88 (0.01) & 1.00 (0.00) & 1.00 (0.00) & 0.96 (0.01) & 1.01 (0.00)\\
   \hline
     \end{tabular}}
     \caption{Estimates of rescaled parameters involved in fire spread from neighbours for the Australia wildfire. Each cell gives the multiplicative change in the rate, with its standard error in parentheses; 1.00 denotes no effect. For all variables except soil moisture the change is the mean absolute daily change; for soil moisture it is 1 kg/m$^2$. Altitude does not vary in time, so its entry is 1.00 by construction.}
     \label{tab:spread_scaled_aus}
 \end{table}

\begin{table}[H]
     \centering
     \resizebox{\textwidth}{!}{
     \begin{tabular}{|c|cccccc|}
     \hline

      Time & Precipitation & Rel hum & Altitude & Soil moisture & Temperature & Pressure\\ \hline
     Nov 1- Nov 14 & 1.24 (0.01) & 1.12 (0.02) & 1.00 (0.00) & 0.97 (0.01) & 0.79 (0.02) & 1.01 (0.00)\\
     Nov 1-Nov 21 & 1.20 (0.01) & 1.12 (0.01) & 1.00 (0.00) & 0.97 (0.01) & 0.89 (0.02) & 1.01 (0.00)\\
     Nov 1-Nov 28 & 1.09 (0.00) & 1.15 (0.01) & 1.00 (0.00) & 0.95 (0.01) & 0.94 (0.01) & 1.00 (0.00)\\
     Nov 1-Dec 5 & 1.09 (0.00) & 1.12 (0.01) & 1.00 (0.00) & 0.97 (0.01) & 1.00 (0.01) & 1.01 (0.00)\\
     Nov 1-Dec 12 & 1.07 (0.00) & 1.10 (0.01) & 1.00 (0.00) & 0.98 (0.00) & 0.98 (0.01) & 1.01 (0.00)\\
     Nov 1-Dec 19 & 1.07 (0.00) & 1.11 (0.01) & 1.00 (0.00) & 0.98 (0.00) & 0.97 (0.01) & 1.01 (0.00)\\
     Nov 1-Dec 26 & 1.07 (0.00) & 1.13 (0.01) & 1.00 (0.00) & 0.98 (0.00) & 0.96 (0.01) & 1.01 (0.00)\\
     Nov 1-Jan 2 & 1.07 (0.00) & 1.12 (0.01) & 1.00 (0.00) & 0.98 (0.00) & 0.94 (0.01) & 1.01 (0.00)\\
     Nov 1-Jan 9 & 1.08 (0.00) & 1.17 (0.00) & 1.00 (0.00) & 0.95 (0.00) & 0.88 (0.01) & 1.01 (0.00)\\
   \hline
     \end{tabular}}
     \caption{Estimates of rescaled parameters involved in recovery from fire for the Australia wildfire. Each cell gives the multiplicative change in the rate, with its standard error in parentheses; 1.00 denotes no effect. For all variables except soil moisture the change is the mean absolute daily change; for soil moisture it is 1 kg/m$^2$. Altitude does not vary in time, so its entry is 1.00 by construction. }
     \label{tab:recovery_scaled_aus}
 \end{table}

\begin{table}[H]
     \centering
     \resizebox{\textwidth}{!}{
     \begin{tabular}{|c|cccccc|}
     \hline

      Time & Precipitation & Rel hum & Altitude & Soil moisture & Temperature & Pressure\\ \hline
     Nov 1- Nov 14 & 0.88 (0.05) & 0.92 (0.04) & 1.00 (0.00) & 1.03 (0.02) & 1.25 (0.06) & 0.96 (0.02)\\
     Nov 1-Nov 21 & 0.90 (0.05) & 0.91 (0.04) & 1.00 (0.00) & 1.03 (0.02) & 1.04 (0.04) & 0.95 (0.02)\\
     Nov 1-Nov 28 & 0.94 (0.02) & 0.91 (0.03) & 1.00 (0.00) & 1.07 (0.02) & 1.06 (0.03) & 0.97 (0.01)\\
     Nov 1-Dec 5 & 0.92 (0.02) & 0.92 (0.02) & 1.00 (0.00) & 1.08 (0.01) & 1.08 (0.03) & 0.95 (0.01)\\
     Nov 1-Dec 12 & 0.95 (0.02) & 0.95 (0.02) & 1.00 (0.00) & 1.06 (0.01) & 1.08 (0.02) & 0.96 (0.01)\\
     Nov 1-Dec 19 & 0.95 (0.02) & 0.98 (0.02) & 1.00 (0.00) & 1.05 (0.01) & 1.08 (0.02) & 0.96 (0.01)\\
     Nov 1-Dec 26 & 0.95 (0.01) & 1.01 (0.02) & 1.00 (0.00) & 1.04 (0.01) & 1.04 (0.02) & 0.96 (0.01)\\
     Nov 1-Jan 2 & 0.95 (0.01) & 1.00 (0.02) & 1.00 (0.00) & 1.03 (0.01) & 1.02 (0.02) & 0.97 (0.01)\\
     Nov 1-Jan 9 & 0.95 (0.01) & 1.02 (0.02) & 1.00 (0.00) & 1.02 (0.01) & 1.02 (0.02) & 0.97 (0.01)\\
   \hline
     \end{tabular}}
     \caption{Estimates of rescaled parameters involved in intrinsic fire ignition for the Australia wildfire. Each cell gives the multiplicative change in the rate, with its standard error in parentheses; 1.00 denotes no effect. For all variables except soil moisture the change is the mean absolute daily change; for soil moisture it is 1 kg/m$^2$. Altitude does not vary in time, so its entry is 1.00 by construction.}
     \label{tab:intrinsic_scaled_aus}
 \end{table}

\begin{figure}[H]
     \centering
     \begin{subfigure}[b]{0.42\linewidth}
         \centering
         \includegraphics[width=\textwidth]{Images/aus_obs_nov_15.png}
         \caption{Observed fire on Nov 15, 2019}
         \label{fig:aus_nov_15_obs}
     \end{subfigure}
     \hfill
     \begin{subfigure}[b]{0.42\linewidth}
         \centering
         \includegraphics[width=\textwidth]{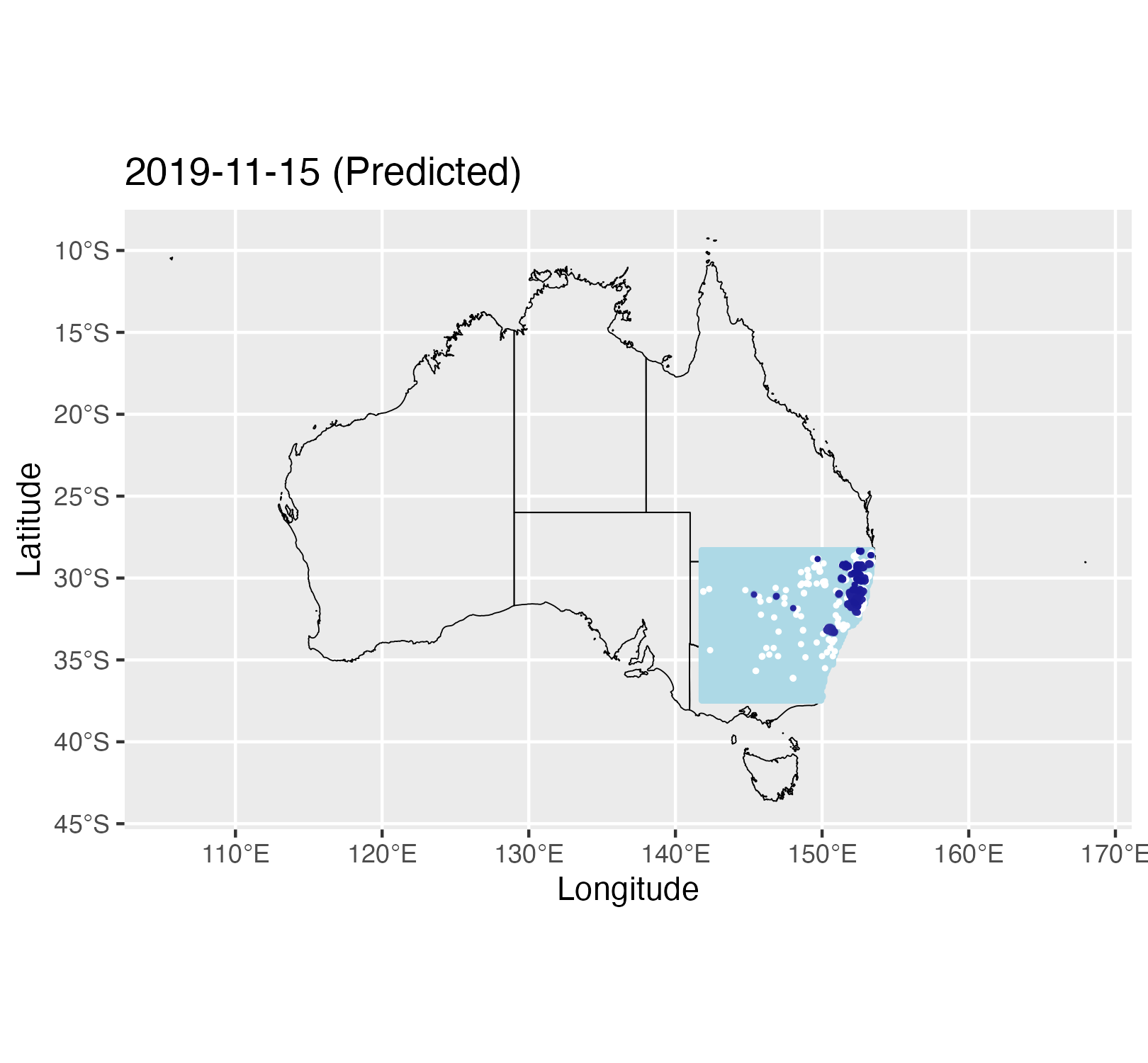}
         \caption{Predicted fire on Nov 15, 2019}
         \label{fig:aus_nov_15_pred}
     \end{subfigure}
      \begin{subfigure}[b]{0.42\linewidth}
         \centering
         \includegraphics[width=\textwidth]{Images/aus_obs_nov_16.png}
         \caption{Observed fire on Nov 16, 2019}
         \label{fig:aus_nov_16_obs}
     \end{subfigure}
     \hfill
     \begin{subfigure}[b]{0.42\linewidth}
         \centering
         \includegraphics[width=\textwidth]{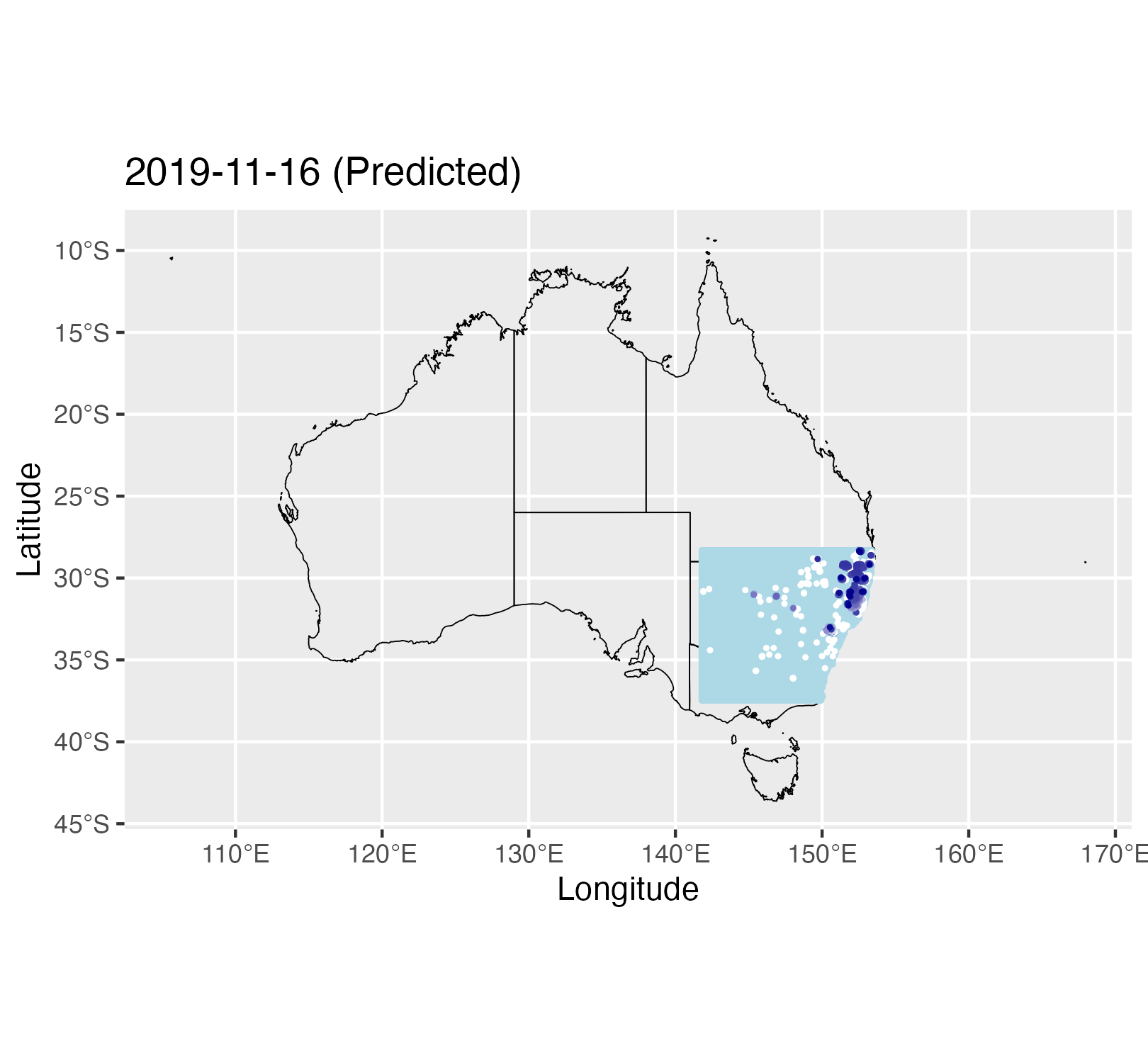}
         \caption{Predicted fire on Nov 16, 2019}
         \label{fig:aus_nov_16_pred}
     \end{subfigure}
      \begin{subfigure}[b]{0.42\linewidth}
         \centering
         \includegraphics[width=\textwidth]{Images/aus_obs_nov_17.png}
         \caption{Observed fire on Nov 17, 2019}
         \label{fig:aus_nov_17_obs}
     \end{subfigure}
     \hfill
     \begin{subfigure}[b]{0.42\linewidth}
         \centering
         \includegraphics[width=\textwidth]{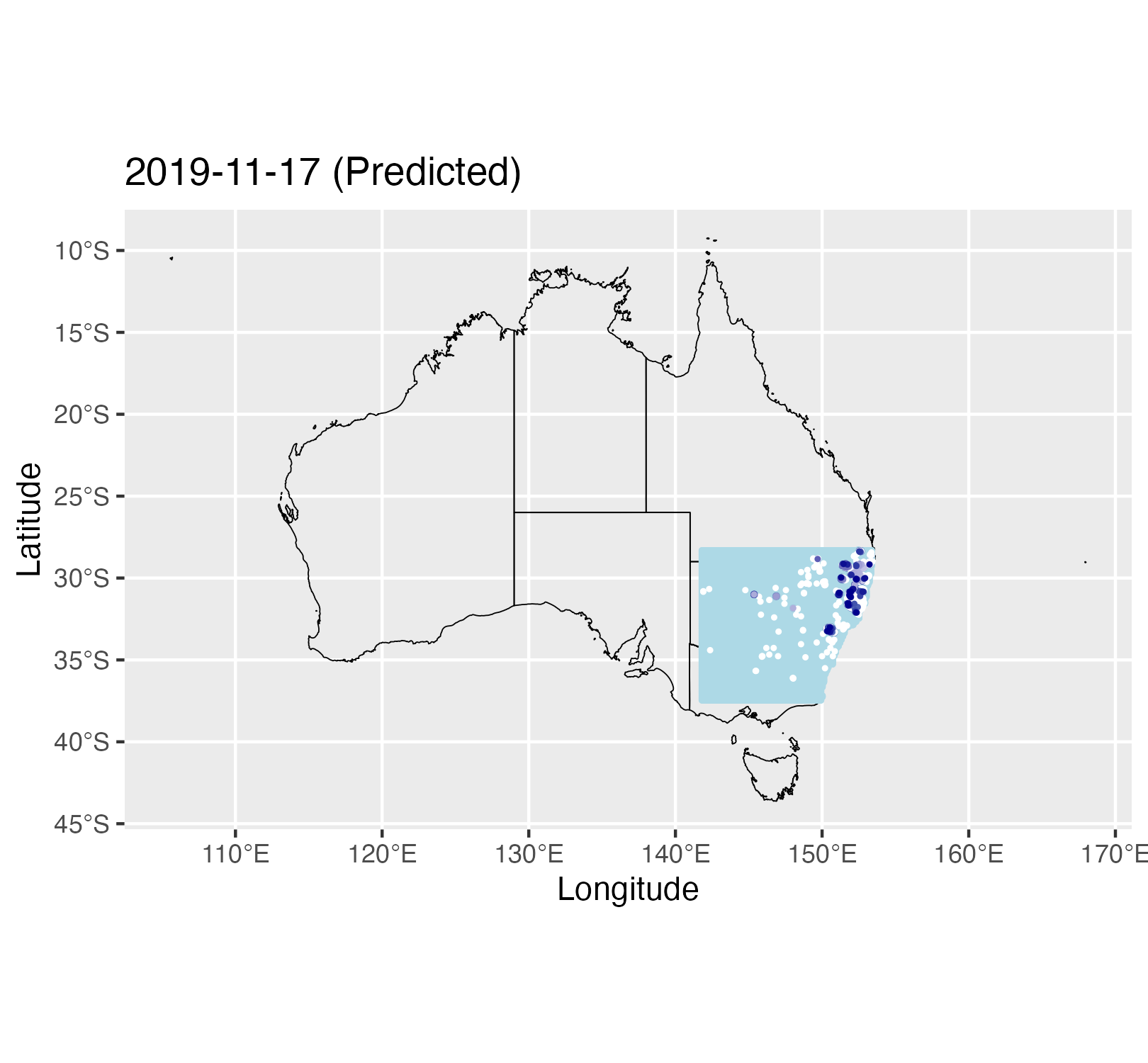}
         \caption{Predicted fire on Nov 17, 2019}
         \label{fig:aus_nov_17_pred}
     \end{subfigure}
     \caption{Observed vs 1, 2 and 3 day predictions}
     \label{fig:nov_15_16_17}
\end{figure}

\begin{figure}[H]
     \centering
     \begin{subfigure}[b]{0.42\linewidth}
         \centering
         \includegraphics[width=\textwidth]{Images/aus_obs_nov_22.png}
         \caption{Observed fire on Nov 22, 2019}
         \label{fig:aus_nov_22_obs}
     \end{subfigure}
     \hfill
     \begin{subfigure}[b]{0.42\linewidth}
         \centering
         \includegraphics[width=\textwidth]{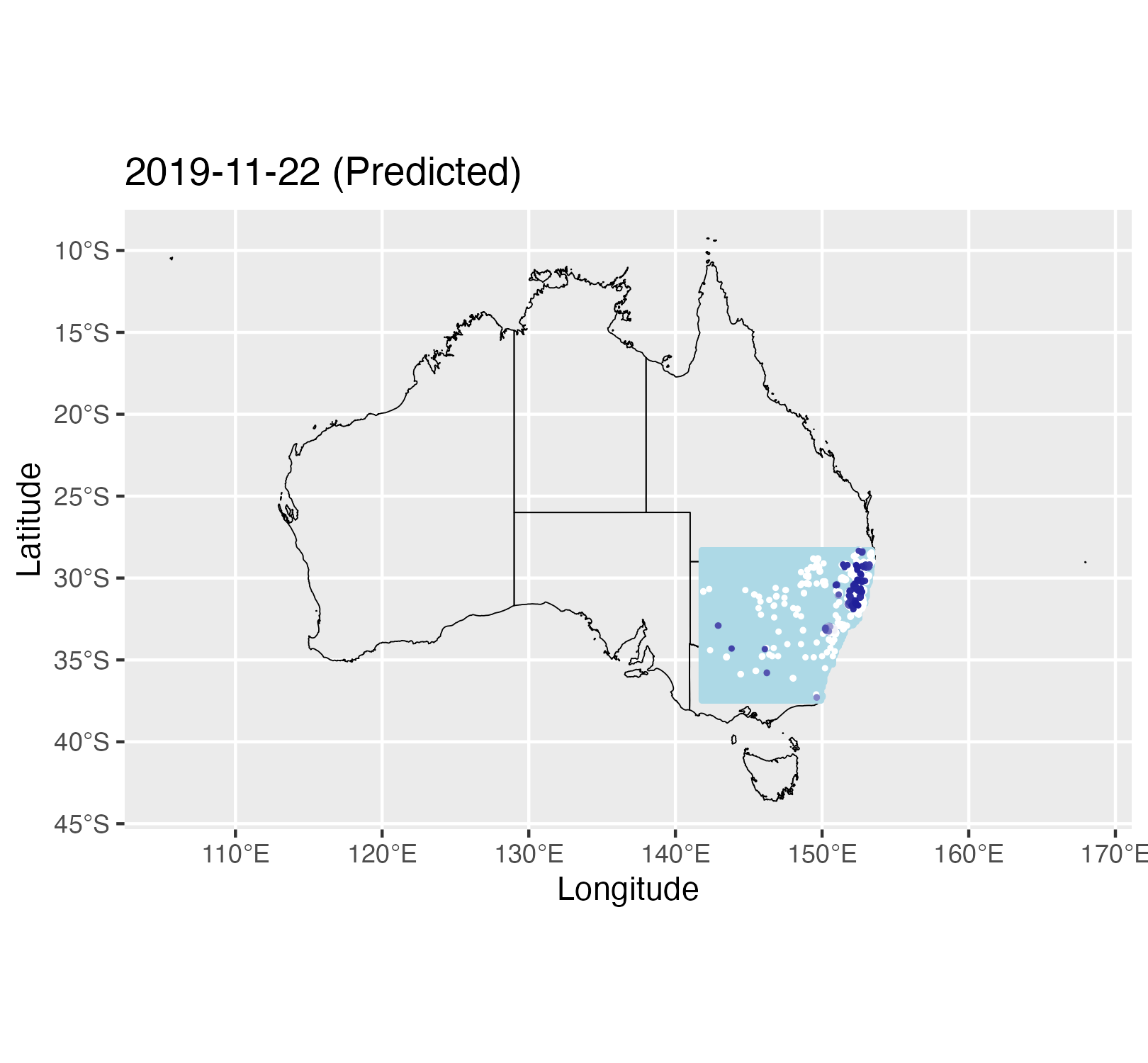}
         \caption{Predicted fire on Nov 22, 2019}
         \label{fig:aus_nov_22_pred}
     \end{subfigure}
      \begin{subfigure}[b]{0.42\linewidth}
         \centering
         \includegraphics[width=\textwidth]{Images/aus_obs_nov_23.png}
         \caption{Observed fire on Nov 23, 2019}
         \label{fig:aus_nov_23_obs}
     \end{subfigure}
     \hfill
     \begin{subfigure}[b]{0.42\linewidth}
         \centering
         \includegraphics[width=\textwidth]{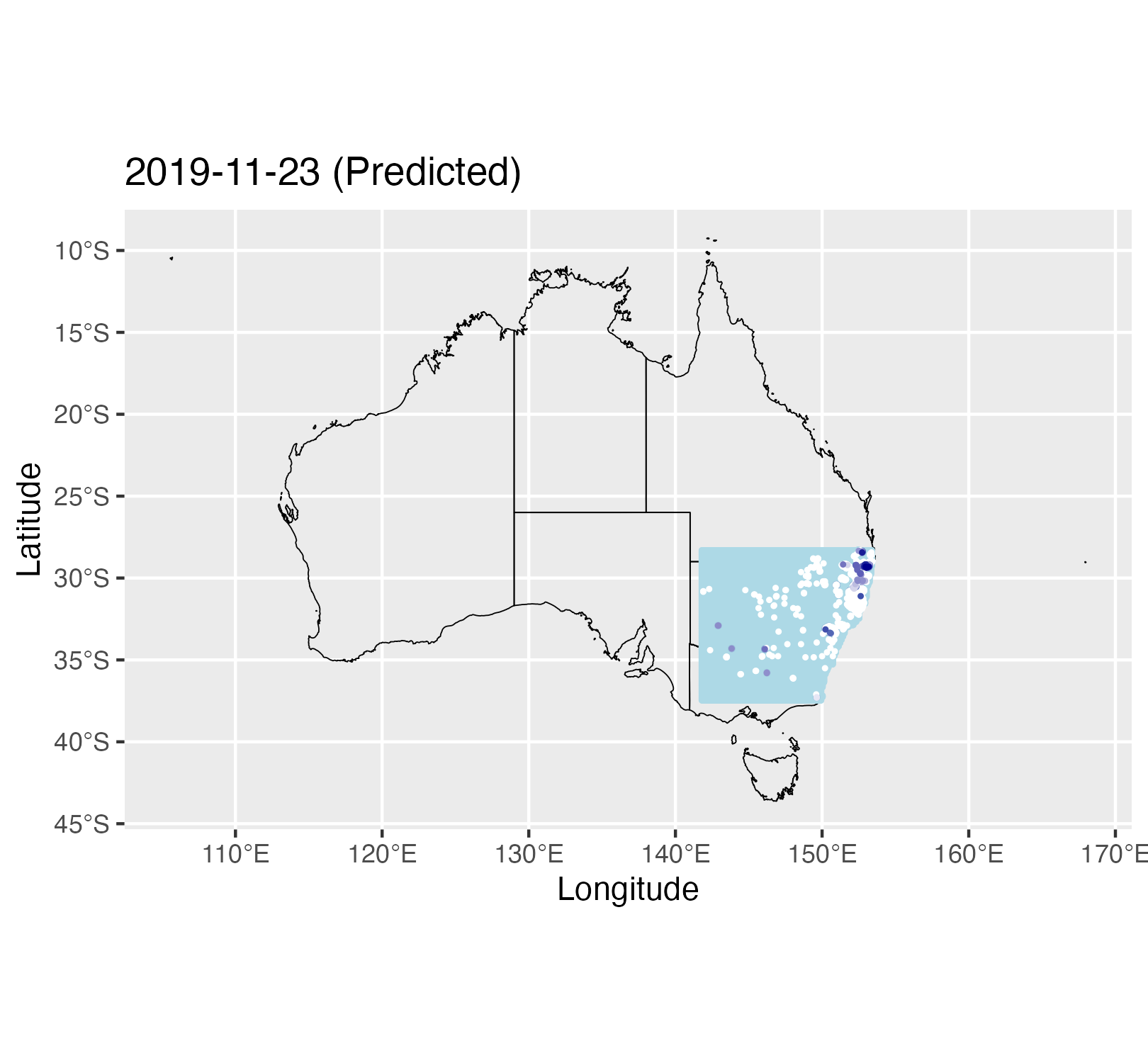}
         \caption{Predicted fire on Nov 23, 2019}
         \label{fig:aus_nov_23_pred}
     \end{subfigure}
      \begin{subfigure}[b]{0.42\linewidth}
         \centering
         \includegraphics[width=\textwidth]{Images/aus_obs_nov_24.png}
         \caption{Observed fire on Nov 24, 2019}
         \label{fig:aus_nov_24_obs}
     \end{subfigure}
     \hfill
     \begin{subfigure}[b]{0.42\linewidth}
         \centering
         \includegraphics[width=\textwidth]{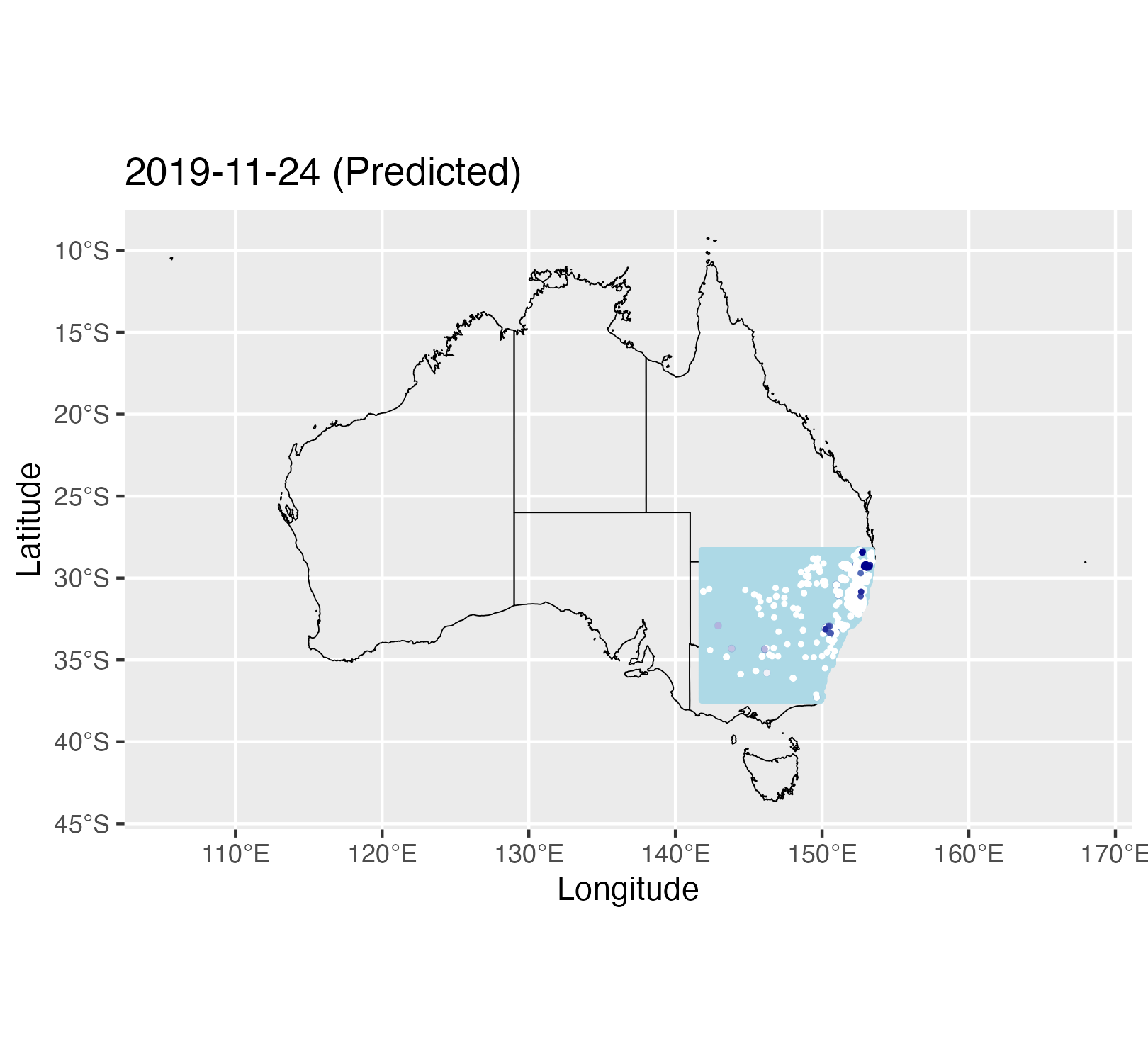}
         \caption{Predicted fire on Nov 24, 2019}
         \label{fig:aus_nov_24_pred}
     \end{subfigure}
     \caption{Observed vs 1, 2 and 3 day predictions}
     \label{fig:nov_22_23_24}
\end{figure}

\begin{figure}[H]
     \centering
     \begin{subfigure}[b]{0.42\linewidth}
         \centering
         \includegraphics[width=\textwidth]{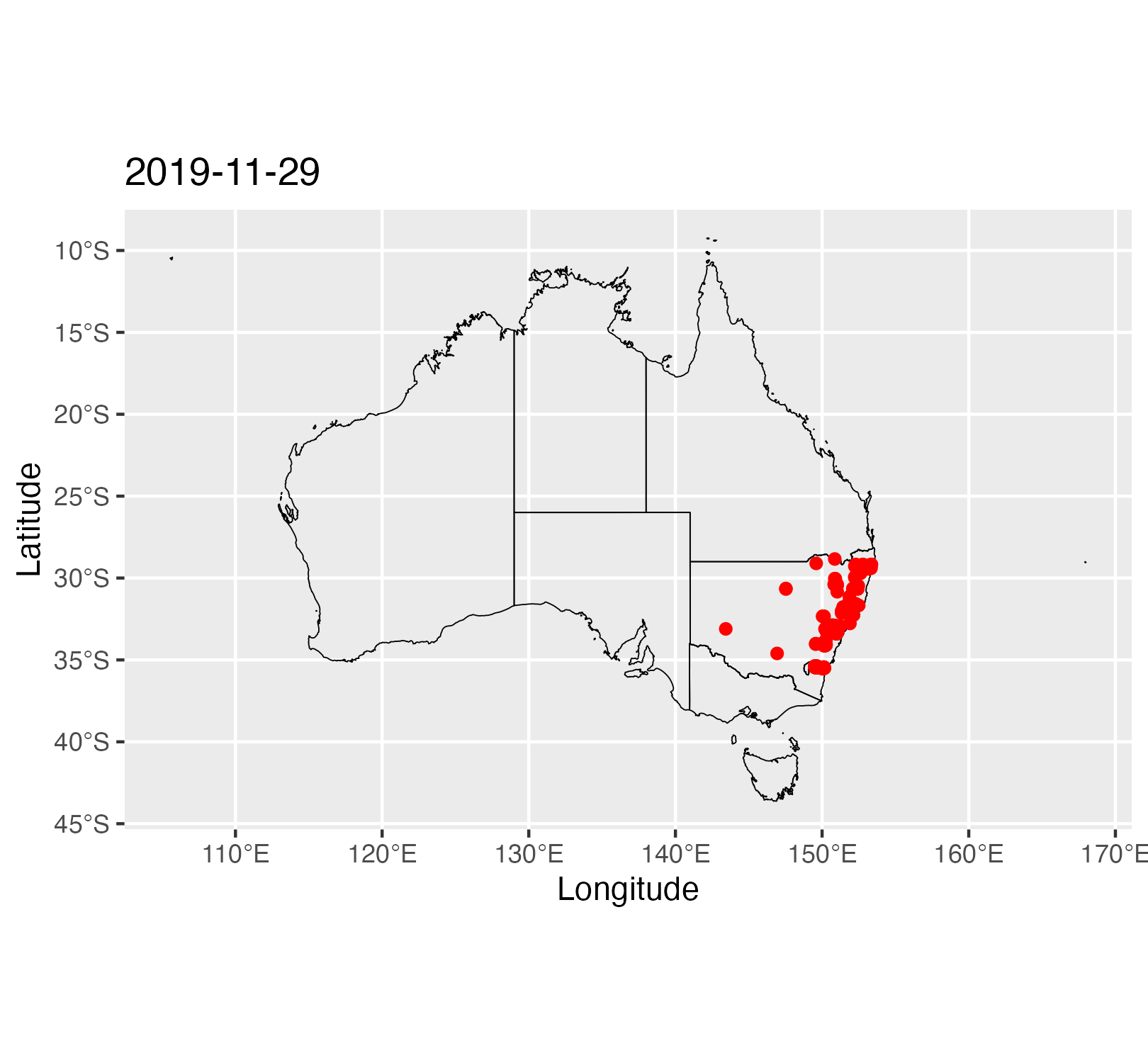}
         \caption{Observed fire on Nov 29, 2019}
         \label{fig:aus_nov_29_obs}
     \end{subfigure}
     \hfill
     \begin{subfigure}[b]{0.42\linewidth}
         \centering
         \includegraphics[width=\textwidth]{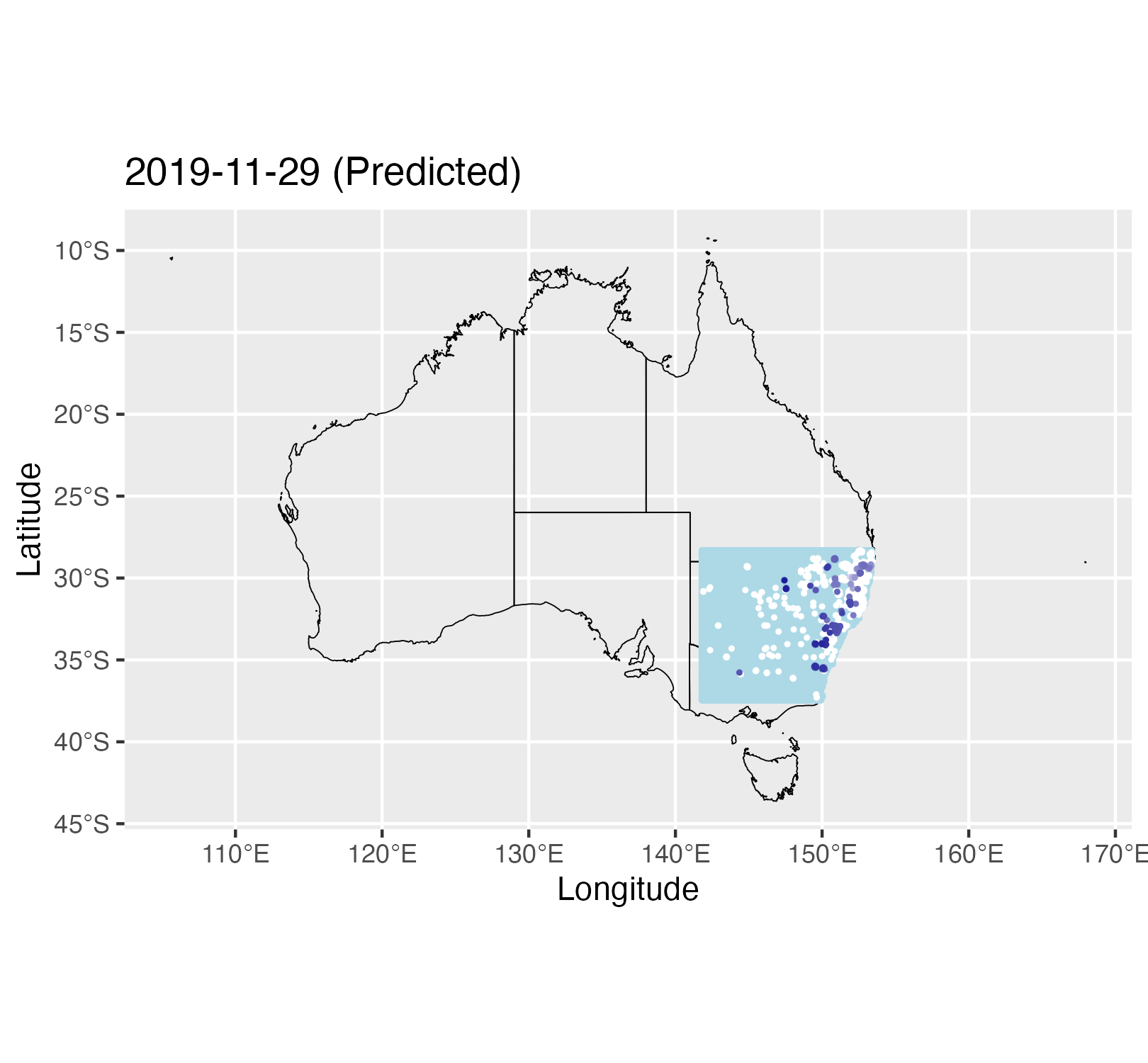}
         \caption{Predicted fire on Nov 29, 2019}
         \label{fig:aus_nov_29_pred}
     \end{subfigure}
      \begin{subfigure}[b]{0.42\linewidth}
         \centering
         \includegraphics[width=\textwidth]{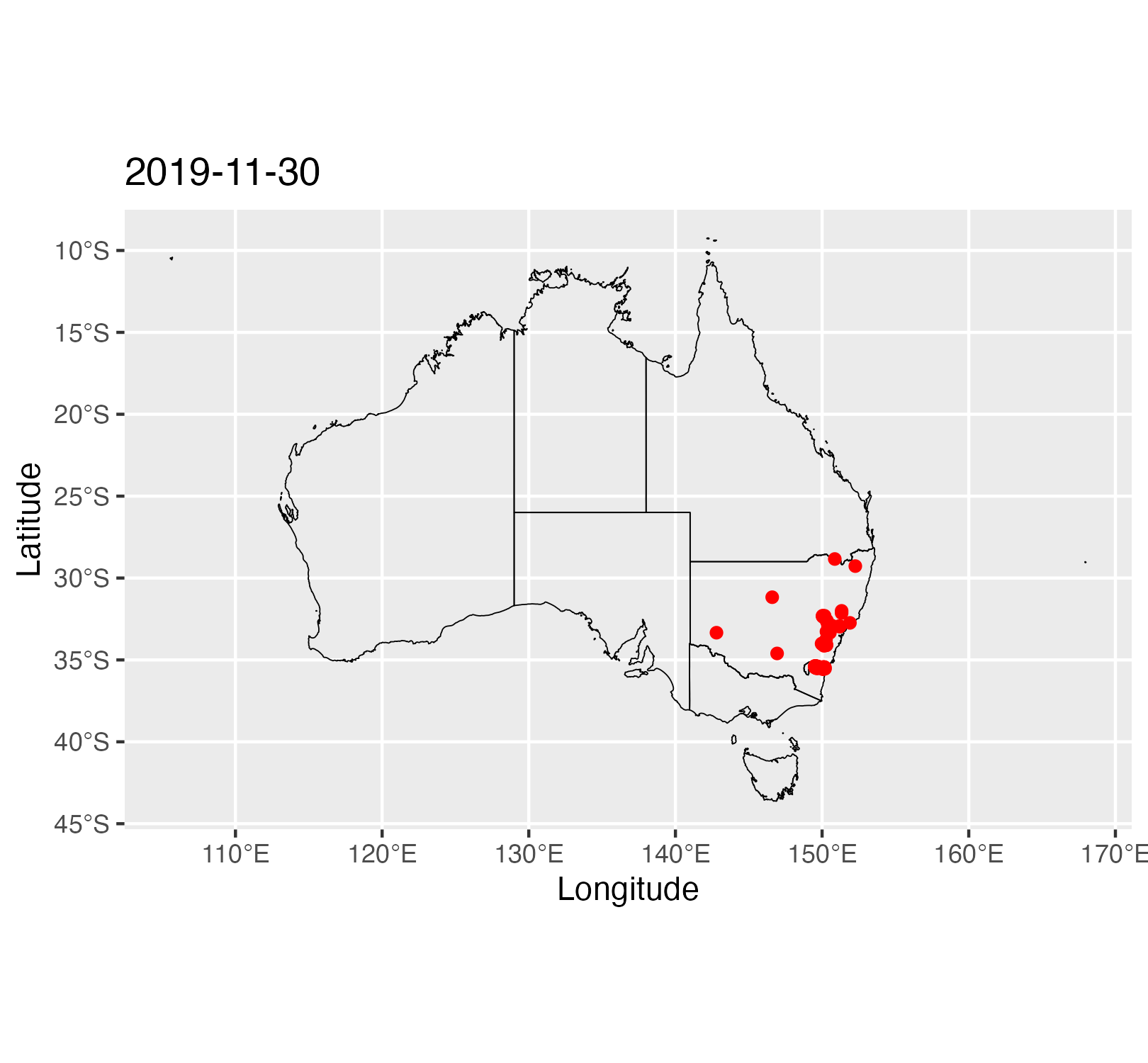}
         \caption{Observed fire on Nov 30, 2019}
         \label{fig:aus_nov_30_obs}
     \end{subfigure}
     \hfill
     \begin{subfigure}[b]{0.42\linewidth}
         \centering
         \includegraphics[width=\textwidth]{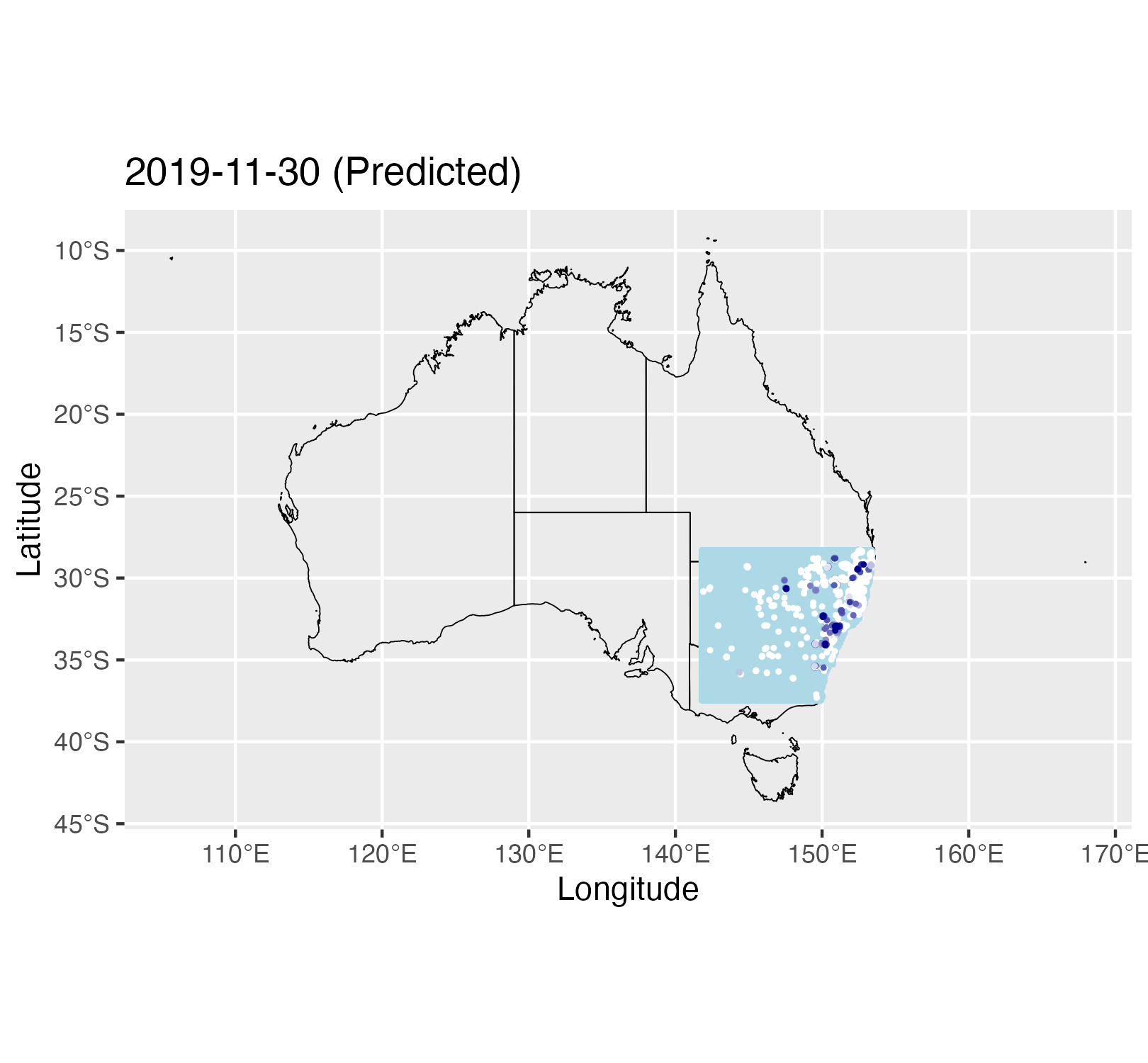}
         \caption{Predicted fire on Nov 30, 2019}
         \label{fig:aus_nov_30_pred}
     \end{subfigure}
      \begin{subfigure}[b]{0.42\linewidth}
         \centering
         \includegraphics[width=\textwidth]{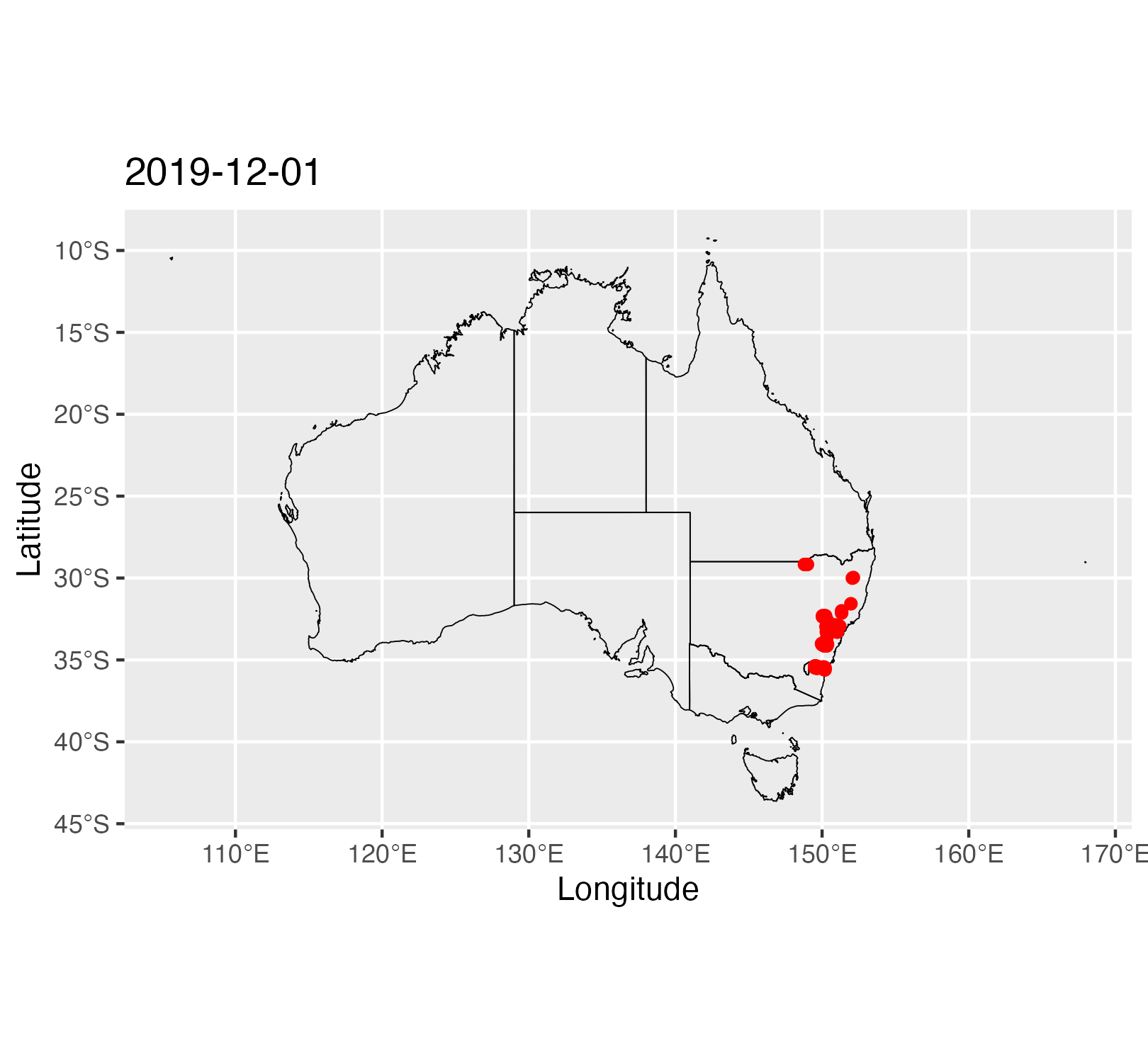}
         \caption{Observed fire on Dec 01, 2019}
         \label{fig:aus_dec_01_obs}
     \end{subfigure}
     \hfill
     \begin{subfigure}[b]{0.42\linewidth}
         \centering
         \includegraphics[width=\textwidth]{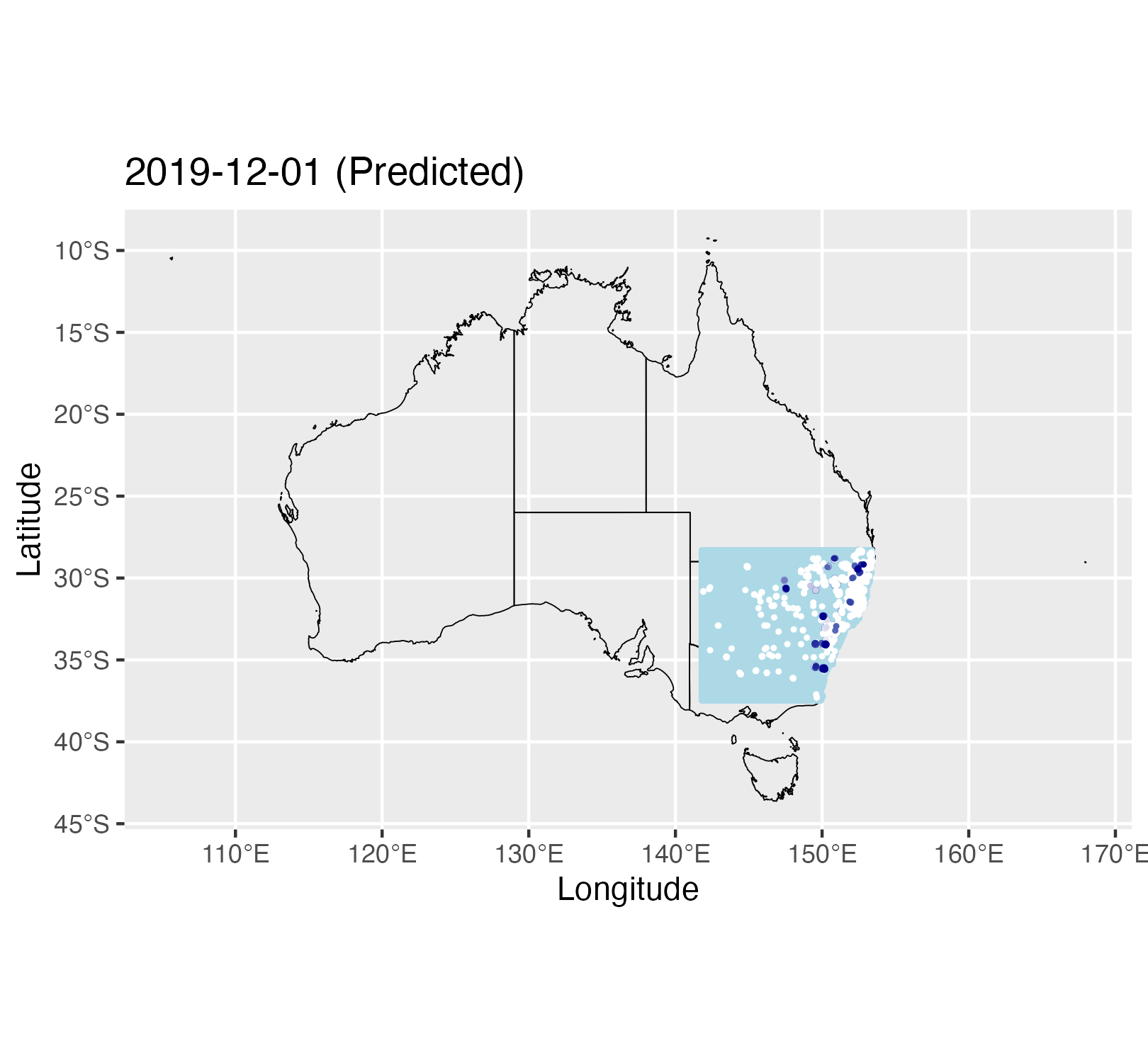}
         \caption{Predicted fire on Dec 01, 2019}
         \label{fig:aus_dec_01_pred}
     \end{subfigure}
     \caption{Observed vs 1, 2 and 3 day predictions}
     \label{fig:nov_29_30_dec_01}
\end{figure}

\begin{figure}[H]
     \centering
 \begin{subfigure}[b]{0.42\linewidth}
         \centering
         \includegraphics[width=\textwidth]{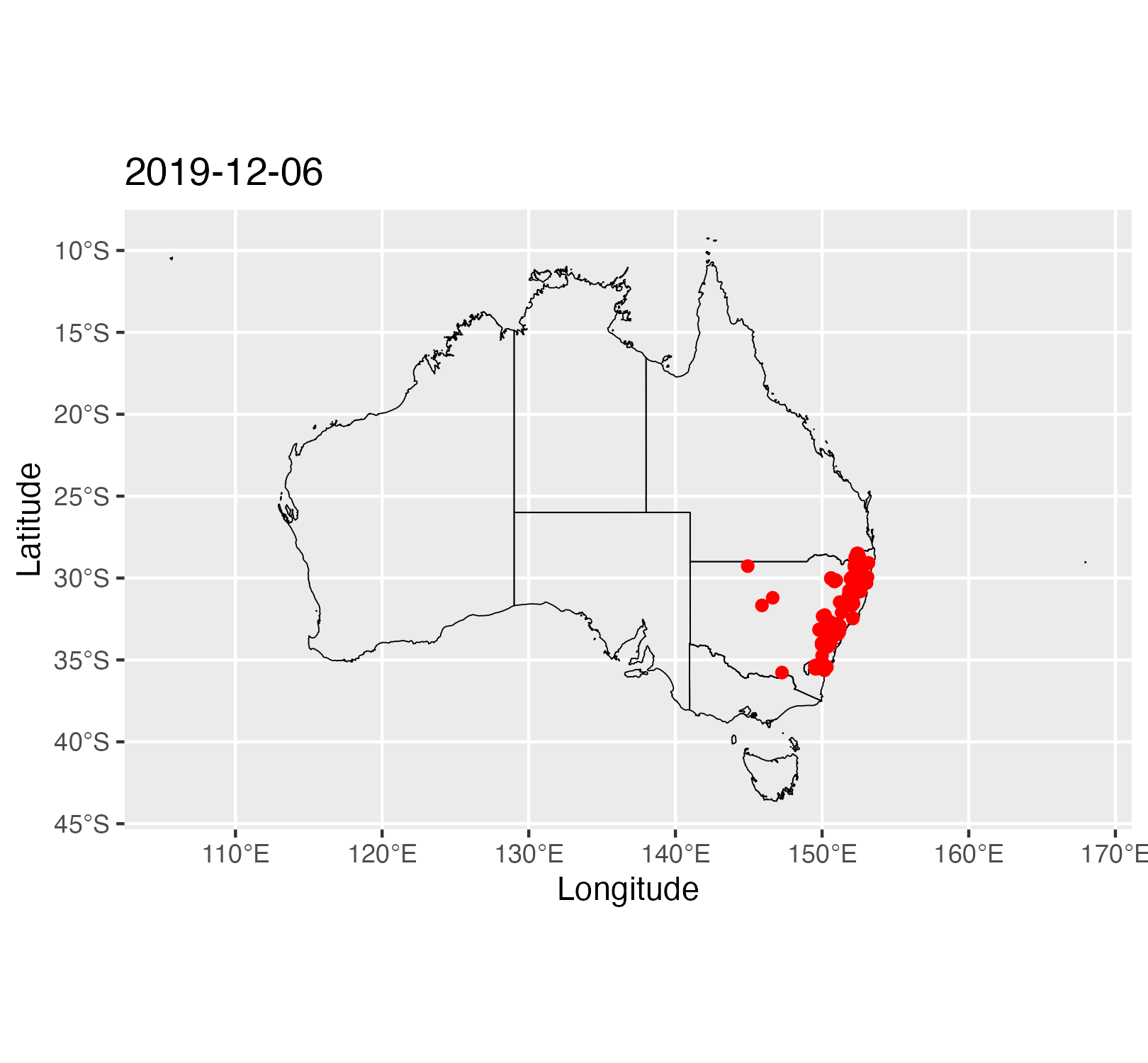}
         \caption{Observed fire on Dec 06, 2019}
         \label{fig:aus_dec_06_obs}
     \end{subfigure}
     \hfill
 \begin{subfigure}[b]{0.42\linewidth}
         \centering
         \includegraphics[width=\textwidth]{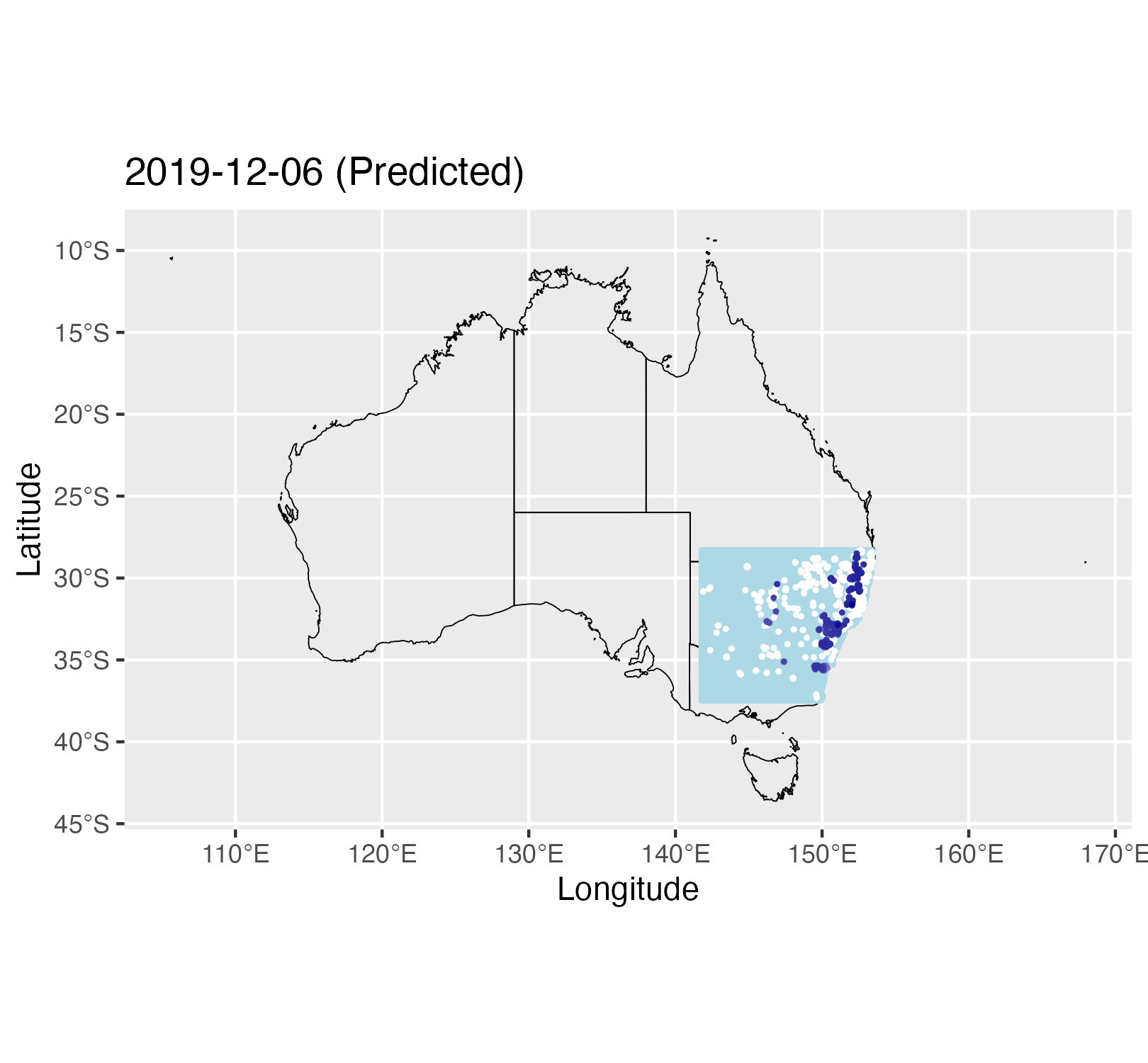}
         \caption{Predicted fire on Dec 06, 2019}
         \label{fig:aus_dec_06_pred}
     \end{subfigure}
 \begin{subfigure}[b]{0.42\linewidth}
         \centering
         \includegraphics[width=\textwidth]{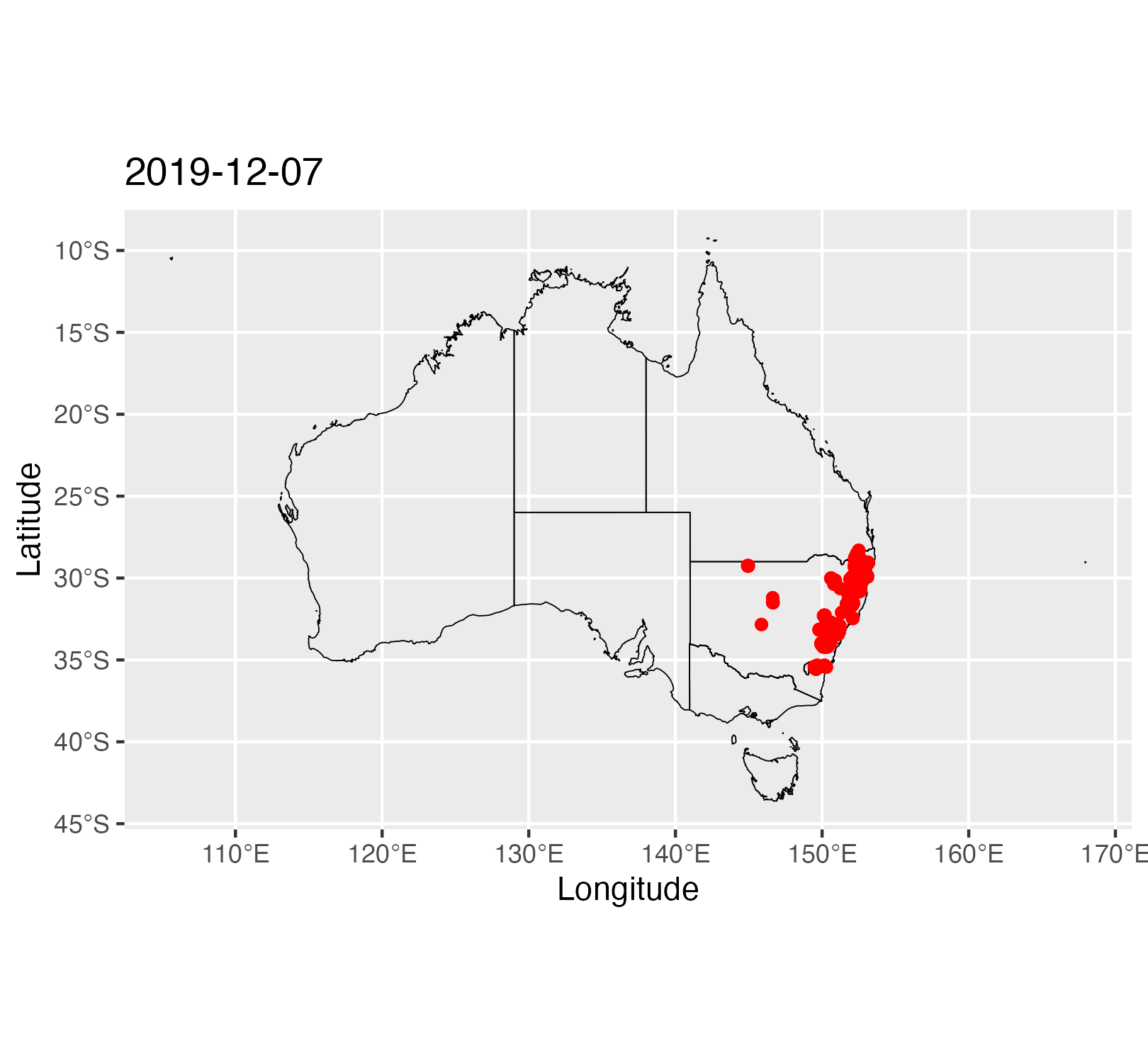}
         \caption{Observed fire on Dec 07, 2019}
         \label{fig:aus_dec_07_obs}
     \end{subfigure}
     \hfill
 \begin{subfigure}[b]{0.42\linewidth}
         \centering
         \includegraphics[width=\textwidth]{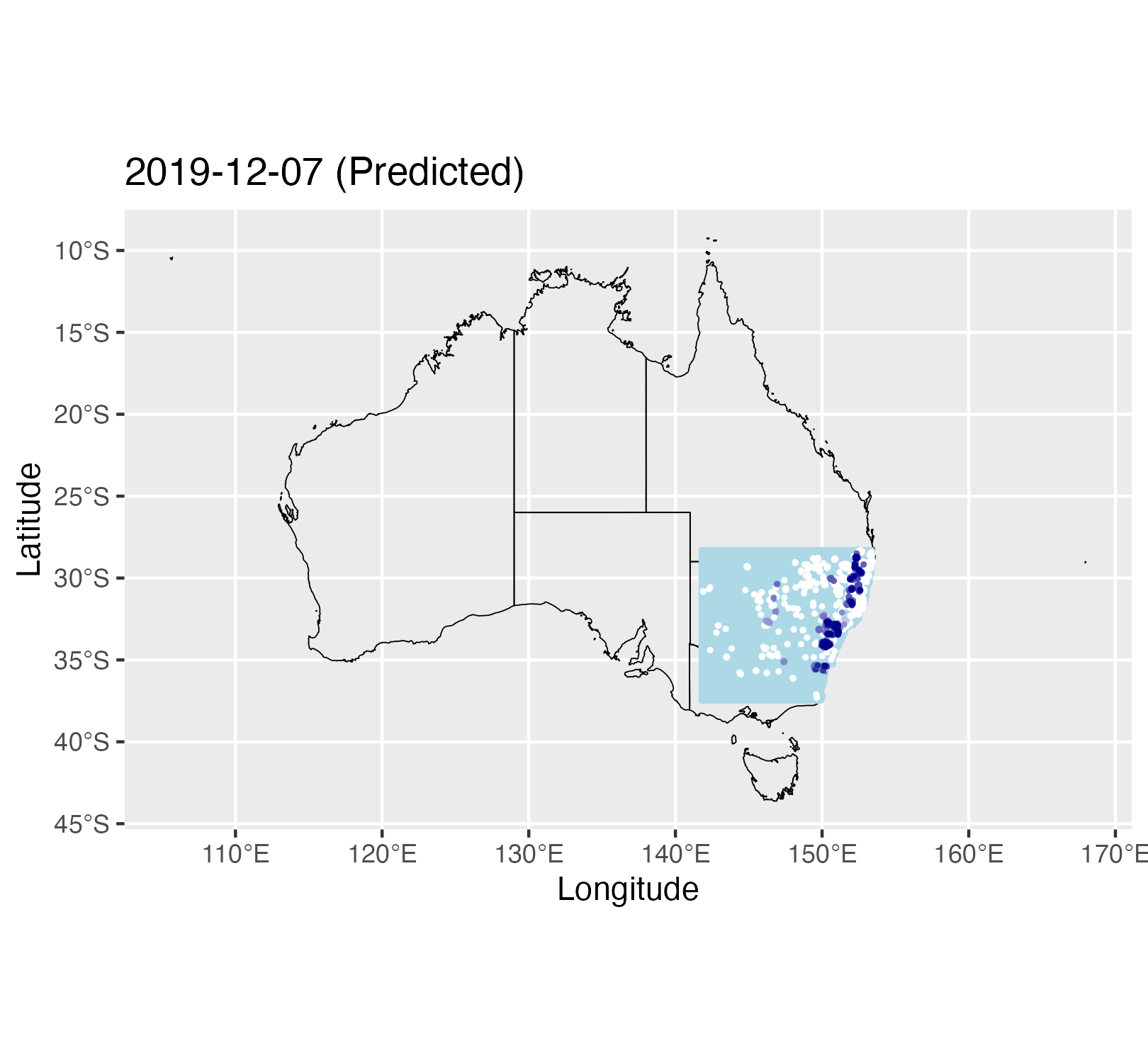}
         \caption{Predicted fire on Dec 07, 2019}
         \label{fig:aus_dec_07_pred}
     \end{subfigure}
 \begin{subfigure}[b]{0.42\linewidth}
         \centering
         \includegraphics[width=\textwidth]{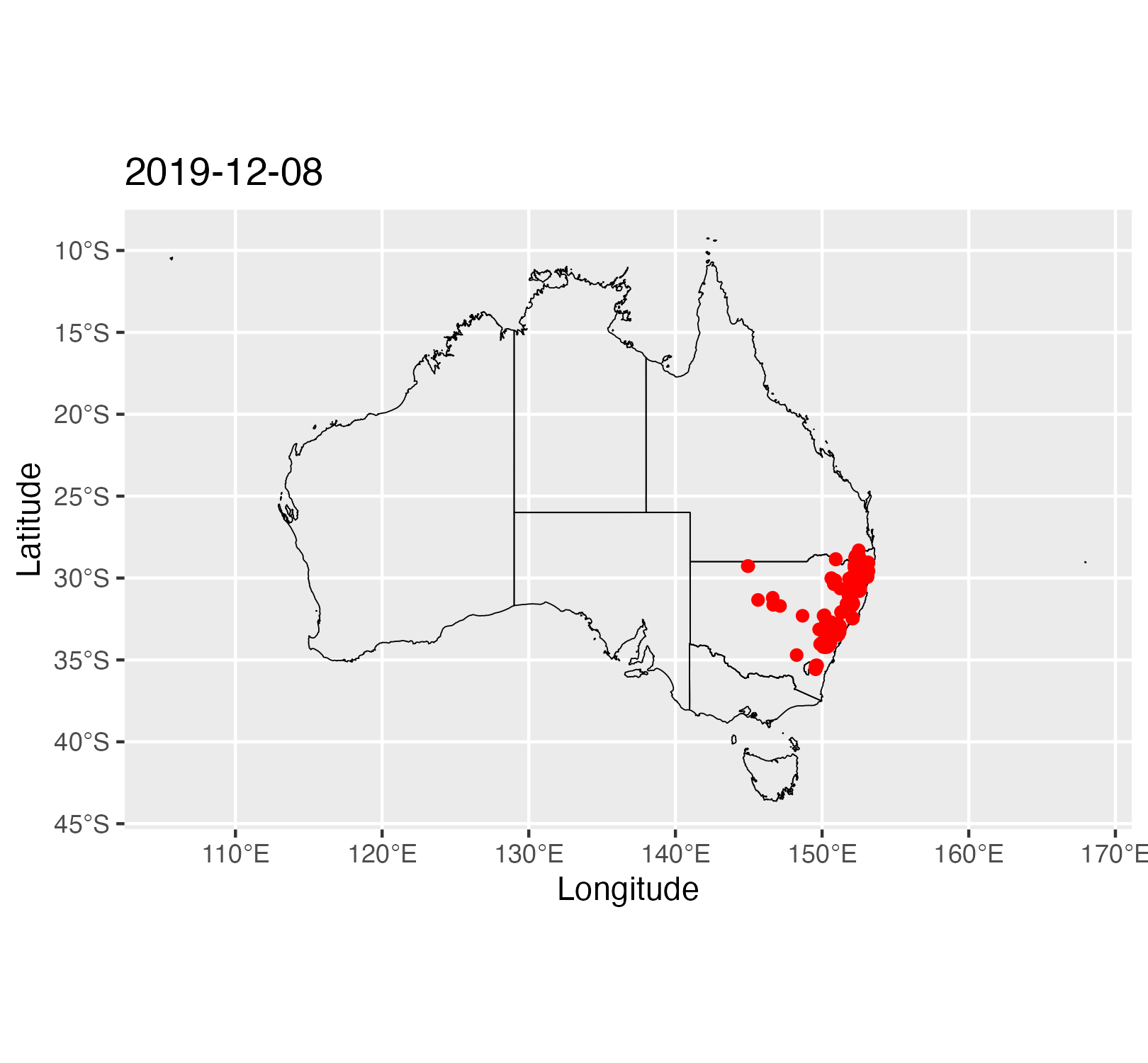}
         \caption{Observed fire on Dec 08, 2019}
         \label{fig:aus_dec_08_obs}
     \end{subfigure}
     \hfill
 \begin{subfigure}[b]{0.42\linewidth}
         \centering
         \includegraphics[width=\textwidth]{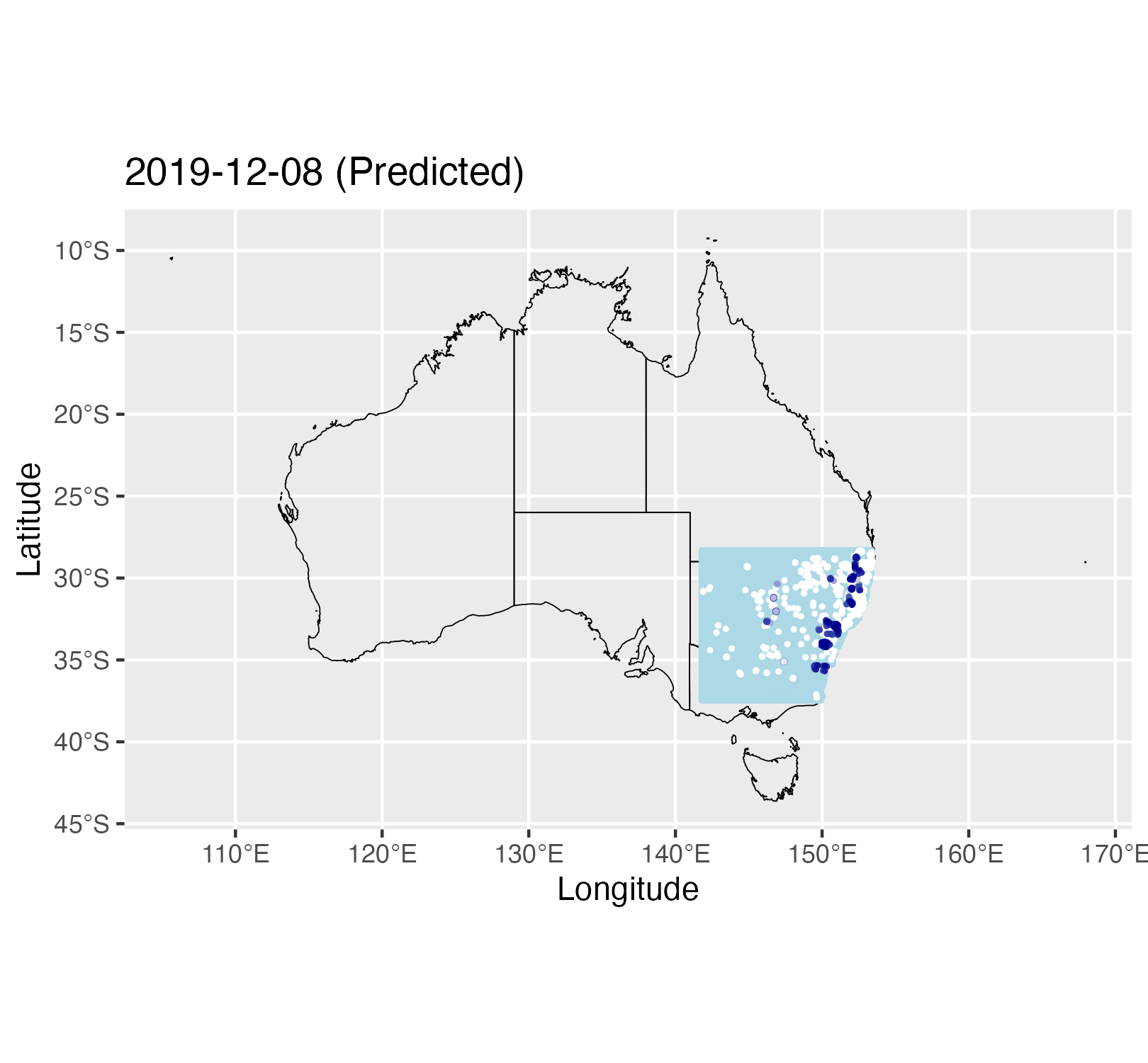}
         \caption{Predicted fire on Dec 08, 2019}
         \label{fig:aus_dec_08_pred}
     \end{subfigure}
     \caption{Observed vs 1, 2 and 3 day predictions}
     \label{fig:dec_06_07_08}
\end{figure}

\begin{figure}[H]
     \centering
 \begin{subfigure}[b]{0.42\linewidth}
         \centering
         \includegraphics[width=\textwidth]{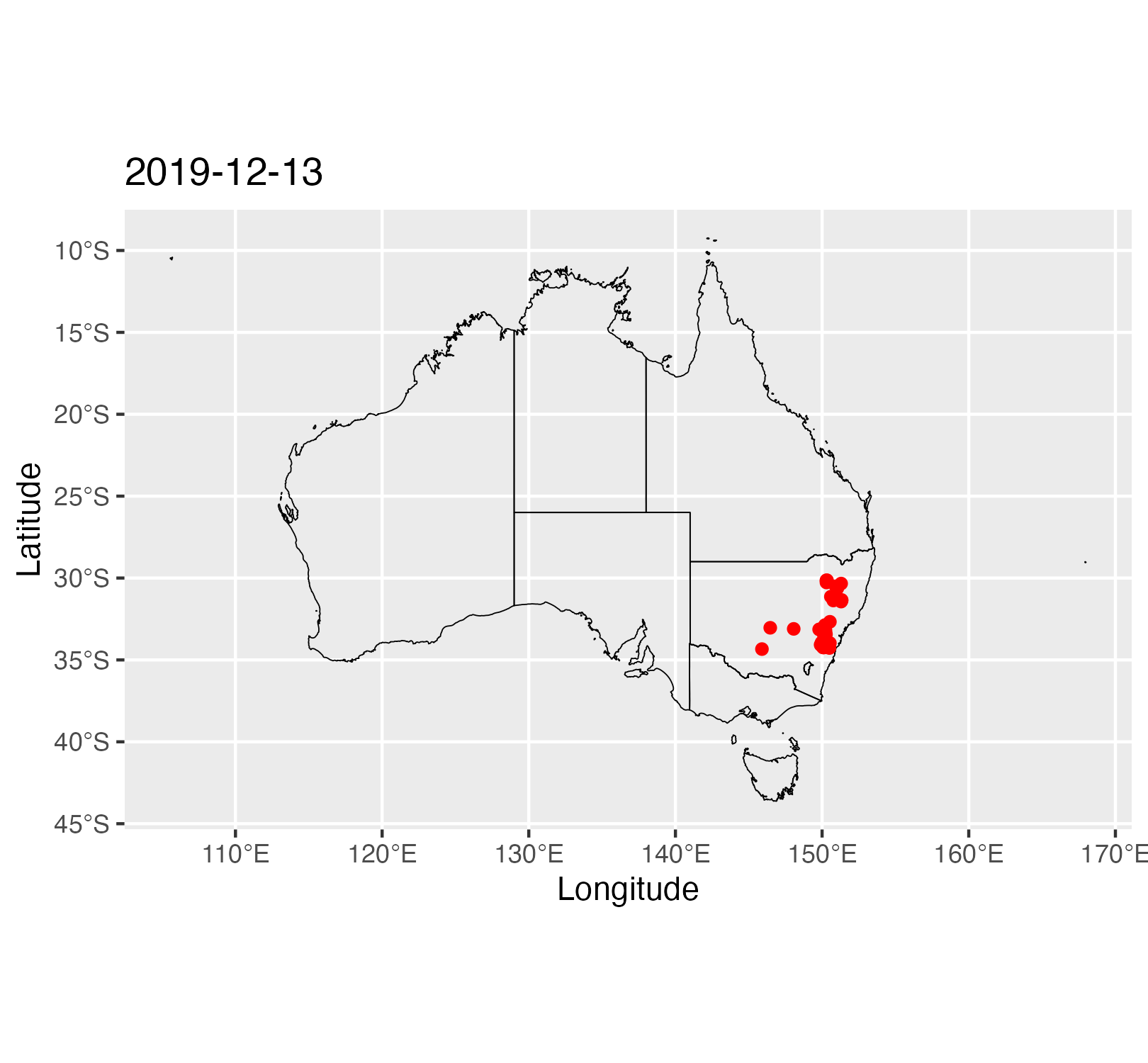}
         \caption{Observed fire on Dec 13, 2019}
         \label{fig:aus_dec_13_obs}
     \end{subfigure}
     \hfill
 \begin{subfigure}[b]{0.42\linewidth}
         \centering
         \includegraphics[width=\textwidth]{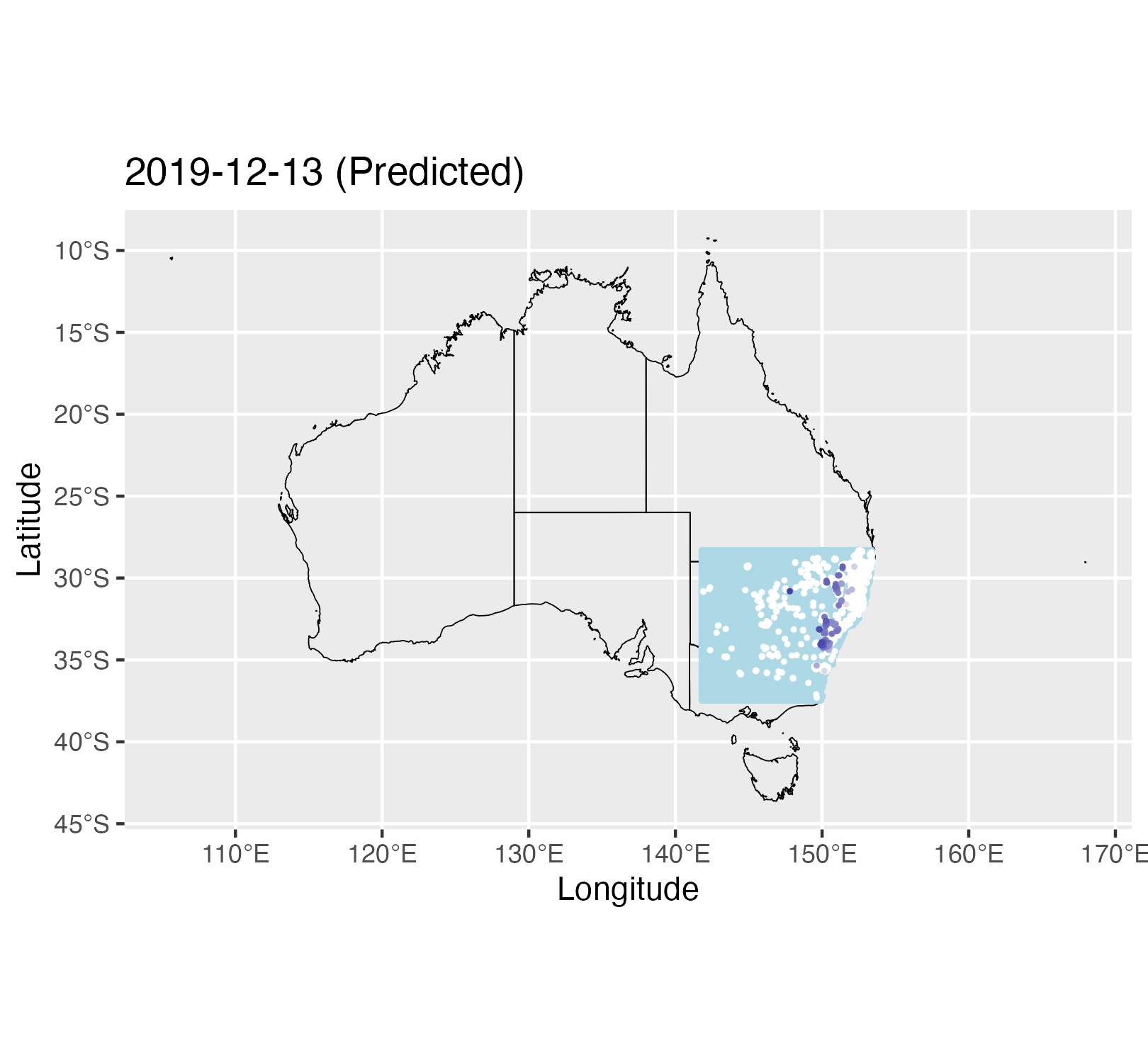}
         \caption{Predicted fire on Dec 13, 2019}
         \label{fig:aus_dec_13_pred}
     \end{subfigure}
 \begin{subfigure}[b]{0.42\linewidth}
         \centering
         \includegraphics[width=\textwidth]{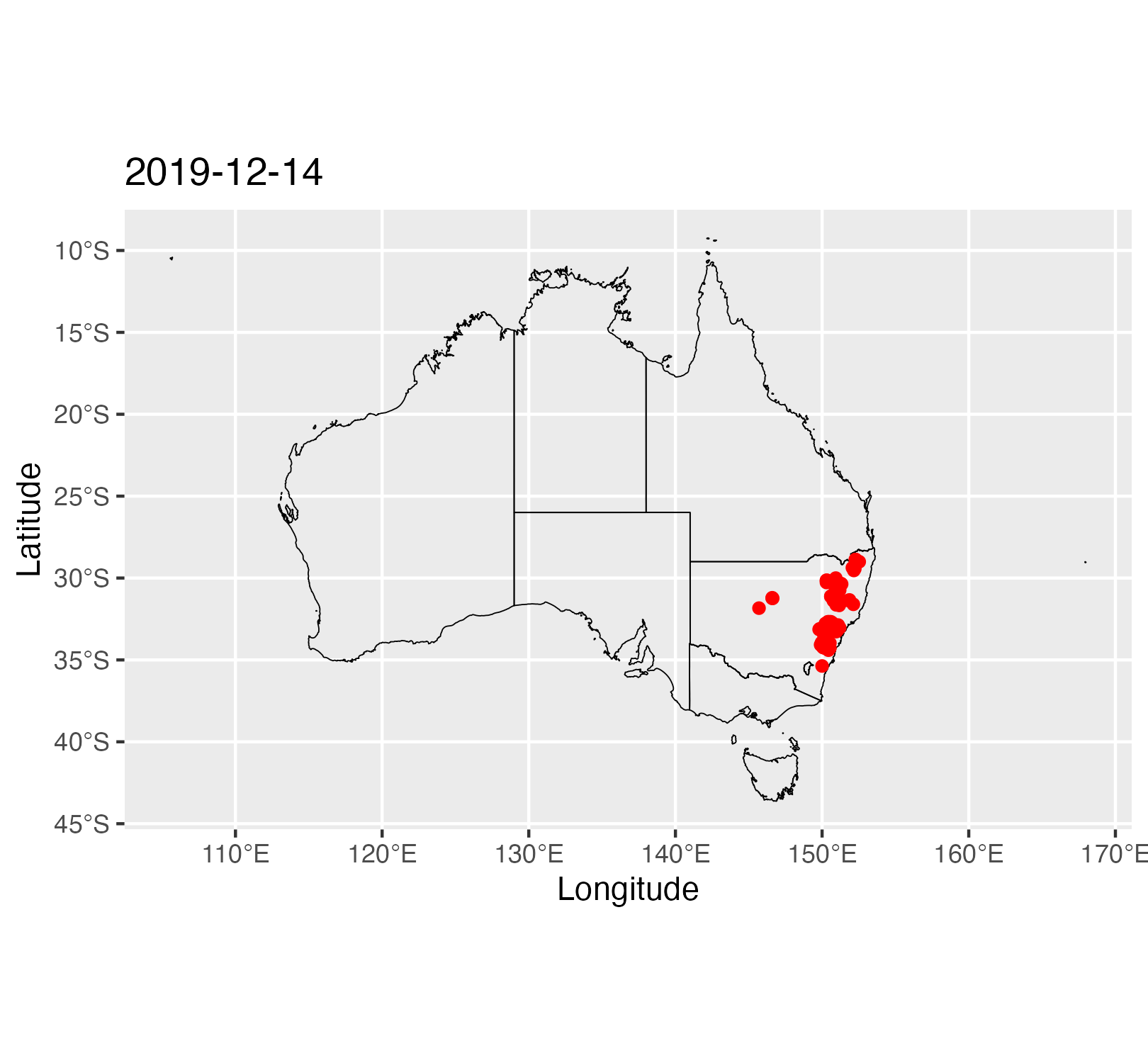}
         \caption{Observed fire on Dec 14, 2019}
         \label{fig:aus_dec_14_obs}
     \end{subfigure}
     \hfill
 \begin{subfigure}[b]{0.42\linewidth}
         \centering
         \includegraphics[width=\textwidth]{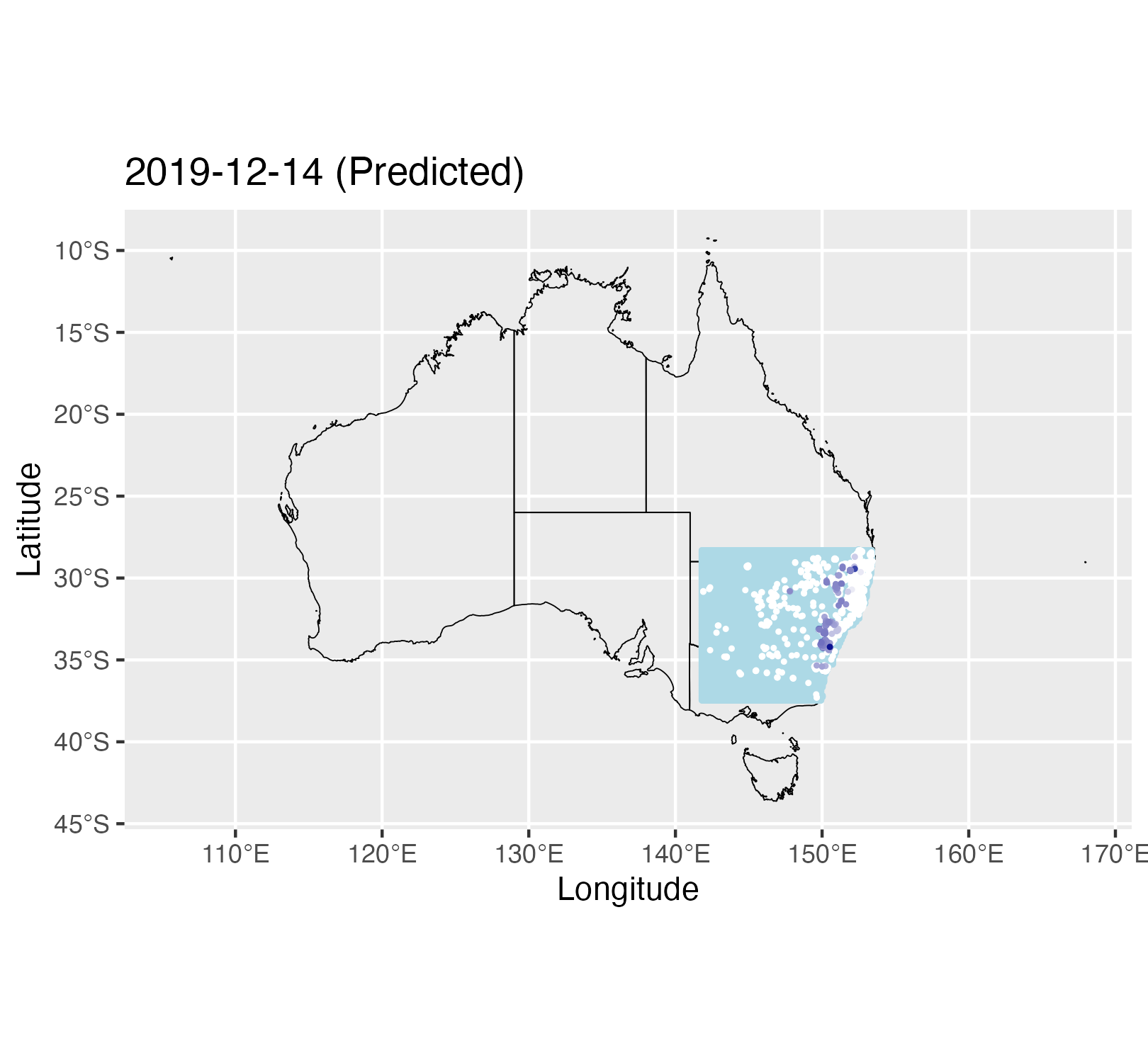}
         \caption{Predicted fire on Dec 14, 2019}
         \label{fig:aus_dec_14_pred}
     \end{subfigure}
 \begin{subfigure}[b]{0.42\linewidth}
         \centering
         \includegraphics[width=\textwidth]{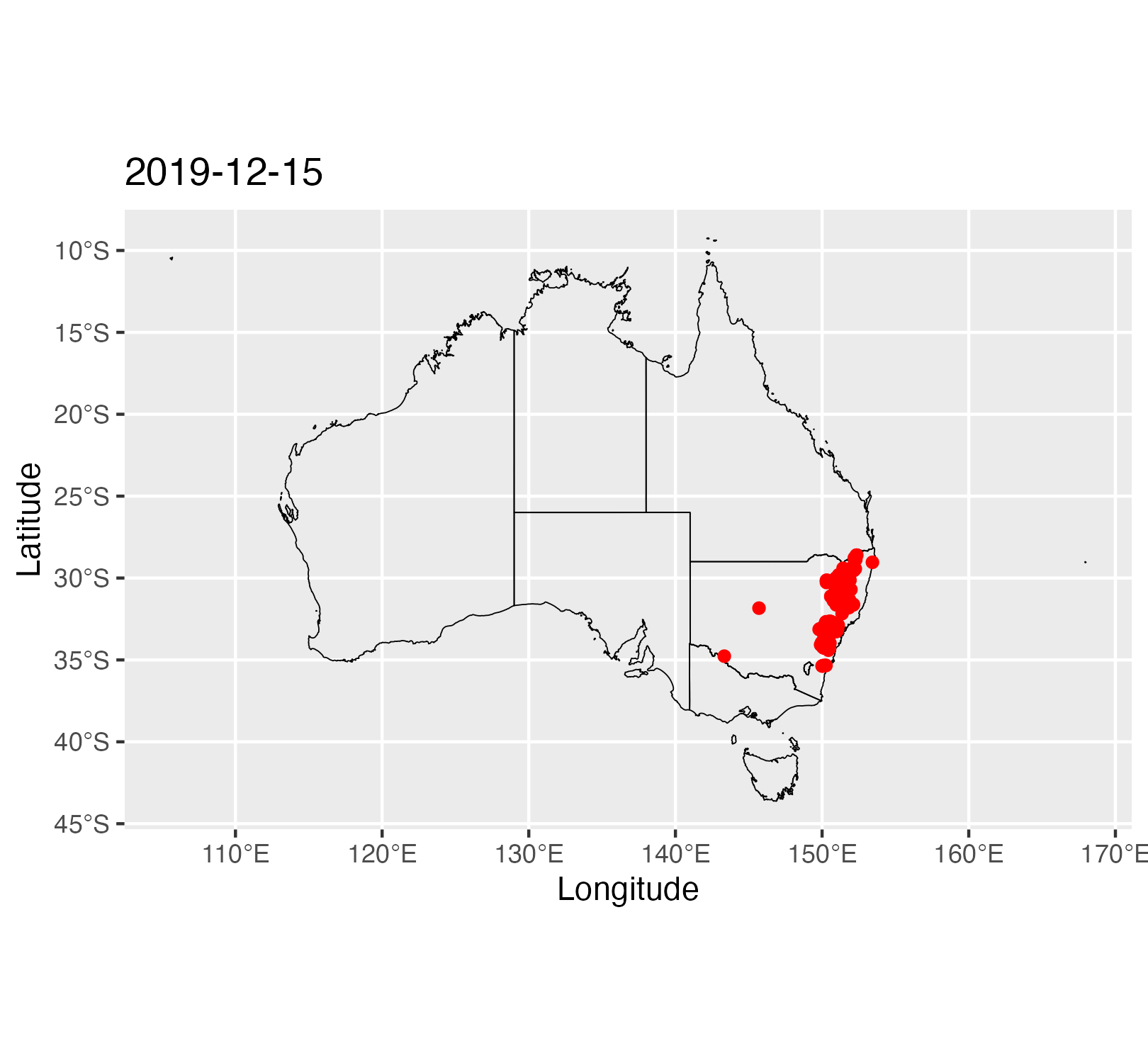}
         \caption{Observed fire on Dec 15, 2019}
         \label{fig:aus_dec_15_obs}
     \end{subfigure}
     \hfill
 \begin{subfigure}[b]{0.42\linewidth}
         \centering
         \includegraphics[width=\textwidth]{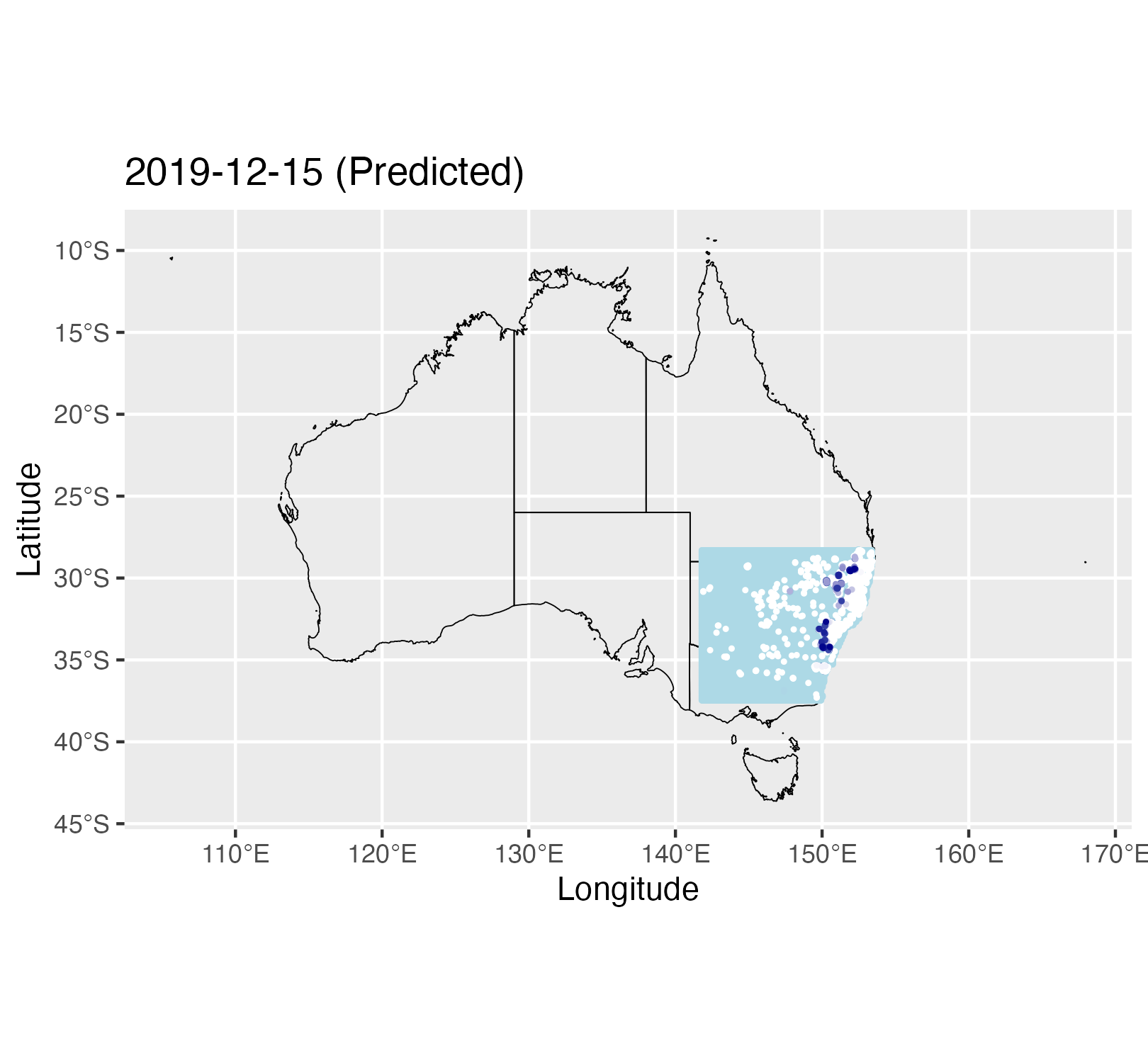}
         \caption{Predicted fire on Dec 15, 2019}
         \label{fig:aus_dec_15_pred}
     \end{subfigure}
     \caption{Observed vs 1, 2 and 3 day predictions}
     \label{fig:dec_13_14_15}
\end{figure}

\begin{figure}[H]
     \centering
 \begin{subfigure}[b]{0.42\linewidth}
         \centering
         \includegraphics[width=\textwidth]{Images/aus_obs_dec_20.png}
         \caption{Observed fire on Dec 20, 2019}
         \label{fig:aus_dec_20_obs}
     \end{subfigure}
     \hfill
 \begin{subfigure}[b]{0.42\linewidth}
         \centering
         \includegraphics[width=\textwidth]{Images/aus_pred_dec_20.png}
         \caption{Predicted fire on Dec 20, 2019}
         \label{fig:aus_dec_20_pred}
     \end{subfigure}
 \begin{subfigure}[b]{0.42\linewidth}
         \centering
         \includegraphics[width=\textwidth]{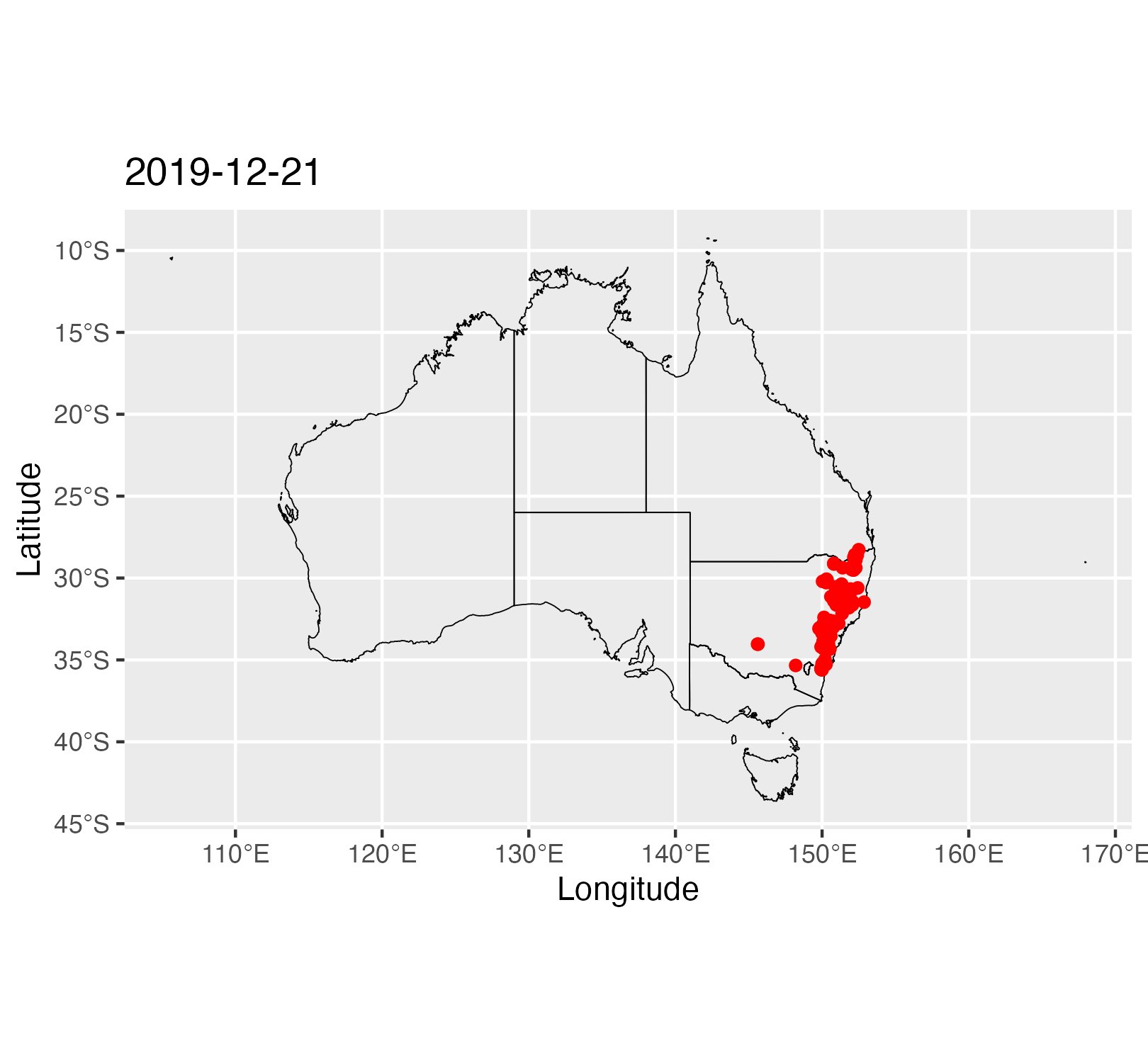}
         \caption{Observed fire on Dec 21, 2019}
         \label{fig:aus_dec_21_obs}
     \end{subfigure}
     \hfill
 \begin{subfigure}[b]{0.42\linewidth}
         \centering
         \includegraphics[width=\textwidth]{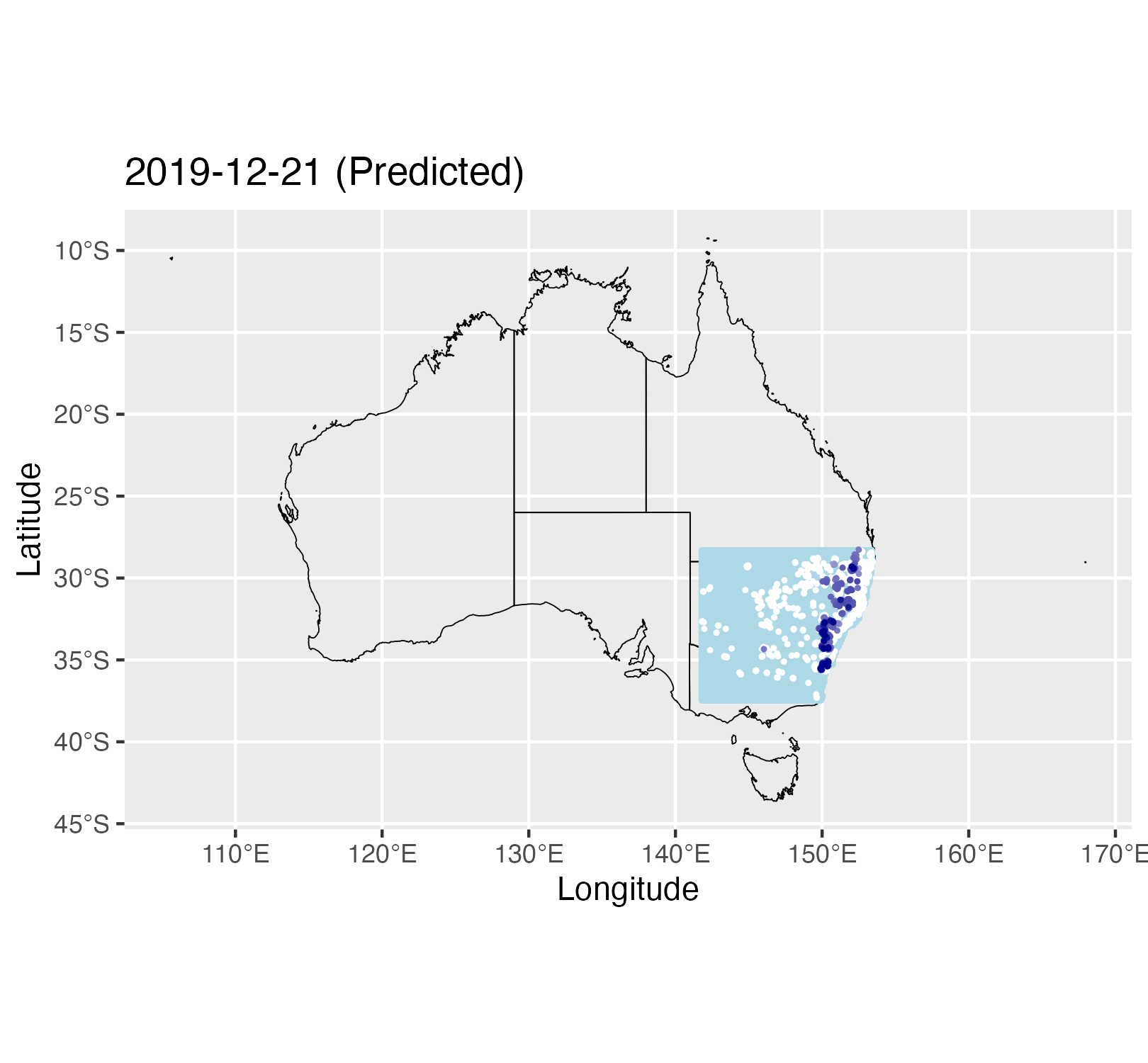}
         \caption{Predicted fire on Dec 21, 2019}
         \label{fig:aus_dec_21_pred}
     \end{subfigure}
 \begin{subfigure}[b]{0.42\linewidth}
         \centering
         \includegraphics[width=\textwidth]{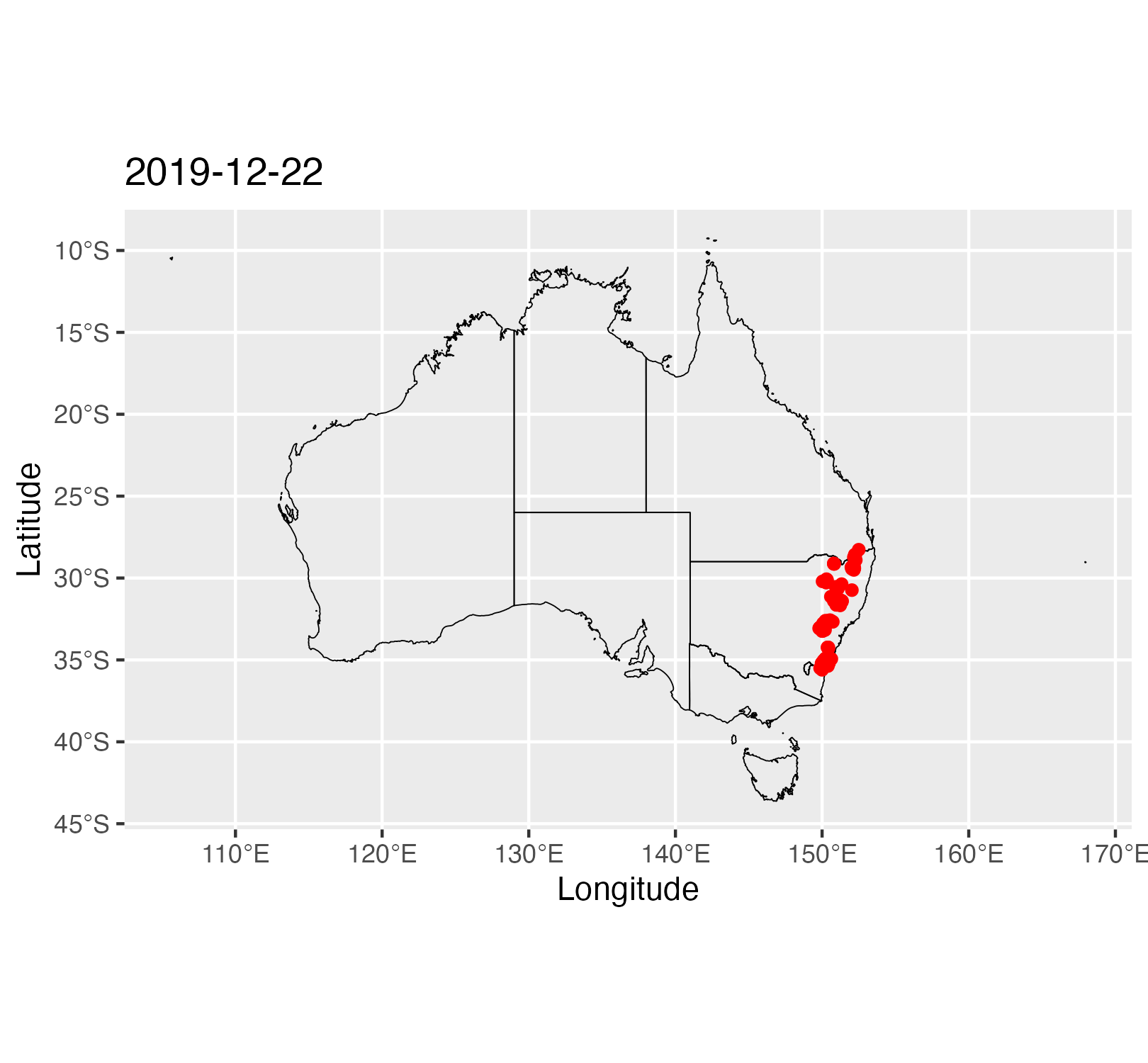}
         \caption{Observed fire on Dec 22, 2019}
         \label{fig:aus_dec_22_obs}
     \end{subfigure}
     \hfill
 \begin{subfigure}[b]{0.42\linewidth}
         \centering
         \includegraphics[width=\textwidth]{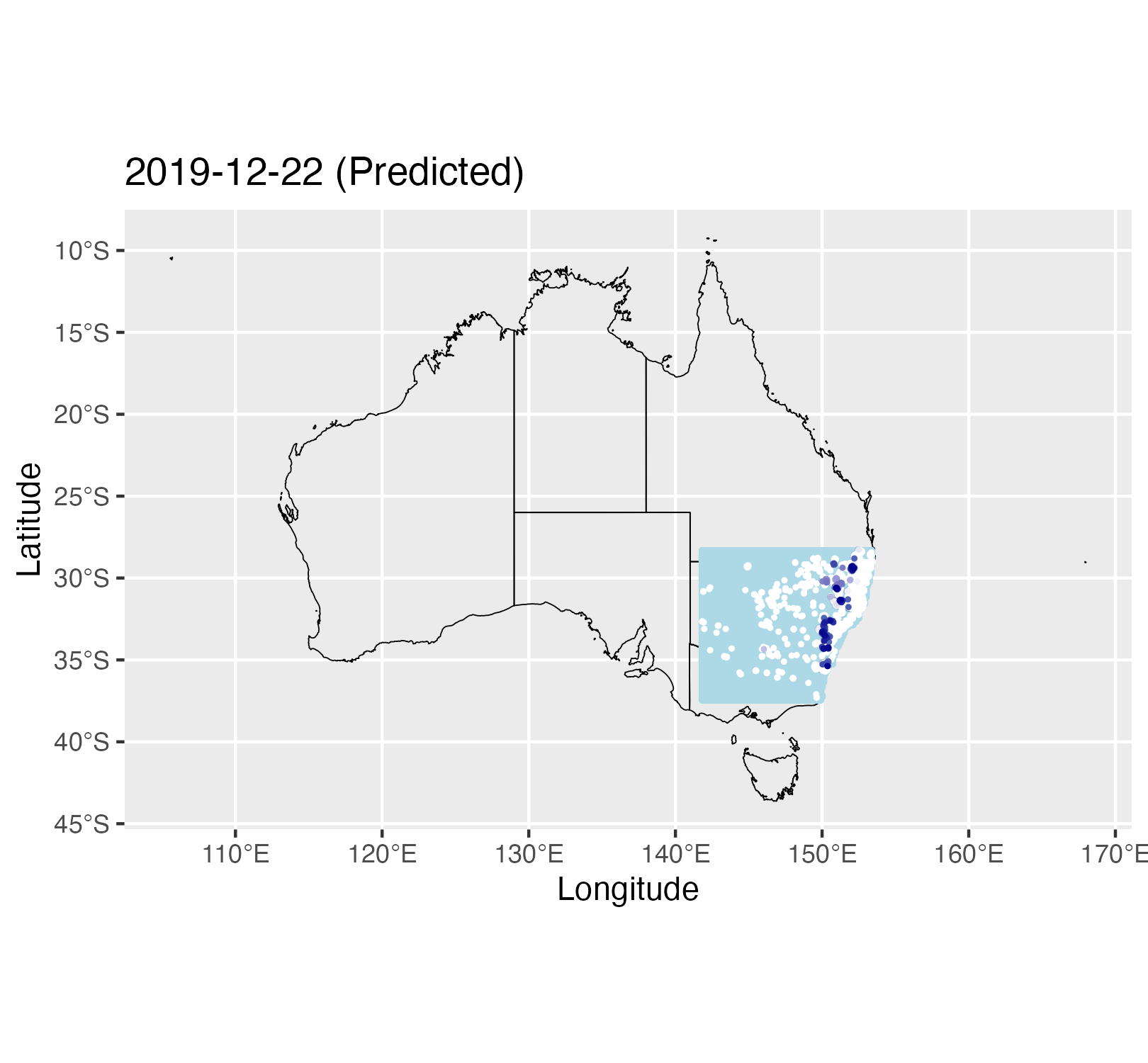}
         \caption{Predicted fire on Dec 22, 2019}
         \label{fig:aus_dec_22_pred}
     \end{subfigure}
     \caption{Observed vs 1, 2 and 3 day predictions}
     \label{fig:dec_20_21_22}
\end{figure}

\begin{figure}[H]
     \centering
 \begin{subfigure}[b]{0.42\linewidth}
         \centering
         \includegraphics[width=\textwidth]{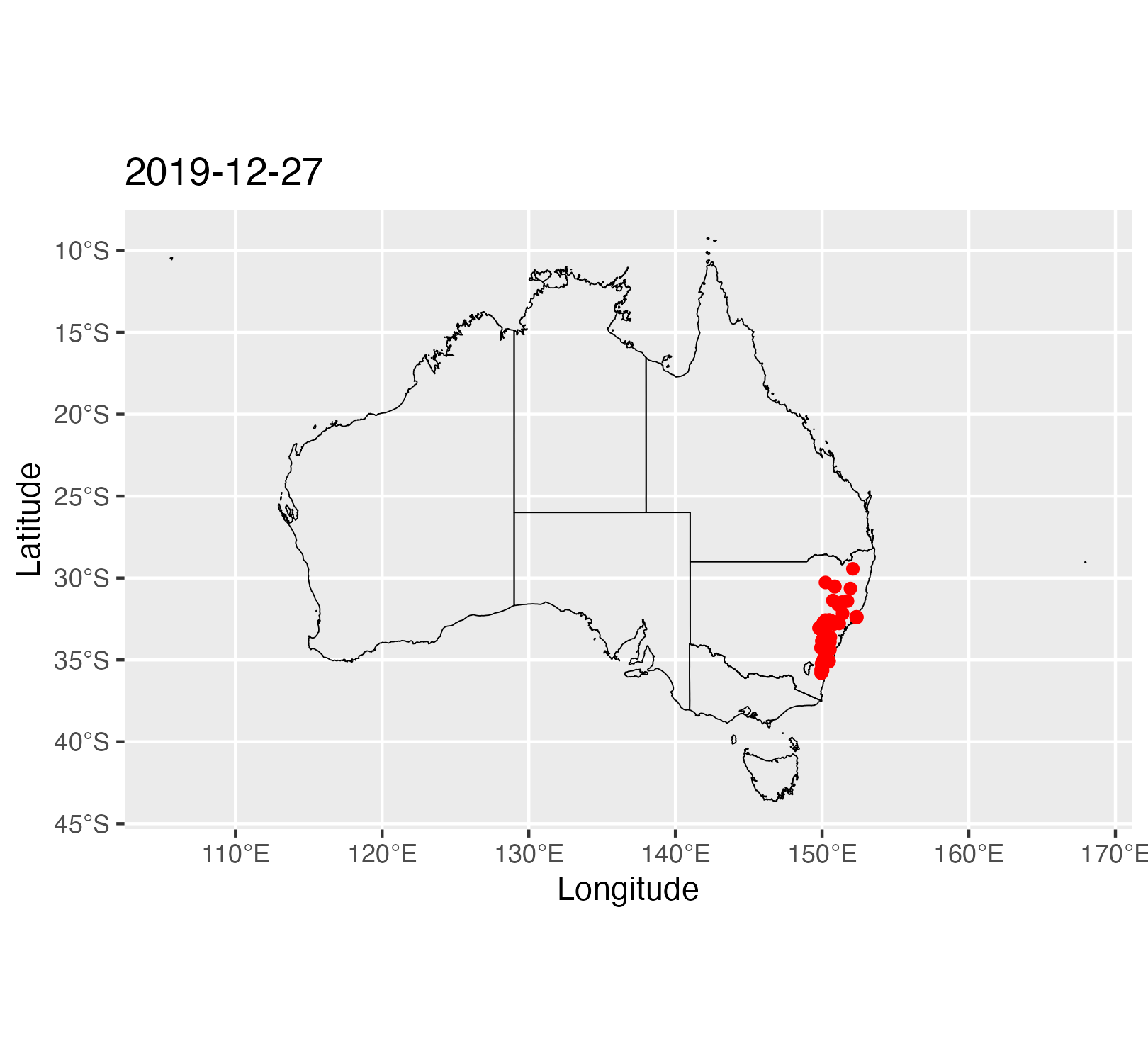}
         \caption{Observed fire on Dec 27, 2019}
         \label{fig:aus_dec_27_obs}
     \end{subfigure}
     \hfill
 \begin{subfigure}[b]{0.42\linewidth}
         \centering
         \includegraphics[width=\textwidth]{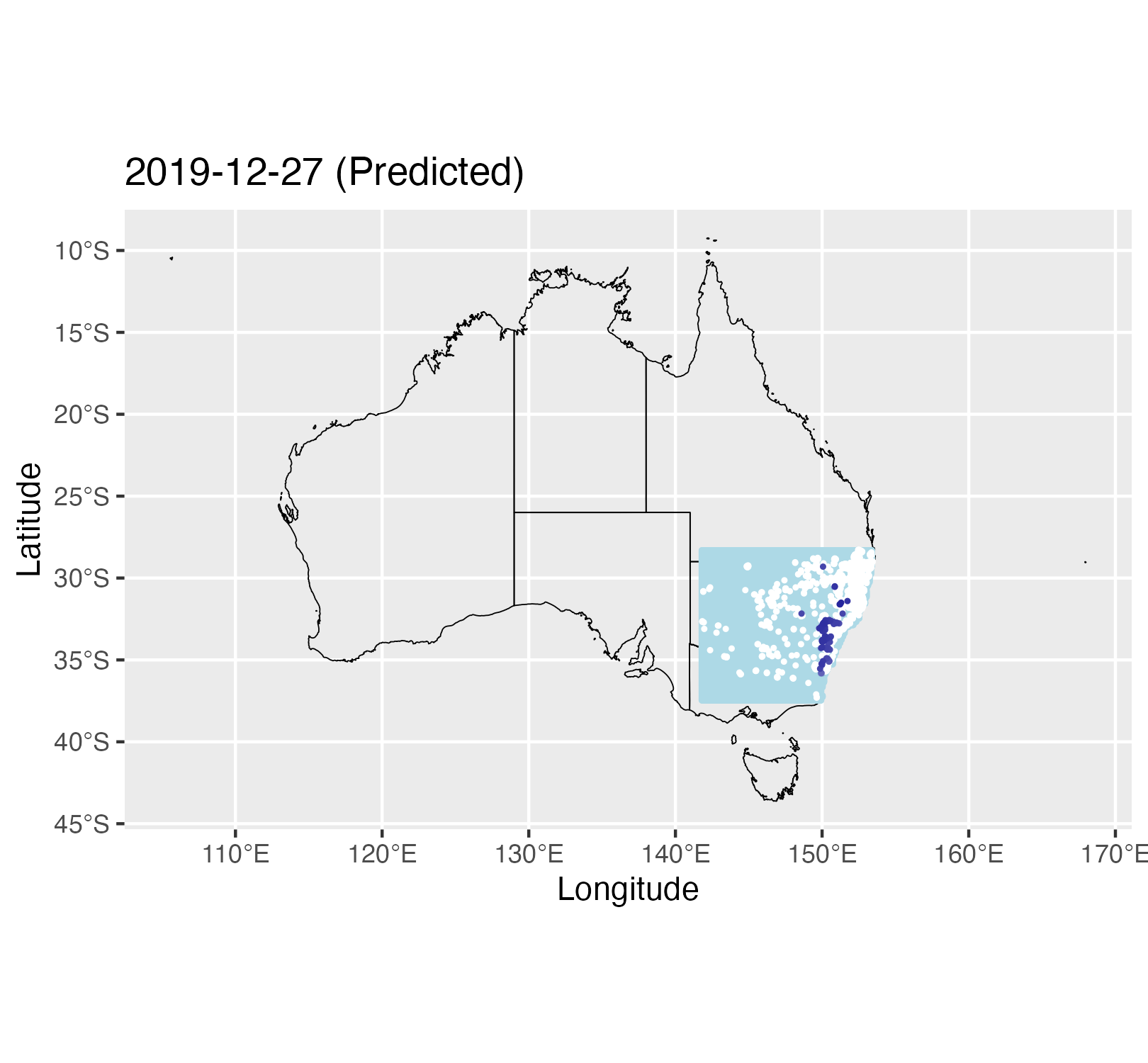}
         \caption{Predicted fire on Dec 27, 2019}
         \label{fig:aus_dec_27_pred}
     \end{subfigure}
 \begin{subfigure}[b]{0.42\linewidth}
         \centering
         \includegraphics[width=\textwidth]{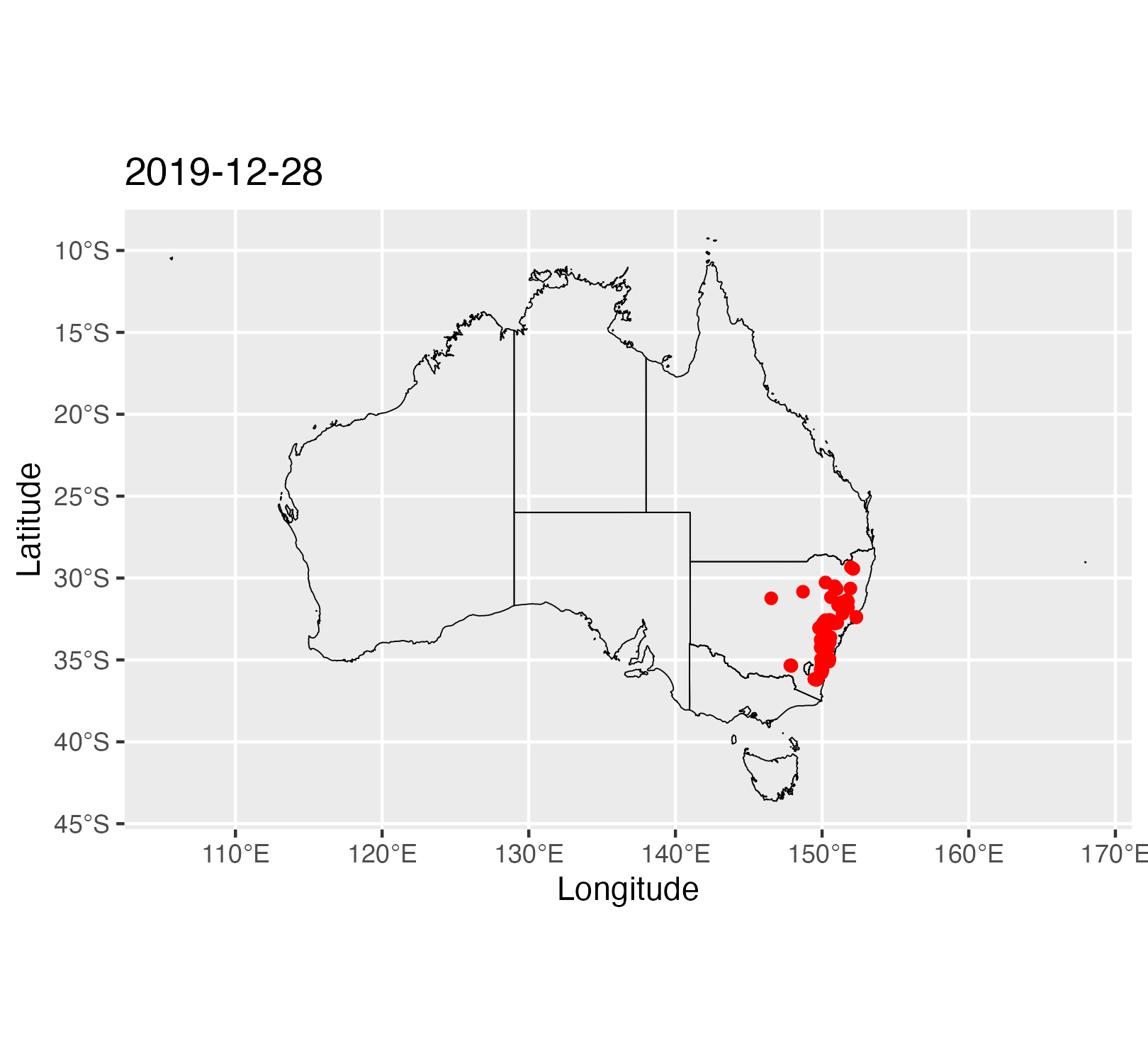}
         \caption{Observed fire on Dec 28, 2019}
         \label{fig:aus_dec_28_obs}
     \end{subfigure}
     \hfill
 \begin{subfigure}[b]{0.42\linewidth}
         \centering
         \includegraphics[width=\textwidth]{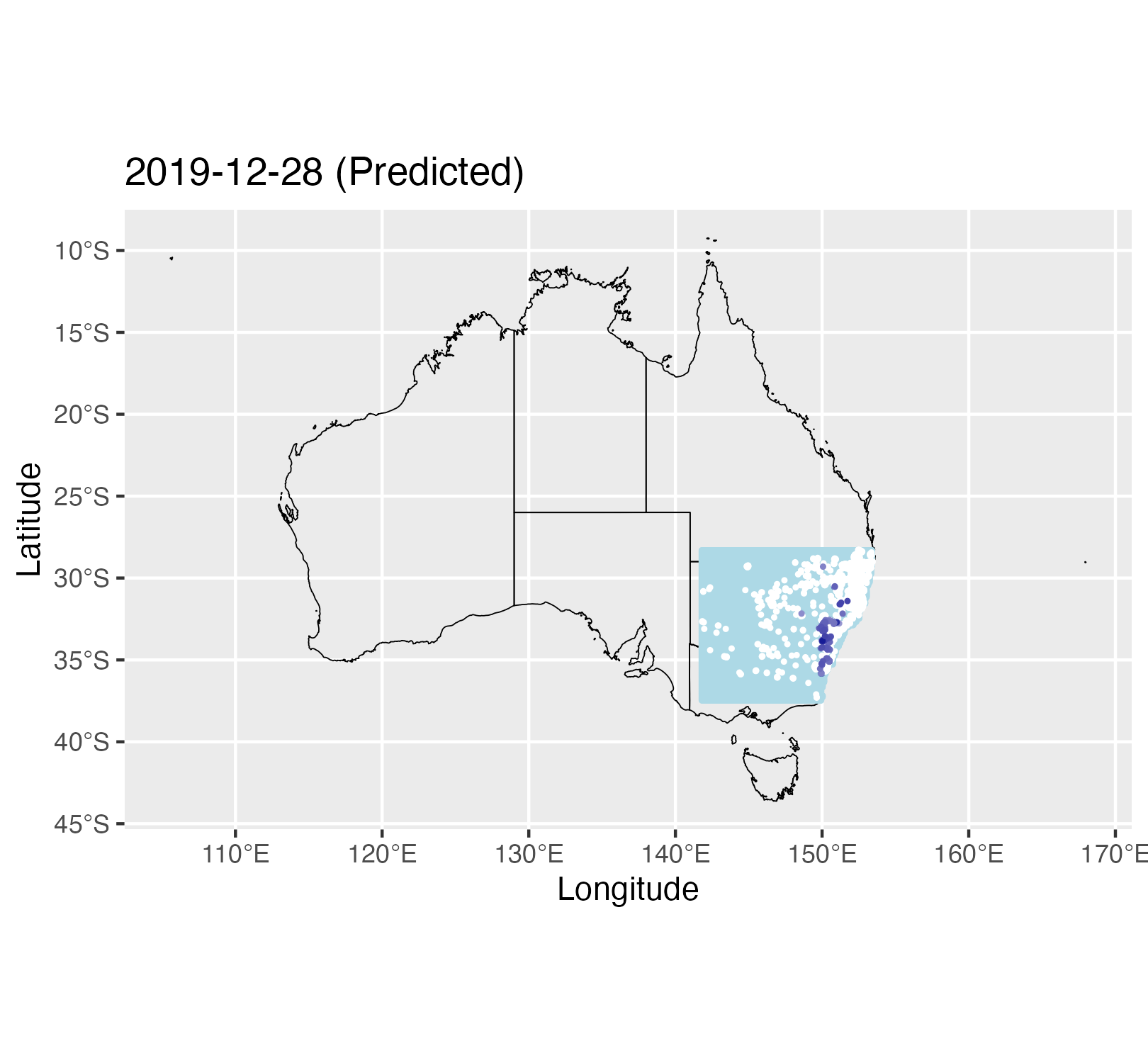}
         \caption{Predicted fire on Dec 28, 2019}
         \label{fig:aus_dec_28_pred}
     \end{subfigure}
 \begin{subfigure}[b]{0.42\linewidth}
         \centering
         \includegraphics[width=\textwidth]{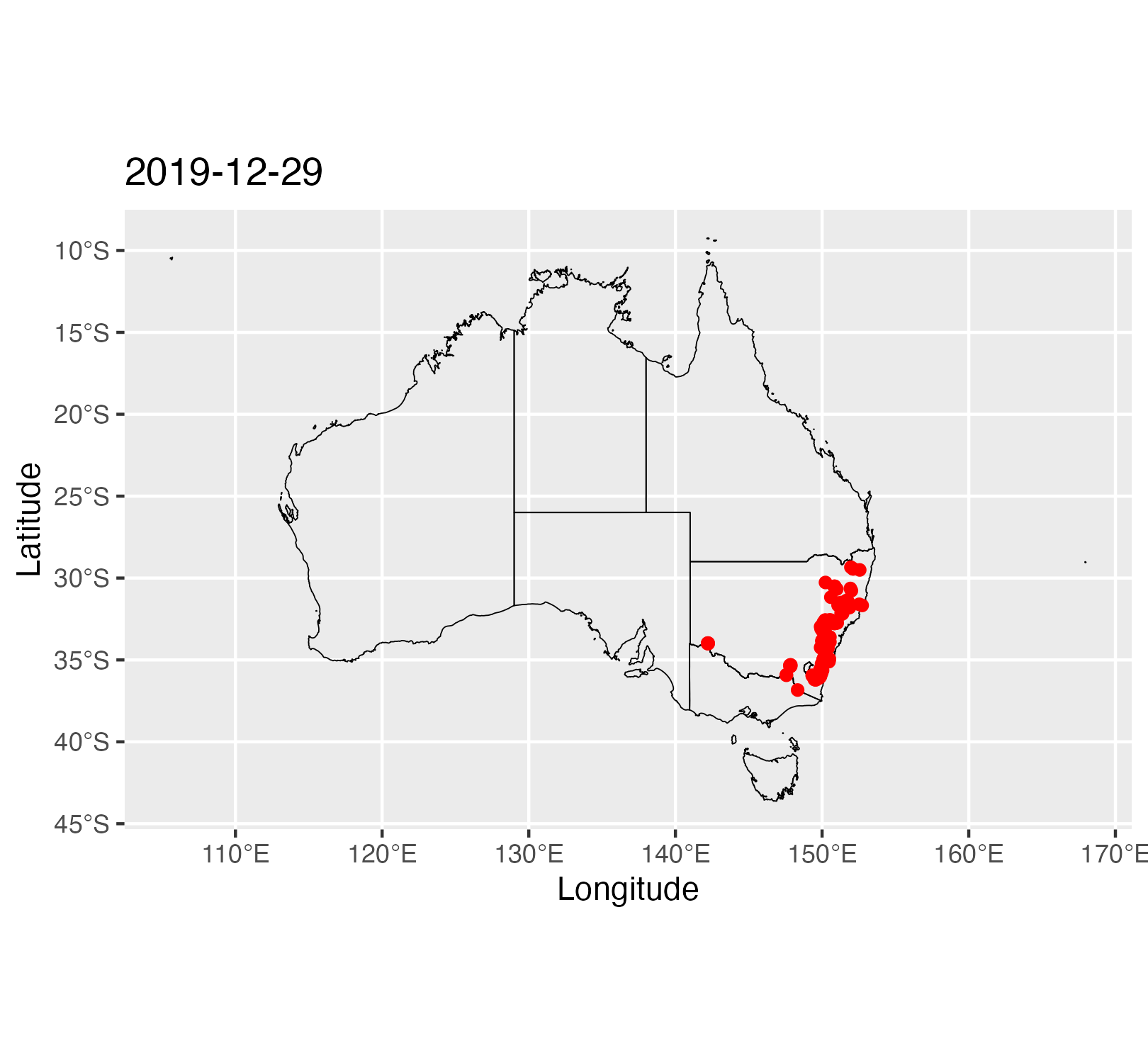}
         \caption{Observed fire on Dec 29, 2019}
         \label{fig:aus_dec_29_obs}
     \end{subfigure}
     \hfill
 \begin{subfigure}[b]{0.42\linewidth}
         \centering
         \includegraphics[width=\textwidth]{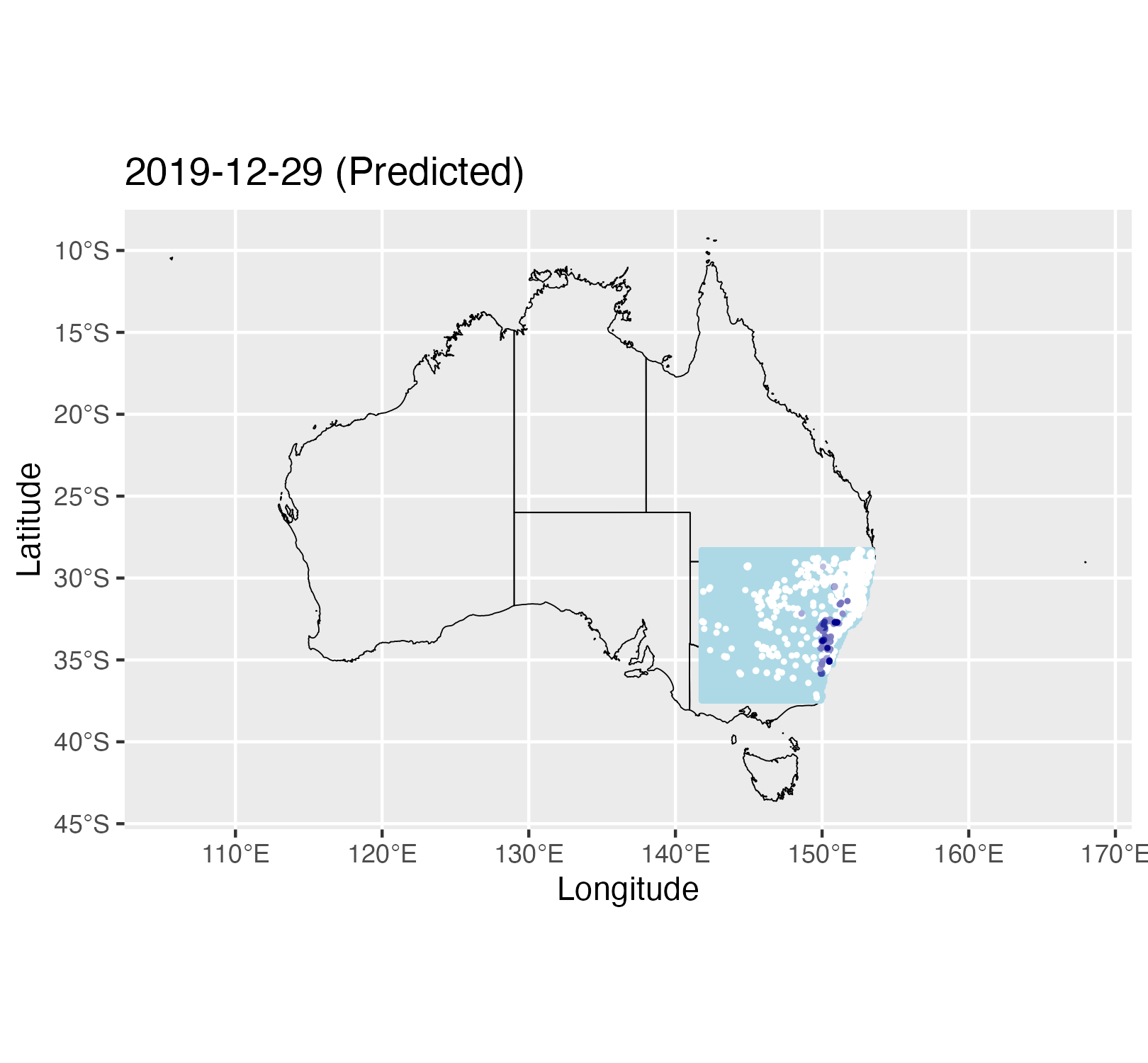}
         \caption{Predicted fire on Dec 29, 2019}
         \label{fig:aus_dec_29_pred}
     \end{subfigure}
     \caption{Observed vs 1, 2 and 3 day predictions}
     \label{fig:dec_27_28_29}
\end{figure}

\begin{figure}[H]
     \centering
 \begin{subfigure}[b]{0.42\linewidth}
         \centering
         \includegraphics[width=\textwidth]{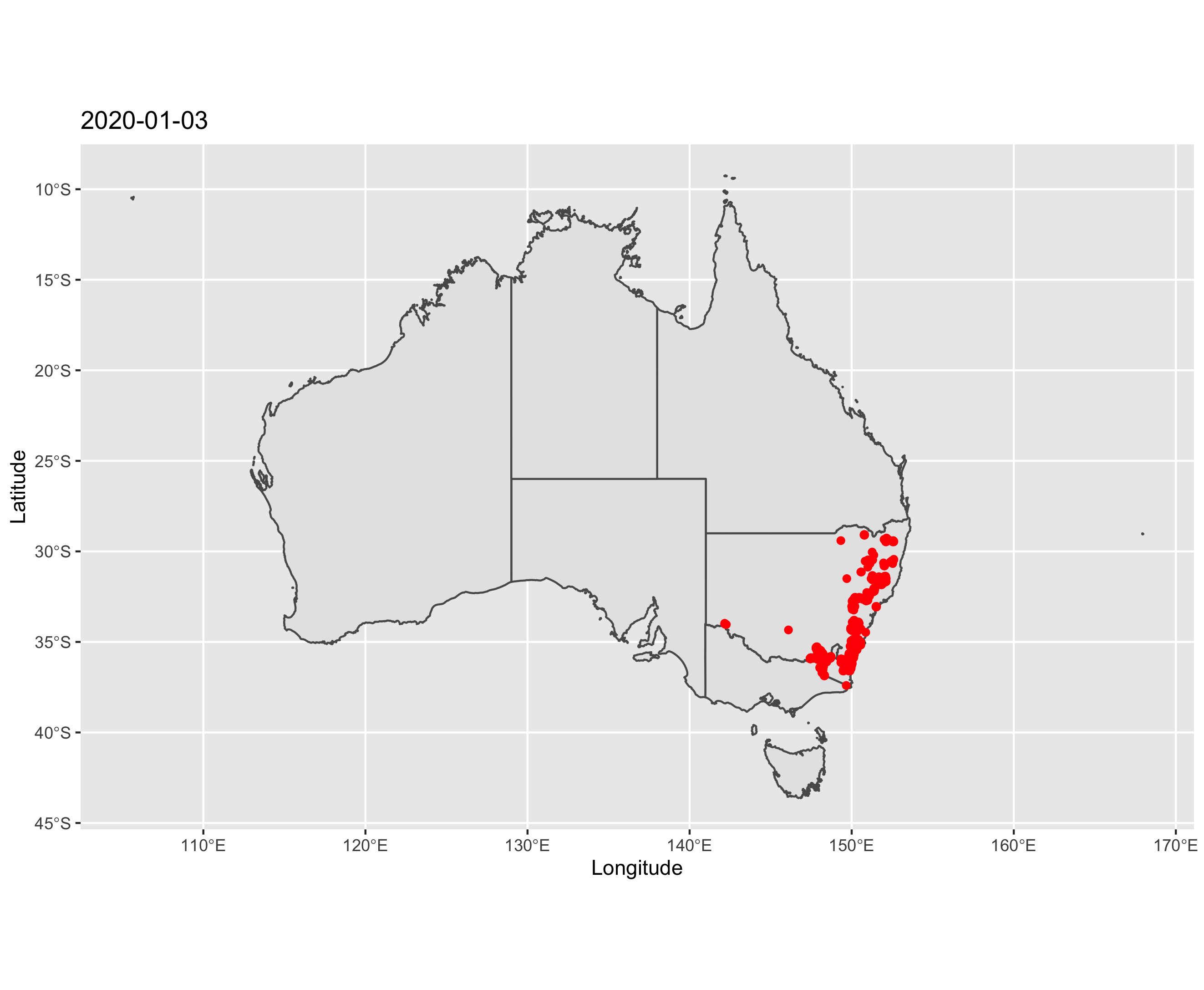}
         \caption{Observed fire on Jan 3, 2020}
         \label{fig:aus_jan_3_obs}
     \end{subfigure}
     \hfill
 \begin{subfigure}[b]{0.42\linewidth}
         \centering
         \includegraphics[width=\textwidth]{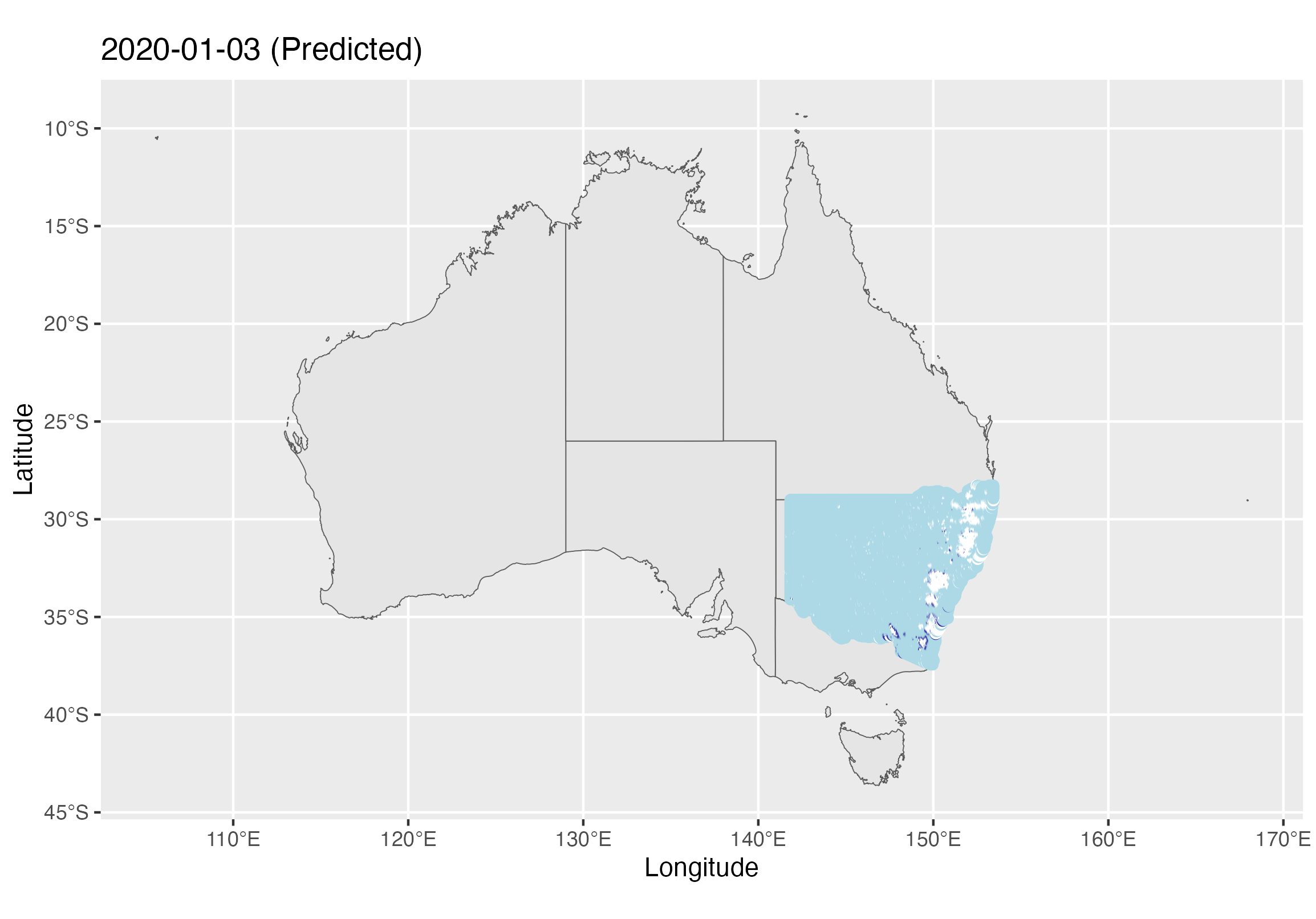}
         \caption{Predicted fire on Jan 3, 2020}
         \label{fig:aus_jan_3_pred}
     \end{subfigure}
 \begin{subfigure}[b]{0.42\linewidth}
         \centering
         \includegraphics[width=\textwidth]{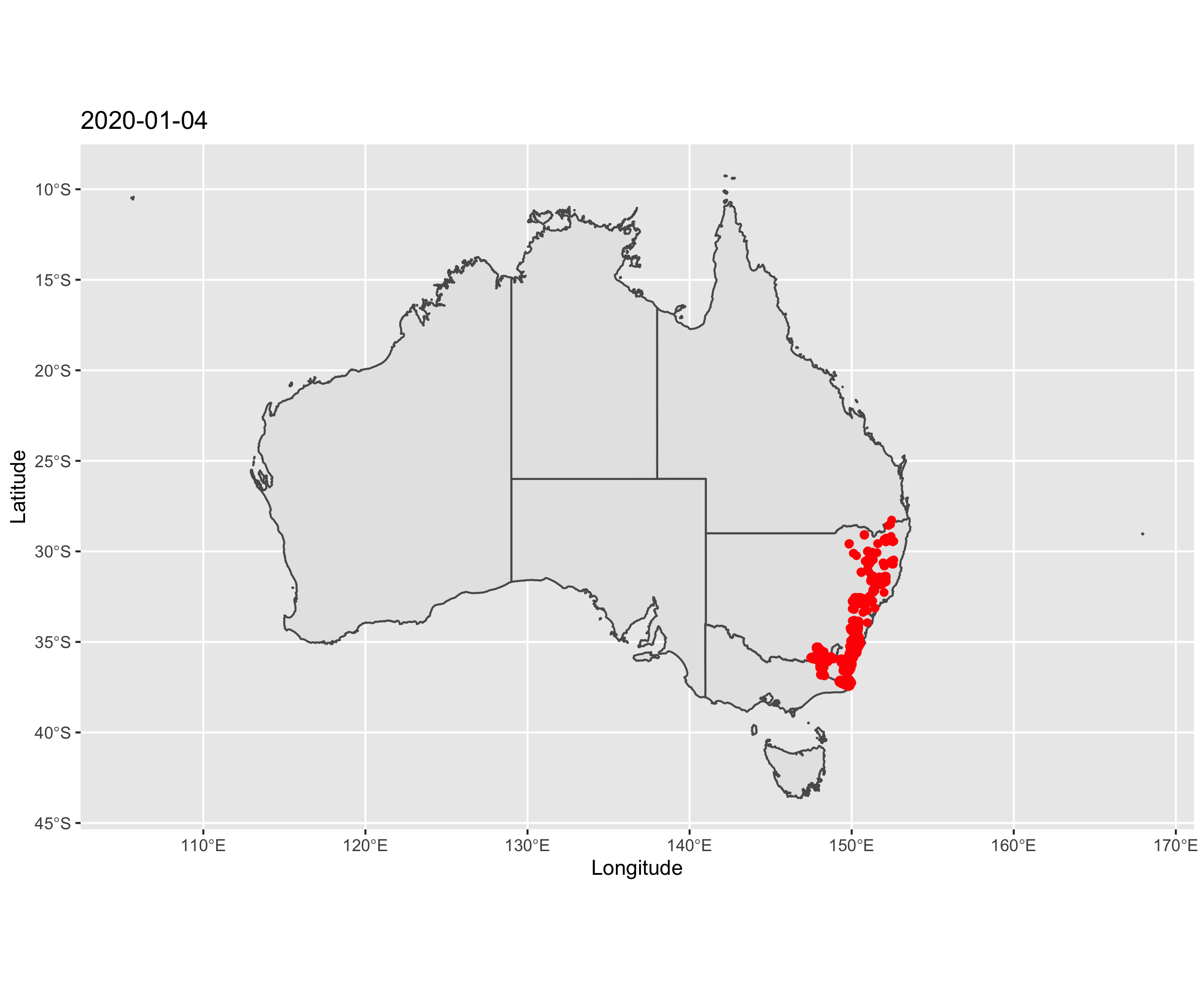}
         \caption{Observed fire on Jan 4, 2020}
         \label{fig:aus_jan_4_obs}
     \end{subfigure}
     \hfill
 \begin{subfigure}[b]{0.42\linewidth}
         \centering
         \includegraphics[width=\textwidth]{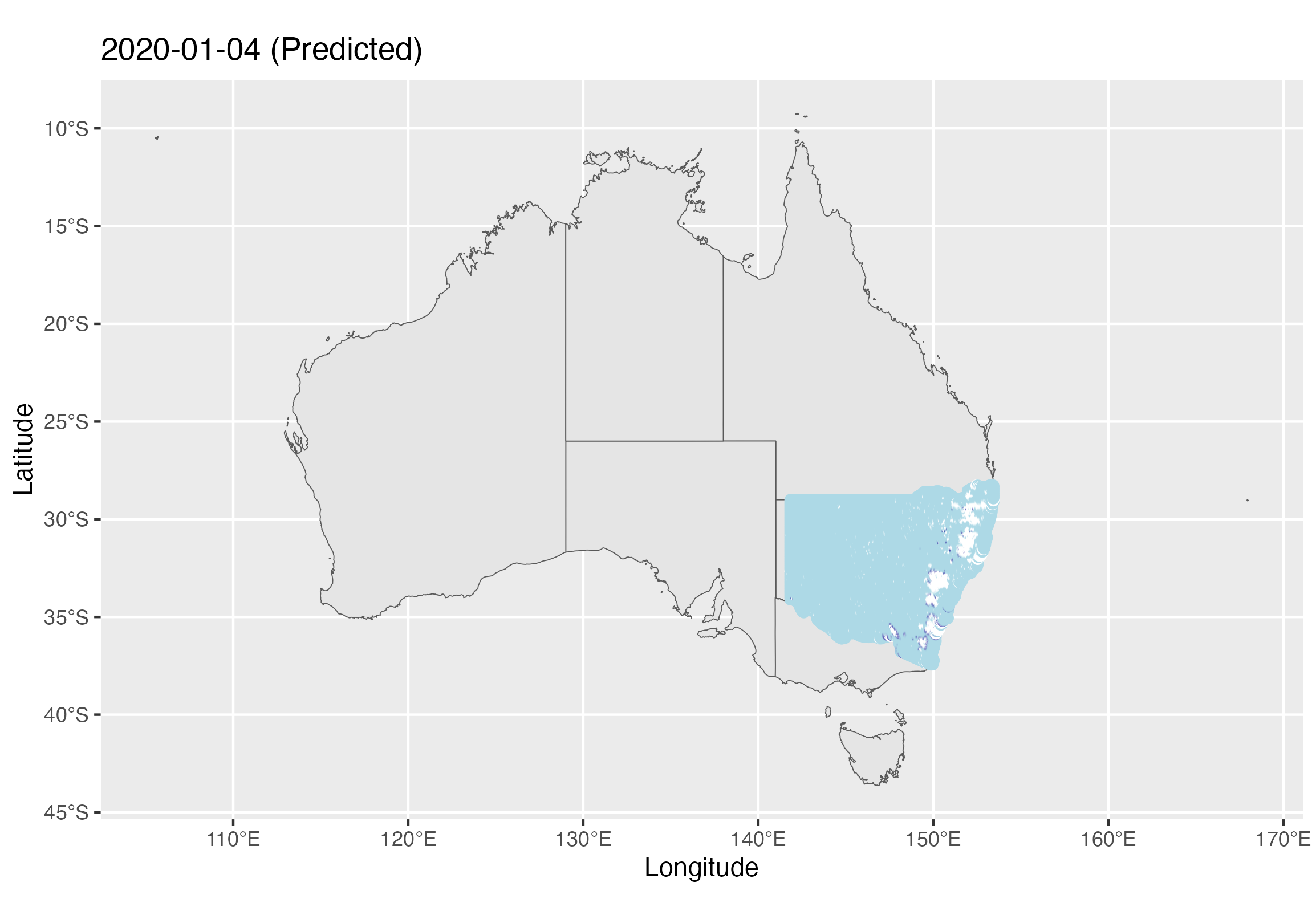}
         \caption{Predicted fire on Jan 4, 2020}
         \label{fig:aus_jan_4_pred}
     \end{subfigure}
 \begin{subfigure}[b]{0.42\linewidth}
         \centering
         \includegraphics[width=\textwidth]{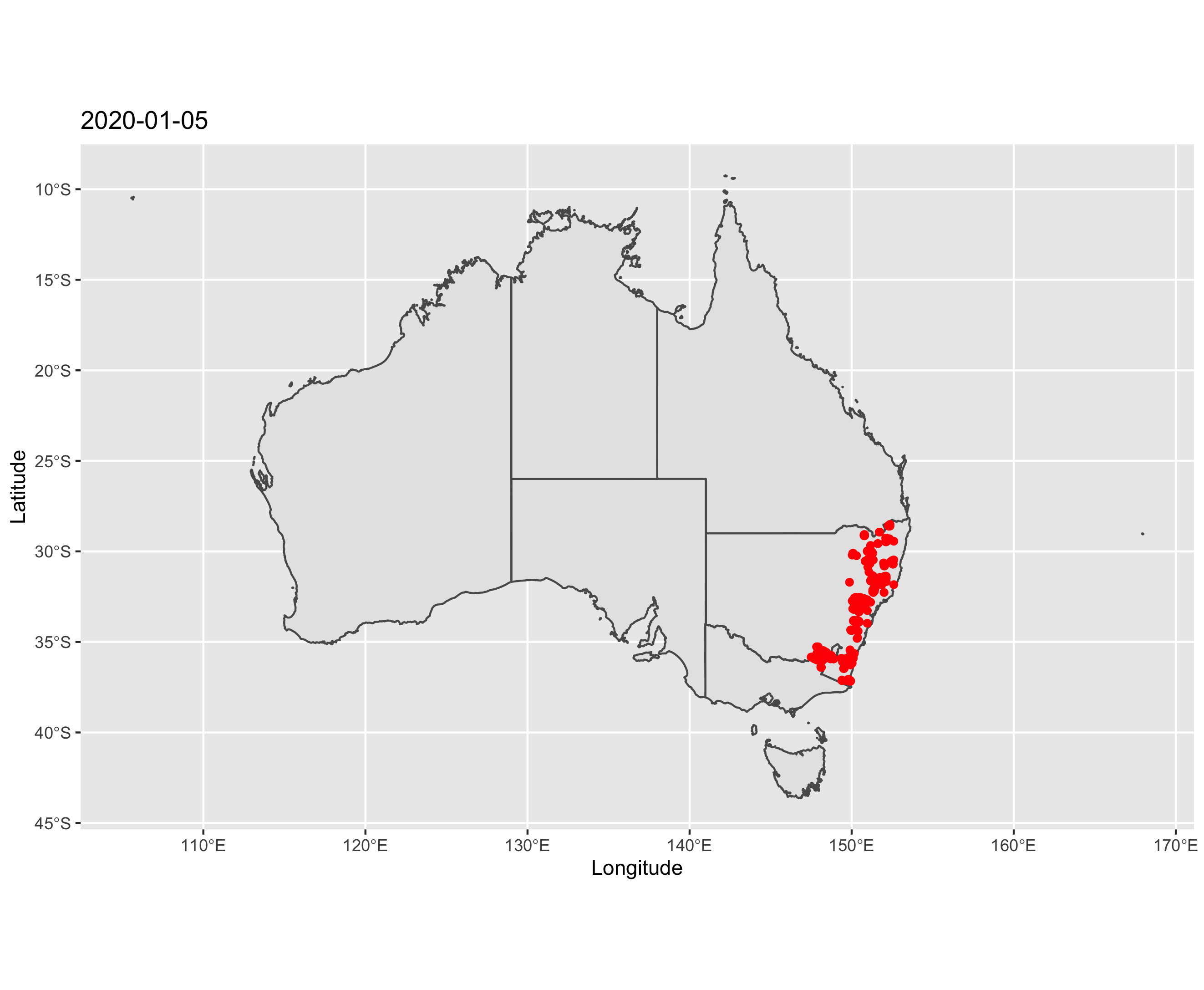}
         \caption{Observed fire on Jan 5, 2020}
         \label{fig:aus_jan_5_obs}
     \end{subfigure}
     \hfill
 \begin{subfigure}[b]{0.42\linewidth}
         \centering
         \includegraphics[width=\textwidth]{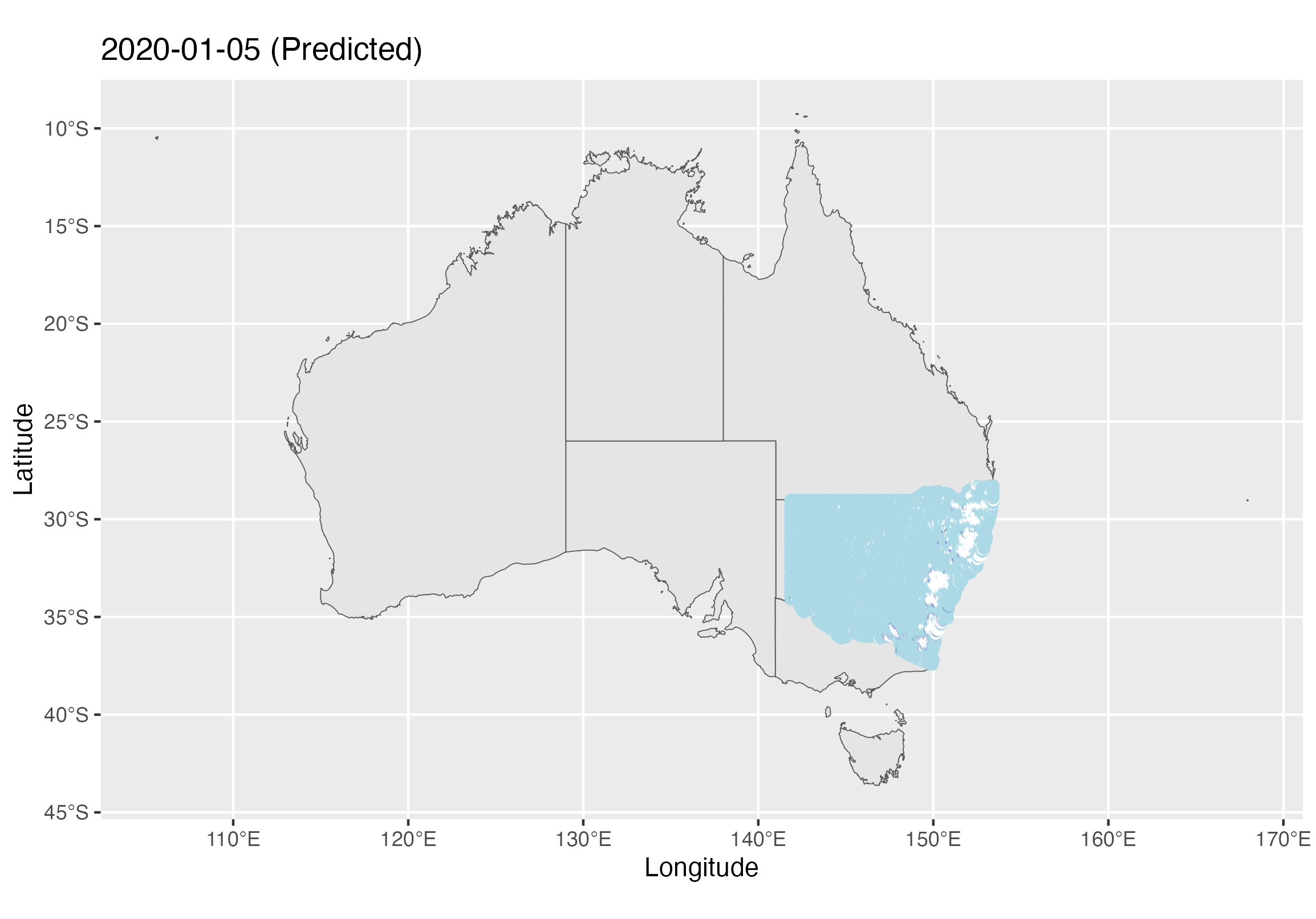}
         \caption{Predicted fire on Jan 5, 2020}
         \label{fig:aus_jan_5_pred}
     \end{subfigure}
     \caption{Observed vs 1, 2 and 3 day predictions}
     \label{fig:jan_3_4_5}
\end{figure}

\begin{figure}[H]
     \centering
 \begin{subfigure}[b]{0.42\linewidth}
         \centering
         \includegraphics[width=\textwidth]{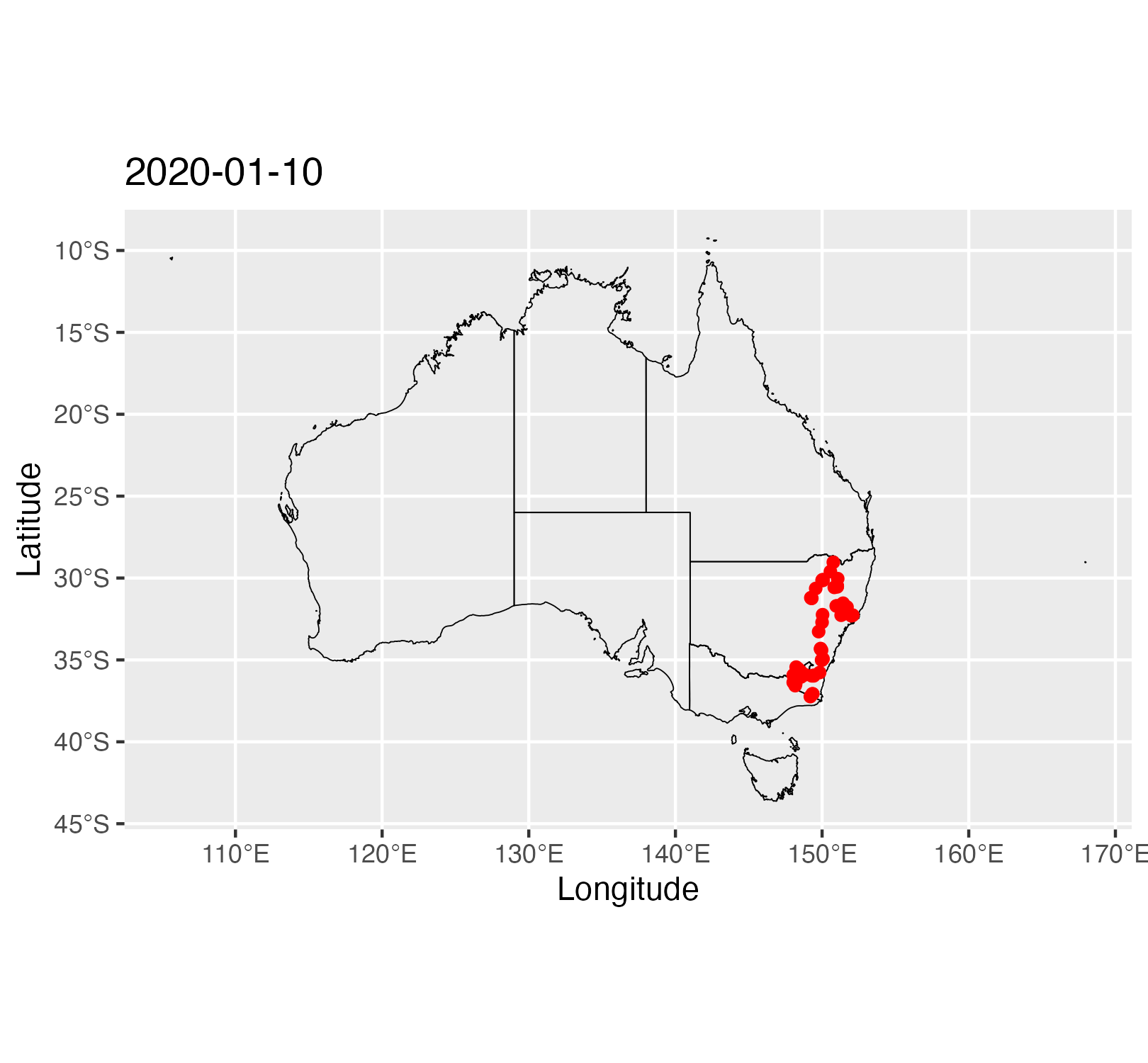}
         \caption{Observed fire on Jan 10, 2020}
         \label{fig:aus_jan_10_obs}
     \end{subfigure}
     \hfill
 \begin{subfigure}[b]{0.42\linewidth}
         \centering
         \includegraphics[width=\textwidth]{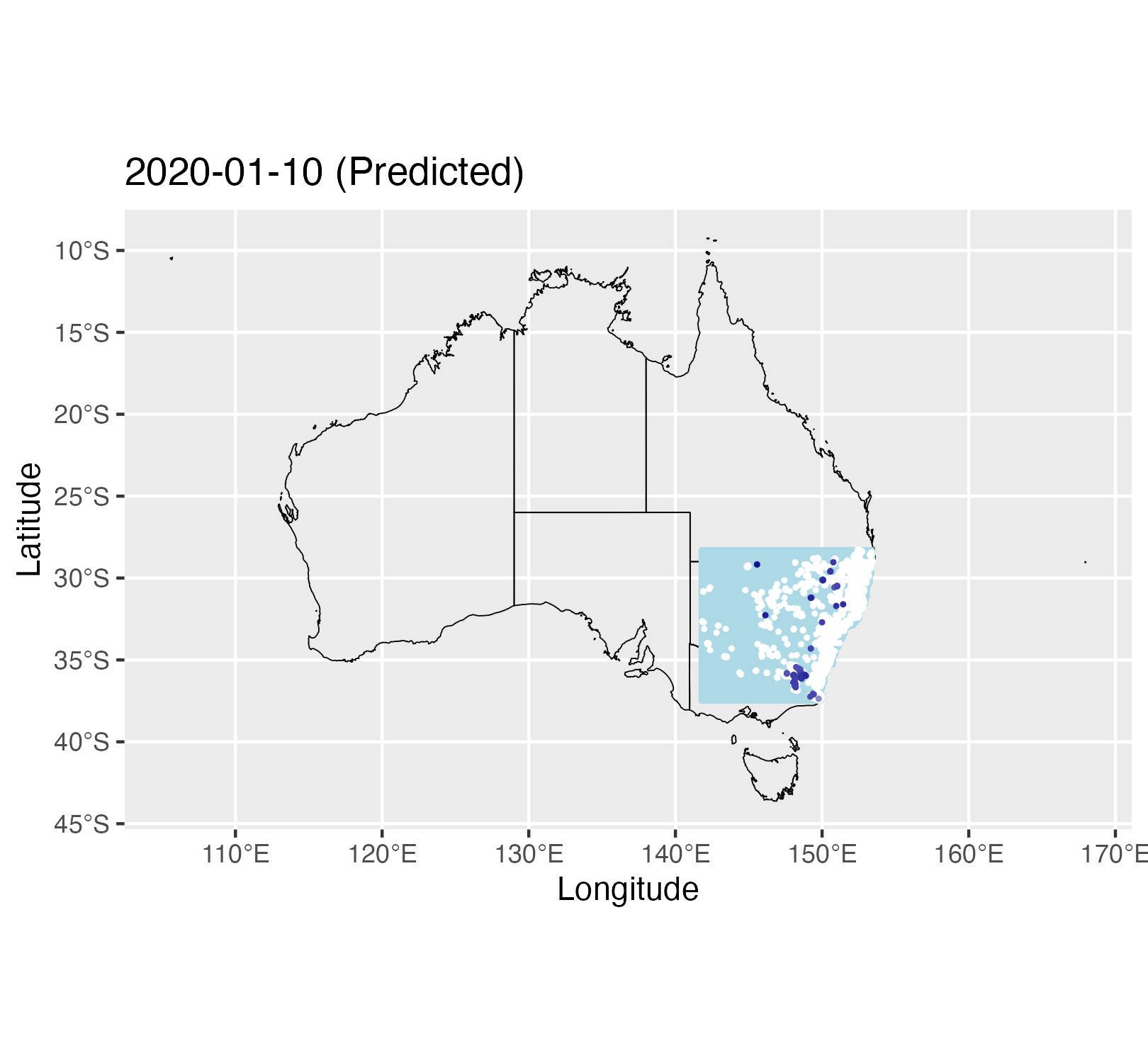}
         \caption{Predicted fire on Jan 10, 2020}
         \label{fig:aus_jan_10_pred}
     \end{subfigure}
 \begin{subfigure}[b]{0.42\linewidth}
         \centering
         \includegraphics[width=\textwidth]{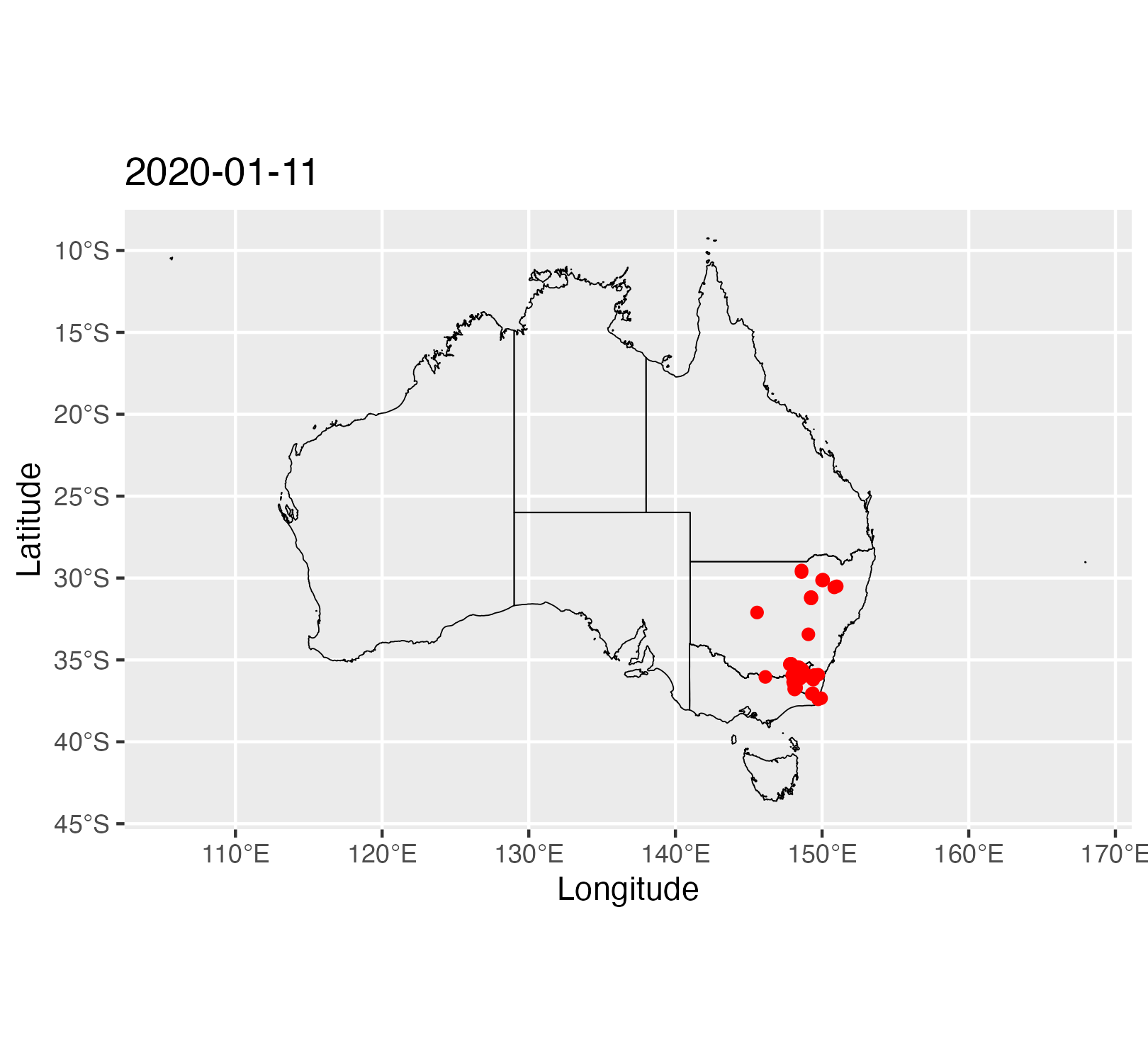}
         \caption{Observed fire on Jan 11, 2020}
         \label{fig:aus_jan_11_obs}
     \end{subfigure}
     \hfill
 \begin{subfigure}[b]{0.42\linewidth}
         \centering
         \includegraphics[width=\textwidth]{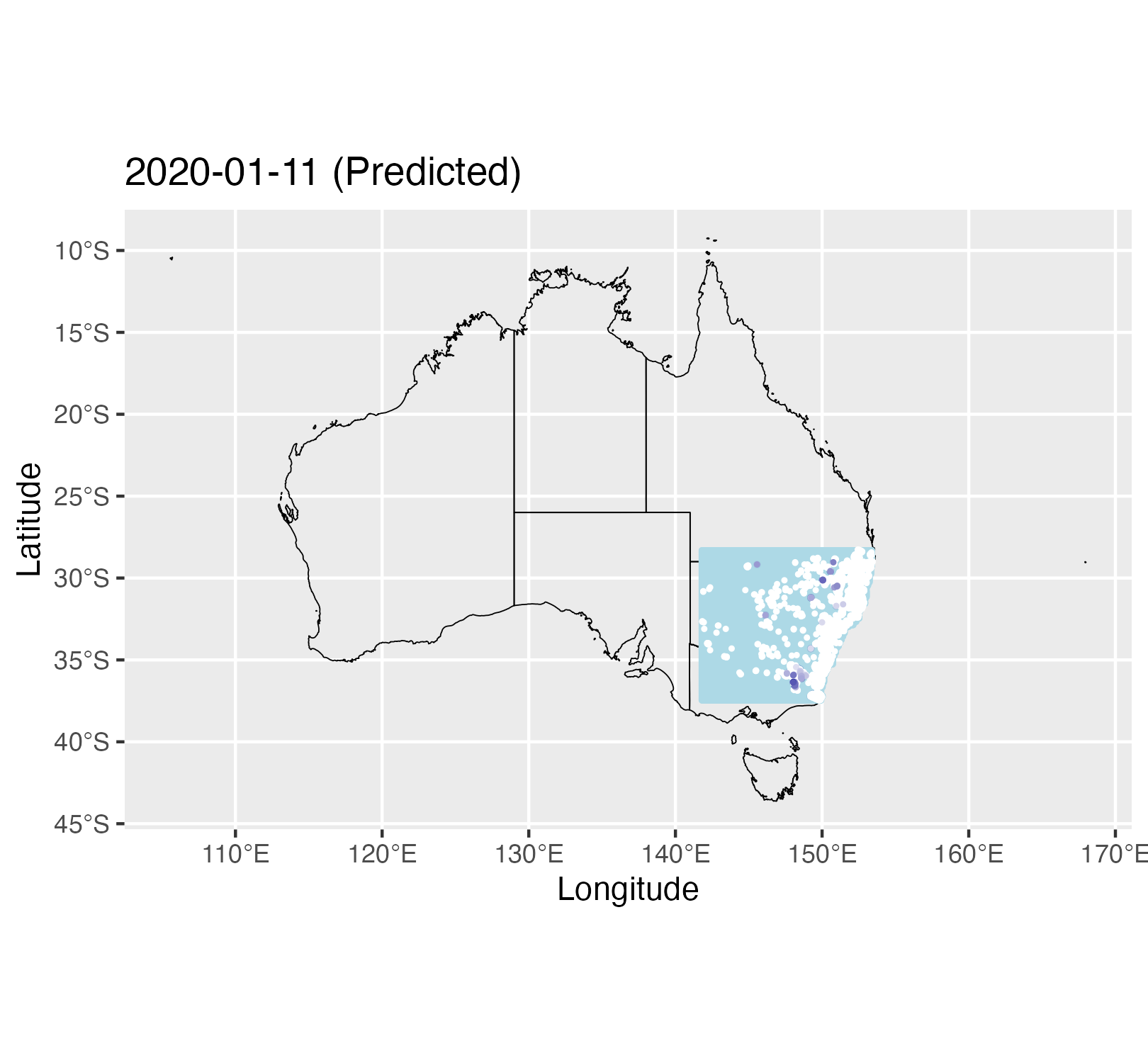}
         \caption{Predicted fire on Jan 11, 2020}
         \label{fig:aus_jan_11_pred}
     \end{subfigure}
 \begin{subfigure}[b]{0.42\linewidth}
         \centering
         \includegraphics[width=\textwidth]{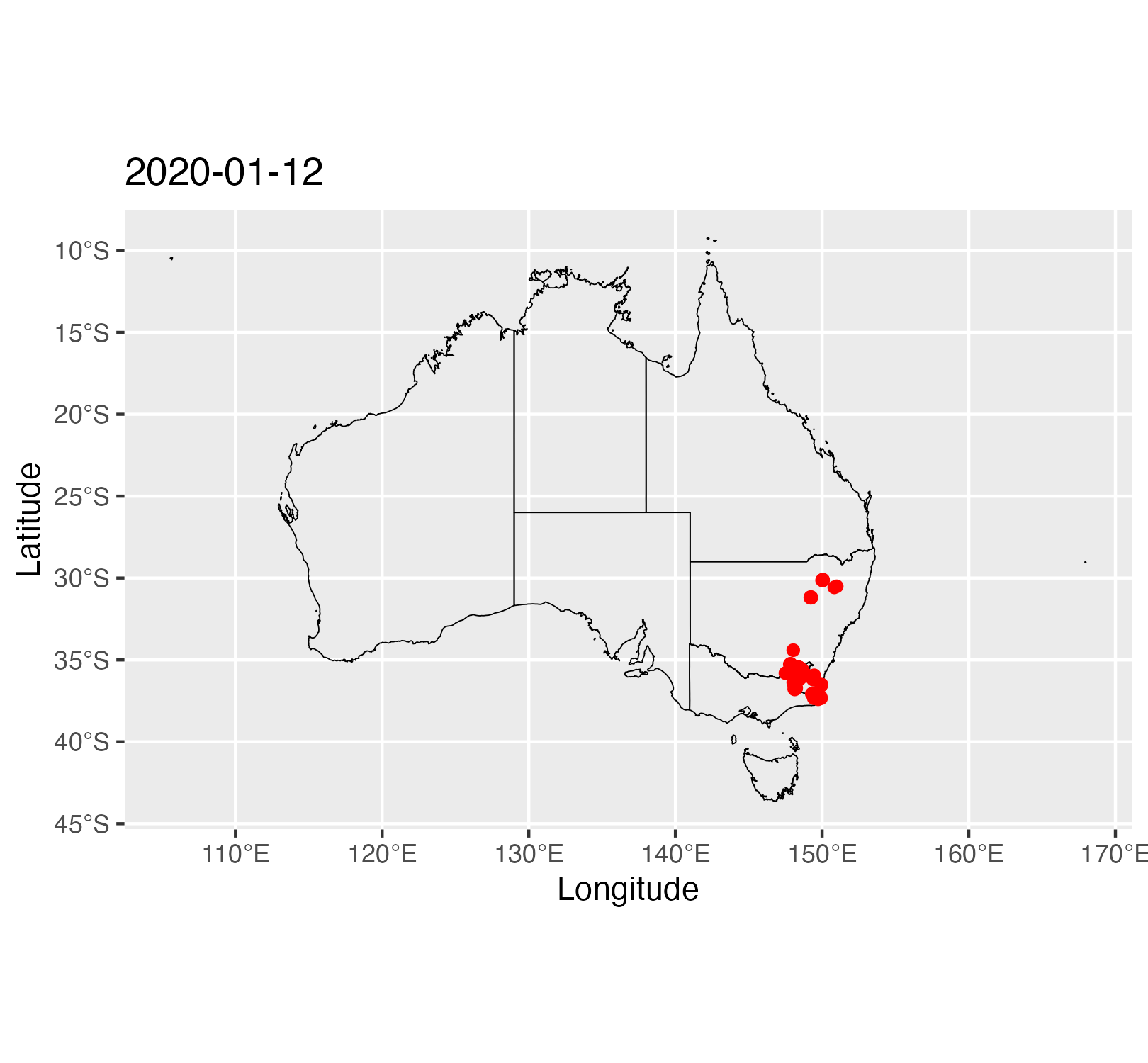}
         \caption{Observed fire on Jan 12, 2020}
         \label{fig:aus_jan_12_obs}
     \end{subfigure}
     \hfill
 \begin{subfigure}[b]{0.42\linewidth}
         \centering
         \includegraphics[width=\textwidth]{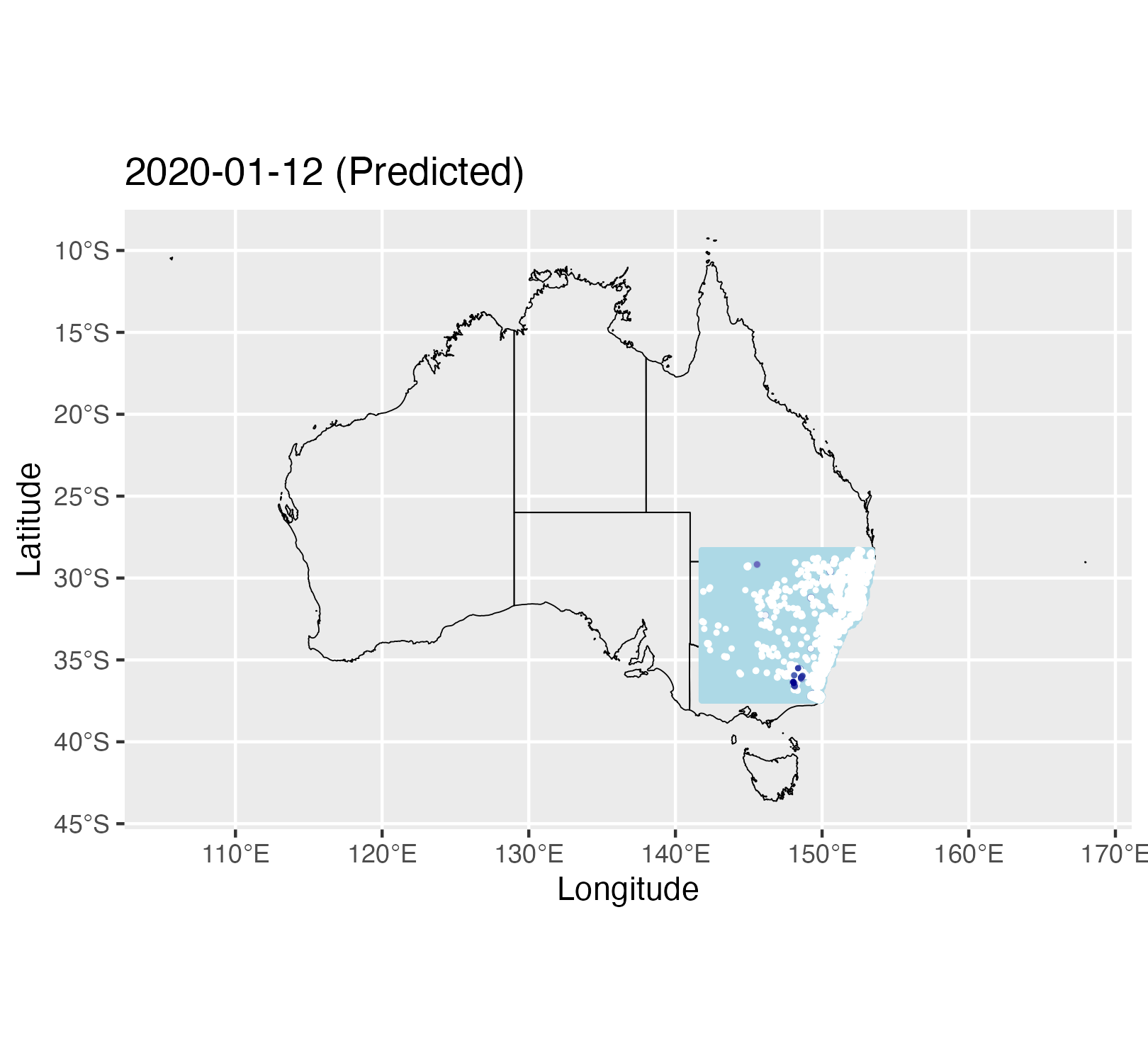}
         \caption{Predicted fire on Jan 12, 2020}
         \label{fig:aus_jan_12_pred}
     \end{subfigure}
     \caption{Observed vs 1, 2 and 3 day predictions}
     \label{fig:jan_10_11_12}
\end{figure}

\clearpage

\bibliography{ref}